\documentclass[11pt]{article}
\pdfoutput=1
\usepackage[utf8]{inputenc}
\usepackage{multirow}
\usepackage{svg}
\usepackage{helvet}
\usepackage{subcaption}
\usepackage{array}
\usepackage{amsmath, amsfonts, amssymb}
\usepackage{caption}
\usepackage{comment}
\usepackage{mathtools}
\usepackage{graphicx}
\usepackage{psfrag}
\usepackage{float}
\usepackage{amsthm}
\usepackage{xcolor}
\usepackage{enumerate}
\usepackage{arydshln}
\usepackage{soul}
 \usepackage{slashed}

\usepackage{mathrsfs}
 \usepackage{a4wide}
  \usepackage{tikz}
  \usepackage{tikz-cd}
  \usetikzlibrary{shapes.geometric}
\usepackage[most,many,breakable]{tcolorbox}
\usepackage{wrapfig}
  \usepackage{color}
  \definecolor{dark-gray}{gray}{0.20}
  \definecolor{gray}{gray}{0.30}
  \definecolor{light-gray}{gray}{0.80}
  \definecolor{dark-red}{rgb}{0.7,0,0}
  \definecolor{dark-green}{rgb}{0.1,0.4,0}
  \definecolor{dark-blue}{rgb}{0.3,0.3,0.7}
  \definecolor{light-blue}{rgb}{0.8,0.8,1}
  \definecolor{swamp}{RGB}{240, 199, 197}
\colorlet{mydarkred}{red!50!black}
\colorlet{mydarkblue}{blue!50!black}
  \usepackage{pifont}
\usetikzlibrary{shadings}
\usepackage{newunicodechar} 

\usepackage{xcolor}

\definecolor{dustyPlum}{HTML}{8E5378}
\definecolor{agedRose}{HTML}{86385B}
\definecolor{wineBrick}{HTML}{863B43}
\definecolor{clayRust}{HTML}{86564D}
\definecolor{burntOchre}{HTML}{865E3E}
\definecolor{antiqueGold}{HTML}{86773B}
\definecolor{mutedTeal}{HTML}{438667}
\definecolor{deepAqua}{HTML}{338681}
\definecolor{stormBlue}{HTML}{366E8E}
\definecolor{softLemon}{HTML}{F0F497}
\usepackage{setspace}

\usepackage{soul}
\sethlcolor{softLemon!70} 

\usepackage{tikz-3dplot}
\usetikzlibrary{decorations.markings,decorations.pathmorphing}
\usetikzlibrary{intersections}
\usetikzlibrary{shapes.misc}
\tdplotsetmaincoords{70}{110}
\tikzset{cross/.style={cross out, draw=black, minimum size=2*(#1-\pgflinewidth), inner sep=0pt, outer sep=0pt},
cross/.default={1pt}}
\usepackage{enumitem}
\usepackage{varwidth}

\tcbuselibrary{skins}
\newtcolorbox{remark}[1][]{%
	enhanced jigsaw,
	colback=gray!20!white,%
	colframe=gray!80!black,
	size=small,
	boxrule=1pt,
	title=\textbf{!},
	halign title=flush center,
	coltitle=black,
	breakable,
	drop shadow=black!50!white,
	attach boxed title to top left={xshift=1cm,yshift=-\tcboxedtitleheight/2,yshifttext=-\tcboxedtitleheight/2},
	minipage boxed title=1.5cm,
	boxed title style={%
			colback=white,
			size=fbox,
			boxrule=1pt,
			boxsep=2pt,
			underlay={%
					\coordinate (dotA) at ($(interior.west) + (-0.5pt,0)$);
					\coordinate (dotB) at ($(interior.east) + (0.5pt,0)$);
					\begin{scope}
						\clip (interior.north west) rectangle ([xshift=3ex]interior.east);
						\filldraw [white, blur shadow={shadow opacity=60, shadow yshift=-.75ex}, rounded corners=2pt] (interior.north west) rectangle (interior.south east);
					\end{scope}
					\begin{scope}[gray!80!black]
						\fill (dotA) circle (2pt);
						\fill (dotB) circle (2pt);
					\end{scope}
				},
		},
	#1,
}

\usetikzlibrary{arrows,calc,shadows.blur}
\tcbuselibrary{skins}
\newtcolorbox{pending}[1][]{%
	enhanced jigsaw,
	colback=red!20!white,%
	colframe=red!80!black,
	size=small,
	boxrule=1pt,
	title=\textbf{?},
	halign title=flush center,
	coltitle=black,
	breakable,
	drop shadow=black!50!white,
	attach boxed title to top left={xshift=1cm,yshift=-\tcboxedtitleheight/2,yshifttext=-\tcboxedtitleheight/2},
	minipage boxed title=1.5cm,
	boxed title style={%
			colback=white,
			size=fbox,
			boxrule=1pt,
			boxsep=2pt,
			underlay={%
					\coordinate (dotA) at ($(interior.west) + (-0.5pt,0)$);
					\coordinate (dotB) at ($(interior.east) + (0.5pt,0)$);
					\begin{scope}
						\clip (interior.north west) rectangle ([xshift=3ex]interior.east);
						\filldraw [white, blur shadow={shadow opacity=60, shadow yshift=-.75ex}, rounded corners=2pt] (interior.north west) rectangle (interior.south east);
					\end{scope}
					\begin{scope}[red!80!black]
						\fill (dotA) circle (2pt);
						\fill (dotB) circle (2pt);
					\end{scope}
				},
		},
	#1,
}

\newtcbtheorem[number within=section]{Definition}{Definition}{enhanced,
	before skip=2mm,after skip=2mm, colback=olive!5,colframe=olive!80!black,boxrule=0.5mm,
	attach boxed title to top left={xshift=1cm,yshift*=1mm-\tcboxedtitleheight}, varwidth boxed title*=-3cm,
	boxed title style={frame code={
					\path[fill=tcbcolback]
					([yshift=-1mm,xshift=-1mm]frame.north west)
					arc[start angle=0,end angle=180,radius=1mm]
					([yshift=-1mm,xshift=1mm]frame.north east)
					arc[start angle=180,end angle=0,radius=1mm];
					\path[left color=tcbcolback!60!black,right color=tcbcolback!60!black,
						middle color=tcbcolback!80!black]
					([xshift=-2mm]frame.north west) -- ([xshift=2mm]frame.north east)
					[rounded corners=1mm]-- ([xshift=1mm,yshift=-1mm]frame.north east)
					-- (frame.south east) -- (frame.south west)
					-- ([xshift=-1mm,yshift=-1mm]frame.north west)
					[sharp corners]-- cycle;
				},interior engine=empty,
		},
	fonttitle=\bfseries,
	title={#2},#1}{def}
\newtcbtheorem[number within=section]{definition}{Definition}{enhanced,
	before skip=2mm,after skip=2mm, colback=olive!5,colframe=olive!80!black,boxrule=0.5mm,
	attach boxed title to top left={xshift=1cm,yshift*=1mm-\tcboxedtitleheight}, varwidth boxed title*=-3cm,
	boxed title style={frame code={
					\path[fill=tcbcolback]
					([yshift=-1mm,xshift=-1mm]frame.north west)
					arc[start angle=0,end angle=180,radius=1mm]
					([yshift=-1mm,xshift=1mm]frame.north east)
					arc[start angle=180,end angle=0,radius=1mm];
					\path[left color=tcbcolback!60!black,right color=tcbcolback!60!black,
						middle color=tcbcolback!80!black]
					([xshift=-2mm]frame.north west) -- ([xshift=2mm]frame.north east)
					[rounded corners=1mm]-- ([xshift=1mm,yshift=-1mm]frame.north east)
					-- (frame.south east) -- (frame.south west)
					-- ([xshift=-1mm,yshift=-1mm]frame.north west)
					[sharp corners]-- cycle;
				},interior engine=empty,
		},
	fonttitle=\bfseries,
	title={#2},#1}{def}

\newcommand{\be}{\begin{equation}}
\newcommand{\ee}{\end{equation}}

\newcommand{\ket}[1]{\vert #1 \rangle}

\def\be{\begin{equation}}
\def\ee{\end{equation}}
\def\bea{\begin{eqnarray}}
\def\eea{\end{eqnarray}}

\theoremstyle{definition}

\newcommand{\R}{\mathbb{R}}

\newcommand{\Z}{\mathbb{Z}}

\newcommand{\ang}[1]{\langle #1 \rangle}
\definecolor{antiquefuchsia}{rgb}{0.57, 0.36, 0.51}

\def\simleq{\; \raise0.3ex\hbox{$<$\kern-0.75em
      \raise-1.1ex\hbox{$\sim$}}\; }
   \def\simgeq{\; \raise0.3ex\hbox{$>$\kern-0.75em
      \raise-1.1ex\hbox{$\sim$}}\; }

\numberwithin{equation}{section}

\usepackage{jheppub}
\usepackage{hyperref}
\usepackage{cleveref}
\usepackage{braket}

 \usepackage{color}

\definecolor{dark-blue2}{RGB}{0, 114,178}

\hypersetup{
	colorlinks=true,
	linkcolor=stormBlue,
	citecolor=stormBlue,
	urlcolor=stormBlue,
	linktoc=section
}

\newtcolorbox{symmetrybox}[2][]{%
  colback=#2!20,        
  colframe=#2!70!black, 
  boxrule=0.8pt,
  arc=4pt,
  left=8pt,
  right=8pt,
  top=8pt,
  bottom=8pt,
  enhanced,
  breakable,
  #1                    
}
\usepackage{tocloft}

\theoremstyle{remark}

\crefname{appendix}{Appendix}{Appendices}

\definecolor{darklavender}{rgb}{0.45, 0.31, 0.59}
\definecolor{darkrlavender}{rgb}{0.3375, 0.2325, 0.4425}
\definecolor{darkerlavender}{rgb}{0.225, 0.155, 0.295}
\def\({\left(}
\def\){\right)}
\def\[{\left[}
\def\]{\right]}

\title{{\color{stormBlue} Gauge It} \textbf{or} {\color{stormBlue}Break It}{\textbf{:}}\\ {\Large Lectures on Global Symmetries and the Cobordism Conjecture}}

\author{\small Matilda Delgado,$^{1,2}$}
\author{\small Corinne Englert,$^{3}$}
\author{\small Guillermo Mera \'Alvarez,$^{4}$}
\author{\small Eduardo Velasco-Aja$^{5,6}$}

\usetikzlibrary{calc}
\usetikzlibrary{intersections}
\usetikzlibrary{positioning} 
\usetikzlibrary{arrows}
\usetikzlibrary{shapes.misc}
\tikzstyle{spring}=[line width=0.8,blue!7!black!80,snake=coil,segment amplitude=4.25,segment length=4.75,line cap=round]
\usetikzlibrary{patterns,snakes}
\tikzstyle{spring}=[line width=0.8,blue!7!black!80,snake=coil,segment amplitude=4.25,segment length=4.75,line cap=round]
\usetikzlibrary{decorations.pathmorphing}
\usetikzlibrary{decorations.markings}
\usetikzlibrary{arrows.meta}
\usetikzlibrary{matrix} 
\usetikzlibrary{positioning} 
\usetikzlibrary{arrows}
\tikzset{>=latex} 
\tikzstyle{radiation}=[-{Latex[length=2,width=1.5]},stormBlue!70!black!50,opacity=0.7,very thin,decorate,
                       decoration={snake,amplitude=0.7,segment length=2,post length=2}]

\usepackage[normalem]{ulem} 
\usepackage{booktabs} 
\newcommand\pt{\mathrm{pt}}

\newcommand\SL{\mathrm{SL}}

\newcommand\SO{\mathrm{SO}}

\newcommand\UU{\mathrm{U}}
\newcommand\SU{\mathrm{SU}}

\newcommand\Spin{\mathrm{Spin}}

\affiliation{$^1$Jefferson Physical Laboratory, Harvard University\\ 17 Oxford St, Cambridge, MA 02138, United States of America}
\affiliation{$^2$Max-Planck-Institut f\"ur Physik (Werner-Heisenberg-Institut)\\
Boltzmannstr. 8, 85748 Garching, Germany}
\affiliation{$^3$Institute for Theoretical Physics and Astrophysics, Julius-Maximilians-Universit\"at W\"urzburg,
Am Hubland, D-97074 W\"urzburg, Germany}
\affiliation{$^4$Instituut voor Theoretische Fysica and Leuven Gravity Institute, KU Leuven,\\ Celestijnenlaan 200D, B-3001 Leuven, Belgium}
\affiliation{$^5$Instituto de F\'isica Te\'orica IFT-UAM/CSIC  C/ Nicol\'as Cabrera 13-15, Campus de Cantoblanco, 28049 Madrid, Spain}
\affiliation{$^6$Departamento de F\' \i sica Te\'orica, Universidad Aut\'onoma de Madrid, Campus de Cantoblanco, 28049 Madrid, Spain}

\emailAdd{matildadelgado@fas.harvard.edu}
\emailAdd{corinne.englert@stud-mail.uni-wuerzburg.de}
\emailAdd{guillermo.meraalvarez@kuleuven.be}
\emailAdd{eduardo.velasco@uam.es}

\abstract{These lecture notes introduce the No Global Symmetries Conjecture in the context of the Swampland Program, with particular emphasis on the Cobordism Conjecture. We begin with a review of ordinary and higher-form symmetries, their realization by extended operators, and mechanisms for gauging or breaking them. We then discuss the no global symmetries principle, including evidence from black hole physics, string perturbation theory, and AdS/CFT, and its relation to other Swampland constraints. A central theme is the reformulation of symmetry constraints in topological terms: we introduce bordism groups, with special focus on spin bordism, and explain how they provide a natural framework for organizing would-be global symmetries in gravitational theories. With this at hand, we discuss the Cobordism Conjecture, which proposes that nontrivial bordism classes are eliminated by dynamical processes in quantum gravity. We conclude with examples and a discussion of dynamical aspects involving defects, boundaries, and related open questions. These notes are based on lectures delivered by Matilda Delgado at the 21st Modave Summer School on Mathematical Physics. A similar series of lectures was also presented at the KMPB School on D-branes in Physics and Mathematics.
}

\makeatletter
\newcounter{subsubsubsection}[subsubsection]
\renewcommand\thesubsubsubsection{\thesubsubsection.\arabic{subsubsubsection}}

\newcommand{\subsubsubsection}[1]{%
  \refstepcounter{subsubsubsection}%
  \paragraph{\thesubsubsubsection\quad #1}%
}

\makeatother
\begin{document}
\hypersetup{pageanchor=false}
\makeatletter
\let\old@fpheader\@fpheader
\preprint{\begin{tabular}[t]{@{}r@{}}
MPP-2026-196\\
IFT-UAM/CSIC-26-138
\end{tabular}}

\makeatother

\notoc 
\maketitle
\newpage
\newpage
\hrule
\tableofcontents
\newpage

\section*{Summary and reader's guide}

\paragraph{In a nutshell.} The \emph{No Global Symmetries Conjecture} states that every global symmetry of a consistent theory of quantum gravity is either gauged or broken. This includes generalized symmetries: higher-form symmetries acting on extended objects. The \emph{Cobordism Conjecture} is a refinement of this principle that befits the topology of spacetime itself: a non-trivial \emph{bordism class} of backgrounds would define a conserved topological charge, and hence a global symmetry, so all bordism groups of quantum gravity must be trivial. Any background can then be deformed into any other and in particular into ``nothing''.

We discuss the evidence and the assumptions behind these conjectures, drawing on black holes, perturbative string theory and AdS/CFT. We then show how the Cobordism Conjecture postdicts D-branes and O-planes, and predicts new non-supersymmetric objects in string theory. We close with open problems, in particular the dynamics and tensions of non-supersymmetric defects, which set the scale at which the associated symmetries are broken.

These notes are a pedagogical introduction aimed at advanced undergraduate and graduate students. They require only basic general relativity and quantum field theory, plus a rudimentary understanding string theory and holography.

\paragraph{Reader's guide.}
\begin{itemize}
    \item Section~\ref{sec:intro} gives motivation and context, and can be skipped by readers familiar with the Swampland Program (see \cite{Palti:2019pca,vanBeest:2021lhn,Agmon:2022thq} for reviews).
    \item Section~\ref{sec:GlobalSYms101} introduces ordinary and higher-form global symmetries, their gauging and breaking, and gives some explicit examples. Generalized symmetries are used throughout the rest of the notes; for complete treatments see \cite{BottiniR,Gomes:2023ahz,Brennan:2023mmt}.
    \item Section~\ref{sec:No_Global_Symmetries} presents the No Global Symmetries Conjecture along with the evidence for it. The black hole argument (Section~\ref{sec:heuristic}) is essential; the arguments from string perturbation theory and AdS/CFT can be reads independently of each other.
    \item Section~\ref{sec:Bordisms101} is a self-contained primer on bordism groups for the layman. Section~\ref{sec:bordismgeneralities} and the examples of spin bordism of a point are essential; the discussion of the Atiyah--Hirzebruch spectral sequence (Section~\ref{sec:spectralsequences}) can be skipped on a first reading.
    \item Section~\ref{sec:coco} states the Cobordism Conjecture and explains how bordism classes give global symmetries, then applies it to string theory. The dynamical aspects in Section~\ref{sec:dynamicalaspects} onwards are closer to current research.
\end{itemize}
Readers mainly interested in the Swampland may follow Sections~\ref{sec:intro}, \ref{sec:No_Global_Symmetries} and \ref{sec:coco}, returning to Sections~\ref{sec:GlobalSYms101} and \ref{sec:Bordisms101} as needed. More mathematically inclined readers may prefer to go through Sections~\ref{sec:GlobalSYms101}, \ref{sec:Bordisms101} and~\ref{sec:coco}.

\newpage
\section{Motivation \& Context}\label{sec:intro}

In these lecture notes we introduce the No Global Symmetries Conjecture as a  constraint on consistent theories of quantum gravity, discuss its role within the Swampland Program, and place particular emphasis on its formulation via the Swampland Cobordism Conjecture.\footnote{Every time we mention the Cobordism Conjecture in these lecture notes, we are referring to the Swampland Cobordism Conjecture of \cite{McNamara:2019rup}, not to the mathematical conjecture known as the Cobordism Hypothesis/Conjecture.} We begin with a short review of what is meant by quantum gravity, together with a discussion of effective field theories (EFTs) that arise from string theory and the broader perspective provided by the Swampland Program. This material is included primarily for context and motivation, and is not intended as a comprehensive review (for reviews on the Swampland Program in general, see e.g.~\cite{Palti:2019pca,Grana:2021zvf,vanBeest:2021lhn,Agmon:2022thq}). Readers already familiar with string phenomenology or the Swampland literature may wish to skip this initial section and proceed directly to the discussion of global symmetries, bordism, and the Cobordism Conjecture.

Gravity was originally understood with respect to a fixed spacetime background, as in Newtonian mechanics, where space and time provide a rigid stage for particles to move. General Relativity (GR), however, revolutionized this view: spacetime is no longer fixed, but dynamical, curving in response to matter and energy according to Einstein's equations,
\begin{equation}
G_{\mu\nu} + \Lambda g_{\mu\nu} = \frac{8 \pi G_N}{c^4} T_{\mu\nu}.
\end{equation}   
The key feature of GR is its \hl{background independence}. The theory does not assume any fixed spacetime background against which gravitational phenomena occur. Instead, the geometry of spacetime is determined dynamically by the distribution of matter and energy through Einstein's equations. These equations are notoriously difficult to solve, as they are inherently non-linear: the gravitational field itself carries energy and momentum and therefore acts as a source of curvature. Consequently, gravitational perturbations self-interact and solutions of GR cannot, in general, be superposed. 
Although Einstein’s field equations are nonlinear, their linearization about a background solution admits propagating gravitational waves. Such waves have been directly detected by LIGO/Virgo \cite{LIGOScientific:2016aoc}, providing remarkable direct evidence for the radiative degrees of freedom of spacetime. At the same time, their observed waveform also probes the nonlinear dynamics of the binary black hole merger that produced them.

In contrast, the three other fundamental forces are described by Quantum Field Theory (QFT) on a fixed background spacetime. Indeed, in QFT spacetime is just the rigid stage on top of which fields interact. The fields themselves have no effect on the geometry of spacetime. This is not a contradiction with GR: particles are generally so light that they effectively do not backreact on the spacetime. In other words, QFT is the fixed-background approximation that neglects the gravitational backreaction.

It is possible, however, to imagine a number of scenarios where the strength of gravity becomes comparable to that of the other forces and this effective description breaks down. For instance, what if we consider a particle that is very massive? Say, as massive as a planet? How is spacetime curved by a supermassive quantum particle? Einstein's equations tell us the geometry of spacetime should be dictated by the stress-energy tensor of this particle, but what should we use if the particle is in a mixed quantum state? Even for light particles, gravity can become as strong as the other forces near sufficiently small black holes or at the beginning of the universe. These are scenarios in which the semiclassical approximation to spacetime is expected to break down, since it treats the geometry classically while sourcing it with the expectation value of a quantum stress-energy tensor.

This is where we \textit{need} a theory of Quantum Gravity: the issue is not fitting gravity into the framework of particle physics, but rather making general relativity and quantum theory mutually consistent. You cannot consistently couple a quantum theory (the Standard Model) to a classical one (GR) without running into contradictions at energies where the quantum matter starts behaving in a quantum way. These are the same contradictions that arise in electromagnetism when coupling electrons to a classical electromagnetic background (see e.g.~\cite{Belenchia:2018szb} and references therein). At low energies this tension is resolved by quantizing gravity perturbatively, treating gravitons as an effective field theory on a fixed background. This description, however, breaks down at high energies, where a genuinely non-perturbative theory is required. While GR treats spacetime as a smooth, continuous manifold, in this non-perturbative regime the notion of a smooth, macroscopic spacetime breaks down and its fabric becomes subject to quantum effects: superpositions of states, quantum tunnelling, etc. These fluctuations would manifest as \hl{uncertainty in the geometry and topology of spacetime}. This has drastic consequences. The causal structure of spacetime itself can exist in superposition, blurring the notion of fixed light cones and making strict locality only an approximate concept. The idea that a quantum gravity theory should account for the possibility of changes in the topology of spacetime implies a radical change of paradigm that will play a fundamental role in these lectures.\footnote{Note that this view is not universally shared. In asymptotic safety, for instance, the gravitational path integral is typically defined at fixed topology \cite{Basile:2025zjc}. Other approaches, such as loop quantum gravity and spin foams, naturally accommodate topology change (see e.g.~\cite{Perez:2012wv} for a review). Here we take the view that a quantum gravity theory should account for the possibility of changes in the topology of spacetime, and mostly discuss it in the context of string theory.}

It is fair to say that we do not have a full understanding of such quantum gravitational phenomena. However, what we can say for sure is that, at low energies below the Planck scale, gravity should become classical again and be governed by the laws of GR.  This is the concept of \hl{emergence of spacetime}. At the fundamental level, we would like our theory of quantum gravity to make no reference to any specific spacetime background, which means that the geometry of spacetime should not be imposed from the outside but should rather be an emergent property of the theory itself. In this view, spacetime and gravitational dynamics are regarded as low-energy collective phenomena of the underlying quantum degrees of freedom, similar to how fluid dynamics arises from the collective behaviour of molecules, or to how a classical electromagnetic field comes from the collective behaviour (a coherent state) of photons. In our case, macroscopic spacetime emerges from the collective behaviour (a coherent state) of gravitons. This means that, no matter what quantum gravity mess is hiding at high energies, physics should reduce to some theory of QFT + GR at low energies (IR).

We have just described some features that a theory of quantum gravity should have. We saw that the notion of a background spacetime on which one can define quantum field theories and interactions should arise as a low energy description, at scales well below the Planck mass $M_\mathrm{P}$. This leads to a physical theory that one can compare to the laws of our universe, and perhaps even attempt to make predictions.

In the next section, we explain how these ideas are realized within a concrete candidate theory of quantum gravity: string theory. String theory provides a consistent framework for studying quantum gravity. We will emphasize how familiar low-energy effective theories of the form QFT + GR emerge from the underlying string dynamics.

\subsection{Stringy EFTs}

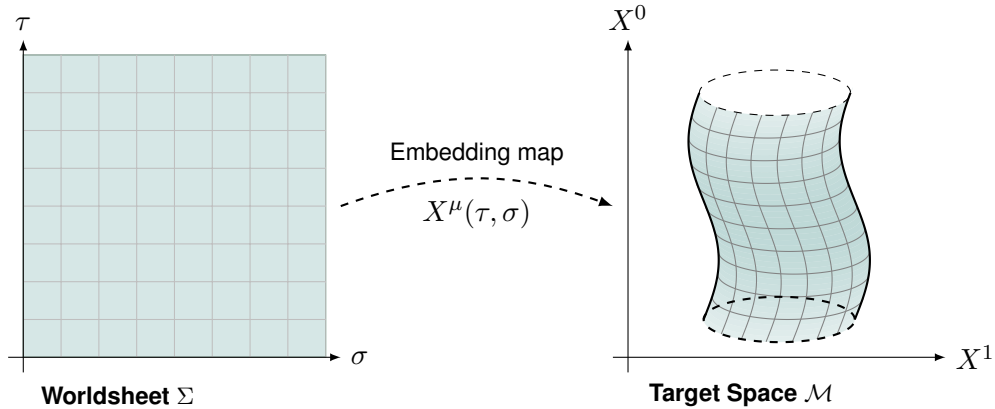
\begin{figure}[h]
    \centering
        \begin{tikzpicture}
    \begin{scope}[shift={(0,0)}]
        \draw[fill=deepAqua!20, draw=mutedTeal] (0,0) -- (4,0) -- (4,4) -- (0,4) -- cycle;
        \draw[lightgray, thin, step=0.5] (0,0) grid (4,4);
        \draw[->] (-0.2,0) -- (4.2,0.) node[right] {$\sigma$};
        \draw[->] (0,-0.2) -- (0,4.2) node[above] {$\tau$};
        \node at (1.25,-0.5) {\fontfamily{phv}\selectfont \footnotesize\textbf{Worldsheet} $\Sigma$};
    \end{scope}
    \draw[->, thick, dashed] (4.2,2) to[out=20,in=160] (7.8,2); 
    \node[above] at (6,2.4) {\fontfamily{phv}\selectfont \footnotesize Embedding map}; 
    \node[] at (6,1.9) {$X^\mu (\tau, \sigma)$};
    \begin{scope}[shift={(10,0.5)}]
        \def\R{1}
        \def\r{0.3}
        \def\h{3}      
        \def\A{0.2}   
        \def\k{2}     

        \def\n{40}
        \foreach \i in {0,...,\n} {
            \pgfmathsetmacro{\z}{\h*\i/\n}
            \pgfmathsetmacro{\znext}{\h*(\i+1)/\n}
            \pgfmathsetmacro{\shade}{70*(0.4 + 0.5*sin(180*\i/\n))}
            \fill[deepAqua!\shade!white, opacity=0.5]
                plot[domain=180:360,samples=40]
                    ({\R*cos(\x) + \A*sin(\k*\z r)}, {\r*sin(\x) + \z})
                --
                plot[domain=360:180,samples=40]
                    ({\R*cos(\x) + \A*sin(\k*\znext r)}, {\r*sin(\x) + \znext})
                -- cycle;}

        \foreach \t in {220,240,...,320} {
            \draw[gray!70]
                plot[domain=0:\h,samples=60]
                ({\R*cos(\t) + \A*sin(\k*\x r)}, {\r*sin(\t) + \x});}
        \foreach \z in {0.3,0.6,...,3} {
            \draw[gray!70]
                plot[domain=180:360,samples=100]
                ({\R*cos(\x) + \A*sin(\k*\z r)}, {\r*sin(\x) + \z});}
        \draw[dashed,thick]
            plot[domain=0:360,samples=100]
            ({\R*cos(\x)}, {\r*sin(\x)});
        \draw[dashed, fill=white]
            plot[domain=0:360,samples=100]
            ({\R*cos(\x) + \A*sin(\k*\h r)}, {\r*sin(\x) + \h});
        \draw[thick]
            plot[domain=0:\h,samples=100]
            ({-\R + \A*sin(\k*\x r)}, {+\x});
        \draw[thick]
            plot[domain=0:\h,samples=100]
            ({\R + \A*sin(\k*\x r)}, {+\x});
        \draw[->] (-2.2,-0.5) -- (2.2,-0.5) node[right] {$X^1$};
        \draw[->] (-2,-0.7) -- (-2,3.7) node[above] {$X^0$};
        \node at (-0.5,-1)  {\fontfamily{phv}\selectfont \footnotesize\textbf{Target Space} $\mathcal{M}$};
        
    \end{scope}

\end{tikzpicture}
    \caption{
Similar to how one can associate a worldline to the path of a relativistic point particle in spacetime, the one-dimensional string sweeps out a two-dimensional surface, the worldsheet. By embedding functions one can map from the worldsheet of the string to the target space (spacetime) in which the (closed) string moves.}
    \label{fig:WStoTS}
\end{figure}

String theory is a theory of quantum gravity with renowned success. In its essence, it posits that the fundamental building blocks of the universe are not point-like particles but rather one-dimensional entities known as strings. String theory thus only has one fundamental scale, the string length $l_s$. These strings are assumed to propagate in a $d$-dimensional spacetime, sweeping out a worldsheet (the analog of a particle worldline for a string), as depicted in Figure \ref{fig:WStoTS} for a closed string. Consistency of the worldsheet theory combined with Lorentz invariance constrains $d$ to take specific values. In perturbative string theory, we assume that spacetime is flat Minkowski space and we quantize the oscillatory modes of the string and find its spectrum. All of the modes of the string describe particle states in spacetime. The massless modes lead (at the very least) to a $d$-dimensional graviton, a 2-form gauge field and a scalar field $\phi$, the \textit{dilaton}, that parametrizes the coupling of the string $g_s= e^\phi$. All of the heavy modes of the string have masses at the string scale, so they decouple from physics at energy scales much lower than $m_s = l_s^{-1}$. We therefore obtain a $d$-dimensional low-energy EFT that describes these three fields propagating on a flat spacetime.

\begin{figure}[h!]
    \centering
    \includegraphics[scale=0.35]{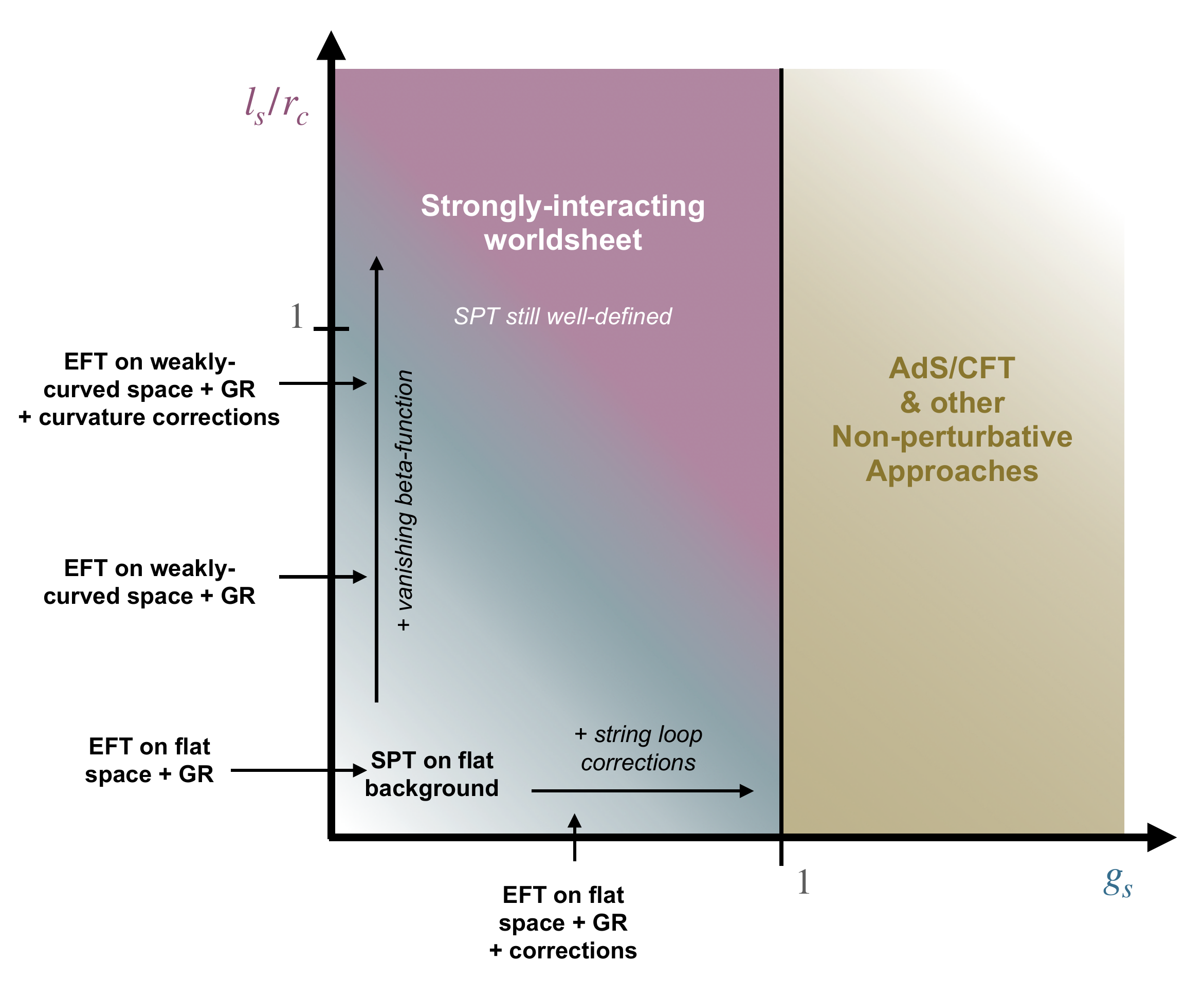}
    \caption{A diagram describing our understanding of string theory according to the value of the string coupling $g_s$ and the curvature of spacetime relative to the string scale. At weak coupling and low curvatures, we have 10d EFTs of string theory, describing various particles moving on the background described by general relativity. At strong curvatures, the worldsheet theory becomes strongly coupled but string perturbation theory is still well defined (the path integral over the moduli space of Riemann surfaces still makes sense). At strong coupling, string perturbation theory completely breaks down and we have to resort to other methods.  }
    \label{fig:stringtheory}
\end{figure}

\noindent We can incorporate weakly curved spacetime into this discussion as a perturbation around the flat background. The big surprise that is responsible for some of the fame of string theory is that, in order for the worldsheet theory to remain consistent in curved spacetime, the metric has to satisfy Einstein's equations. In this way, GR arises naturally in the \hl{low-energy limit} of string theory. Therefore, at low energies $E\ll l_s^{-1}$ and as long as spacetime is weakly curved, string theory is described by a low-energy effective theory of a 2-form gauge field, the dilaton and all other massless modes of the string propagating on a background spacetime dictated by GR. In this limit, string theories thus reduce to theories that look like our world, though in a higher number of dimensions and possibly with a lot more supersymmetry. 

\paragraph*{Emergent Spacetime.} At low energies $E\ll l_s^{-1}$, a massless oscillatory mode of the string gives rise to a graviton in spacetime, which satisfies the equations of motion of GR in weakly curved spacetime. When one goes up in energies and considers string theory at the string scale, this effective description completely breaks down and one needs to take into account all of the massive oscillatory modes of the string. That is, you cannot consider only the massless excitations which lead to familiar particle states in spacetime, you need to consider the string itself as the primary object of interest. In this sense, spacetime and GR only emerge at low energies. Importantly, at the string scale, you consider strings themselves moving in a flat ``background'', but in no shape or form is this theory describing GR. The better description would be in terms of some gas of strings, perhaps in the spirit of \cite{Brandenberger:1988aj}, or in terms of string field theory \cite{Witten:1992qy}. Even in the limit $g_s \to 0$, one would have to consider a non-interacting theory of strings moving around in flat space. When talking about the emergence of a  \textit{spacetime}, we always refer to the one that is governed by GR and that only exists in the low-energy limit of string theory.

\begin{figure}[h!]
    \centering
    \includegraphics[width=1.1\textwidth]{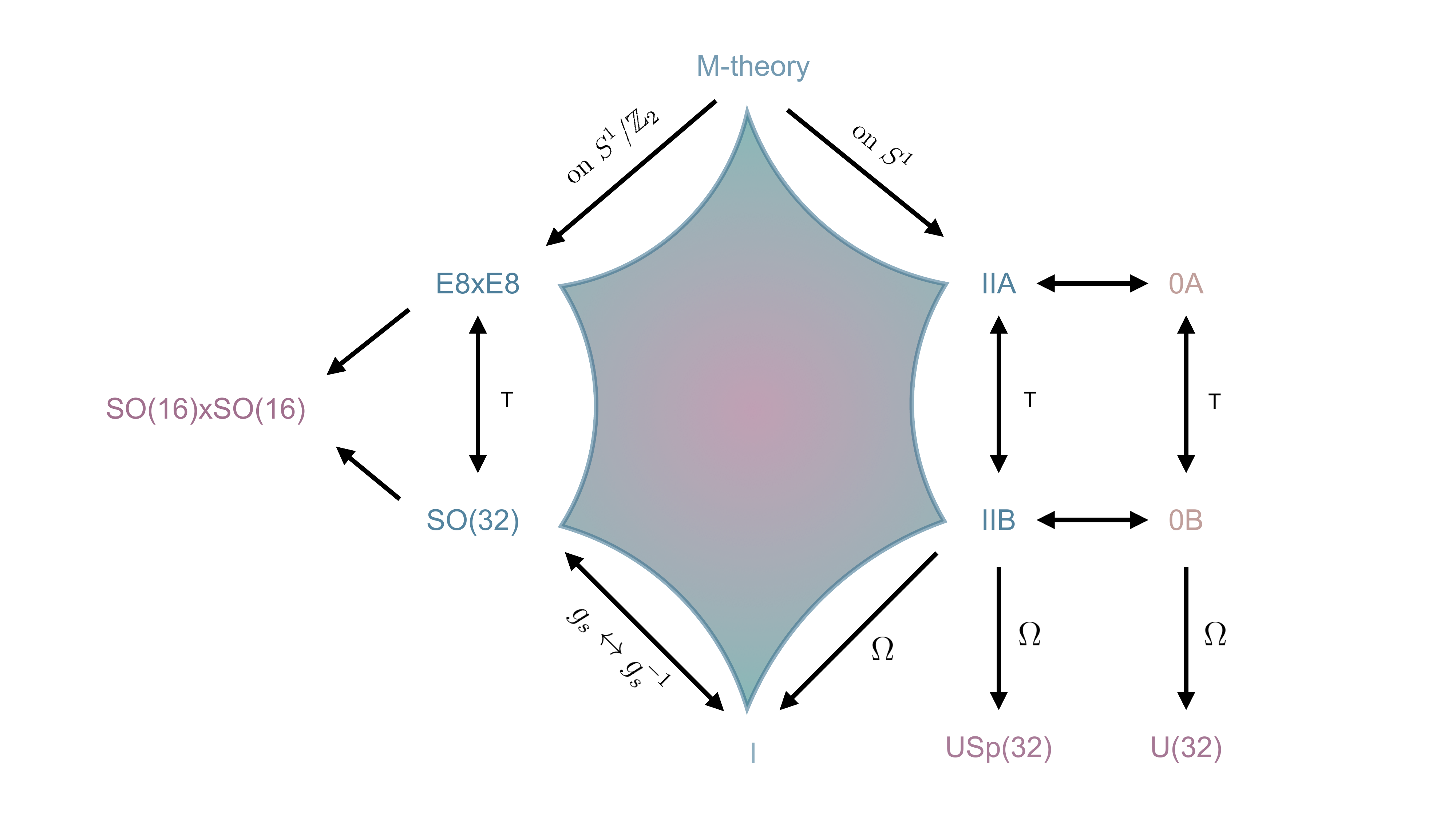}
    \caption{A diagram representing the eight ten-dimensional non-tachyonic string theories and some arrows demonstrating how they are related to each other, to M-theory and to tachyonic type 0 string theories. }
    \label{fig:dualitystar}
\end{figure}

For the simplicity of this discussion, we will mostly consider the eight known 10d non-tachyonic string theories, five of which lead to supersymmetric spacetimes \cite{Polchinski:1998rr,Alvarez-Gaume:1986ghj,Dixon:1986iz,Sagnotti:1995ga,Sagnotti:1996qj,Sugimoto:1999tx}. There are other consistent string theories but they always have a tachyon in their spectrum, which signals an instability of the vacuum. All eight of these theories, shown in Figure \ref{fig:dualitystar}, reduce at low energies to ten-dimensional EFTs with matter and gauge fields minimally coupled to gravity. These theories are all related to M-theory whose low-energy limit is eleven-dimensional supergravity.

\paragraph*{Background Independence in String Theory.}\label{par:BI_in_ST}
Above, we discussed how we quantized the oscillations of the string on a flat background and showed at low energies that this could be considered to be a flat (or weakly-curved) spacetime. This was for practical reasons, as quantizing the string on curved backgrounds is a very daunting task. From the perspective of the worldsheet, it is not obvious that string theory is background independent: because we picked a background to quantize it, background independence of the quantum theory is not manifest. However, in the low-energy EFTs, there are subtle ways in which the background independence of the full theory can be seen. For instance, there are \hl{dualities} between string theories that map a string theory on a background X to another one (possibly the same one) on a different background Y, as is schematically depicted in Figure \ref{fig:dualitystar}. The duality states that the spectra of the two theories are exactly the same. This hints at the fact that the choice of background may be somewhat irrelevant. On the other hand there are holographic dualities, which equate the partition function of certain string theories on curved spacetime with the partition function of a QFT (so no gravity) in one dimension less. The most famous example is the \hl{AdS/CFT correspondence} \cite{Maldacena:1997re,Witten:1998qj,Gubser:1998bc}, which relates string theories on asymptotically AdS space to conformal field theories (CFTs) living at the asymptotic boundary of spacetime, with the global conformal symmetry of the CFT matching the asymptotic symmetry of the bulk (see e.g.~\cite{Ammon:2015wua,Aharony:1999ti} for detailed reviews on gauge-gravity duality). The identification of the partition functions allows one to recast all of the gravity in AdS in terms of CFT correlators on flat space. A crucial aspect of holography is that the geometry of the extra (holographic) dimension is not meant to be fixed, and one technically has to sum over all geometries (even topologically different ones) that are compatible with the asymptotic boundary conditions. This versatility suggests that the essential features of the correspondence are not tied to any specific background geometry but rather reflect the background independence of the gravitational theory in a profound way. Finally, in non-perturbative formulations of string theory, such as string field theory, background independence could play a more prominent role (see e.g.~\cite{Sen:1994kx}).

\paragraph*{The String Landscape.} The string theories discussed above and shown in Figure \ref{fig:dualitystar} live in ten spacetime dimensions. In order to make contact with our four-dimensional world, we can compactify some of the extra dimensions. This comes down to wrapping some of them on a compact manifold of size $l_{\rm KK}$ and making $l_{\rm KK}^{-1} \gg E$, where $E$ here represents the energy-scale of the physical process under consideration. Decomposing all of the ten-dimensional fields into their Fourier modes, the massless modes survive the compactification and appear in the lower-dimensional theory whilst the heavy modes, often called Kaluza-Klein modes or replicas, decouple from the lower-dimensional physics due to their mass $m_{\rm KK} = l_{\rm KK}^{-1} \gg E$. During the process of compactification, the parts of the ten-dimensional fields that propagate in the internal dimensions lead to new dynamical scalars in the lower-dimensional theory, called \textit{moduli}. They parametrize the geometry of the compact manifold.
\begin{wrapfigure}{r}{0.5\textwidth} 
    \centering
 \includegraphics[width=0.5\textwidth]{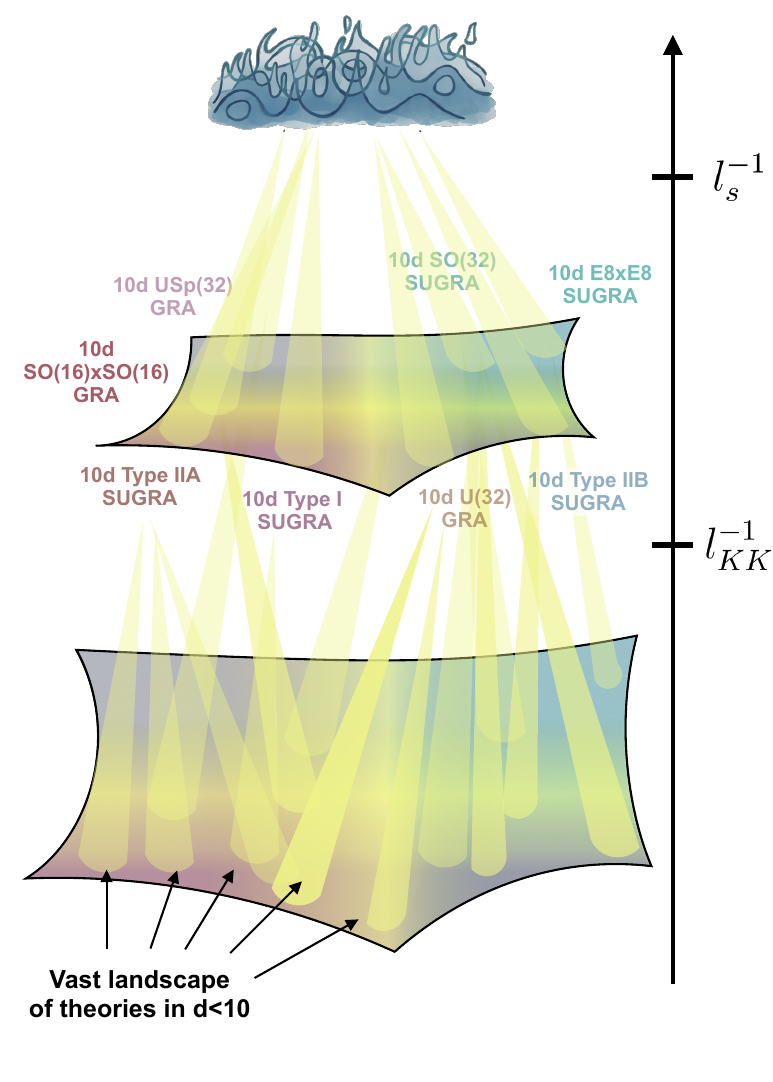} 
    \caption{At energies well below the string scale, string theories are described by their ten-dimensional low-energy effective actions. One can further compactify these EFTs and obtain a myriad of theories in 4d, with varying amounts of supersymmetry.}
    \label{fig:stringpheno}
\end{wrapfigure}
Through this process, the ten-dimensional EFTs that arise as the low-energy limits of string theories give rise to four-dimensional EFTs containing gravity, gauge and matter fields, together with an array of scalars: the dilaton and the moduli of the compact space. Compactification is therefore more than a way of ``getting rid of'' the extra dimensions: it can generate a much richer field content and phenomenology than is present in ten dimensions.

Geometric compactification is, however, not the only route to four dimensions. A four-dimensional string vacuum only requires a consistent internal worldsheet CFT with the appropriate central charge, and this CFT need not admit a geometric interpretation. Some asymmetric orbifolds~\cite{Narain:1986qm, Baykara:2024tjr}, for instance, are intrinsically four-dimensional, and many of them are tachyon-free and supersymmetric. Recently, 4d Minkowski vacua have also been constructed using Landau-Ginzburg models \cite{Becker:2024ayh,Rajaguru:2024emw,Chen:2025rkb}. Alternatively, one can construct subcritical string theories directly in four dimensions by turning on a linear dilaton background, but at the cost of breaking four-dimensional Lorentz invariance and typically introducing tachyonic instabilities. Compactification nonetheless remains the dominant approach in string phenomenology, owing to the versatility of the four-dimensional physics it allows one to engineer. This is summarized schematically in Figure~\ref{fig:stringpheno}. The big question is then: do these 4d theories look like our world, and can we use them to make predictions from string theory?

The fact that there seem to be many different ways of carrying out this compactification, depending on the choice of compact manifold and on the additional ingredients introduced along the way, makes it difficult to extract unique low-energy predictions from string theory without some further guiding principle. Indeed, different choices generally lead to very different four-dimensional low-energy theories: although all of them contain GR in their gravitational sector, they can differ in their particle content, vacuum energy, coupling constants, and symmetries. The collection of such theories is said to form the \hl{string Landscape}. In practice, while the basic idea of compactifying string theory to four dimensions is conceptually well understood, carrying it out in complete mathematical detail is extremely difficult, except for some isolated cases that preserve a lot of (super)symmetries.\footnote{Type II strings on a six-torus yield $4d$ $\mathcal{N}=8$ theories, while compactifications on Calabi--Yau threefolds yield $4d$ $\mathcal{N}=2$ theories. Further ingredients, such as orientifolds, D-branes and fluxes, reduce this to $4d$ $\mathcal{N}=1$ vacua, where quantum corrections are already only partially under control. Generic six-manifolds, on the other hand, break supersymmetry completely, and the resulting landscape of $4d$ $\mathcal{N}=0$ vacua remains very poorly understood.} A lot of progress has been made in recent years to find detailed, top-down examples of compactifications that lead to realistic universes, but it is fair to say that very few, if any at all, are under full computational control.

\subsection{The Swampland Program}

Faced with this computational difficulty, one can ask whether or not it is possible to obtain \textit{any} low energy theory from string theory. In recent years, this has been shown not to be true. There are some theories that simply cannot be obtained from string theory at low energies. Identifying the characteristics that separate these theories from the ones that can be obtained has been the whole premise of the Swampland Program \cite{Vafa:2005ui}. The Swampland Program aims to distinguish between effective theories that arise from a consistent theory of quantum gravity ---those in the ``Landscape''--- and those that do not, that are said to be in the ``Swampland''. In other words, the Swampland Program provides criteria that EFTs must satisfy to be considered as low-energy limits of a consistent theory of quantum gravity, such as string theory. EFTs that are in the Swampland are therefore thought to be \hl{inconsistent with fundamental principles} of a UV-complete theory of quantum gravity. The Swampland Program has generated significant interest and research activity within the string theory community and beyond, as it provides valuable insights into the structure of quantum gravity and the landscape of EFTs that truly stem from it.

Swampland conjectures, which form the basis of the program, are the criteria that segregate the EFTs in the Landscape from those in the Swampland. The central strategy of the Swampland Program is to combine insights from \hl{top-down constructions} in string theory with \hl{bottom-up consistency requirements} expected of any theory of quantum gravity. On the one hand, one can study large classes of effective theories that arise from explicit string compactifications and identify common structural patterns among them. On the other hand, one can impose general consistency conditions that any theory coupled to quantum gravity should satisfy, ranging from familiar constraints such as anomaly cancellation to more subtle restrictions suggested by black hole physics. Black holes provide a unique window into the fundamental consistency of quantum gravity because:

\begin{itemize}
  \item \textbf{They are universal in theories with gravity}. A sufficiently energetically dense field configuration in a gravitational theory may collapse into a black hole. Black holes are widely believed to constitute a fundamental part of the high-energy spectrum of the theory and therefore constrain the UV density of states, whose behavior must match the result obtained from the Bekenstein-Hawking formula for the entropy \cite{Bekenstein:1973ur, Hawking:1975vcx,Hawking:1974rv}.
\item\textbf{They are thermodynamic objects with entropy/temperature that radiate}. Hawking evaporation implies that certain black holes might decay into lighter states. As a result, consistency of black hole evaporation places non-trivial constraints on the spectrum of states of the theory. For instance, for charged black holes to be able to shed their charge and decay, there must exist sufficiently light charged states. This is the essence of the Weak Gravity Conjecture~\cite{ArkaniHamed:2006dz}.
 \item \textbf{They mix the UV and the IR}. One can get some intuition for this fact by noting that a high-energy process, whose microscopic description necessarily involves quantum-gravitational degrees of freedom, is described at long distances by a large, smooth, and entirely macroscopic spacetime configuration. Indeed, extremely high-energy (UV) collisions with center of mass energy given, in four spacetime dimensions, by $\sqrt{s} \gg M_{\rm P}$, collapse, at sufficiently small impact parameter, into black holes whose Schwarzschild radius grows with the collision energy, making the outcome of the collision a large infrared (IR) object \cite{Eardley:2002re}. The black hole then emits Hawking radiation with a characteristic wavelength set by its size, which is low-energy (IR). Thus, an ultraviolet (UV) scattering process produces infrared physics, illustrating UV/IR mixing.
       \begin{equation*}
         \begin{tikzpicture}[scale=1, decoration={coil,amplitude=4.25,segment length=4.75}]
     \draw[thick, ->,blue!7!black!80](-2.5,0.4) -- (-1.5,0.4);
    \draw[thick, ->,blue!7!black!80](1.55,0.4) -- (0.55,0.4);
     \draw[spring] (-3.25,0) -- (-1,0);
  
     \draw[spring] (0,0) -- (2.25,0);
     \filldraw [black] (-2.2,-0.2) circle (0pt) node[anchor=north ]{\large${h}_{\rho\sigma}(p^0=\frac{\sqrt{s}}{2}) $};
     \filldraw [black] (1.25,-0.2) circle (0pt) node[anchor=north ]{\large${h}_{\mu\nu}(p^0=\frac{\sqrt{s}}{2})  $};
    \draw[thick, ->,blue!7!black!80](2.55,0) -- (3,0);
    
    \fill[thick,dustyPlum]   (4.5,0) circle (12pt);

  \foreach \ang [evaluate={\angin=\ang-180+10*rand; \r=0.6+0.05*rand; \l=0.6+0.02*rand}] in {10,45,100,140,190,240,290,330}{
    \draw[radiation] ($(4.5,0) + (\ang:\r)$) --++ (\angin:-\l);
  }
 \filldraw [black] (4.5,-1.5) circle (0pt) node[anchor=north ]{$R_S\sim G_N\, {\sqrt{s}}$};
    \end{tikzpicture}
       \end{equation*}

This intuition can be made more precise in the context of the AdS/CFT correspondence \cite{Maldacena:1997re,Witten:1998qj,Gubser:1998bc}. In fact, the correspondence itself is a fantastic realization of UV/IR mixing in that the radial direction in the bulk is associated with the renormalization group (RG)-flow energy scale in the boundary theory \cite{Susskind:1998dq, Peet:1998wn,Polchinski:2010hw}. 
 
\end{itemize}

\noindent These considerations explain why black holes play such an important role in the Swampland Program. They provide a controlled setting in which apparently infrared questions about charges, symmetries, and the low-energy spectrum become tied to the ultraviolet consistency of quantum gravity. In these lectures, we will focus on one of the most robust lessons to emerge from this perspective: a consistent theory of quantum gravity cannot possess exact global symmetries. As we will see in Section~\ref{sec:No_Global_Symmetries}, black hole evaporation provides a particularly simple physical motivation for this statement, while sharper arguments can be obtained from string perturbation theory and holography. These amount to proofs within specific frameworks and under suitable assumptions, whose limitations we also discuss.

\hl{The Cobordism Conjecture} can be understood as a topological formulation of this principle, applying it to charges associated with \hl{bordism invariants} (see \cref{sec:bordismgeneralities}). Non-trivial cobordism classes would divide the space of allowed quantum-gravity configurations into distinct topological sectors that cannot be connected by any physical process, thereby giving rise to conserved global charges. The conjecture asserts that quantum gravity must contain the appropriate objects, defects or mechanisms that allow these sectors to be connected, so that the relevant cobordism groups are trivial. We will therefore begin by reviewing ordinary and generalized global symmetries and the arguments for their absence in quantum gravity, before turning to the Cobordism Conjecture and exploring how this abstract principle leads to concrete predictions for the spectrum of defects and transitions in string theory.

\section{Global Symmetries 101}\label{sec:GlobalSYms101}

In this section we review the notion of global symmetries in (quantum) field theory, with particular emphasis on their modern generalization known as higher-form symmetries. Unlike ordinary symmetries, which are associated with operators acting on local operators, higher-form symmetries are naturally described by topological operators supported on extended submanifolds of spacetime and act on extended objects. In this way, extended objects already play an important role in quantum field theory, independently of string theory.

In these lectures, we will focus on invertible symmetries, fusing according to a group law, though this can be relaxed.\footnote{Details about non-invertible/categorical symmetries, where the group fusion law is replaced by a more general fusion category, can be consulted in e.g.~\cite{NabilR, BottiniR,NamekiR, Cordova:2022ruw,McGreevy:2022oyu,Brennan:2023mmt,Gomes:2023ahz}.
For this reason, we will implicitly assume throughout this section that the term ``symmetry'' refers specifically to an invertible symmetry, fusing according to a group law.} 
We conclude with some examples to clarify the formal definitions and detail the possibility of either gauging or breaking higher-form symmetries.

We will loosely follow the presentation of the topic in \cite{NabilR, BottiniR}, but we will be concise and, when unavoidable, sacrifice technical details and generality for intuition. 
The interested reader is encouraged to consult these sources directly, along with other introductions to the topic, such as \cite{NamekiR, Gaiotto:2014kfa,Cordova:2022ruw, McGreevy:2022oyu, Brennan:2023mmt, Gomes:2023ahz}, and references therein.

\noindent

\subsection{What are Global Symmetries?}  \label{sec:2.1} 
Here, a ``global symmetry'' is simply a \hl{genuine symmetry of the physical theory}. The adjective \emph{global} is used only to distinguish it from a gauge symmetry, which is not a physical symmetry, but rather a redundancy of the theoretical description: gauge-related configurations represent the same physical state.\footnote{Provided they act trivially at the boundary/have trivial asymptotic action.} How does one define a genuine symmetry of a physical theory? A working definition for a \hl{unitary symmetry} in Quantum Mechanics (QM) is the following:

\begin{symmetrybox}{deepAqua}
\textbf{Symmetries in QM:} A time-independent unitary symmetry is represented by a {\bf unitary} operator $U$ that acts on the Hilbert space and commutes with the Hamiltonian $H$,
\begin{equation}
 [H,U]=0\,.
\end{equation}
It acts on observables by conjugation.
In Quantum Field Theory,
\begin{equation}
 U\mathcal{O}(t,\vec{x})U^\dagger:=\mathcal{O}'(t,\vec{x})\,,
\end{equation}  
mapping local operators to local operators.\footnote{More generally, if the unitary operator implements a spacetime symmetry depending on time, such as a Galilean or Lorentz boost, these statements require refinements, see e.g., \cite{Weinberg:QM}. We will, for simplicity, not delve further into such considerations here.}
\end{symmetrybox}
\noindent

\noindent Note that a symmetry should act non-trivially on at least one physical state in the Hilbert space; otherwise, it is not clear in what meaningful sense it constitutes a symmetry. In this respect, gauge symmetries are not genuine symmetries of the physical Hilbert space: by construction, physical states and observables are gauge invariant. We now turn to an example that makes the definition above more concrete.

\paragraph{Example 0: complex scalar with global U$(1)$.}Consider the following action for a local complex scalar field $\phi(x)$ in $d$-dimensional Minkowski space $\mathbb{R}^{1,d-1}$:\footnote{We locally resort to Lorentzian signature in mostly plus convention in this example because we want to make a connection with the usual notion of conservation under time evolution. We shall otherwise work in Euclidean signature all the way to the end of this section. See also our later comment on \hyperlink{Euclidean}{Euclidean vs.~Lorentzian space.}
}

\begin{equation}\label{eq:free_cx_scalar}
 S[\phi]=-\int \, d^dx\, \left(\partial_\mu \phi^\dagger \,\partial^\mu \phi+m^2 \phi^\dagger \phi\right)\;.
\end{equation}
The action is invariant under the global phase transformation
\begin{equation}\label{eq:U1_transf}
  \phi(x)\longrightarrow\, \phi'(x)=e^{i \alpha} \phi(x)\,,
\end{equation}
where $e^{i \alpha} \in \mathrm{U}(1)$ is a global (spacetime-independent) phase. This continuous global symmetry has an associated conserved current $j^\mu$, which can be obtained via the Noether procedure:
\begin{equation}\label{eq:currentphi}
 j^\mu:= i \left( \phi^\dagger \partial^\mu \phi-\phi\partial^\mu \phi^\dagger \right)\;.
\end{equation}
The current satisfies an on-shell conservation law 
\begin{equation}\label{eq:continuity} 
  \partial_\mu\, j^\mu=0\qquad \qquad \mbox{(on shell)} \,, 
\end{equation}
which allows us to define a conserved charge
\begin{equation}
 Q(t):=\int_{\Sigma_t} d^{d-1}x\; j^0(t,\vec{x})\;,
\end{equation}
where the integral is performed over a constant time slice $\Sigma_t\equiv \mathbb{R}^{d-1}$. The conservation of $Q(t)$ follows from the continuity relation in \cref{eq:continuity}: 
\begin{equation}
 \frac{d}{dt}Q(t)=\int_{\mathbb{R}^{d-1}} d^{d-1}x\; \partial_t j^0(t,\vec{x})\;=-\int_{\mathbb{R}^{d-1}} d^{d-1}x\; \partial_{x^i}j^i(t,\vec{x})\;=0 \;,
\end{equation}
where, in the last step, we used the fact that the integral can be recast as a boundary term and that there is no flux at infinity.\footnote{These boundary conditions are part of the definition of the theory.} 

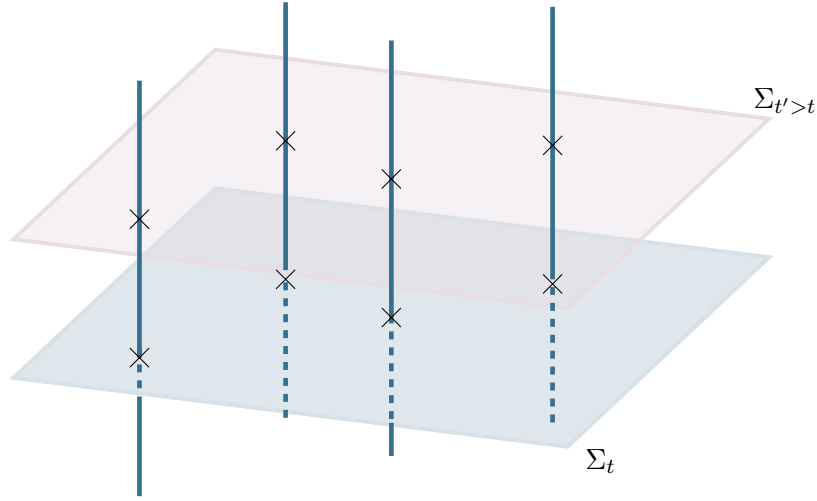
\begin{figure}[htbp]
  \centering
\begin{tikzpicture}[scale=1.3,tdplot_main_coords]
    \filldraw[stormBlue!20, opacity=0.8] (-3,-3,0) -- (3,-3,0) -- (3,3,0) -- (-3,3,0) -- cycle;
    \draw[stormBlue!20, line width=1.5] (-3,-3,0) -- (3,-3,0) -- (3,3,0) -- (-3,3,0) -- cycle;
  \filldraw[dustyPlum!20, opacity=0.4] (-3,-3,1.5) -- (3,-3,1.5) -- (3,3,1.5) -- (-3,3,1.5) -- cycle;
    \draw[dustyPlum!20, line width=1.5] (-3,-3,1.5) -- (3,-3,1.5) -- (3,3,1.5) -- (-3,3,1.5) -- cycle;
\draw[line width=1.75, stormBlue] (0,0,0) -- (0,0,3);
\draw[line width=1.75, stormBlue,dashed] (0,0,-1.1) -- (0,0,0);
\draw[line width=1.75, stormBlue] (0,0,-1.14) -- (0,0,-1.5);
\draw[line width=1.75, stormBlue] (-1.5,1.2,0.05) -- (-1.5,1.2,3);
\draw[line width=1.75, stormBlue,dashed] (-1.5,1.2,-1.5) -- (-1.5,1.2,0);
\draw[line width=1.75, stormBlue] (2,-2,0) -- (2,-2,3);
\draw[line width=1.75, stormBlue,dashed] (2,-2,0.1) -- (2,-2,-0.4);
\draw[line width=1.75, stormBlue] (2,-2,-1.5) -- (2,-2,-0.42);
\draw[line width=1.75, stormBlue] (-.7,-1.4,0.1) -- (-.7,-1.4,3);
\draw[line width=1.75, stormBlue,dashed] (-.7,-1.4,-1.5) -- (-.7,-1.4,0);

\filldraw[stormBlue] (0,0,0) circle (0pt) node[cross=4pt] {};
\filldraw[stormBlue] (-1.5,1.2,0) circle (0pt) node[cross=4pt] {};
\filldraw[stormBlue] (2,-2,0) circle (0pt) node[cross=4pt] {};
\filldraw[stormBlue] (-.7,-1.4,0) circle (0pt) node[cross=4pt] {};
\node at (2.5,2.9,-0.1) [anchor=north west] { $\Sigma_t$ };
\filldraw[stormBlue] (0,0,1.5) circle (0pt) node[cross=4pt] {};
\filldraw[stormBlue] (-1.5,1.2,1.5) circle (0pt) node[cross=4pt] {};
\filldraw[stormBlue] (2,-2,1.5) circle (0pt) node[cross=4pt] {};
 \filldraw[stormBlue] (-.7,-1.4,1.5) circle (0pt) node[cross=4pt] {};
\node at (-2.5,2.9,1.6) [anchor=south west] { $\Sigma_{t'>t}$ };
\end{tikzpicture}
    \caption{\small Charged particle worldlines intersecting a codimension-1 constant $t$ submanifold \\ $\Sigma_t\subset \mathbb{R}^{1,d-1}.$
 The number of intersections is the same for any future or past time slice $\Sigma_{t'\neq t}$. } \label{fig:2.1}
\end{figure}

One simplistic yet effective way to think about this conservation law is the following. The operator $Q(t)$ measures net charge on a given time slice: each particle contributes $+1$ and each antiparticle $-1$. In the free theory (\ref{eq:free_cx_scalar}), the numbers of particles and of antiparticles are separately conserved, so charged worldlines cannot end. A constant time slice therefore intersects the same number of particle and antiparticle worldlines, counted with their respective signs, at all times; hence
\begin{equation}
 Q(t)=Q(t')\qquad \qquad \forall\; t,\,t'\in \mathbb{R}\,.
\end{equation}
This is represented in Figure \ref{fig:2.1}.
This classical result is modified in QFT, which famously allows for particle-antiparticle pair production. 
If we insert this conservation law in a path integral to obtain an expectation value, we recover the corresponding \emph{Ward Identity}, 
\begin{equation}
\begin{split}
     & \partial_\mu\langle j^\mu(x) \phi(x_1)\phi^\dagger(x_2)\cdots \rangle \equiv \frac{1}{Z}\partial_\mu \int\left[\mathcal{D} \phi \right]e^{iS[\phi]}\, j^\mu(x) \phi(x_1)\phi^\dagger(x_2)\cdots\,\\
     & = (\delta^d(x-x_1) - \delta^d(x-x_2) + \ldots) \braket{\phi(x_1)\phi^\dagger(x_2)\cdots} \;.
\end{split}\label{eq:ward1}
\end{equation}
Here, the dots denote operator insertions at non-coincident points, and $\delta$ is the $d$-dimensional Dirac delta distribution. The coefficients in front of each $\delta^d(x-y)$ are related to the infinitesimal transformation law of the corresponding operator insertion (for details on the derivation see e.g.~\cite[\S14.8.1]{Schwartz:2014sze}).
We see that, inside of correlation functions, the divergence of the current vanishes everywhere except perhaps at the insertion points of other local operators, such as $\phi(x_1)$ or $\phi^\dagger(x_2)$. On physical grounds, the local operator $\phi(x_1)$ creates a point-like unit charge at $x_1$, while $\phi^\dagger(x_2)$ creates a point-like charge of opposite sign at $x_2$. The Ward identity shows that this effect of charge creation is directly related to the non-trivial transformation of these two operators under the symmetry, which is given by (\ref{eq:U1_transf}). We thus refer to any operator transforming non-trivially under a global symmetry as a \hl{charged operator} under the symmetry.

\begin{figure}[htbp]
  \centering
\includegraphics[width=0.7\linewidth]{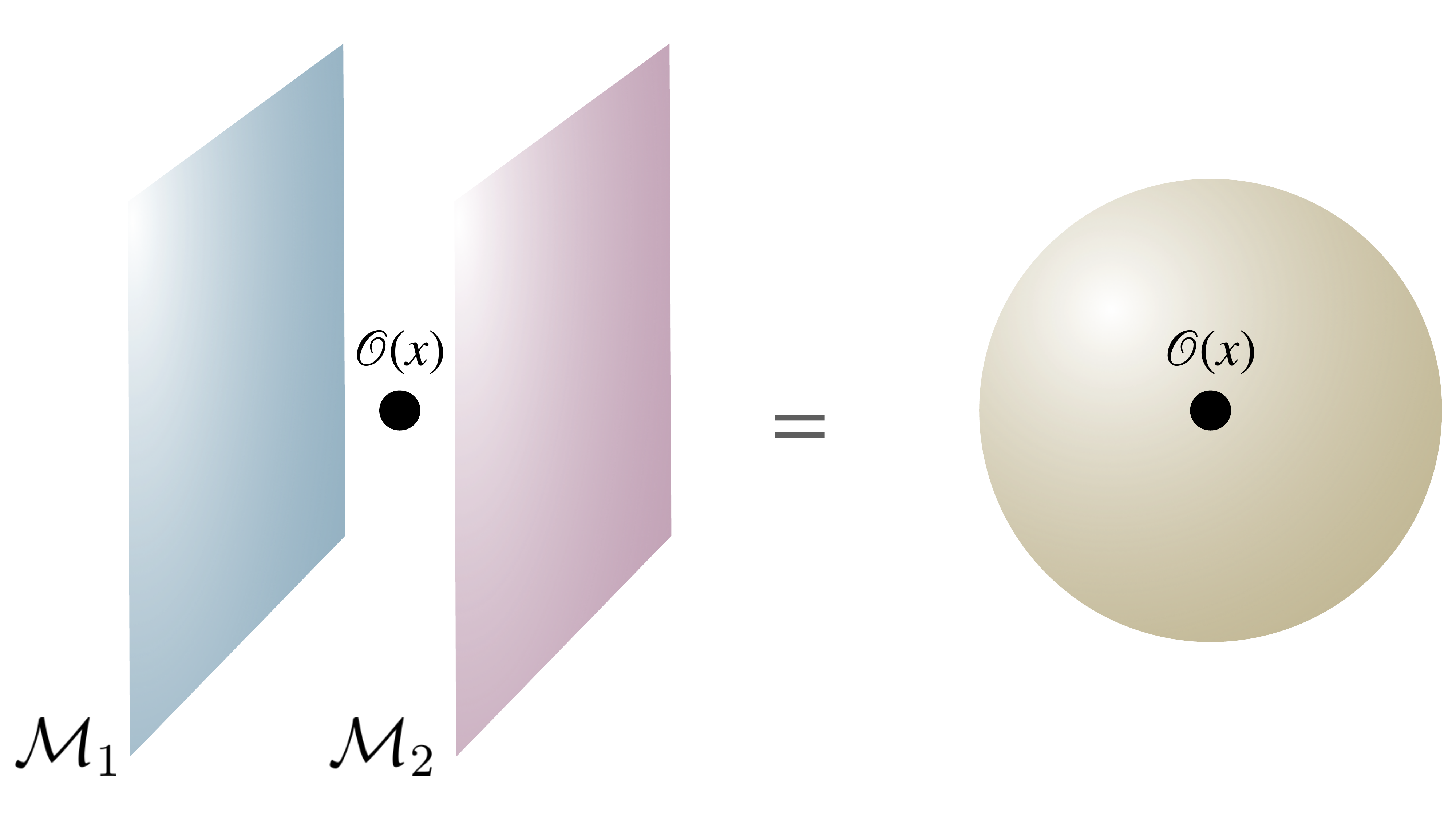}
\caption{\small Deformation of two topological operators $ Q(\mathcal{M}_1)$ and $ Q(\mathcal{M}_2)$ into a third operator $ Q(S^{d-1})$ that is supported in a $S^{d-1}$ and links with the charged operator.} \label{fig:2.2}
\end{figure}

\paragraph{Topological operators and symmetries.}The local field $\phi(x)$ played a central role in our previous example.
It turns out that this is by no means necessary, and that there is an alternative approach in which the topological nature of the associated operator $Q(t)$ is central, instead of the field itself.\footnote{One further advantage of this presentation is that, in avoiding the choice of a \textit{fundamental field}, we are making the notion of symmetry duality-invariant.} Let us write the conservation law \cref{eq:continuity} using differential forms. 
\begin{equation}
    \partial_\mu\, j^\mu=0\qquad\longrightarrow \qquad  d \star \mathbf{j} =0\,, \label{eq:contistar} 
\end{equation} 
with $\mathbf{j}\in \Lambda^1(\mathcal{M})$ the one form current. The Hodge star, which is a map from $p$-forms into $(d-p)$-forms
\begin{equation}
   \star: \Lambda^p(\mathcal{M})\to \Lambda^{d-p}(\mathcal{M})\,,
\end{equation}
allows us to define a $(d-1)$-form $\star \mathbf{j}$ which can be integrated over a general codimension-1 (dimension $d-1$) submanifold $\mathcal{M}_{d-1}$. 
\begin{equation}
 Q(\mathcal{M}_{d-1}):=\int_{\mathcal{M}_{d-1}}d^{d-1}x\;\sqrt{h}\,n_\mu\, j^\mu=\int_{\mathcal{M}_{d-1}}\star \,\mathbf{j}\;.
\end{equation} 
By virtue of the conservation law \cref{eq:contistar} and the Stokes theorem, the integral of the current only depends on the choice of $\mathcal{M}_{d-1}$ through its homology class. Consider for example the situation represented in Figure \ref{fig:2.2}. The action of the two topological operators supported in the two submanifolds $\mathcal{M}_1$ and $\mathcal{M}_2$ is equivalent to the action of a single topological operator supported on a $S^{d-1}$ sphere that links with the operator, that is,
\begin{equation}
 Q(\mathcal{M}_1)-Q(\mathcal{M}_2)=Q(S^{d-1})\;. \label{eq:topo1} 
\end{equation}   
The minus sign in the second term comes from the consistent choice of a normal such that, when deformed to a $S^{d-1}$, gives a vector that is outward-pointing everywhere; equivalently, we have a well-defined orientation so that we can integrate over the $S^{d-1}$ submanifold. We say that $Q$ is a \hl{topological symmetry operator}.\footnote{Note that this operator is sometimes called \emph{charge} operator, since it measures a charge. In order not to confuse it with \emph{charged} operators, we will stick to the \emph{symmetry} operator terminology.} This is precisely what we observed in the scalar field discussion, where the role of the $\mathcal{M}_{d-1}$ was played by constant time slices and we saw that the result did not depend on the particular choice of time.

\noindent
We then make the following suggestive  observation:
\begin{symmetrybox}{deepAqua}
   \textbf{Symmetry operator:} To a global symmetry acting on local operators, there is a corresponding topological symmetry operator supported on a codimension-1 surface $\mathcal{M}_{d-1}$.
\end{symmetrybox}
\noindent
The question that we need to address now is: how does this topological operator act on a charged local operator? In the quantum theory, a local operator $\mathcal{O}_q$ with charge $q$ must obey the {Ward Identity}
\begin{equation}
  \partial_\mu\, j^\mu(x) \mathcal{O}_q(y)= q\, \delta^d(x-y)\mathcal{O}_q(y)\,, \label{eq:continuity2} 
\end{equation}  
which should be understood as an operator equation inside a path integral, similarly to \cref{eq:ward1}.\footnote{We stress that the divergence in \cref{eq:continuity2} should be interpreted as acting on the whole correlation function $\langle j^\mu(x) \mathcal{O}_q(y) \rangle$.} A consequence of this identity is that if we try to make the topological operator $Q$ pass through a charged operator insertion, the charged operator gets modified.\footnote{This is precisely what identifies the local operator as \textit{charged} under the symmetry.}
Let us now write this as an action of $Q$ on a charged operator. Using eq.~(\ref{eq:topo1}) together with \cref{eq:continuity2}, we have
\begin{align}\label{eq:charge_op_through_insertion}
 \left(  Q(\mathcal{M}_1)\right.&-\left.Q(\mathcal{M}_2)\right)\mathcal{O}_q(y) =Q(S^{d-1})\mathcal{O}_q(y)=\oint_{S^{d-1}} \; \star \,\mathbf{j}(x)\; \mathcal{O}_q(y)=\nonumber\\
   &=\int_{B_d}(d\star \mathbf{j})(x)\mathcal{O}_q(y)= q \int_{B_d} d^dx\,\delta^d(x-y)\mathcal{O}_q(y)= q \,\mathcal{O}_q(y)\;,
\end{align}
where $B_d$ is the ball with boundary $S^{d-1}=\partial B_d$.  
We then learn that if we act with the topological operator on a charged operator insertion, the charged operator picks up a contribution proportional to its charge. 

\begin{figure}[htbp]
  \centering
\includegraphics[width=0.7\linewidth]{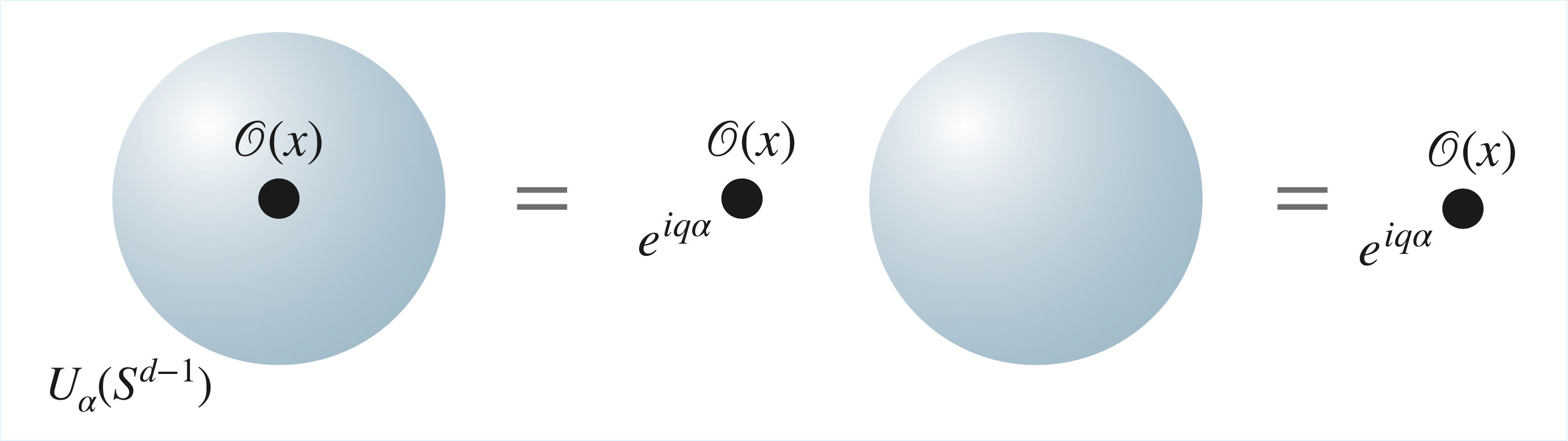}
    \caption{\small The action of a finite charge rotation operator when dragged through a point where a charged operator insertion lies: the operator picks up a $\mathrm{U}(1)$ phase corresponding to the $\alpha$ number defining the finite rotation. Once the topological operator has crossed the charged operator and no longer links with it, it can be contracted to a point, yielding a trivially acting operator. } \label{fig:2.3}
\end{figure}
\noindent
From these results, exponentiating $Q$, we can build the topological operator that is tantamount to a finite charge rotation by an angle $\alpha$, 
\begin{equation}
 U_\alpha(\mathcal{M}_{d-1}):=\operatorname{exp}\left(i \alpha Q(\mathcal{M}_{d-1})\right)\;. \label{eq:finitecharge} 
\end{equation} 
This operator acts as we would expect on charged operators. In the complex scalar example, $\phi$ is a charged operator with $q=+1$ and $\phi^\dagger$ one with $q=-1$,\footnote{In other words, $[Q,\phi]= \phi$ and $[Q,\phi^\dagger]= -\phi^\dagger$, as one can show using \eqref{eq:currentphi}.} so that \cref{eq:linking_action_0-form} below reproduces the transformation \eqref{eq:U1_transf}: 
\begin{equation}\label{eq:linking_action_0-form}
 U_\alpha(S_{x\in \operatorname{int}(B_d)}^{d-1})\,\mathcal{O}_q(x)=e^{i q \alpha} \mathcal{O}_{q}(x)\,;
\end{equation}
this is depicted in Figure \ref{fig:2.3}.

\begin{figure}[htbp]
  \centering
\includegraphics[width=0.9\linewidth]{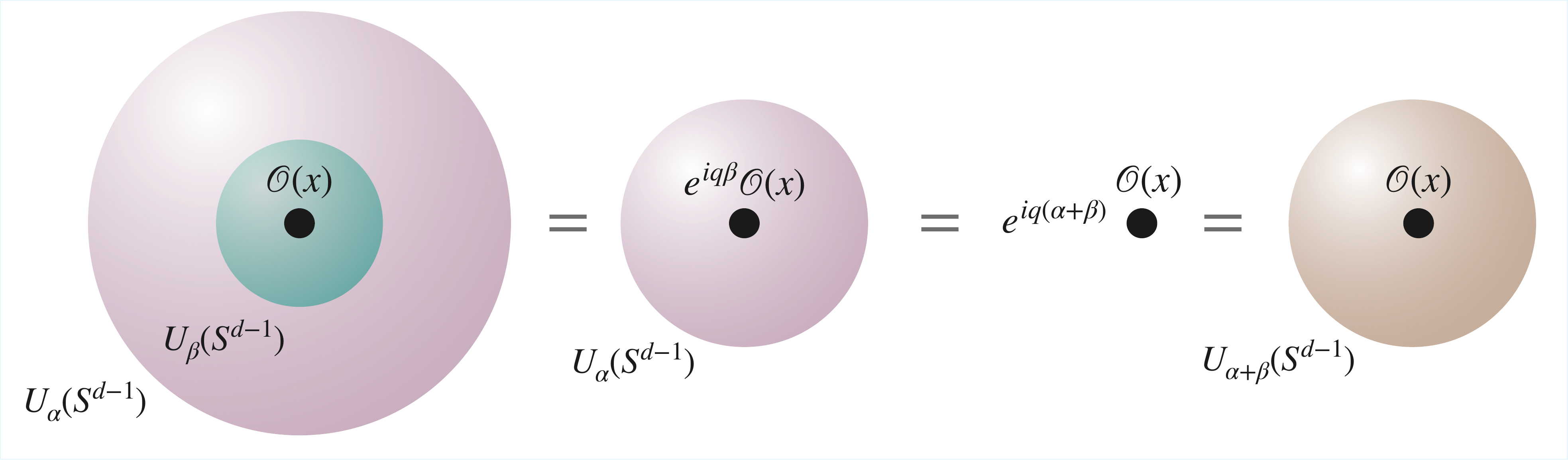}
    \caption{\small The composition of two symmetry operators.  } \label{fig:2.5}
\end{figure}

\noindent
The last ingredient we are missing is to incorporate the composition of several such finite rotations. 
This is the simplest for the $\mathrm{U}(1)$ symmetry case. Imagine acting first with a $U_\beta(S^{d-1}_{\rho_1})$ and then with a second operator $U_\alpha(S^{d-1}_{\rho_2})$, with $\rho_2>\rho_1$ such that it encloses the first topological operator, see Figure \ref{fig:2.5}.\footnote{The order $\rho_1>\rho_2$ or $\rho_2>\rho_1$ is irrelevant in this example because $\mathrm{U}(1)$ is abelian, but becomes important if we replace $\mathrm{U}(1)$ by a non-abelian group.}  
Since there are no insertions between the two topological operators, we can deform their support to a single $S^{d-1}$, 
\begin{equation}
 U_\alpha(S^{d-1})\, U_\beta(S^{d-1})=U_{\alpha+\beta}(S^{d-1})\,.
\end{equation}
This is depicted in Figure \ref{fig:2.5}.

\hypertarget{Euclidean}{\paragraph{Disclaimer: Euclidean vs.~Lorentzian spacetime.}} For a globally hyperbolic Lorentzian spacetime, one can choose a global time function and write the spacetime as $\mathcal{M}= \mathbb{R}_t \times\Sigma_{d-1}$, where the hypersurfaces $\Sigma_{d-1}$ are constant-time slices. No analogous canonical decomposition exists for a generic Euclidean manifold, which is the setting in which higher-form symmetries are often discussed, as above. Let us therefore briefly digress to clarify the role of time in the discussion above.\footnote{We follow \cite{Gaiotto:2014kfa} in this remark.}

When the Euclidean manifold under consideration can be written as $\mathcal{M}=\mathbb{R} \times \Sigma_{d-1}$ or  $\mathcal{M}=S^1\times \Sigma_{d-1}$, it is natural to interpret the $\mathbb{R}$ or $S^1$ factor as the (Euclidean) time $\tau_E$. In this case, if the topological operator is supported on a constant $\tau_E$ submanifold, it can be interpreted as a symmetry operator acting on the Hilbert space.

On the other hand, if the topological operator extends along the natural time direction, it should instead be regarded as a \hl{defect}: rather than acting within the Hilbert space of the theory, it modifies the theory in such a way that one associates a different Hilbert space to the resulting configuration. In the following, \hl{we will work in Euclidean signature} and will, therefore use the terms (topological/symmetry) defect/operator interchangeably.

\paragraph{Summary.}The summary of this discussion is:
\begin{symmetrybox}{deepAqua}
    \textbf{A 0-form symmetry operator:} is a codimension-1 operator that is topological and invertible.
\end{symmetrybox}
\noindent The adjective `0-form' comes from the fact that these symmetries act on local operators $\mathcal{O}(x)$ (localized in a 0-dimensional spacetime point). Furthermore, 0-form symmetries combine to form a \textit{0-form symmetry group} $ G^{(0)}$.
\begin{symmetrybox}{deepAqua}
\textbf{A 0-form symmetry group:} $ G^{(0)}$ is a set of topological codimension-1 operators 
\begin{equation}
 U_g\qquad \forall \,g\in G^{(0)}\,,
\end{equation} 
parametrized by the elements of the group $ G^{(0)}$, with composition of operators described by the group composition operation ``$\cdot$'' 
\begin{equation}
 U_g(\mathcal{M}_{d-1})\, U_{g'}(\mathcal{M}_{d-1})=U_{g\cdot g'}(\mathcal{M}_{d-1})\;.
\end{equation}
\end{symmetrybox}

\noindent If we went back to a Lorentzian signature manifold, and restricted to constant $t$ time slices, then these operators would be the unitary operators $U(t)$ that act on the Hilbert space and commute with the Hamiltonian $H$.\footnote{See, however, our previous `Euclidean vs. Lorentzian' remark.}
These are precisely those we defined at the beginning of our discussion on symmetries in a quantum theory. 
 
\paragraph{Action on a local operator.}In general, a local operator $\mathcal{O}$ will transform in a representation $\Gamma$ of $G^{(0)}$. The action of the symmetry group on $\mathcal{O}$ is then
\begin{equation}
 U_g(S^{d-1})\mathcal{O}_a(x)=(\Gamma(g))^b_{\;\,a}\,\mathcal{O}_b(x) \qquad \forall \,g\in G^{(0)}\,,
\end{equation}
where $S^{d-1}$ is a $(d-1)$-sphere that links with the insertion point $x$.

In the previous $\mathrm{U}(1)$ example, we took $\Gamma$ to be the one-dimensional representation and then $\Gamma(\alpha)=e^{i q \alpha}$. 
For a non-abelian example, if we were considering $\SU(2)$ in the fundamental representation, $\Gamma$ would correspond to exponentials of Pauli matrices. 

\paragraph{A comment on discrete symmetries.}Before discussing the generalization to higher-form symmetries, we briefly comment on discrete symmetries. 
In this case, there is \hl{no conserved current} that we can integrate to straightforwardly construct a topological operator. 
Nevertheless, there is still a set of topological operators of codimension one that implement the action of the corresponding discrete symmetry, 
\begin{equation}
 U_g(S^{d-1})\mathcal{O}_a(x)=(\Gamma(g))^b_{\;\,a}\,\mathcal{O}_b(x)\,.
\end{equation}
For example, a real scalar with just a mass term, 
\begin{equation}
 S[\phi]=\int \, d^dx\,\frac{1}{2} \left( (\partial \phi)^2+m^2 \phi^2 \right)\;,
\end{equation}
is invariant under the $\mathbb{Z}_2$ symmetry $\phi(x)\to -\phi(x)$. 
Then, one can define a topological defect that implements this transformation, i.e.,
\begin{equation}
 U_{\mathbb{Z}_2}(S^{d-1}) \phi(x)=-\phi(x)\,.
\end{equation}
Given the fact that 
\begin{equation}
 \mathbb{Z}_2:=\left\{e,\sigma\right\} \qquad\mbox{with}\qquad \sigma^2=e,
\end{equation}
the only non-trivial element in $\mathbb{Z}_2$ satisfies the following fusion rule:
 \begin{equation}
 U_{\mathbb{Z}_2}(S^{d-1})\, U_{\mathbb{Z}_2}(S^{d-1})=\mathbf{1}\,.
\end{equation}
\noindent
All this is to say that the lesson from the continuous case still applies to discrete 0-form symmetries, even though they lack a conserved current.

\subsection{Higher-form Symmetries}\label{ssec:higher_form}
In the previous section, we introduced codimension-1 topological defects that linked with charged 0-dimensional local operators as a way to describe the action of a symmetry over the latter. This idea, which we referred to as ``0-form symmetries'', admits a straightforward generalization: \hl{higher-form symmetries}. 

\begin{figure}[H]
  \centering
  \begin{tikzpicture}[scale=1.45]
  
      \draw[dustyPlum,line width=3]  (-1.2,1) ellipse (1 and 0.5);
      \draw[line width=7,domain=2:1.2*pi, samples=100, shift={(-1,-.2)},white] plot ({.25*cos(deg(.75*\x))-0.25},0.8*\x-0.25);
      \draw[line width=4,domain=0:1.13, samples=100, shift={(-1,-.2)}] plot ({.25*cos(deg(.75*\x))-0.25},0.8*\x-0.25);
      \draw[line width=4,domain=1.25:1.2*pi, samples=100, shift={(-1,-.2)}] plot ({.25*cos(deg(.75*\x))-0.25},0.8*\x-0.25);
      \node at (-0.5,-0.4) [] { $\;\;L_p(\mathcal{N}_p)$ };
      \node at (0.4,0.8) [dustyPlum] { $\;\;\;\;U(\Sigma_{d-p-1})$ };
  \end{tikzpicture}
    \caption{\small A charged extended operator $L_p(\mathcal{N}_{p})$ that extends along a $p$-dimensional submanifold $\mathcal{N}_p$ and is linked with a topological symmetry operator $U$ that has support on a $(d-p-1)$-dimensional manifold $\Sigma_{d-p-1}$.} \label{fig:2.7}
\end{figure}
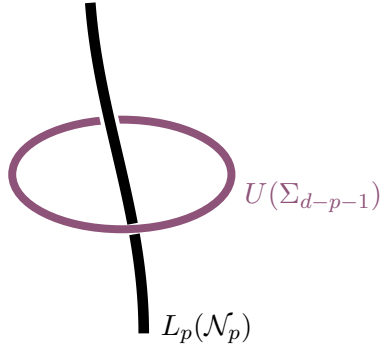

Higher-form symmetries are a formalism to study the properties of extended charged operators. Indeed, if the charged object under study is not a point in space but a higher-dimensional object, the submanifold that links with the charged insertion and over which we define our topological operator needs to have lower dimension, or equivalently, higher codimension, see Figure \ref{fig:2.7}.\footnote{Note that this is different from the situation in Figure \ref{fig:2.1}, where we were representing particle worldlines. The analogous situation in the case in which the extended object is one-dimensional, a line, is discussed below, and depicted in Figure \ref{fig:2.9}.} 

\begin{figure}[H]
    \centering
  \includegraphics[width=0.5\linewidth]{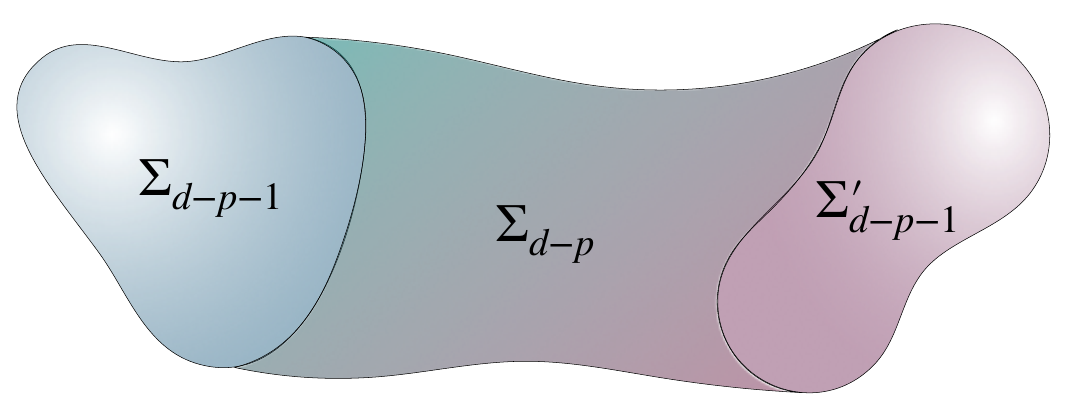}
  \captionof{figure}{If the disjoint union of $\Sigma_{d-p-1}$ and $\Sigma_{d-p-1}'$, with consistent orientations, forms the boundary of a $(d-p)$-dimensional manifold $\Sigma_{d-p}$ that does not cross any charged insertions, then $U(\Sigma_{d-p-1})=U(\Sigma'_{d-p-1})$.}
  \label{fig:2.6}
\end{figure}

\noindent As in the 0-form case, continuous higher-form symmetries have associated conserved currents.\footnote{We will always assume that we are working with a Lagrangian theory. Then, Noether's Theorem applies. } Furthermore, higher-form symmetries are associated with the conservation of extended charges in the same way 0-form symmetries are related to particle charge conservation. In that spirit:
\begin{symmetrybox}{deepAqua}
   \textbf{A $p$-form symmetry operator:} is a codimension-$(p+1)$ symmetry operator $U$ that is topological and invertible.
\end{symmetrybox}
\noindent Generalizing the 0-form case discussed above, such a \hl{higher-form symmetry operator} has the following properties: 
\begin{enumerate}
  \item It is inserted along a general codimension-$(p+1)$ submanifold $\Sigma_{d-p-1}$: $ U(\Sigma_{d-p-1}).$ 

  \item It is \hl{topological} in that \begin{equation}\label{eq:deftop}
 U(\Sigma_{d-p-1})=U(\Sigma'_{d-p-1})\,
  \end{equation}
 provided $\Sigma_{d-p-1}$ and $\Sigma'_{d-p-1}$ are topologically equivalent, i.e., either can be obtained as a topological deformation of the other, by deforming in regions away from charged insertions, which in particular means that they constitute the boundary of a certain codimension $p$ manifold, 
  \begin{equation}
    \Sigma'_{d-p-1}-\Sigma_{d-p-1}=\partial \Sigma_{d-p}\,.
  \end{equation}
 This is shown in Figure \ref{fig:2.6}.
 
   \item It is \hl{invertible}: there is an associated operator $U^{-1}$ such that the composition 
   \begin{equation}
 U(\Sigma_{d-p-1})\,U^{-1}(\Sigma_{d-p-1})=\mathbf{1}\,,
   \end{equation}
 acts as the identity operator.
\end{enumerate}

\noindent With this in hand, we now define the generalization of the \textit{p-form symmetry group}:
\begin{symmetrybox}{deepAqua}
    \textbf{A $p$-form symmetry group}: $ G^{(p)}$ is a set of topological codimension-$(p+1)$ symmetry operators, 
\begin{equation}
 U_g\qquad \forall \,g\in G^{(p)}\,,
\end{equation} 
parametrized by the elements of the group $ G^{(p)}$ and with composition of operators described by the group operation ``$\cdot$'' 
\begin{equation}\label{eq:grouplaw}
 U_g(\mathcal{M}_{d-p-1})\, U_{g'}(\mathcal{M}_{d-p-1})=U_{g\cdot g'}(\mathcal{M}_{d-p-1})\;.
\end{equation}
\end{symmetrybox}

\paragraph{Action of higher-form symmetry operators over extended charged objects.} Having explored the topological nature of higher-form symmetries, we may now study their action on charged operators.
We have seen that a 0-form symmetry acts on 0-dimensional local operators. 
In the topological description, we understood this as the effect of moving the codimension-1 topological operator associated with the 0-form symmetry past the insertion point of the local operator, as shown in Figure \ref{fig:2.3}.

\begin{figure}[htbp]
  \centering
  \begin{tikzpicture}[scale=1.5]
      \draw[mutedTeal,line width=3]  (-1.2,1) ellipse (1 and 0.5);
      \draw[line width=7,domain=2:1.2*pi, samples=100, shift={(-1,-.2)},white] plot ({.25*cos(deg(.75*\x))-0.25},0.8*\x-0.25);
      \draw[line width=4,domain=0:1.13, samples=100, shift={(-1,-.2)}] plot ({.25*cos(deg(.75*\x))-0.25},0.8*\x-0.25);
      \draw[line width=4,domain=1.25:1.2*pi, samples=100, shift={(-1,-.2)}] plot ({.25*cos(deg(.75*\x))-0.25},0.8*\x-0.25);
      \node at (-0.5,-0.4) [] { $L_p(\mathcal{N}_p)$ };
      \node at (-2.8,0.6) [mutedTeal] { $U_g(S^{d-p-1})$ };
      \node at (0.33,1) [] { $=$ };
      \node at (0.95,1.03) [] { $\phi(g)\;\times$ };
      \draw[line width=4,domain=0:1.2*pi, samples=100, shift={(-1,-.2)}] plot ({.25*cos(deg(.75*\x))-0.25+3},0.8*\x-0.25);
      \node at (2.5,-0.4) [] { $L_p(\mathcal{N}_p)$ };
  \end{tikzpicture}
    \caption{\small Action of a topological operator $U_g(S^{d-p-1})$ on a charged extended operator $L_p(\mathcal{N}_{p})$ with support on a $p$-dimensional submanifold $\mathcal{N}_p$. The operator $L_p(\mathcal{N}_{p})$ is assumed to be in an irreducible representation of $G^{(p)}$. Since $G^ {(p)}$ is abelian, its irreducible representations are one-dimensional and hence can be identified with phase functions $\phi(g)$. Note that, in this example, $\mathrm{Link}(S^{d-p-1},\mathcal{N}_p) = 1$.} \label{fig:2.10}
\end{figure}
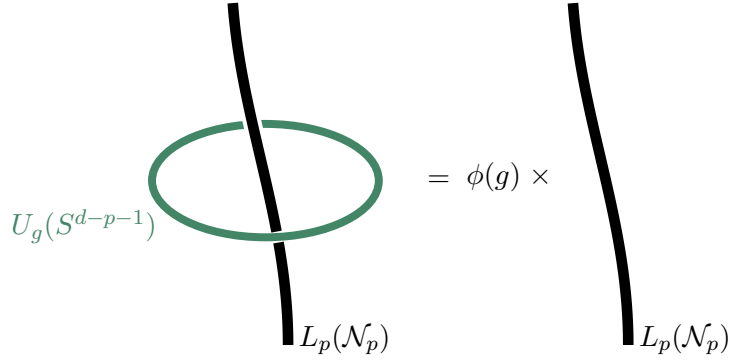

A priori, higher-form symmetry operators may act over charged operators of any dimension, not just local operators. Note however that a topological operator of codimension $(p+1)$ can only link non-trivially with charged operators of dimension $q$, where 
\begin{equation*}
 q\geq p\,.
\end{equation*}
Indeed, if $q < p$, we can deform the support of the topological symmetry operator past the charged operator, showing that it acts trivially on it. Here, we will restrict ourselves to $p=q$; other cases can be consulted, for example, in \cite{BottiniR,Bhardwaj:2023wzd,Bartsch:2023pzl}. In this case, one can show that $p$-dimensional operators transform in representations of the $p$-form symmetry group  $G^{(p)}$, which acts as
\begin{equation}\label{eq:p-form_rep}
    U_g^\dagger(\Sigma_1)\,L_p(\mathcal{N}_p)\,U_g(\Sigma_2) \neq L_p(\mathcal{N}_p) \;,
\end{equation}
see Figure \ref{fig:2.10}. The inequality \eqref{eq:p-form_rep} only expresses that the action is non-trivial. More precisely, since $G^{(p)}$ is abelian for $p\geq1$ (as we show below), a charged operator in an irreducible representation transforms by a phase $\phi(g)$, i.e., a character $\phi\colon G^{(p)}\to\mathrm{U}(1)$, raised to the linking number\footnote{\label{fn:linking_number}Given two closed, disjoint, null-homologous submanifolds $\Sigma_{d-p-1}$ and $\mathcal{N}_p$ of a $d$-dimensional spacetime with $\Sigma_{d-p-1}=\partial D_{d-p}$, the linking number $\mathrm{Link}(\Sigma_{d-p-1},\mathcal{N}_p)\in\mathbb{Z}$ is the number of times $\mathcal{N}_p$ intersects $D_{d-p}$, each intersection counted with sign $\pm1$ according to orientation. Its value does not depend on the choice of $D_{d-p}$.} between the supports of the two operators:
\begin{equation}
    U_g(\Sigma_{d-p-1})\, L_p(\mathcal{N}_p) = \phi(g)^{\operatorname{Link}(\Sigma_{d-p-1},\,\mathcal{N}_p)}\, L_p(\mathcal{N}_p)\,.
\end{equation}
This is the form used in the Maxwell example below (Example~3), where $\phi(e^{i\alpha})=e^{iq\alpha}$. 

\paragraph{$\mathbf{p}$-form symmetry groups are abelian.} A general property of any $p$-form symmetry group $G^{(p)}$ is that, for $p\geq 1$, it must be abelian. This admits a simple geometric explanation, illustrated in Figure \ref{fig:2.8}. Since the topological symmetry operators are supported on submanifolds of codimension greater than one, they can be continuously moved past one another without obstruction. Exchanging their order therefore leaves the configuration unchanged, implying that the corresponding symmetry operators commute.

\begin{figure}[htbp]
  \centering
  \begin{tikzpicture}[scale=1.3]
      \draw[mutedTeal,line width=3]  (-1.2,1) ellipse (0.5 and 0.25);
      \draw[dustyPlum,line width=3]  (-1.2,1) ellipse (1.5 and 0.75);
      \draw[line width=7,domain=2:1.2*pi, samples=100, shift={(-1,-.2)},white] plot ({.25*cos(deg(.75*\x))-0.25},0.8*\x-0.25);
      \draw[line width=4,domain=0:0.82, samples=100, shift={(-1,-.2)}] plot ({.25*cos(deg(.75*\x))-0.25},0.8*\x-0.25);
      \draw[line width=4,domain=0.94:1.44, samples=100, shift={(-1,-.2)}] plot ({.25*cos(deg(.75*\x))-0.25},0.8*\x-0.25);
      \draw[line width=4,domain=1.58:1.2*pi, samples=100, shift={(-1,-.2)}] plot ({.25*cos(deg(.75*\x))-0.25},0.8*\x-0.25);
      \node at (-0.5,-0.4) [] { $L_p(\mathcal{N}_p)$ };
      \node at (-0.45,1) [mutedTeal] { $U_g$ };
      \node at (0.55,0.7) [dustyPlum] { $U_{g'}$ };
      \node at (0.8,1) [] { $=$ };
      \draw[dustyPlum,line width=3]  (2.8,1) ellipse (0.5 and 0.25);
      \draw[mutedTeal,line width=3]  (2.8,1) ellipse (1.5 and 0.75);
      \draw[line width=7,domain=2:1.2*pi, samples=100, shift={(-1,-.2)},white] plot ({.25*cos(deg(.75*\x))-0.25+4},0.8*\x-0.25);
      \draw[line width=4,domain=0:0.82, samples=100, shift={(-1,-.2)}] plot ({.25*cos(deg(.75*\x))-0.25+4},0.8*\x-0.25);
      \draw[line width=4,domain=0.94:1.44, samples=100, shift={(-1,-.2)}] plot ({.25*cos(deg(.75*\x))-0.25+4},0.8*\x-0.25);
      \draw[line width=4,domain=1.58:1.2*pi, samples=100, shift={(-1,-.2)}] plot ({.25*cos(deg(.75*\x))-0.25+4},0.8*\x-0.25);
      \node at (3.5,-0.4) [] { $L_p(\mathcal{N}_p)$ };
      \node at (3.55,1) [dustyPlum] { $U_{g'}$ };
      \node at (4.55,0.7) [mutedTeal] { $U_{g}$ };
  \end{tikzpicture}
    \caption{\small Pictorial argument for the abelian nature of $p\geq1$ higher-form symmetries. All insertions with codimension $\geq2$ can be topologically deformed past each other. Therefore, any two elements in the higher $p$-form group $U_g,\, U_{g'}\in G^{(p)}$ must obey $U_g U_{g'}= U_{g'}U_g$.} \label{fig:2.8}
\end{figure}
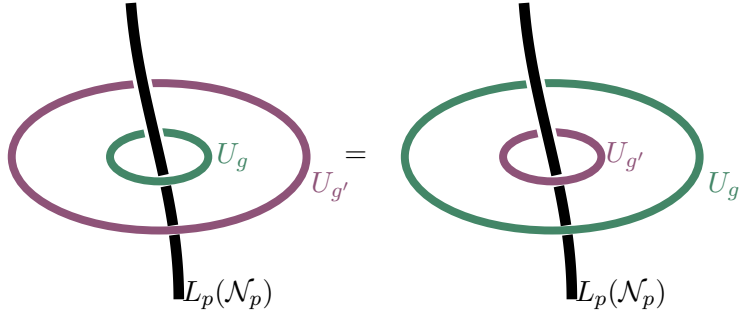
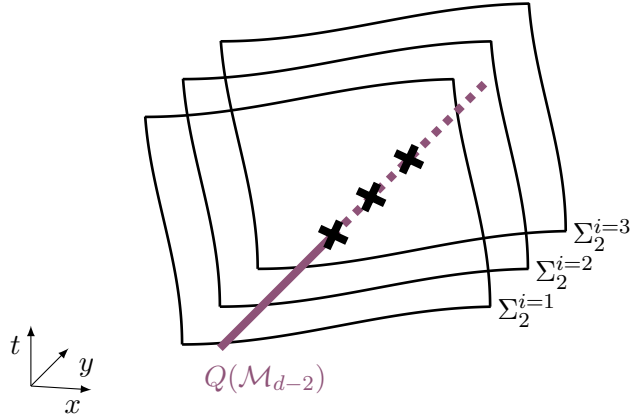
\begin{figure}[h!]
  \centering
  \begin{tikzpicture}[scale=1]
  \draw[domain=0:1.3*pi, samples=100, shift={(-1,-.2)}, line width = 1] plot (\x, {.25*cos(deg(.75*\x-3*pi))});
  \draw[domain=0:1.3*pi, samples=100, shift={(-1,-.2)}, line width = 1] plot (\x-0.49, {.25*cos(deg(.75*\x-3*pi))+3.01});
  \draw[domain=0:1.2*pi, samples=100, shift={(-1,-.2)}, line width = 1] plot ({.25*cos(deg(.75*\x))-0.25},0.8*\x-0.25);
  \draw[domain=0:1.2*pi, samples=100, shift={(-1,-.2)}, line width = 1] plot ({.25*cos(deg(.75*\x))-0.175+4},0.8*\x+0.25);
  \draw[domain=0:1.3*pi, samples=100, shift={(-1,-.2)}, line width = 1] plot (\x+1, {.25*cos(deg(.75*\x-3*pi))+1});
  \draw[domain=0:1.3*pi, samples=100, shift={(-1,-.2)}, line width = 1] plot (\x-0.49+1, {.25*cos(deg(.75*\x-3*pi))+3.01+1});
  \draw[domain=0:1.2*pi, samples=100, shift={(-1,-.2)}, line width = 1] plot ({.25*cos(deg(.75*\x))-0.25+1},0.8*\x-0.25+1);
  \draw[domain=0:1.2*pi, samples=100, shift={(-1,-.2)}, line width = 1] plot ({.25*cos(deg(.75*\x))-0.175+4+1},0.8*\x+0.25+1);
  \draw[domain=0:1.3*pi, samples=100, shift={(-1,-.2)}, line width = 1] plot (\x+0.5, {.25*cos(deg(.75*\x-3*pi))+0.5});
  \draw[domain=0:1.3*pi, samples=100, shift={(-1,-.2)}, line width = 1] plot (\x-0.49+0.5, {.25*cos(deg(.75*\x-3*pi))+3.01+0.5});
  \draw[domain=0:1.2*pi, samples=100, shift={(-1,-.2)}, line width = 1] plot ({.25*cos(deg(.75*\x))-0.25+0.5},0.8*\x-0.25+0.5);
  \draw[domain=0:1.2*pi, samples=100, shift={(-1,-.2)}, line width = 1] plot ({.25*cos(deg(.75*\x))-0.175+4+0.5},0.8*\x+0.25+0.5);
  \draw[line width=3,dustyPlum] (1,1)--(-0.5,-0.5);
  \draw[line width=3,dustyPlum,dashed] (3,3)--(1,1);
  \node at (1,1) [cross=7pt,rotate=20,line width=3] {  };
  \node at (1.5,1.5) [cross=7pt,rotate=20,line width=3] {  };
  \node at (2,2) [cross=7pt,rotate=20,line width=3] {  };
  \node at (0.1,-0.9) [dustyPlum] { $Q(\mathcal{M}_{d-2})$ };
  \node at (3.4,0) [black] { $\;\;\;\Sigma_{2}^{i=1}$ };
  \node at (4.4,1) [black] { $\;\;\;\Sigma_{2}^{i=3}$ };
  \node at (3.9,0.5) [black] { $\;\;\;\Sigma_{2}^{i=2}$ };
  \draw[->] (-3,-1)--(-2.2,-1.05);
  \draw[->] (-3,-1)--(-3,-0.2);
  \draw[->] (-3,-1)--(-2.5,-0.5);
  \node at (-2.2,-1.05) [anchor= north east] { $x$ };
  \node at (-3,-0.2) [anchor=north east] { $t$ };
  \node at (-2.5,-0.5) [anchor=north west] { $y$ };
  \end{tikzpicture}
  \caption{Worldsheets swept by three different line operators $\Sigma^i$. They are counted by a codimension-2 topological operator $Q(\mathcal{M}_{d-2})$ that, in 1+2, corresponds to a line piercing through the worldsheets.}\label{fig:2.9} 
\end{figure}

\paragraph{Example 1: 1-form symmetry.}Let us, for concreteness, consider the case of a $1$-form continuous symmetry to gain some intuition and translate this back to our initial motivation. In this case, the charged operator is 1d, i.e., a line operator, which we will refer to as a string. There is a two-form current:
\begin{equation}
 \mathbf{j}_2\in \Lambda^2(\mathcal{M})\,,\qquad\qquad  \mathbf{j}_2=\frac{1}{2}j_{\mu \nu} dx^\mu\wedge dx^\nu\,,
\end{equation} 
with fully antisymmetric components, 
\begin{equation}
j_{[\mu \nu]}=\frac{1}{2}\left(j_{\mu\nu}-j_{\nu\mu}\right)=j_{\mu \nu}\,.
\end{equation}
We can define the associated topological symmetry operator as the integral of the Hodge dual of the current supported on some codimension two manifold $\mathcal{M}_{d-2}$,
\begin{equation}
 Q(\mathcal{M}_{d-2}):=\;\int_{\mathcal{M}_{d-2}}\star\, \mathbf{j}_2\;.
\end{equation}  
This topological defect is associated with the conservation of the string charges, which in spacetime sweep 2d worldsheets. In order to count the string charge at a given instant of time, it is not necessary to integrate over the entire space. 
Since these strings do not end, it suffices to integrate over a line that traverses the worldsheets; any displacement or deformation of this codimension-2 operator will count the same number of string charges. 
This situation, for the 1+2 case, where both the charged operator and the symmetry operator are 1d, is shown in Figure \ref{fig:2.9} and is to be compared with the particle worldline case in Figure \ref{fig:2.1}.

\noindent Summarizing, we will work with the following definition of a \hl{$p$-form symmetry}: 
\begin{symmetrybox}{deepAqua}
    \textbf{A $p$-form global symmetry:} is a set of topological codimension-$(p+1)$ operators that: 
    \begin{enumerate}
    \item Follow a group law \eqref{eq:grouplaw}.
        \item Act non-trivially on the Hilbert space. For any $g \in G$ not equal to the identity, there exists some operator $L_p(\mathcal{N}_p)$ on which it acts non-trivially \eqref{eq:p-form_rep}.
        \item Commute with the Hamiltonian, more precisely, leave the stress tensor invariant. For any $g \in G$ and $x$, we have
\begin{equation}
U_g^\dagger(\Sigma)\,T_{\mu\nu}(x)\,U_g(\Sigma)
= T_{\mu\nu}(x)\,,
\end{equation}
where $T_{\mu\nu}$ is the stress tensor of the theory.
     \item Map $p$-dimensional charged operators $L_p(\mathcal{N}_p)$ into $p$-dimensional charged operators. 
    \end{enumerate}
\end{symmetrybox}
\noindent Note that the first condition can be relaxed, which leads to the notion of non-invertible symmetries; we will not discuss these here. Moreover, if one considers $(-1)$-form symmetries, there is no need for a group law (see \hyperlink{Example4}{Example 4} below). Finally, as mentioned above, a $p$-form symmetry may also act non-trivially on operators of dimension greater than $p$, but we will ignore this possibility in these lectures.

\subsection{Gauging/Breaking Global Symmetries}\label{sec:gaugebreaking}

\noindent
In this section, we describe the explicit breaking and gauging of global $p$-form symmetries. As we will see, this can be done purely in terms of the topology of the symmetry operators. 

\paragraph{Explicit breaking.}

\begin{figure}[htbp]
  \centering
\begin{tikzpicture}[scale=1.5]
      \draw[dustyPlum,line width=3]  (-1.2,1.3) ellipse (1 and 0.5);

      \def\a{0.70}   
      \def\b{0.99}   

      \draw[line width=7,domain=2:1.2*pi, samples=100, shift={(-1,-0.2)},white]
        plot ({.25*cos(deg(.75*\x))-0.25},0.8*\x-0.25);
      \draw[line width=4,domain=0:\a, samples=100, shift={(-1,-.2)}]
        plot ({.25*cos(deg(.75*\x))-0.25},0.8*\x-0.25);

      \draw[line width=4,domain=\b:1.43, samples=100, shift={(-1,-0.2)}]
        plot ({.25*cos(deg(.75*\x))-0.25},0.8*\x-0.25);
      \draw[line width=4,domain=1.25:1.2*pi, samples=100, shift={(-1,-0.2)}]
        plot ({.25*cos(deg(.75*\x))-0.25},0.8*\x-0.25);

      \fill[deepAqua] ({-1 + (.25*cos(deg(.75*\a))-0.25)},{-0.2 + (0.8*\a-0.25)}) circle (2.2pt);
      \fill[deepAqua] ({-1 + (.25*cos(deg(.75*\b))-0.25)},{-0.2 + (0.8*\b-0.25)}) circle (2.2pt);
      
      \node[deepAqua] at ({-0.9 + (.25*cos(deg(.75*\a))-0.25)},{-0.2 + (0.8*\a-0.25)}) [anchor=west] { $\phi^\dagger(\partial\mathcal{N}_p)$ }; 
      \node[deepAqua] at ({-1.1 + (.25*cos(deg(.75*\a))-0.25)},{ + (0.8*\a-0.25)}) [anchor=east] { $\phi(\partial\mathcal{N}_p)$ };

      \node at (-0.5,-0.4) [] { $L_p(\mathcal{N}_p)$ };
      \node at (-2.8,0.6) [dustyPlum] { $U_g(S^{d-p-1})$ };
      \node at (0.43,1) [] { $\Longrightarrow$ };

      \draw[line width=4,domain=0:\a, samples=100, shift={(-1,-.2)}]
        plot ({.25*cos(deg(.75*\x))-0.25+3},0.8*\x-0.25);
      \draw[line width=4,domain=\b:1.2*pi, samples=100, shift={(-1,-.2)}]
        plot ({.25*cos(deg(.75*\x))-0.25+3},0.8*\x-0.25);

      \fill[deepAqua] ({-1 + (.25*cos(deg(.75*\a))-0.25+3)},{-0.2 + (0.8*\a-0.25)}) circle (2.2pt);
      \fill[deepAqua] ({-1 + (.25*cos(deg(.75*\b))-0.25+3)},{-0.2 + (0.8*\b-0.25)}) circle (2.2pt);

      \node[deepAqua] at ({-0.9 + (.25*cos(deg(.75*\a))-0.25)+3},{-0.2 + (0.8*\a-0.25)}) [anchor=west] { $\phi^\dagger(\partial\mathcal{N}_p)$ }; \node[deepAqua] at ({-1.1 + (.25*cos(deg(.75*\a))-0.25)+3},{ + (0.8*\a-0.25)}) [anchor=east] { $\phi(\partial\mathcal{N}_p)$ };  

      \draw[dustyPlum,line width=3]  (4.0,1) ellipse (1 and 0.5);

      \node at (2.5,-0.4) [] { $L_p(\mathcal{N}_p)$ };
\end{tikzpicture}
    \caption{\small If the charged operator can end, the symmetry is explicitly broken. Symmetry operators $U_g$ are no longer \textit{topological}.} \label{fig:unlinking}
\end{figure}
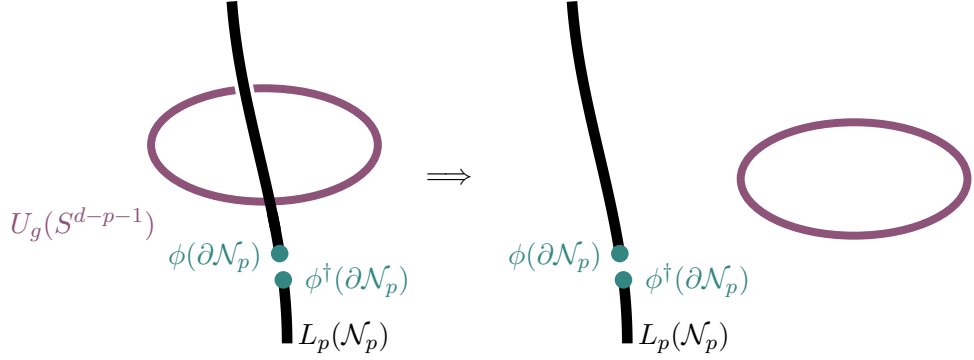

Consider a $p$-form global symmetry with a conserved $(p+1)$-form current $\mathbf{j}_{p+1}$,
\begin{equation}
d \star \mathbf{j}_{p+1} = 0 \; .
\end{equation}
Introducing dynamical charged matter that violates the current conservation law \hl{explicitly breaks} the corresponding generalized symmetry.
Specifically, in the presence of such dynamical elements, the current conservation identity is violated,
\begin{equation}\label{eq:non-conservation}
d \star \mathbf{j}_{p+1} \propto \delta_{d-p} \, ,
\end{equation}
where $\delta_{d-p}$ is a \textit{delta-function} $(d-p)$-form supported on the $p$-dimensional worldvolume of the dynamical charged objects.\footnote{For a closed, oriented $k$-dimensional submanifold $\Sigma \subset M_d$, the Poincar\'e dual $\delta_\Sigma$ is the distribution-valued $(d-k)$-form defined by $\int_{M_d} \omega_k \wedge \delta_\Sigma = \int_\Sigma \omega_k$ for every $k$-form $\omega_k$.}
Given $V$ such that $\partial V=\Sigma-\Sigma'$ the rhs of
\begin{equation}
    Q(\Sigma)-Q(\Sigma')=\int_Vd \star \mathbf{j}_{p+1}\,,
\end{equation}
can be nonzero. This in turn implies that the charge depends on the deformation from $\Sigma$ to $\Sigma'$, losing its ``topological'' behavior. 
 In terms of operators, the explicit breaking of a global symmetry can be interpreted as the fact that the charged operator is allowed to end. The charged operator can thus unlink from the symmetry operator (see Figure \ref{fig:unlinking}), which consequently ceases to be topological.
\begin{symmetrybox}{antiqueGold}
    \begin{center}
        \textbf{Lesson:} A $p$-form symmetry is explicitly broken if its charged operators can end. 
    \end{center}
\end{symmetrybox}
\noindent In other words, charged $p$-dimensional operators $L_p(\mathcal{N}_p)$ are no longer required to form closed manifolds; they may end on $(p-1)$-dimensional operators.

Another form of expressing the conservation law is by coupling the current to a background $(p+1)$-form field $B_{p+1}$, which acts as a source and is not integrated over in a path integral formulation. In Euclidean signature:
\begin{equation}
S \;\supset\; i \int B_{p+1} \wedge \star \mathbf{j}_{p+1} \, .\label{eq:bckg}
\end{equation}
In this language, in the absence of 't Hooft anomalies, conservation of the global symmetry is equivalent to the invariance of the generating functional under background gauge transformations,
\begin{equation}\label{eq:Bjjj}
    B_{p+1}\longmapsto B_{p+1}+d \Lambda_{p}\;.
\end{equation}
If the symmetry is explicitly broken, \eqref{eq:Bjjj} is no longer a symmetry and the Ward identity associated to the background field $B_{p+1}$ reflects the non-conservation of the higher-form charge.

\begin{figure}[htbp]
\vspace{0.5cm}
\centering
\begin{tikzpicture}[scale=1.2]

  \def\rS{0.90}     
  \def\yB{-0.60}    
  \def\amp{0.045}   
  \def\kw{320}      

  \def\xL{-1.4}
  \def\xM{2.0}
  \def\xR{5.8}

  \pgfmathsetmacro\xGauge{0.5*((\xL+\rS)+(\xM-\rS))}
  \pgfmathsetmacro\xStokes{0.5*((\xM+1.35)+(\xR-0.85))}


  \def\yL{1.0}

  \fill[dustyPlum, opacity=0.13]
    (\xL,\yL) circle (\rS);

  \draw[dustyPlum,line width=3]
    (\xL,\yL) circle (\rS);

  \draw[dustyPlum,line width=1.4pt,opacity=0.75]
    (\xL,\yL) ellipse [x radius=\rS cm, y radius=0.30cm];

  \fill[black]
    (\xL,\yL) circle (3pt);

  \node[black,anchor=west]
    at ({\xL+0.10},{\yL+0.05})
    {$\phi(x)$};

  \node[dustyPlum,anchor=east]
    at ({\xL-\rS-0.10},{\yL+0.25})
    {$U_g(S^{d-1})$};

  \node at (\xL,\yB)
    {$\text{global }0\text{-form symmetry}$};


  \node at (\xGauge,1)
    {$\xRightarrow{\text{gauge}}$};


  \def\yM{1.0}
  \pgfmathsetmacro\yP{\yM+\amp*sin(\kw*\rS)}   

  \fill[dustyPlum, opacity=0.13]
    (\xM,\yM) circle (\rS);

  \draw[dustyPlum,line width=3,opacity=0.6]
    (\xM,\yM) circle (\rS);

  \fill[white]
    ({\xM+\rS},\yP) ellipse [x radius=0.07, y radius=0.20];

  \draw[dustyPlum,line width=1.4pt,opacity=0.45]
    ({\xM+\rS},{\yP+0.20})
    arc[start angle=90, end angle=270, x radius=0.07, y radius=0.20];

  \draw[
    deepAqua,
    line width=2.0,
    samples=80,
    smooth,
    domain=0:1.25
  ]
    plot ({\xM+\x},{\yM+\amp*sin(\kw*\x)});

  \draw[dustyPlum,line width=1.4pt]
    ({\xM+\rS},{\yP-0.20})
    arc[start angle=-90, end angle=90, x radius=0.07, y radius=0.20];

  \fill[black]
    (\xM,\yM) circle (3pt);

  \node[black,anchor=east]
    at ({\xM-0.10},{\yM+0.08})
    {$\phi(x)$};

  \node[deepAqua,anchor=south]
    at ({\xM+1.25},{\yM+0.06})
    {$\gamma$};

  \node[dustyPlum,anchor=south]
    at (\xM,{\yM+\rS+0.08})
    {$S^{d-1}\setminus D_\epsilon$};



  \node at (\xStokes,1)
    {$\xRightarrow{\text{Stokes}}$};


  \def\yR{1.0}
  \pgfmathsetmacro\xRing{\xR+1.125}   

  \draw[dustyPlum,line width=2pt,opacity=0.4]
    (\xRing,{\yR+0.25})
    arc[start angle=90, end angle=270, x radius=0.09, y radius=0.25];

  \draw[
    deepAqua,
    line width=2.0,
    samples=100,
    smooth,
    domain=0:1.6
  ]
    plot ({\xR+\x},{\yR+\amp*sin(\kw*\x)});

  \draw[dustyPlum,line width=2pt]
    (\xRing,{\yR-0.25})
    arc[start angle=-90, end angle=90, x radius=0.09, y radius=0.25];

  \fill[black]
    (\xR,\yR) circle (3pt);

  \node[black,anchor=east]
    at ({\xR-0.10},{\yR+0.08})
    {$\phi(x)$};

  \node[deepAqua,anchor=south]
    at ({\xR+1.6},{\yR+0.30})
    {$\displaystyle \exp\Big(iq\int_{\gamma} A\Big)$};

  \node[dustyPlum,anchor=north]
    at (\xRing,{\yR-0.25})
    {$\displaystyle\frac{i}{g^2} \int_{S^{d-2}}\star F = q$};

  \node at ({\xR+0.8},\yB)
    {$\text{gauged }0\text{-form symmetry}$};

\end{tikzpicture}

\caption{\small
Left: in the ungauged theory, a charged local operator $\phi(x)$ is detected by the topological symmetry operator $U_g(S^{d-1})$ on a sphere surrounding it.
Middle: after gauging, $\phi(x)$ must be dressed by a Wilson line $\gamma$ extending to infinity, which pierces the sphere. On the closed $S^{d-1}$ the operator is now trivial (Gauss law), but removing a small disk $D_\epsilon$ around the piercing point leaves a surface whose boundary is a codimension-2 sphere $S^{d-2}$ linking $\gamma$.
Right: by Stokes' theorem, the charge is measured by the electric flux through this $S^{d-2}$.}
\label{fig:gauged-zero-form}
\end{figure}
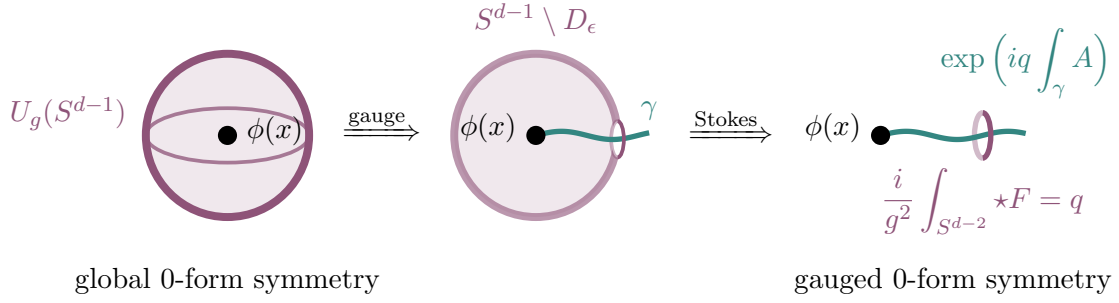

\paragraph{Gauging.} Gauging the $p$-form symmetry consists of promoting the background field $B_{p+1}$ in \cref{eq:bckg} to a \hl{dynamical} gauge field and summing over its configurations. The action contains now the kinetic term along with the coupling to the current
\begin{equation}
S \,\supset\, \frac{1}{2 g^2} \int d B_{p+1}\wedge\star\, dB_{p+1} + i\int B_{p+1} \wedge \star\, \mathbf{j}_{p+1} \,.
\end{equation}
The equation of motion for $B_{p+1}$ sets
\begin{equation}
\frac{1}{g^2}\, d \star d B_{p+1} = i\,(-1)^{p+1}\star \mathbf{j}_{p+1}\, ,
\end{equation}
so that $\star \,\mathbf{j}_{p+1}$ becomes not only \emph{closed} but \emph{exact} in the gauged theory. As a consequence, the symmetry operator $Q$ is no longer an independent observable: the transformation it generates has become a gauge redundancy with a Gauss-law constraint $Q(\mathcal{M}_{d-p-1}) = 0$ (for closed $\mathcal{M}_{d-p-1}$). Correspondingly, the original global symmetry current is absorbed into the dynamical gauge sector. Indeed, upon gauging, the operators that were charged under the $p$-form symmetry are no longer gauge invariant and therefore do not, by themselves, define physical observables of the gauged theory. Instead, they become \hl{relative} operators/defects, appearing as the boundaries of dynamical $(p+1)$-dimensional gauge operators/defects.\footnote{For specifics on the concept of \textit{genuine vs non-genuine} operators, see \cite{NabilR}.} A familiar example is ordinary electromagnetism: the electron operator is not gauge invariant by itself, but must be dressed by a Wilson line in order to define a gauge-invariant object. The symmetry operators, in turn, acquire the interpretation of operators implementing the Gauss-law constraint $\tfrac{1}{g^2}\int_{\mathcal{M}_{d-1}} d\star F = 0$ (for closed $\mathcal{M}_{d-1}$). When integrated over an appropriate region, the Gauss law relates them to a flux operator on the boundary of that region, which measures the enclosed gauge charge, or equivalently the charge seen at infinity. This is illustrated in Figure \ref{fig:gauged-zero-form}.

\begin{symmetrybox}{antiqueGold}
    \begin{center}
        \textbf{Lesson:} A $p$-form symmetry is gauged if its charged operators become the boundary of $(p+1)$-dimensional gauge defects. They are no longer genuine observables.  
    \end{center}
\end{symmetrybox}

\noindent In summary, explicit breaking is characterized by the ability of charged operators to end, while gauging is characterized by the fact that charged operators become boundaries of $(p+1)$-dimensional operators of the dynamical gauge field. 

Note that if two global symmetries share a mixed anomaly, gauging one of them can explicitly break the other. This is an important phenomenon in gauge theory. We will see a canonical example of this in Example~3.

\paragraph{An invitation to higher-form spontaneous symmetry breaking.} 
We have discussed explicit breaking and gauging of higher-form symmetries. One may also wonder whether higher-form symmetries can be spontaneously broken. This brief aside provides an answer to that question.

In the usual discussions about the spontaneous symmetry breaking of a 0-form symmetry, one is concerned with the expectation value of some local (charged) operator that acts as an order parameter 
\begin{equation}
    \braket{\Phi(x)}\sim\left\lbrace
\begin{array}{ll}
0 \qquad{\rm symmetric}\;{\rm  phase} \\
\Phi_0\neq 0 \qquad {\rm SSB}\;{\rm  phase}
\end{array}
\right.\;.
\end{equation}
In case of a $p$-form symmetry, rather than that of a local operator, the order parameter is the expectation value of an operator extended over a submanifold $\mathcal C$ of dimension $p$, $\langle W(\mathcal C) \rangle$.

The phases are now organized according to the scaling of this expectation value for large closed $\mathcal{C}$. One can distinguish:
\begin{enumerate}
    \item \textbf{Unbroken phase:}
    \begin{equation}
        \braket{W(\mathcal C)}\sim\operatorname{exp}\left(-T \operatorname{vol}(\mathcal{X}) \right),
    \end{equation}
    where $\mathcal X$ refers to the minimal manifold satisfying $ \partial \mathcal X=\mathcal C\,,$ and $T$ is some physical parameter.
    In non-abelian gauge theories, this \textit{``area'' scaling} is a sign of \hl{confinement.}\footnote{This language for SSB of a higher form symmetry is reminiscent of the prescription for QCD confinement in \cite{Wilson:1974sk, tHooft:1977nqb,tHooft:1978pzo}. These shaped the modern notion of confinement/deconfinement corresponding to the $1$-form center symmetry being conserved/spontaneously broken. For a crash course, see Section 2.2 of \cite{Witten:1998zw}.}
\item \textbf{Broken phase:}
    \begin{equation}
        \braket{W(\mathcal C)}\sim\operatorname{exp}\left(-m \operatorname{vol}(\mathcal{C}) \right).
    \end{equation}
    This \textit{``perimeter'' law} is a sign of \hl{deconfinement} in non-abelian gauge theories. Since the perimeter term can be removed ($m\to0$) by a local counterterm on $\mathcal{C}$, the broken phase is equivalently characterized by $\langle W(\mathcal{C})\rangle$ approaching a non-zero constant for large $\mathcal{C}$, in analogy with $\langle\Phi\rangle\neq0$.
\end{enumerate}
A great deal of the interest in higher-form SSB comes from the following generalization of the \hl{Goldstone Theorem}: when the \textit{perimeter law} holds for a continuous $p$-form symmetry, there is a gapless mode in the spectrum of the theory \cite{Kapustin:2014gua, Hofman:2018lfz, Lake:2018dqm,NabilR}. This Goldstone mode corresponds to a $p- $form Gauge field.

For the particular case of pure 4-dimensional Maxwell theory, which we will discuss in Example~3 below, both the electric and the magnetic one-form symmetries turn out to be spontaneously broken in the unique vacuum of the theory. This follows from the computation of the Wilson line along a circular loop. 't Hooft lines display a similar behaviour, as can be argued from electric-magnetic duality. The 4d photon can then be seen as the soft mode of these broken higher-form symmetries \cite{Hofman:2018lfz, NabilR}.\footnote{The fact that the two broken 1-form symmetries do not produce two independent photon fields, but both are realized by the photon, can be seen as a manifestation of electric-magnetic duality in pure Maxwell.}

The discussion so far has been somewhat abstract; we now turn to concrete examples that illustrate these general principles.

\subsection{Examples}\label{ssec:globsym_examples}
We now present four additional examples of global symmetries. The first three are continuous: the 0-form symmetry associated to the constant shifts of an axion, the two higher-form global symmetries of free Maxwell theory, and the $(-1)$-form symmetries associated with coupling constants. The fourth example, $\Z_N$ gauge theory, illustrates discrete higher-form symmetries. We will also discuss how these global symmetries arise and how they can be gauged and/or broken.

\paragraph{Example 2: axion field 0-form symmetry.} Consider the action for a (dimensionless) axion $a(x)$ with a continuous shift symmetry,
\begin{equation}
  S[a(x)]=\frac{f^2}{2}\int d^dx\, (\partial_\mu a(x) )(\partial^\mu a(x))\,. 
\end{equation} 
This action is invariant under constant shifts, 
\begin{equation}
  a(x)\mapsto a(x)+c\, \quad \mbox{with} \quad c\in\mathbb{R}.
\end{equation}
By standard Noether methods on the Euclidean action, we have a conserved current proportional to
$\frac{\partial \mathcal{L}}{\partial (\partial_\mu a(x))}=f^2\, \partial^\mu a(x)$. 
In form language, and to match our conventions, 
\begin{equation}
  \mathbf{j}_{1}=i f^2 \,d a(x)\,,
\end{equation}
such that $d \star\mathbf{j}_{1}=0$. Therefore, we can define topological symmetry operators with support on codimension-1 manifolds 
\begin{equation}
  U_\alpha(\mathcal{M}_{d-1}):=\operatorname{exp}\left(i \alpha \int_{\mathcal{M}_{d-1}}  \star \,\mathbf{j}_{1}\right) \,.
\end{equation}
In this case, charged local operators under this symmetry can be explicitly constructed as 
\begin{equation}
  \mathcal{O}_q(x):=\operatorname{exp} \left(i q\, a(x)\right) \,,
\end{equation}
which the reader can show, using the same argument as in \eqref{eq:charge_op_through_insertion}, satisfy
\begin{equation}
  U_\alpha(S^{d-1})\mathcal{O}_q(x)=e^{i q \alpha} \mathcal{O}_q(x) U_\alpha(S^{d-1})=e^{i q \alpha} \mathcal{O}_q(x)
\end{equation}
for a defect $U_\alpha(S^{d-1})$ that links with the point $x$, as in Figure \ref{fig:2.3}.\footnote{Note that in this case, $q\in \mathbb{R}$. Details about the U(1) construction and the dual $(p-1)$-form symmetry can be consulted in e.g.~\cite{Brennan:2023mmt}.}

We now consider gauging this global 0-form symmetry. In order to gauge the symmetry, we couple the current to a dynamical gauge field $A$ by adding the term
\begin{equation}
    i\int A\wedge \star \,\mathbf{j}_{1} = -f^2 \int A\wedge \star \,d a
\end{equation}
to the action. The gauging exchanges the partial derivatives in $ S[a(x)]$ to gauge-covariant derivatives:
\begin{equation}
   S[a(x)]=\frac{f^2}{2}\int d^dx\, (D_\mu a(x) )(D^\mu a(x))\,,
\end{equation}
with $ \partial_\mu\to D_\mu$ with $D_\mu a:=\partial_\mu a - A_\mu\,.$
Adding a kinetic term for the gauge field, we arrive at the action:
\begin{equation}
S[a(x),A_\mu(x)]=\frac{f^2}{2}\int D a\wedge\star\,D a + \frac{1}{2g^2}\int F\wedge \star F\,, \label{eq:gaugedaxion} 
\end{equation}
where $F:= dA$ is the field strength.\footnote{Note in this case that the field strength is simply given by the exterior derivative of the gauge potential because the shift symmetry has one single generator.}  
The action given by \cref{eq:gaugedaxion} is gauge invariant under the transformations
\begin{equation}
  a(x)\mapsto a(x) + c(x)\,,\qquad A(x)\mapsto A(x) + dc(x)\,.
\end{equation}
Once the shift symmetry is gauged, the charged local operators $\mathcal{O}_q(x)=e^{i q a(x)}$ are no longer observable, as they are not gauge invariant. Indeed, under
\begin{equation}
  a(x)\mapsto a(x)+c(x)
\end{equation}
the operator $\mathcal{O}_q(x)$ transforms as
\begin{equation}
  \mathcal{O}_q(x)\mapsto e^{i q c(x)}\,\mathcal{O}_q(x)\,.
\end{equation}
The only way of getting a gauge invariant operator out of $\mathcal{O}_q(x)$ is attaching it to a Wilson line of $A$ ending at the point $x$. 
Explicitly, for a curve $\gamma$ extending from $x$ to infinity (or to another insertion of opposite charge), the composite operator
\begin{equation}
  \widetilde{\mathcal{O}}_q(x)
  :=
  \exp\!\left(i q a(x)\right)
  \exp\!\left(i q \int_{\gamma} A \right)
\end{equation}
is gauge invariant.
It is in this sense that the original operator is `not genuinely local' in the gauged theory: it must instead be regarded as the endpoint of a line operator. 
Equivalently, $\mathcal{O}_q(x)$ defines the boundary of a one-dimensional defect, no longer ``topological'', reflecting the fact that the 0-form symmetry has been promoted to a gauge redundancy. 

By contrast, had the shift symmetry been explicitly broken rather than gauged, the operator $\mathcal{O}_q(x)$ would remain a genuine local operator, but it would no longer be charged under any conserved symmetry. 
Explicit breaking could arise, for example, by adding a potential term such as
\begin{equation}
  V(a)=\frac{f^2}{2}\, m^2 a^2 \,,
\end{equation}
which violates the continuous shift symmetry. 
In that case the Noether current would no longer be conserved,
\begin{equation}
  d \star \mathbf{j}_1 \neq 0 \, ,
\end{equation}
and the symmetry operators $U_\alpha(\mathcal{M}_{d-1})$ would cease to be topological: their correlation functions would depend on deformations of $\mathcal{M}_{d-1}$. 
Thus, under explicit breaking, charged operators remain local and genuine, but the symmetry operators lose their topological character; in contrast, upon gauging, the symmetry defects become unphysical redundancies, while the charged operators must be attached to gauge Wilson lines and therefore cease to be genuine local observables.

\paragraph{Example 3: Maxwell U(1) 1-form symmetries.} Consider a $\mathrm{U}(1)$ pure gauge theory in $d$ dimensions. 
\begin{equation}
    S[A]=\frac{1}{2 e^2}\int {\;F\wedge\star F}=\frac{1}{4 e^2}\int d^dx {\;F_{\mu\nu} F^{\mu\nu}}\,.
\end{equation}
If the gauge connection is given by the 1-form $A=A_\mu dx^\mu$, its field strength is the 2-form $F$ locally defined by
\begin{equation}
  F= dA \,.
\end{equation} 
This 2-form satisfies the Bianchi identity, 
\begin{equation}
  dF= 0 \,.
\end{equation} 
Using this along with the equations of motion in the absence of sources,
\begin{equation}
  d \star F= 0 \,,
\end{equation} 
we can identify two conserved currents, proportional to $F$ and $\star F$. In general dimension, one is a $2$-form current corresponding to a $1$-form symmetry and the other is a $(d-2)$-form current $\mathbf{j}_m$ giving rise to a $(d-3)$-form symmetry. For $g\coloneqq e^{i \alpha}\in \mathrm{U}(1)$, the corresponding topological operators are
\begin{equation}\label{eq:U1e&m}
      U^{(m)}_g(\Sigma_2) =\operatorname{exp}  \left( \frac{i \alpha}{2\pi} \int_{\Sigma_2}\; F\right)\,, 
      \qquad 
      U^{(e)}_g(\Sigma_{d-2}) =\operatorname{exp}  \left(- \frac{\alpha}{e^2}  \int_{\Sigma_{d-2}}\; \star F\right) \,, \ 
\end{equation}
where $\Sigma_p$ is a dimension-$p$ submanifold. In the specific $d=4$ case, both are 1-form symmetries.
The superscripts $(m)$ and $(e)$ in \cref{eq:U1e&m} come from the respective names magnetic/electric 1-form symmetry, which is the standard way to refer to these two symmetries. 

Let us now discuss the nature of charged objects under these higher-form symmetries. 

\begin{enumerate}
  \item For the electric $\mathrm{U}^{(e)}(1)$, the charged objects are \emph{Wilson loops}, corresponding to the closed worldline of a electrically-charged test particle:
  \begin{equation}
  W_q(\gamma):=\operatorname{exp}\left( i \,q\oint_\gamma A\right).
  \end{equation}  
  The action of the symmetry operator over the Wilson loop corresponds to
  \begin{equation}
        U^{(e)}_g(\Sigma_{d-2})\;W_q(\mathcal{L}_1)=\, e^{iq\alpha\operatorname{Link}(\Sigma_{d-2},\,\mathcal{L}_1)\;} W_q(\mathcal{L}_1)\,.
  \end{equation}
Here $\operatorname{Link}(\Sigma_{d-2},\,\mathcal{L}_1)\in\mathbb{Z}$ denotes the linking number between the symmetry defect and the Wilson loop; see Footnote \ref{fn:linking_number}.

  \item For the magnetic $\mathrm{U}^{(m)}(1)$, the  characterization is more involved. 
  In $d=4$, $\mathrm{U}^{(m)}(1)$ is also a 1-form symmetry and by electric-magnetic duality the charged operators correspond to the closed worldlines of charged magnetic monopoles, i.e., to \textit{'t Hooft loops}, $T_p(\gamma)$. 
  In the original electric description in terms of $A_\mu$, the ’t Hooft loop cannot be constructed locally from $A_\mu$. Instead, it is defined by prescribing a monopole-like singularity along $\gamma$. Upon electromagnetic duality, however, the same operator becomes an ordinary Wilson loop of the dual gauge potential $\widetilde A_\mu$:
    \begin{equation}
  T_p(\gamma):=\operatorname{exp}\left(i \,p\oint_\gamma \widetilde {A}\right).
  \end{equation} 
  In general, these are $(d-3)$-dimensional operators, which link with the $(d-3)$-form magnetic symmetry operator above as: 
  \begin{equation}
        U^{(m)}_g(\Sigma_{2})\;T_p(\mathcal{L}_{d-3})=\, e^{i p\alpha \operatorname{Link}(\Sigma_{2},\,\mathcal{L}_{d-3})\;} T_p(\mathcal{L}_{d-3})\,.
  \end{equation}

  \end{enumerate}

\noindent As a direct application of the general discussion around eq.~\eqref{eq:non-conservation} and Figure~\ref{fig:unlinking}, both symmetries are explicitly broken once dynamical charged objects are added. Electrically charged matter with current $\mathbf{J}_e$ couples as $i\int A\wedge\star\mathbf{J}_e$. It modifies the Euclidean Maxwell equation to
\begin{equation}
  \frac{1}{e^2}\, d\star F = -i\star\mathbf{J}_e
  \qquad\Longleftrightarrow\qquad
  d\star\mathbf{j}_e = \star\mathbf{J}_e \,,
  \label{eq:electric-breaking}
\end{equation}
so Wilson lines can end on charged fields and $U^{(e)}_g$ is no longer topological.
In the same way, dynamical monopoles with current $\mathbf{J}_m$ source the Bianchi identity,
\begin{equation}
  dF = 2\pi\star\mathbf{J}_m
  \qquad\Longleftrightarrow\qquad
  d\star\mathbf{j}_m = \star\mathbf{J}_m \,,
  \label{eq:magnetic-breaking}
\end{equation}
so 't Hooft lines can end on monopoles and $U^{(m)}_g$ also stops being topological. 

One may also attempt to gauge these higher-form symmetries by coupling their conserved currents to background gauge fields 
\begin{align}
   S^{(e)}_{\rm coupling}[F,B^{(e)}_{2}]&=i\int \;B^{(e)}_{2}\,\wedge \star \,\mathbf{j}_e=-\frac{1}{e^2}\int\;B^{(e)}_{2}\,\wedge \star F\;,\label{eq:gaugeleectric}\\
   S^{(m)}_{\rm coupling}[F,B^{(m)}_{d-2}]&= i\int\;B^{(m)}_{d-2}\,\wedge\,\star\,\mathbf{j}_m=\frac{i}{2 \pi} \int\;B^{(m)}_{d-2}\,\wedge\,F\;,
\end{align}
and promoting those to dynamical fields in a path integral. However, the electric and magnetic higher-form symmetries possess a mixed ’t Hooft anomaly \cite{Gaiotto:2014kfa,Brennan:2023mmt}.
Because of this mixed anomaly, both higher-form symmetries cannot be gauged simultaneously in a local, gauge-invariant manner. 
The action cannot be simultaneously invariant under the background transformations
\begin{equation}
    B^{(e)}_{2}\to B^{(e)}_{2}+d\, \Lambda_{1}\;,\qquad \qquad  B^{(m)}_{d-2}\to B^{(m)}_{d-2}+d\, \Lambda_{d-3} \,.
\end{equation}
Hence, gauging one necessarily obstructs gauging the other. In fact, gauging the full electric $\mathrm{U}^{(e)}(1)$ explicitly breaks the magnetic $\mathrm{U}^{(m)}(1)$. To see this, note that the linear coupling \eqref{eq:gaugeleectric} is by itself not invariant under $B^{(e)}_2\to B^{(e)}_2+d\Lambda_1$. Gauge invariance requires $A$ to shift as well, $A\to A+\Lambda_1$, and the coupling to be completed into the minimal coupling
\begin{equation}
    S[A,B^{(e)}_2]=\frac{1}{2e^2}\int \mathcal{F}\wedge\star\mathcal{F}\,,\qquad \mathcal{F}\equiv dA-B^{(e)}_2\,,
\end{equation}
whose cross term reproduces \eqref{eq:gaugeleectric}. The gauge-invariant field strength is now $\mathcal{F}$ rather than $dA$, and it no longer satisfies the Bianchi identity:
\begin{equation}
    d\mathcal{F} = -\,dB^{(e)}_2 \,.
\end{equation}
As long as $B^{(e)}_2$ is a fixed background with $dB^{(e)}_2=0$, nothing changes. Once $B^{(e)}_2$ is promoted to a dynamical field, however, it is summed over in the path integral, and its field strength $H_3=dB^{(e)}_2$ is generically non-zero. The magnetic current $\star\mathbf{j}_m\propto\mathcal{F}$ is then no longer closed, and the magnetic $\mathrm{U}^{(m)}(1)$ symmetry is explicitly broken. 

\hypertarget{Example4}{\paragraph{Example 4: global $(-1)$-form symmetries.}}
Given its relevance for \cref{sssec:BabyUniv}, let us explain how the notion of $p$-form symmetry extends to $p=-1$. According to the definition above, the conserved current associated to a global $(-1)$-form symmetry is a 0-form $\mathbf{j}_0$, whose conservation is trivial because $\star \mathbf{j}_0$ is a top-form, $d \star \mathbf{j}_0 = 0$. The corresponding topological operators are supported on a closed codimension-0 manifold, i.e., on the entire spacetime manifold $M_d$,
\begin{equation}\label{eq:(-1)-form_symm}
    U_\alpha(M_d) \coloneqq \exp\left(i\alpha \int_{M_d} \star \mathbf{j}_0\right)\,.
\end{equation}
Two spacetime-filling defects cannot be fused in the usual way; they should rather be regarded as theory-changing operations. Indeed, inside of a path integral, $U_\alpha(M_d)$ has the same effect as adding a term $-i\alpha\int\star \mathbf{j}_0$ to the action.\footnote{No local operator can take a state in the Hilbert space of one of the theories to the Hilbert space of another one, so it is also often said that $(-1)$-form symmetries map different superselection sectors of a larger theory.} 

Recall that a global $p$-form symmetry can be probed by coupling its conserved $(p+1)$-form current to a background $(p+1)$-form field $B_{p+1}$. In the case of a $(-1)$-form symmetry, that background field is a 0-form, i.e., an ordinary scalar $\lambda(x)$, so the coupling is simply
\begin{equation}
    S  \;\rightarrow\; S + i \int B_0 \wedge \star \mathbf{j}_0 \,\equiv\, S + i \int d^dx\sqrt{g}\;\lambda(x) j_0(x) \,.
\end{equation}
Setting a constant background $\lambda(x) \equiv \bar{\lambda}$ leads to a deformation:
\begin{equation}
    S + \bar{\lambda}\int d^dx\sqrt{g}\; \mathcal{O}(x) \,,
\end{equation}
with $\mathcal{O}(x) \equiv ij_0(x)$.\footnote{Note that, because the Euclidean path integral is weighted as $e^{-S}$, acting with the symmetry operator in \eqref{eq:(-1)-form_symm} corresponds to shifting $\bar{\lambda} \to \bar{\lambda}-\alpha$.} In this sense, a fixed coupling constant can be seen as a constant background for a $(-1)$-form symmetry. As we will see, the statement that there cannot be any free parameters in quantum gravity (in dimension $d>3$) can be seen as the No Global Symmetries Conjecture applied to $(-1)$-form symmetries \cite{McNamara:2020uza}. Moreover, the symmetry must be gauged: every coupling constant must be the vacuum expectation value of a dynamical field.

\paragraph{Example 5: $\Z_N$ gauge theory and discrete higher-form symmetries.}  Discrete symmetries play a central role in what follows, since many non-trivial bordism groups are torsion (= every element has finite order). The simplest QFT with discrete higher-form symmetries is $\Z_N$ gauge theory, which in $d=4$ can be written as a BF theory\footnote{The name stands for fields $B$ and $F\sim dA_1$.} \cite{Banks:2010zn,Kapustin:2014gua,Gaiotto:2014kfa,Brennan:2023mmt},
\begin{equation}
    S = -i\frac{N}{2\pi}\int_{M_4} B_2\wedge dA_1\,,
\end{equation}
with path integral weight $e^{-S}$, where $A_1$ is a $\mathrm{U}(1)$ 1-form gauge field and $B_2$ a $\mathrm{U}(1)$ 2-form gauge field. This theory arises, for instance, as the low-energy limit of a $\mathrm{U}(1)$ gauge theory Higgsed by a scalar of charge $N$, where $B_2$ is dual to the phase of the Higgs field \cite{Banks:2010zn}. The equations of motion, $N\,dA_1=0$ and $N\,dB_2=0$, force both fields to be locally flat. Furthermore, the theory projects out all field configurations except those with $\mathbb{Z}_N$-valued holonomies. The gauge-invariant operators are generated by
\begin{equation}
    W_q(\gamma)=\exp\Big(iq\oint_\gamma A_1\Big)\,,\qquad V_m(\Sigma)=\exp\Big(im\oint_\Sigma B_2\Big)\,,\qquad q,m\in\Z_N\,,
\end{equation}
which have non-trivial correlations only through their linking,
\begin{equation}\label{eq:WV-link}
    \big\langle V_m(\Sigma)\,W_q(\gamma) \big\rangle = \exp\Big(\frac{2\pi i}{N}\,q\,m\,\operatorname{Link}(\gamma,\Sigma)\Big)\,,
\end{equation}
up to an orientation convention. This follows directly from the path integral. Inserting $V_m(\Sigma)$ adds the term $-i m \int B_2 \wedge \delta_\Sigma$ to the action, where $\delta_\Sigma$ is the 2-form delta function supported on $\Sigma$. Since $B_2$ appears linearly, integrating it out imposes
\begin{equation}
  \frac{N}{2\pi}\, dA_1 = - m\, \delta_\Sigma ,
  \label{eq:flux-tube}
\end{equation}
so $A_1$ is flat away from $\Sigma$ but has non-trivial holonomy around it; $V_m(\Sigma)$ inserts a thin tube of $\mathbb{Z}_N$ flux along $\Sigma$. Writing $\gamma = \partial D$, the Wilson line then measures
\begin{equation} \label{eq:holonomy}
  \oint_\gamma A_1 = \int_D dA_1 = -\frac{2\pi m}{N}\, \#(D \cap \Sigma)
  = -\frac{2\pi m}{N}\, \mathrm{Link}(\gamma,\Sigma) ,
\end{equation}
where $\#(D\cap\Sigma)$ is the signed intersection number. This reproduces \eqref{eq:WV-link}, with the overall sign fixed by the orientation convention.

In the language of Section~\ref{ssec:higher_form}, the surfaces $V_m$ are codimension-2 topological operators generating a $\mathbb{Z}_N^{(1)}$ 1-form symmetry, under which the Wilson line $W_q$ carries charge $q \bmod N$. Indeed, by flatness of $A_1$, $\Sigma$ can be freely deformed without changing any correlator as long as it does not cross $\gamma$. The linking number changes (by $\pm 1$) only when $\Sigma$ is swept through $\gamma$, and the correlator then picks up a phase $e^{\pm 2\pi i q m/N}$. In particular, take $\Sigma = S^2$ to be a small sphere linking $\gamma$ once. It cannot be shrunk to a point without crossing $\gamma$, and doing so gives, inside correlation functions,
\begin{equation}
  V_m(S^2)\, W_q(\gamma) = e^{2\pi i q m/N}\, W_q(\gamma) .
  \label{eq:charge-action}
\end{equation}
The roles can be exchanged: inserting $W_q(\gamma)$ and integrating out $A_1$ gives $\frac{N}{2\pi}\, dB_2 \propto q\, \delta_\gamma$, so $B_2$ acquires a $\mathbb{Z}_N$ holonomy on surfaces linking $\gamma$. Hence the lines $W_q$ are codimension-3 topological operators generating a $\mathbb{Z}_N^{(2)}$ 2-form symmetry, under which $V_m$ carries charge $m \bmod N$. As for the electric and magnetic symmetries of Maxwell theory, the two symmetries have a mixed 't~Hooft anomaly. The physical picture is Aharonov-Bohm \cite{AharonovBohm}: a charge-$q$ particle transported around a vortex string carrying $m$ units of $\Z_N$ flux picks up the phase $e^{2\pi i qm/N}$.

Following Section~\ref{sec:gaugebreaking}, the 1-form symmetry is broken by dynamical particles of charge $q\neq0$ mod $N$, on which Wilson lines can end, while the 2-form symmetry is broken by dynamical vortex strings, on which the surfaces $V_m$ can end. In quantum gravity, the $\Z_N$ gauge symmetry itself is perfectly allowed, but its associated higher-form symmetries must be broken or gauged. This requires that both charged particles and strings carrying $\Z_N$ flux be present in the spectrum, a discrete version of the completeness hypothesis \cite{Banks:2010zn}. For more details on discrete gauge theories and their generalized symmetries, see e.g.~\cite{BottiniR,Gomes:2023ahz,Brennan:2023mmt}. 

\section{No Global Symmetries in Quantum Gravity}
\label{sec:No_Global_Symmetries}

 In this section we aim to describe the idea that a consistent theory of quantum gravity cannot admit exact global symmetries. We begin by carefully formulating the conjecture and clarifying what is meant by a global symmetry in a gravitational context in Section \ref{sec:statement}. We then review the arguments supporting the conjecture in increasing order of rigor: starting with heuristic considerations based on black holes, and culminating in more precise results in perturbative string theory and in AdS/CFT. We finish by briefly commenting on how this conjecture is related to other Swampland conjectures in Section \ref{sec:relation}.

\subsection{Statement}\label{sec:statement}
The idea that consistent theories of quantum gravity do not admit exact global symmetries is both old and influential. Early precursor arguments \cite{MISNER1957525} already suggested that conserved charges in a gravitational theory should be encoded in dynamical fields or spacetime geometry, rather than exist merely as abstract globally conserved labels. Later, this expectation was explored and sharpened using further ideas from the Hawking evaporation of black holes \cite{Hawking:1975vcx}, and string theory \cite{Polchinski2003Monopoles}. Today, the \emph{No Global Symmetries} Conjecture occupies a central place in the Swampland Program: it has been formulated precisely and can even be proven in certain settings:

\begin{symmetrybox}{dustyPlum}
    \textbf{The No Global Symmetries Conjecture:}\;
In any consistent theory of quantum gravity, there are no exact global symmetries. Every symmetry is either gauged or broken.
\end{symmetrybox}

\noindent The basic philosophy behind the No Global Symmetries Conjecture is the following. Consider a theory that we believe admits a consistent ultraviolet completion including quantum gravity, like, for instance, a supergravity theory arising as the low-energy limit of string theory, or even our own universe. Such theories are well described at long distances by an effective field theory coupled to gravity. 
The conjecture asserts that in any such theory, \hl{exact} global symmetries cannot exist. From an effective field theory perspective, global symmetries may appear as \hl{accidental or approximate symmetries in the infrared.} 
However, once gravity is included and one probes sufficiently high energies, these symmetries must either be gauged or explicitly broken. 
In other words, what looks like a perfectly good global symmetry in the IR cannot remain exact in a consistent UV completion to quantum gravity.

Importantly, in the case of a broken symmetry, the conjecture places no a priori restriction on how weak the breaking is at low energies. Approximate global symmetries are therefore perfectly allowed, and in fact ubiquitous. Baryon and lepton number in the Standard Model are accidental symmetries of the renormalizable Lagrangian, violated only by higher-dimensional operators suppressed by a high scale $\Lambda$ \cite{PhysRevLett.43.1566} (and non-perturbatively by electroweak sphalerons \cite{PhysRevLett.37.8}). Axion shift symmetries are typically broken only by instanton effects of order $e^{-S_{\rm inst}}$ (see Section \ref{sec:spt}). Schematically, an approximate symmetry is one whose breaking in the IR is controlled by a small parameter, such as $(E/\Lambda)^n$ or $e^{-S}$, where $\Lambda$ is the scale at which charged objects or symmetry-violating operators appear. The sharp question is therefore quantitative: how weak can the breaking be, and how high can $\Lambda$ be compared to the Planck scale? Refinements of this type underlie several other Swampland conjectures (Section~\ref{sec:relation}), and they are also what makes the dynamical side of the Cobordism Conjecture subtle (Section~\ref{sec:charact}). 

There are several heuristic reasons to expect exact global symmetries to be incompatible with quantum gravity. One is their apparent tension with consistent black hole evaporation, which we review in detail in Section \ref{sec:heuristic}. More broadly, their absence is closely tied to the principle of background independence introduced in Section~\ref{sec:intro}. Recall that background independence is the idea that there should be no preferred choice of background in quantum gravity. The expectation is that different spacetime backgrounds should correspond to dynamical configurations of the same underlying theory, and there should in principle exist physical processes that interpolate between them. Exact global charges would obstruct this picture by assigning conserved labels to backgrounds that are neither gauge redundancies nor asymptotic observables. They would therefore divide the space of backgrounds into superselection sectors that cannot be connected by any physical process; see also Section \ref{sssec:BabyUniv}.

Over time, these heuristic arguments have been supplemented by more systematic analyses in concrete frameworks such as perturbative string theory \cite{Banks:1988yz} and the AdS/CFT correspondence \cite{Harlow:2018jwu,Harlow:2018tng}. While a completely general theorem is still lacking, no counterexamples are known, and in a number of well-defined settings the absence of exact global symmetries can be established with considerable precision.

Before we move on to the arguments in favor of this conjecture, it is worth noting that, even outside of quantum gravity, the absence of global symmetries has also played a useful role as a guiding principle in quantum field theory model building. In particular, it is often considered preferable for global symmetries to arise accidentally (as approximate symmetries) rather than to postulate nearly exact global symmetries without an underlying explanation (see \cite{Reece:2023TASI} and references therein). 

\subsection{The Black Hole Argument}\label{sec:heuristic}
We now summarize the black-hole-based arguments for the No Global Symmetries Conjecture. Although heuristic, these arguments capture much of the underlying intuition and motivate the more precise arguments we will review in the following sections.

\begin{figure}
    \centering
    \includegraphics[width=1\linewidth]{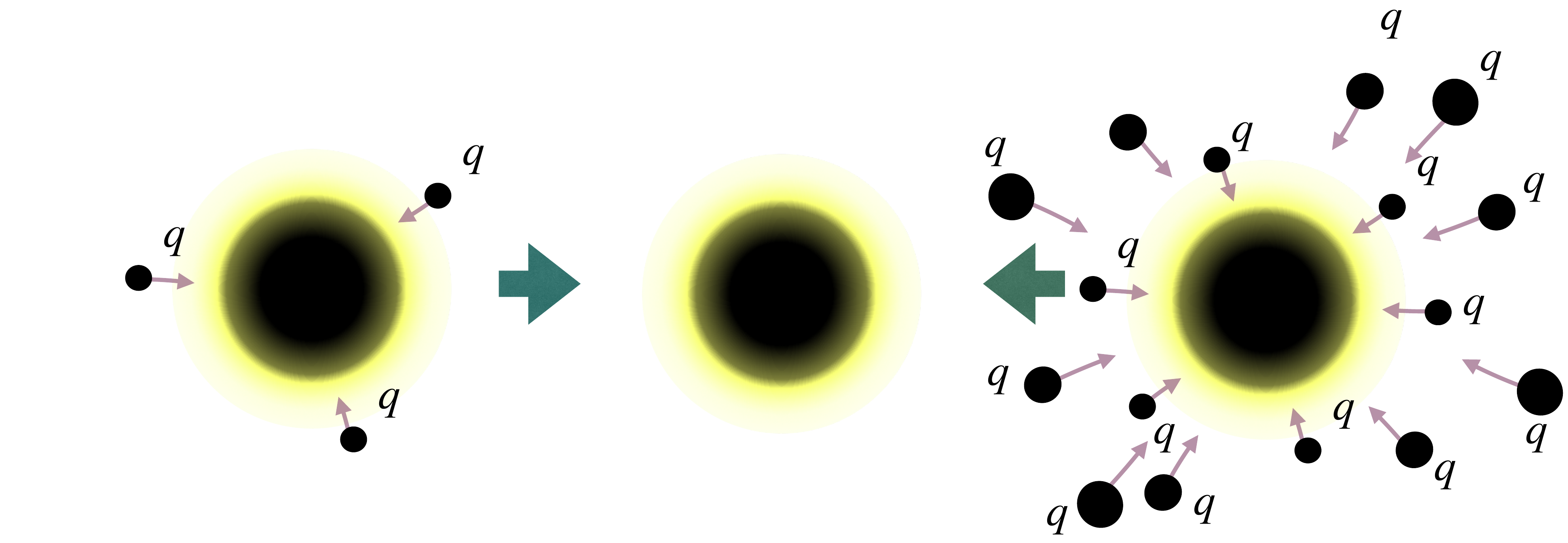}
    \caption{A black hole's horizon is completely insensitive to global charge (no-hair theorem). One can imagine constructing many configurations of the same black hole --as described from its asymptotic charges-- with different global charges. If semiclassical evaporation can bring black holes carrying arbitrarily many distinct global charges to a common final geometry, these global charges label an unbounded number of different states belonging to the same black hole macrostate. This number of sates may, in principle, exceed $\operatorname{area/4G_N}$ and thus contradict the area law for black hole entropy \eqref{eq:BHentropy}.}
    \label{fig:basicblackholeargument}
\end{figure}

The basic observation is that global symmetries are not gauge symmetries and therefore do not admit charges that are measurable at infinity. By contrast, black hole solutions are believed to be completely characterized by their asymptotic charges, according to the no-hair theorems, see e.g.~\cite{Heusler1996} (but note the caveats in \hyperlink{higherdBH}{footnote 5}). Consequently, any global charge carried by the matter that collapses to form a black hole is not reflected among the parameters specifying the resulting exterior black hole geometry. This leads to a tension once the black hole evaporates \cite{Hawking:1974rv,Hawking:1975vcx}. Since   the Hawking radiation is insensitive to the global charge of the black hole, the complete evaporation process naively appears to violate global charge conservation.\footnote{This is too quick as we are extrapolating the evaporation beyond semiclassical control. } For continuous global symmetries, a sharper tension arises within the semiclassical regime from the finiteness of the Bekenstein-Hawking entropy \cite{bekensetin2,Hawking:1974rv}:
\begin{equation}\label{eq:BHentropy}
S_{\rm BH} = \frac{A}{4 G_N},
\end{equation}
where $A$ is the area of the event horizon. If the horizon geometry is insensitive to the amount of global charge hidden behind it, then arbitrarily many states with different global charge could evaporate to the same black hole macrostate, in stark contradiction with \eqref{eq:BHentropy}. This is shown schematically in Figure \ref{fig:basicblackholeargument}. Inspired by the Bekenstein bound \cite{bekensetinbound}, the development of the holographic principle \cite{tHooft:1993dmi,Susskind:1994vu} led to the \hl{Covariant Entropy Bound} \cite{Bousso1999,Bousso2002}, which allowed one to formulate a more precise argument against global symmetries in quantum gravity. Indeed, in \cite{Banks:2010zn}, Banks and Seiberg examined the status of, and presented arguments in support of, the following three conjectures:
\begin{enumerate}
    \item \textbf{No Global Symmetries in quantum gravity}: 
    \\[0.5em]
    All symmetries must be gauged or broken in a theory of quantum gravity.
    \item \textbf{Compactness of the gauge group}:
    \\[0.5em]
    All continuous gauge symmetries must have compact gauge group in a theory of quantum gravity.   
    \item \textbf{Completeness}:
    \\[0.5em]
    The entire lattice of charged states must be populated in a theory of quantum gravity.
\end{enumerate}
They argued these three conjectures follow from the consistency of black hole evaporation with the bounds mentioned above, and furthermore are not logically independent. In what follows, we revisit their arguments and summarize the main conclusions that can be drawn from them. Our primary focus will be on the first of the three conjectures, with a presentation that has been largely inspired by \cite{Reece:2023TASI}.

\paragraph{The No Global Symmetries argument.}There are two main pillars that lie at the heart of the argument. The first one is: 
\begin{symmetrybox}{dustyPlum}
    \textbf{1. The No Hair Theorem:} 
    Black hole solutions in GR are determined by a small number of parameters that are related to asymptotic symmetries; typically, mass $M$, angular momenta $J^i$, and charges related to gauge fields $Q_e,\,Q_m,\, \ldots$\hypertarget{higherdBH}{\footnote{In $d\geq5$ dimensions, black objects with different horizon topologies can share the same $\{M,\, J^i, Q_e, Q_m\}$ (see e.g.~\cite{Emparan:2001wn}), so these asymptotic quantities no longer single out a unique solution. Still, one can regard the horizon topology as additional discrete data, so the spirit of the no-hair theorem survives.}}
\end{symmetrybox}
\noindent
In particular, solutions are insensitive to charges that are not associated with a gauge bundle that can be read in the asymptotic conditions of the solution, see, e.g.~\cite{Heusler1996} for a self-contained review. 

The second pillar is the idea that any region of spacetime cannot carry more entropy than a black hole of the same size:
\begin{equation}
    S(A) \leq S_{\rm BH}= \frac{A}{4 G_N}\,,\label{eq:3.2}
\end{equation} where $S(A)$ is the entropy of a region in spacetime enclosed by a sphere of area $A$ \cite{tHooft:1993dmi,Susskind:1994vu}. For weakly gravitating systems, this is implied by Bekenstein's bound \cite{bekensetinbound}, for which a precise formulation and proof were given in \cite{Casini:2008cr}. The bound in (\ref{eq:3.2}) was adapted to strongly gravitating or cosmological settings, by the Covariant Entropy Bound \cite{Bousso1999,Bousso2002}: 

\begin{symmetrybox}{dustyPlum}
    \textbf{2. The Covariant Entropy Bound (CEB):}

    Let $\partial Y$ be a spacelike codimension-two surface and let $L$ be a
    \emph{light-sheet} of $\partial Y$.\footnote{A null hypersurface generated by the orthogonal
    null geodesics along which the expansion is nowhere positive.} Then the entropy on $L$ obeys
    \begin{equation}
        S[L] \;\leq\; \frac{A(\partial Y)}{4\, G_N}\,.
    \end{equation}
    Taking $\partial Y$ to be a cross-section of the black hole's horizon, the number
    of distinguishable internal states of the black hole is therefore at most
    $\exp\!\left[A(\partial Y)/4G_N\right]=e^{S_{BH}}$.
\end{symmetrybox}
\noindent
Again, the maximum entropy of a region $Y$, with boundary $\partial Y$ is given by the entropy of the black hole that would have as its horizon $\mathcal{H}=\partial Y$. 

Let us see how these two properties of black holes can be used to argue for the three conjectures above. 

If we had a global symmetry, we could build a black hole transforming in a non-trivial way under the symmetry by collapsing identical particles transforming  in a representation $\mathbf{R}$ of the symmetry group. The resulting black hole would transform in a representation $\Gamma\subset \mathbf{R}\otimes\mathbf{R}\otimes\mathbf{R}\otimes \cdots $. 
According to the no-hair theorem, the resulting black hole geometry will have no imprint of this non-trivial transformation property under the global symmetry. If we follow Hawking's computation for the evaporation, the black hole will evaporate semiclassically until it reaches a size, $r_{\mathcal{H}}\geq \ell_{\rm P}$, at which point we can no longer trust the semiclassical computation for the evaporation,
\begin{equation}
    r_{\mathcal{H}}= X \ell_{\rm P}\sim \mathcal{O}(\ell_{\rm P}) \quad\Longrightarrow \quad S_{\mathcal{H}}\coloneqq\frac{A_{\mathcal{H}}}{4\, G_N}\sim \pi X^2\sim\mathcal{O}\left((\ell_{\rm P})^0\right),
\end{equation}
with $X>0$ a finite numerical factor,\footnote{It is sometimes assumed that $X$ is order 1 such that the black holes are Planck-sized remnants. For non-supersymmetric black holes, the naive thermodynamic picture may drastically change near extremality \cite{Rakic:2023vhv, Iliesiu:2022onk}; see \cite{Turiaci:2023wrh} for a lightning introduction and further references. In spite of these subtleties, for violating the Covariant Entropy Bound, even in the Schwarzschild case, it suffices to obtain infinitely many black hole states below a finite energy scale, which we can separate from the \textit{breakdown} scale at which these derivations in the extremal case become relevant, $E_{\rm break}$. } parametrizing the scale at which the semiclassical computation of Hawking breaks down. In this fashion, we can build arbitrarily many different black holes with an exterior geometry corresponding to a black hole of radius $r_{\mathcal{H}}= X \ell_{\rm P}$. The tension with the Covariant Entropy Bound is then immediate: since all such black holes have the same exterior geometry, their entropy is bounded by $S_{\mathcal H}\sim\pi X^2$ which is finite by construction. Therefore, the number of independent black hole states is bounded by $\dim\mathcal H_{\rm BH}\lesssim e^{\pi X^2}$. An exact global symmetry, however, would allow the collapsing matter to carry arbitrarily many distinct global charges, requiring an arbitrarily large number of distinguishable black hole states of size $r_\mathcal{H}$ and therefore violating this finite entropy bound.

It is important to note that the argument above does not work for charges associated to gauge symmetries. On the one hand, gauge charge is part of the allowed black-hole hair, since it can be measured by an observer at asymptotic infinity using Gauss's law, meaning that the exterior geometry ---and therefore the event horizon--- \emph{is} sensitive to the amount of this charge. In particular, an extremality bound $M_{\rm BH} \geq M_{\rm ext}(|Q|)$ typically prevents the formation of the many black holes of size $r_{\mathcal{H}} \sim \ell_{\rm P}$ with arbitrarily large $|Q|$ which are required by the argument. On the other hand, as opposed to what happens with global charge, the spectrum of Hawking radiation is sensitive to gauge charge, favouring the emission of particles that carry away such charge. Thus, a black hole that carries gauge charge can in principle discharge simply by emission of Hawking quanta.

\paragraph{Compactness of Gauge Group.}The same line of thought allows one to argue against non-compact gauge groups.\footnote{The argument presented here should be supplemented by charge completeness, which shows that the three conjectures that we are discussing here are in fact deeply interconnected.} Suppose, for instance, that a single U$(1)$ theory contains two abelian charges whose elementary charge units are incommensurate,
\begin{equation}
\frac{e_2}{e_1}\notin \mathbb{Q}.
\end{equation}
For concreteness one may take $e_2=\sqrt{2} \,e_1$, although the particular irrational number is immaterial. Black holes assembled from particles carrying these two charges can then have total charge
\begin{equation}
Q_{\rm BH}=n_1 e_1+n_2 e_2\,,
\qquad (n_1,n_2)\in\mathbb Z^2 .
\end{equation}
Because $e_2/e_1$ is irrational, the set of such charges is dense in $\mathbb R$. In particular, any finite interval of charge contains infinitely many distinct allowed values of $Q_{\rm BH}$. Hence, even restricting to black holes whose mass, horizon area, and charge remain within fixed finite ranges, one obtains arbitrarily many black holes that could subsequently Hawking-radiate into a black hole state with $r_{\mathcal{H}}=X\ell_P$. This is clearly in tension with the Covariant Entropy Bound, which assigns only a finite-dimensional Hilbert space to bounded regions of spacetime. Thus, a dense, non-quantized charge spectrum is incompatible with finite black hole entropy. This motivates the requirement that abelian gauge groups in quantum gravity be compact, so that electric charges lie on a discrete lattice rather than densely populating $\mathbb R$. We thus see that the same argument that excludes global symmetries in quantum gravity excludes non-compact gauge groups.

\paragraph{Limitations.}Of course, these arguments are not air-tight. A first subtlety concerns whether the black hole states used in the counting are stable, which depends on the lightest particle carrying global charge. If this particle has mass $m>0$, a black hole with $M<m\,|Q|$ cannot shed its charge, by energy conservation alone. Such states are kinematically protected, independently of the dynamics: they eventually settle into a static configuration, a light-sheet from a surface enclosing them captures all their distinguishable states, and the Covariant Entropy Bound applies directly. If instead the charged particle is \emph{massless}, discharge is always kinematically allowed. Semiclassically it does not happen, since the exterior is insensitive to the global charge and the charge merely random-walks. However, the semiclassical analysis breaks down precisely when the black hole becomes small, $r_{\mathcal H}\sim X\ell_{\rm P}$ with $X\sim\mathcal{O}(1)$, so quantum gravity effects could discharge it. There is then no guaranteed static end state, and the bound must be applied to an evaporating horizon, whose light-sheets need not contain the collapsed matter. The argument then requires assuming a statistical interpretation of $S_{\rm BH}$ rather than deriving it from the Covariant Entropy Bound.

A second limitation is that the argument relied on the possibility of building arbitrarily many different states corresponding to the same black hole in order to reach a tension with the Covariant Entropy Bound. This makes it difficult to apply to finite symmetry groups. Indeed, for a finite symmetry group $G$ the charge of a black hole can only take finitely many values, one for each irreducible representation of $G$. The number of charge sectors is then bounded, and can easily be accommodated within finite entropy: the counting argument above does not go through for discrete symmetries. It would depend on quantities such as $X$ in $S_{\mathcal{H}}\sim\pi X^2$, that are related to black hole evaporation processes in regimes in which they are not under full semiclassical control. Moreover, discrete \emph{gauge} charge produces no long-range classical field, so the classical no-hair theorem does not distinguish it from discrete global charge; it can nonetheless be measured at infinity via Aharonov--Bohm interference with cosmic strings, making it a form of quantum hair \cite{Krauss:1988zc,Coleman:1991ku}, albeit a more subtle one. Arguments against discrete global symmetries therefore require different tools, such as the Harlow--Ooguri argument of Section~\ref{sssec:harlow-ooguri} or those collected in \cite{Banks:2010zn,Hellerman:2006tx} for discrete symmetries being gauged in string theory.

\paragraph{Completeness of Charge Spectrum.}Finally, we turn to the third conjecture: completeness of the charge spectrum. Note that in the argument for compactness of the gauge group we implicitly assumed that the completeness hypothesis holds, namely that all charges allowed by the gauge structure are actually realized by physical states. The expectation that every allowed charge sector is populated is a statement that is intrinsically tied to quantum gravity and black holes. Indeed, it needs not to hold in ordinary quantum field theory, where one can take the mass of a charged particle to infinity and effectively decouple that state from the low-energy spectrum while keeping the gauge theory otherwise well defined. In a theory of quantum gravity, however, this decoupling argument eventually breaks down. As the mass of a charged object is increased, its gravitational backreaction becomes increasingly important, and beyond some point the appropriate description of the state is no longer that of a particle excitation, but rather that of a charged black hole. In this sense, gravity prevents one from simply removing an entire allowed charge sector by sending the mass of its lightest state to infinity: sufficiently massive charged configurations are themselves gravitational states. For more details on this, we refer the reader to a beautiful discussion in this direction by Polchinski \cite{Polchinski2003Monopoles}, where, inspired by perturbative string theory examples and consistency of the Dirac quantization condition, he argues in favour of the completeness hypothesis. In more recent years, more precise arguments have been put forward to show that completeness of charge spectrum follows from No Global (Non-Invertible) Symmetries \cite{Heidenreich:2021xpr,McNamara:2021cuo}.

\paragraph{Final comments.}
The main lesson is that our current understanding of black hole physics and black hole thermodynamics provides evidence for the three conjectures stated above. The heuristic arguments reviewed above also admit more concrete realizations in perturbative string theory, where related conclusions have been known for a long time, see e.g.~\cite{Banks:1988yz}. We review some of these examples in the next section.

These arguments excluding continuous $0$-form global symmetries can be extended to continuous higher-form global symmetries. A useful way to see this is by compactification. If the theory is compactified on a torus, for instance
\begin{equation*}
\mathbb{T}^n=S^1\times\cdots\times S^1,
\end{equation*}
then a higher-form symmetry can give rise, upon wrapping the corresponding symmetry operators on suitable cycles of the compactification manifold, to an ordinary $0$-form symmetry in the lower-dimensional theory. The absence of exact continuous $0$-form global symmetries in quantum gravity then implies the absence of the corresponding exact continuous higher-form global symmetry in the higher-dimensional theory. In this sense, the black-hole arguments discussed above extend naturally to generalized global symmetries.

One could also try to formulate the same argument directly in the higher-dimensional theory using charged black branes, whose entropy is extensive along the brane worldvolume but has a finite entropy density, so that the same tension with an arbitrarily large number of conserved higher-form charge sectors can be phrased per unit brane volume.

\subsection{Evidence} In the previous section, we described a heuristic argument in favour of there being no global symmetries in quantum gravity. In this section, we present more robust evidence. We will first go through the proof that there are no global symmetries in string perturbation theory. Then, we will present an argument against global symmetries in AdS/CFT. For both, we will discuss the extent and limitations of these proofs. We note in passing that although we have not discussed them in these lectures, the conjecture can be proven explicitly for \textit{non-invertible} symmetries \cite{Heckman:2024obe} in string theory.
\subsubsection{In String Perturbation Theory}\label{sec:spt}

We will now review the argument that perturbative string theory cannot give rise to exact global symmetries in target space. This argument was originally presented in \cite{Banks:1988yz} and is also discussed in Section 18.1 of Polchinski's \textit{String Theory} \cite{Polchinski:1998rr}. The derivation relies on a few basic facts about string perturbation theory, which we briefly summarize below. For a more detailed treatment, we refer the reader to \cite{Polchinski:1998rr}.
 
\paragraph{Perturbative string theory in a nutshell.} One starts from a two-dimensional theory describing the embedding of the string worldsheet into spacetime, with the worldsheet metric serving as the gauge field for $\operatorname{Diff}\ltimes\operatorname{Weyl}$ invariance. After gauge fixing and quantization, consistency of Weyl invariance at the quantum level requires the worldsheet theory to be a two-dimensional conformal field theory (CFT) of the appropriate central charge. More precisely, the total central charge, including the ghost sector introduced by gauge fixing, must vanish; equivalently, the matter CFT has central charge $c=26$ in the bosonic string and $c=15$ in the RNS superstring.

Suppressing the ghost factors, physical string states are represented by vertex operators. For closed bosonic strings, these are local primaries $\mathcal O$ of left- and right-moving conformal weights $(h,\tilde h)=(1,1)$, so that their integrated form
\begin{equation}
V_{\mathcal O}=\int_{\Sigma} d^2z\,\mathcal O(z,\bar z)\,,
\end{equation}
is invariant under the worldsheet gauge symmetries. Massless vertex operators correspond to spacetime fields. In backgrounds admitting a geometric description, turning on their spacetime-dependent couplings deforms the worldsheet theory into a nonlinear sigma model whose couplings encode the target-space background. Requiring quantum Weyl invariance of the deformed theory, equivalently the vanishing of the worldsheet $\beta$-functions, yields the spacetime equations of motion.

The string path integral also includes a sum over worldsheet topologies. The Euler-characteristic coupling
\begin{equation}
\frac{\lambda}{4\pi}\int_\Sigma d^2\sigma\,\sqrt g\,R_g=\lambda\,\chi(\Sigma)\,,
\end{equation}
weights a worldsheet of Euler characteristic $\chi$ by a power of the string coupling $g_s=e^\lambda$, thereby generating the genus expansion. Perturbative string theory is applicable when this expansion is under control, which in a conventional weakly coupled background requires $g_s\ll1$.

When the worldsheet CFT arises from a nonlinear sigma model, there is in addition a useful spacetime derivative expansion controlled schematically by $\alpha'\mathcal R$, where $\mathcal R$ denotes the characteristic curvature scale of the target-space background and $\alpha'$ is inversely related to the string tension. In the regime $\alpha'\mathcal R\ll1$, the spacetime equations extracted from the worldsheet $\beta$-functions reduce at leading order to those of (super-)gravity, with higher-derivative corrections suppressed by powers of $\alpha'\mathcal R$. This low-curvature condition is therefore required for a conventional supergravity description, but not for perturbative string theory itself. 

Now that we have clarified what we mean by \emph{string perturbation theory}, we turn to the question of whether its target-space backgrounds admit global symmetries. The argument below is formulated entirely within the perturbative worldsheet description and therefore constrains symmetries acting on the perturbative string spectrum. By itself, it does not exclude the logical possibility of a symmetry that exists only at a genuinely strong-coupling point, or that acts only on intrinsically nonperturbative states. Such possibilities must be addressed by additional, nonperturbative arguments; when a strong-coupling regime admits a dual weakly coupled description, however, the perturbative argument may be applied in that duality frame.

Naively, one would expect that any global symmetry of the spacetime theory comes from a global symmetry of the worldsheet theory. We will initially work under that assumption and postpone the assessment of its validity to a later point in this section.

\paragraph{The Banks-Dixon argument.} Our first goal will be to show that \emph{continuous} global symmetries of the worldsheet theory do not give rise to global symmetries in target space; they produce gauge redundancies instead. The proof originally appeared in \cite{Banks:1988yz} and is also covered in Section 18.1 of Polchinski's \textit{String Theory} book \cite{Polchinski:1998rr}. The argument is simplest in bosonic string theory. We will consider a Euclidean worldsheet and use complex coordinates $(z,\Bar{z})$ in which the flat metric takes the form $ds^2 = dz d \Bar{z}$. By Noether's theorem, a continuous global symmetry of the worldsheet CFT will give rise to a conserved charge $Q$, which may be expressed in terms of the associated worldsheet current $j_\alpha$ as
\begin{equation}
    Q = \frac{1}{2\pi} \oint \star \mathbf{j} = \frac{1}{2\pi i} \oint (dz j_z - d\Bar{z} j_{\Bar{z}}) \,.
\end{equation}
To be a well-defined symmetry of the physical spectrum, $Q$ must be conformally invariant, which in turn requires $j_z$ and $j_{\Bar{z}}$ to transform as primaries of respective weights $(1,0)$ and  $(0,1)$. 
Then, recalling that for the worldsheet bosons $X^\mu(z,\bar z)$, which describe the embedding of the worldsheet into spacetime (cf.~Figure~\ref{fig:WStoTS}), $\partial X^\mu$ is primary of weights $(1,0)$ and $\Bar{\partial}X^\mu$ is primary of weights $(0,1)$, we may construct the vertex operators
\begin{equation}\label{eq:vertices_Banks-Dixon}
    j_z \Bar{\partial} X^\mu e^{ik \cdot X}\,, \quad \partial X^\mu j_{\Bar{z}} e^{ik \cdot X} \quad\text{with}\quad k^2=0\,,
\end{equation}
which couple to massless spacetime vectors $A_\mu$, and $\Tilde{A}_\mu$ respectively. Let's focus for now on the first of these vertices. The associated integrated vertex operator is
\begin{equation}
    V[A] \equiv \int d^2z\, j_z \Bar{\partial} X^\mu A_\mu(X)\,, \quad A_\mu(X) = \int d^D k\, \epsilon_\mu(k) e^{ik \cdot X}\,,
\end{equation}
where $D$ is the dimension of spacetime. We shall momentarily show that this operator is invariant under a gauge transformation $A \mapsto A + d\Lambda$. Indeed, letting 
\begin{equation}\label{eq:conservation_to_gauge_1}
\begin{split}
    V[A + d\Lambda] = \int d^2z\, j_z \Bar{\partial} X^\mu \int d^Dk\,(\epsilon_\mu(k) + ik_\mu \lambda(k))\, e^{ik \cdot X} \\
    = V[A] + \int d^2z \int d^Dk\, \lambda(k) j_z \Bar{\partial} e^{ik \cdot X} \;.
\end{split}
\end{equation}
A primary operator of weight $\tilde{h}=0$, like $j_z$, must be holomorphic;\footnote{For any primary operator $\mathcal{O}$ of weights $(h,\tilde{h})$, $$\braket{\partial\mathcal{O} | \partial\mathcal{O}} = \lVert L_{-1}\ket{\mathcal{O}} \rVert^2 = 2h\braket{\mathcal{O}|\mathcal{O}}\,, \quad \braket{\Bar{\partial}\mathcal{O} | \Bar{\partial}\mathcal{O}} = \lVert \tilde{L}_{-1}\ket{\mathcal{O}} \rVert^2 = 2\tilde{h}\braket{\mathcal{O}|\mathcal{O}}\;,$$ so in a unitary CFT $\Tilde{h}=0$ implies $\Bar{\partial}\mathcal{O} = 0$ and similarly for $h$, $\partial\mathcal{O}$.} hence $\Bar{\partial} j_z = 0$. Then, after integration by parts in the worldsheet coordinates,
\begin{equation}\label{eq:conservation_to_gauge_2}
    V[A + d\Lambda] - V[A] = - \int d^2z \int d^Dk\, \lambda(k) (\Bar{\partial} j_z) \, e^{ik \cdot X} = 0 \,,
\end{equation}
which proves the gauge invariance of $V[A]$. From a Hilbert space point of view, the transformation shifts the state by a BRST-exact term, which does not change its gauge-equivalence class. The same applies to the other vertex operator in (\ref{eq:vertices_Banks-Dixon}). We thus conclude that the worldsheet global symmetry generated by $Q$ translates into a \hl{gauge redundancy in spacetime}, and not into a global symmetry.

A similar argument may be formulated in superstring theory using superspace. The conserved charge associated to a continuous global symmetry on the worldsheet of a type I or type II string can be expressed in terms of a conserved superfield current $J, \tilde{J}$ as
\begin{equation}
    Q = \frac{1}{2\pi i} \oint(dzd\theta J - d\Bar{z} d\Bar{\theta} \Tilde{J}) \;.
\end{equation}
This charge must be invariant under superconformal transformations, which forces $J$ and $\Tilde{J}$ to transform as $(\frac{1}{2},0)$ and $(0,\frac{1}{2})$ tensor superfields, respectively.\footnote{A tensor superfield of weight $(h,\Tilde{h})$ transforms under a general superconformal transformation $\mathbf{z} = (z,\theta) \mapsto \mathbf{z'} = (z',\theta')$ as $(D_\theta\theta')^{2h}(D_{\Bar{\theta}}\Bar{\theta}')^{2\Tilde{h}} \mathbf{\phi'(z',\Bar{z}')} = \mathbf{\phi(z,\Bar{z})}\;.$ The super-Jacobian, meanwhile, transforms as $dz' d\theta' = dz d\theta\, D_\theta\theta'\,.$ } These superfields can then be respectively combined with the anticommuting worldsheet fields $\Tilde{\psi}^\mu$ and $\psi^\mu$, which have complementary superconformal weights, to give gauge boson vertex operators.\footnote{Superspace vertex operators must be tensor superfields of weight $(\frac{1}{2}, \frac{1}{2})$ in order to give a superconformally invariant integrated operator. This follows from the transformation law of the super-Jacobian.} Hence, once more, the global symmetry of the worldsheet manifests as a gauge redundancy in spacetime. The same is true for the heterotic string, by application of the bosonic argument on one side of the string and the supersymmetric argument on the other.

Let us now reverse the argument to show that the gauge redundancy associated with any such massless spacetime vector gives rise to a continuous global symmetry on the worldsheet, focusing on bosonic string theory. The subspace of massless vector states is spanned by\footnote{Here we assume that the worldsheet CFT contains a flat $D$-dimensional spacetime factor described by free bosons $X^\mu$, together with an otherwise arbitrary internal CFT.
}
\begin{equation}\label{eq:states_massless_vector}
    \alpha_{-1}^\mu \ket{k}_L \otimes \ket{\chi}_R \quad\text{and}\quad \ket{\chi}_L \otimes \Tilde{\alpha}_{-1}^\mu\ket{k}_R
\end{equation}
where $k^2=0$ and $\ket{\chi}_{L,R}$ have conformal weight 1 and no spacetime indices. Let $j(z),\,\bar j(\bar z)$ be the image under the state-operator correspondence of $\ket{\chi}_{L,R}$; then, the states in (\ref{eq:states_massless_vector}) will correspond to the operators
\begin{equation}
    j(z)\Bar{\partial}X^\mu e^{ik \cdot X} \quad\text{and}\quad \partial X^\mu \Tilde{j}(\Bar{z}) e^{ik \cdot X} \,.
\end{equation}
Then, turning (\ref{eq:conservation_to_gauge_1})--(\ref{eq:conservation_to_gauge_2}) around, the gauge redundancy associated to these modes implies that $j(z)$ and $\Tilde{j}(\Bar{z})$ are holomorphic/antiholomorphic, which in turn implies that $j_\alpha = (j(z), \Tilde{j}(\Bar{z}))$ is conserved:
\begin{equation}
    \partial^\alpha j_\alpha = 2(\Bar{\partial} j + \partial \Tilde{j}) = 0 \;.
\end{equation}

We stress that the argument, as presented here, applies to \hl{geometric backgrounds}, for which the worldsheet CFT is a nonlinear sigma model: this provides a Lagrangian description, weakly coupled in the $\alpha'$ expansion, in which the symmetry acts on the worldsheet fields and Noether's theorem can be applied. For more exotic worldsheet theories, such as those describing non-geometric compactifications, it is not known in general that every continuous symmetry of an abstract two-dimensional CFT comes with a conserved Noether current.

\paragraph{Briefly on open strings.}The Banks--Dixon argument is typically formulated for closed string worldsheets, and an analogous argument applies to open strings in the language of boundary conformal field theory (BCFT). Symmetries descending from bulk worldsheet currents preserved by the boundary conditions act on both sectors and are gauged by the corresponding closed-string vectors. Symmetries acting only on the open-string sector, such as those rotating Chan--Paton labels, are instead generated by boundary operators. These can be gauged by dressing the generator with $\partial_t X^\mu$, where $\partial_t$ is the derivative tangent to the worldsheet boundary, which gives the vertex operator of a massless open-string vector that lives on the worldvolume of a D-brane.

Symmetries of the open sector can become approximately global in the low-energy limit where closed strings decouple; an example is the $\mathrm{SU}(4)$ R-symmetry in the decoupling limit of the D3-brane solution. The relationship between the symmetries of the closed and the open-string sector is best understood in holographic setups and will essentially be the topic of the next section. For a more complete discussion of the role of open strings and D-branes, see Section 2.2.2 of \cite{Burgess:2008ri}.

\paragraph{Caveats?} Earlier we took for granted that any global symmetry of the spacetime theory is visible on the worldsheet. This is not always the case and in fact an exception was already identified in the original paper by Banks and Dixon \cite{Banks:1988yz}: the continuous shift symmetry associated with a massless axion-like scalar field $a$,
\begin{equation}
a \mapsto a + \text{const.} ,
\end{equation}
Axionic shift symmetries give rise to a vanishing Noether charge on the worldsheet, so they do not have a two-dimensional realization \cite{Banks:1988yz}. One would then be tempted to conclude that such symmetries invalidate the hypothesis that quantum gravity has no global symmetries. However, perturbation theory is \emph{not} a complete picture of string theory as a theory of quantum gravity. In general, continuous axion shift symmetries are broken by non-perturbative effects. A simple example is provided by the RR axion $C_0$ of type IIB string theory. At the level of perturbative type IIB supergravity, the axion enjoys a continuous shift symmetry $C_0\mapsto C_0+c$, as part of the SL$(2,\mathbb{R})$ symmetry of the classical theory. D$(-1)$-brane instantons break this continuous symmetry to the discrete shift $C_0\mapsto C_0+n$, $n\in\mathbb{Z}$, or, more generally, break SL$(2,\mathbb{R})$ to SL$(2,\mathbb{Z})$. Thus, the continuous axion shift is not a global symmetry of full string theory; it is only a symmetry of its perturbative description. This illustrates the dangers of purely perturbative considerations.

What, then, is left of the discrete shift symmetry $C_0\mapsto C_0+1$? This transformation is not an ordinary global symmetry of type IIB string theory. Rather, it is the $T$-generator of the SL$(2,\mathbb{Z})$ duality group, acting on the axiodilaton $\tau=C_0+i e^{-\phi}$ as $\tau\mapsto\tau+1$. In quantum gravity, such duality transformations are understood as gauge identifications: backgrounds related by an SL$(2,\mathbb{Z})$ transformation correspond to physically equivalent configurations, rather than distinct configurations related by a global symmetry. Thus, the discrete remnant of the perturbative axion shift is not a surviving global symmetry, but rather a discrete gauge symmetry contained in the duality group of type IIB string theory.

A second `exception' that is mentioned in the original Banks-Dixon paper \cite{Banks:1988yz} is the rigid Lorentz invariance of Minkowski spacetime (do not confuse with local Lorentz frame rotations in the vielbein formalism, which are gauge redundancies). In that case, although the worldsheet current as a whole is conserved, its two components are not separately holomorphic/antiholomorphic, so the vertex operators of (\ref{eq:vertices_Banks-Dixon}) are not BRST-invariant, i.e., correspond to unphysical states. In the more modern language of the post-AdS/CFT world (to be fully rolled out in the next section), rigid Lorentz transformations are the asymptotic symmetries associated to the `long-range gauge symmetry' \cite{Harlow:2018tng} of diffeomorphism invariance on asymptotically-flat spacetimes. Such symmetries are not considered global symmetries and are thus allowed in a quantum gravity theory ---another example is the asymptotic conformal symmetry of asymptotically-AdS spacetimes. Hence rigid Lorentz symmetry is not a counterexample to the conjecture either.

\paragraph{Discrete global symmetries.}
The argument above does not directly extend to discrete symmetries, since these have no associated conserved currents. Discrete global symmetries in quantum gravity were discussed in detail by Banks and Seiberg~\cite{Banks:2010zn}. Their analysis is not a general proof, but a collection of examples supporting the expectation that exact discrete global symmetries are also absent in quantum gravity.

As in the continuous case, one expects that a discrete symmetry that is exact in the full theory should be interpretable as a gauge symmetry in spacetime. There is a simple way to motivate this from the worldsheet~\cite{Hellerman:2006tx}. Suppose the sigma model has a discrete symmetry $G$, and compactify one additional direction on a circle of radius $R$. Consider a background in which the fields come back to themselves only up to the action of some $g \in G$ when transported around the circle, i.e.\ a compactification with $G$-monodromy. From the spacetime point of view, this is precisely a non-trivial background for a discrete gauge field: its field strength vanishes, but it has a non-trivial holonomy around the circle.

The natural worldsheet realization of this background is an orbifold by $g$ combined with a shift along the circle~\cite{Dixon:1986jc}. This amounts to gauging $\langle g \rangle$ on the worldsheet, so it is consistent if and only if the orbifold by $g$ alone is, i.e.\ if $\langle g \rangle \subset G$ is free of anomalies~\cite{Freed:1987qk,Cheng:2026abk,Delgado:2026qvy}. Otherwise, level matching and modular invariance fail. Take $G = \mathbb{Z}_N$ for concreteness. In the standard orbifold, momentum along the circle is quantized in units of $1/(NR)$. When $\langle g \rangle$ is anomalous, it was shown in~\cite{Hellerman:2006tx} that one can instead construct a modified CFT in which momentum is quantized in units of $1/(N^2 R)$, and that this restores level matching and modular invariance. Therefore, for any discrete global symmetry of the worldsheet theory, there is a consistent CFT describing the background with holonomy $g$ around the circle: the standard orbifold if $\langle g \rangle$ is anomaly-free, and the modified construction otherwise. In either case, the worldsheet global symmetry gives rise to a discrete \emph{gauge} symmetry in spacetime.

\subsubsection{In AdS/CFT}\label{sssec:harlow-ooguri}

One of the major pieces of evidence for the No Global Symmetries Conjecture is found in AdS/CFT theories of quantum gravity \cite{Maldacena:1997re,Witten:1998qj,Gubser:1998bc}. In this context, under reasonable assumptions about the field theory dual, Harlow and Ooguri have shown \cite{Harlow:2018tng} (see \cite{Harlow:2018jwu} for a letter version of this paper) that any global symmetry in the bulk of an AdS/CFT pair would lead to a \hl{contradiction in the boundary theory}. The objective of this section is to present a summary of the Harlow-Ooguri argument, its consequences and its limitations. At the heart of the argument are the concepts of entanglement wedge reconstruction and splittability, which we briefly review below.

\begin{figure}
    \centering
    \includegraphics[width=0.3\linewidth]{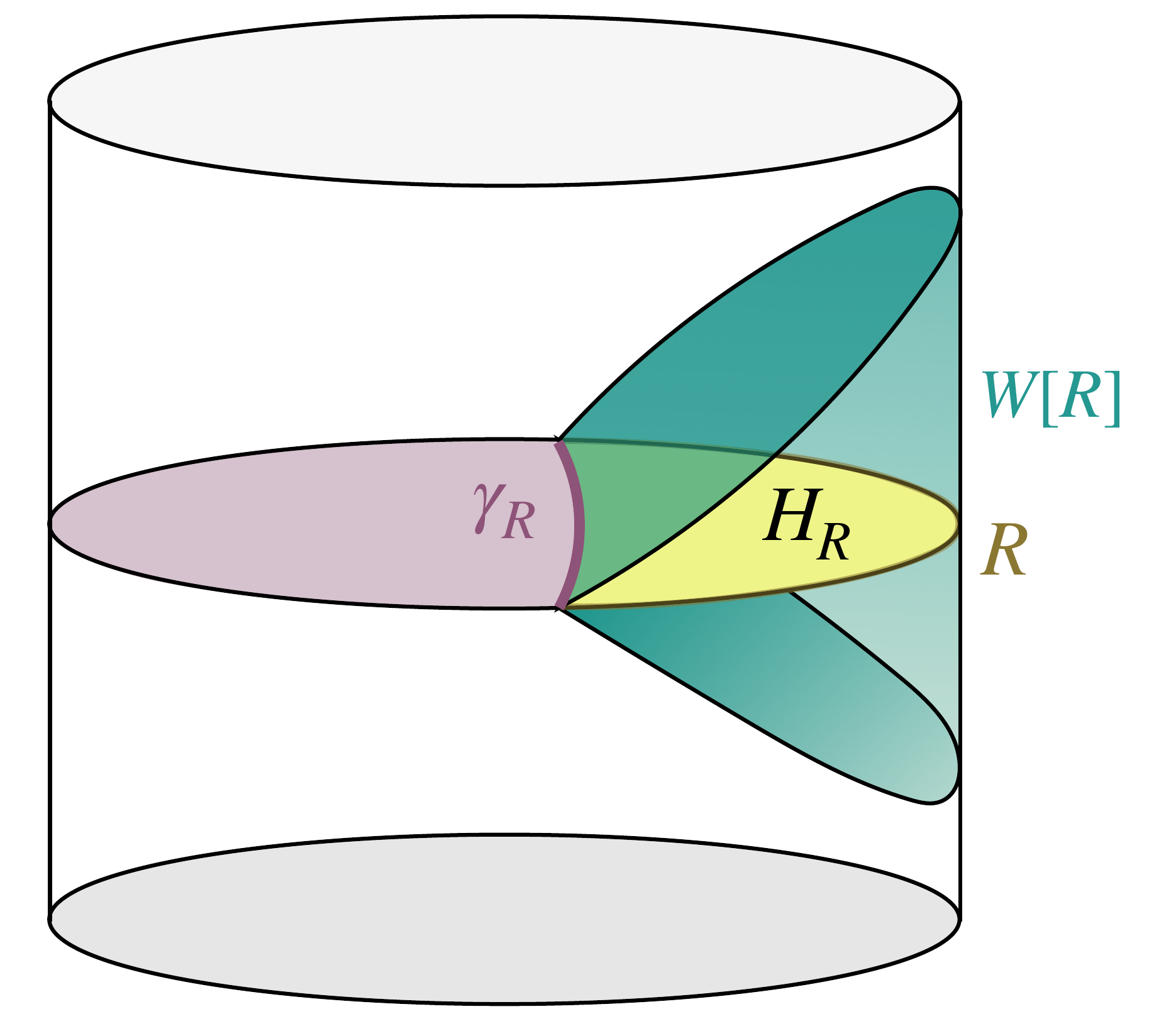}
    \caption{The Hubeny-Rangamani-Takayanagi surface $\gamma_R$ of the boundary subregion $R$ is homologous to $R$ via a bulk spatial surface $H_R$ (in yellow). The entanglement wedge of $R$, $W[R]$, is then defined as the domain of dependence of $H_R$. In the picture it would correspond to the spacetime region bounded by the two blue codimension-1 surfaces. Adapted from \cite{Harlow:2018tng}.}
    \label{fig:entanglement_wedge}
\end{figure}

\paragraph*{Entanglement wedge reconstruction.}In AdS/CFT, knowledge of the quantum state $\Psi$ of the boundary CFT entirely determines the quantum state of the bulk. Suppose instead that a boundary observer has access not to $\Psi$ but only to the density matrix $\rho_R$ that describes measurements in a certain spatial region, $R$. How much information about the quantum state of the bulk does this observer have? Giving a detailed answer to this question would be beyond the scope of these notes; here we will just state what the answer is and point out the main hint that led to it. The interested reader is referred to \cite{Harlow:TASI} or Section 9 of \cite{Witten:BH_lectures} for more information.

The answer was provided by a series of works that culminated in the proofs of \cite{Dong:2016eik, Harlow:EWR_proof}. There it was shown that such an observer would only be able to deduce the quantum state of the bulk in a region $W[R]$ called the \emph{entanglement wedge} of $R$, but has no knowledge of what happens beyond $W[R]$. The act of deducing the contents of $W[R]$ from the density matrix $\rho_R$ of $R$ is called \emph{entanglement wedge reconstruction}. To give a geometric definition of the entanglement wedge, we first need to introduce the associated concept of the Hubeny-Rangamani-Takayanagi surface. Given a spatial subregion $R$ of the boundary CFT, a codimension-2 bulk surface $\gamma_R$ is a \emph{Hubeny-Rangamani-Takayanagi (HRT) surface} for $R$ if all of the following are true:
\begin{enumerate}[label=(\roman*)]
    \item $\gamma_R$ is anchored to $\partial R$, i.e., $\partial\gamma_R = \partial R$.
    \item $\gamma_R$ has extremal area among the surfaces which obey condition (i).
    \item There exists an achronal codimension-1 bulk surface $H_R$ such that $\partial H_R = \gamma_R \cup R$, i.e., $\gamma_R$ is homologous to $R$ via $H_R$.
    \item There is no other surface obeying (i)-(iii) whose area is less than that of $\gamma_R$.
\end{enumerate}
Typically there will be only one such surface. The entanglement wedge of $R$ is then given by the domain of dependence of any bulk surface $H_R$ satisfying the criteria of (iii) for some HRT surface $\gamma_R$ of $R$:
\begin{equation}
    W[R] \equiv D(H_R) \,.
\end{equation}
This procedure is illustrated in Figure \ref{fig:entanglement_wedge}.

The main hint for entanglement wedge reconstruction came from the (one-loop corrected) Ryu-Takayanagi formula, which says that the von Neumann entropy\footnote{For a gentle introduction to von Neumann entropy and its relevance for black hole thermodynamics, see Section 7 of \cite{Witten:BH_lectures}.} of the CFT density matrix for the region $R$ is
\begin{equation}\label{eq:Ryu-Takayanagi}
    S_{\text{vN}}(\rho_R) = \frac{\operatorname{Area}(\gamma_R)}{4G_N} + S_{\text{vN}}(\rho_{H_R}) \,,
\end{equation}
where $\gamma_R$ is the HRT surface of $R$ and $\rho_{H_R}$ is the density matrix of the bulk quantum fields for the region $H_R$. Moreover, the Ryu-Takayanagi formula is somewhat reminiscent of Bekenstein's original proposal for the generalized entropy of a black hole spacetime,
\begin{equation}
    S_{\text{gen}} = \frac{A}{4G_N} + S_{\text{out}} \,,
\end{equation}
where $A$ is the horizon area and $S_{\text{out}}$ is the ordinary entropy of matter and radiation outside the horizon. The HRT surface $\gamma_R$ plays the same role in (\ref{eq:Ryu-Takayanagi}) as the event horizon in Bekenstein's formula, which points to the fact that (in the time slice where $R$ is contained) $\gamma_R$ separates the regions of the bulk that can and cannot be accessed by our `external' observer.
 
Let us stress the aspect of entanglement wedge reconstruction that will be most important for the Harlow-Ooguri argument: CFT operators localized in $R$ should not be sensitive to bulk objects in the complement of the entanglement wedge of $R$.
\paragraph*{Splittability.}Splittability is a quantum field theory concept that generalizes the characteristic way in which continuous global symmetries are implemented across spacetime to other types of global symmetries, like discrete symmetries. The former are expected to give rise to a Noether current $J_a^\mu$, where $a$ is a Lie algebra index. Then, if the symmetry group is also connected and compact, we may define operators
\begin{equation}\label{eq:splittable_op_cont}
    U(e^{i\epsilon^a T_a}, R) \equiv \exp\Big(i \epsilon^a \int_R \star J_a\Big),
\end{equation}
which implement the symmetry on operators localized on a spatial subregion $R$ and (by microcausality) do nothing on operators entirely localized in the complement of $R$. Inspired by this observation, we may say that a global symmetry of a quantum field theory which is preserved on a spacetime $\mathbb{R}_t \times \Sigma$ is \emph{splittable on $\Sigma$} if for every open spatial subregion $R \subset \Sigma$ and every $g \in G$ there is a unitary operator $U(g,R)$ such that
\begin{equation}\label{eq:def_splittable}
    U^\dagger(g,R)\,\mathcal{O}\,U(g,R) = \left\{\begin{array}{ll}
         U^\dagger(g,\Sigma)\,\mathcal{O}\,U(g,\Sigma) &\quad\text{if }\; \mathcal{O} \in \mathcal{A}[R] \ \\
         \mathcal{O} &\quad\text{if }\; \mathcal{O} \in \mathcal{A}[\operatorname{Int}(\Sigma \setminus R)]
    \end{array}\right. \,,
\end{equation}
where $U(g, \Sigma)$ are the topological operators that implement the symmetry on the whole spatial manifold $\Sigma$ and $\mathcal{A}[R]$ denotes the algebra of operators localized in the domain of dependence of $R$. Note that the action of $U(g,R)$ on operators which are localized right at the boundary of $R$ is left unspecified. The condition \eqref{eq:def_splittable} always holds for an operator of the form \eqref{eq:splittable_op_cont}, so a (compact, connected) global symmetry with a Noether current is always splittable on any spatial geometry for which it is preserved.

Splittability is related to the local structure of degrees of freedom of the quantum field theory under consideration. This is illustrated by the fact that all global symmetries are splittable on the lattice \cite[Thm.~2.1]{Harlow:2018tng}, where the total Hilbert space of any QFT is a tensor product of the Hilbert space on each site\footnote{A constrained gauge-theory Hilbert space need not factorize over lattice sites. }. The algebraic quantum field theory community has in fact come up with an axiom, called the \emph{split property}, which if obeyed by a QFT on a spatial manifold $\Sigma$ is sufficient to ensure that any global symmetry is splittable on $\Sigma$; see Def.~2.4 in \cite{Harlow:2018tng} and references therein. The split property is usually taken for granted for QFTs on $\Sigma = \R^{d-1}$, since no counterexamples are known in that case, but may fail for reasonable QFTs on nontrivial manifolds. It is thus possible to have a non-splittable global symmetry on such manifolds. We give a few examples below, all of which are discussed at length in \cite{Harlow:2018tng}:
\begin{itemize}
    \item (\textit{Free theory with a non-compact gauge group}) The global $\mathrm{U}(1)$ symmetry of pure gauge theory with gauge group $G = \R \times \R$ is splittable on $\Sigma = \R^{d-1}$ and $\Sigma = S^{d-1}$, but not on more complicated topologies like $\Sigma = S^1 \times S^{d-1}$. 
    \item (\textit{Free theory with a compact gauge group}) Similarly for the global $\Z_2$ of pure gauge theory with $G = \mathrm{U}(1)$, also called charge conjugation symmetry: it is splittable on $\Sigma = \R^{d-1}$ and $\Sigma = S^{d-1}$, but not on more complicated topologies like $\Sigma = S^1 \times S^{d-2}$. 
    \item (\textit{Interacting theory}) Consider the theory of $N_f$ massless Dirac fermions in 3+1 dimensions,
    \begin{equation}
        S = i \int d^4x \sum_{i=1}^{N_f}\Bar{\psi}_i \slashed{\partial} \psi_i \,,
    \end{equation}
    which has a global chiral symmetry $\mathrm{U_L}(N_f) \times \mathrm{U_R}(N_f)$ which consists in separately rotating the left- and right-handed parts of the fermions. This chiral symmetry has two important $\mathrm{U}(1)$ subgroups: the vector subgroup $\mathrm{U_V}(1)$ and the axial subgroup $\mathrm{U_A}(1)$. The axial subgroup cannot be gauged consistently because it suffers from a 't Hooft anomaly. We may nonetheless gauge the vector subgroup by coupling a dynamical gauge field $A^{\rm V}_\mu$ to the vector current $j_{\rm V}^\mu = \sum_i \overline{\psi}_i\gamma^\mu\psi_i$, and we might then expect that the global $\mathrm{U_A}(1)$ symmetry will still be present in the resulting theory. That, however, turns out not to be entirely the case: some of the new operators constructed out of the gauge field ---'t Hooft loops--- do not transform locally\footnote{Besides requiring global symmetries to map local operators to local operators, Harlow and Ooguri also require them to map the operator algebra $\mathcal{A}[R]$ of any spatial region $R$ to itself. Extended operators localized in the domain of dependence of $R$ are included in this algebra, so they must be acted upon locally, too.} under all of $\mathrm{U_A}(1)$, only under its $\Z_{N_f}$ subgroup. As in the two previous examples, one can show that this remaining $\Z_{N_f}$ symmetry is not splittable on $\Sigma = S^2 \times S^1$.  
\end{itemize}

The obstruction to splittability in all the above examples can be tied to the presence of an additional 1-form global symmetry. Indeed, in these cases there exist line operators which cannot end ---they would have to do so on local operators carrying gauge charge, but none of those exist. It turns out that breaking this 1-form symmetry ---by, for instance, introducing some heavy charges in the UV where the lines can end--- restores splittability, at least in the second and third examples. Not so in the first one, which we expect to be more pathological due to the non-compact gauge group \cite{Harlow:2018tng}.

In any case, the only assumption that we will have to make here is that any global symmetry of any CFT on $\R \times S^{d-1}$ is splittable, since that is what is actually used in the AdS/CFT argument below. For a CFT in $d \geq 2$, splittability on $S^{d-1}$ follows from splittability on $\R^{d-1}$, for which many more arguments ---and no counterexamples--- exist. This is because $\R \times S^{d-1}$ and flat space are conformally equivalent. Since we know exactly how to map a CFT on one geometry to the other, the operators $U(g,R)$ that split the symmetry on $\R^{d-1}$ can be translated into splitting operators on $S^{d-1}$.

\paragraph*{} We may now proceed with the argument.

\begin{figure}
    \centering
    \includegraphics[width=0.4\linewidth]{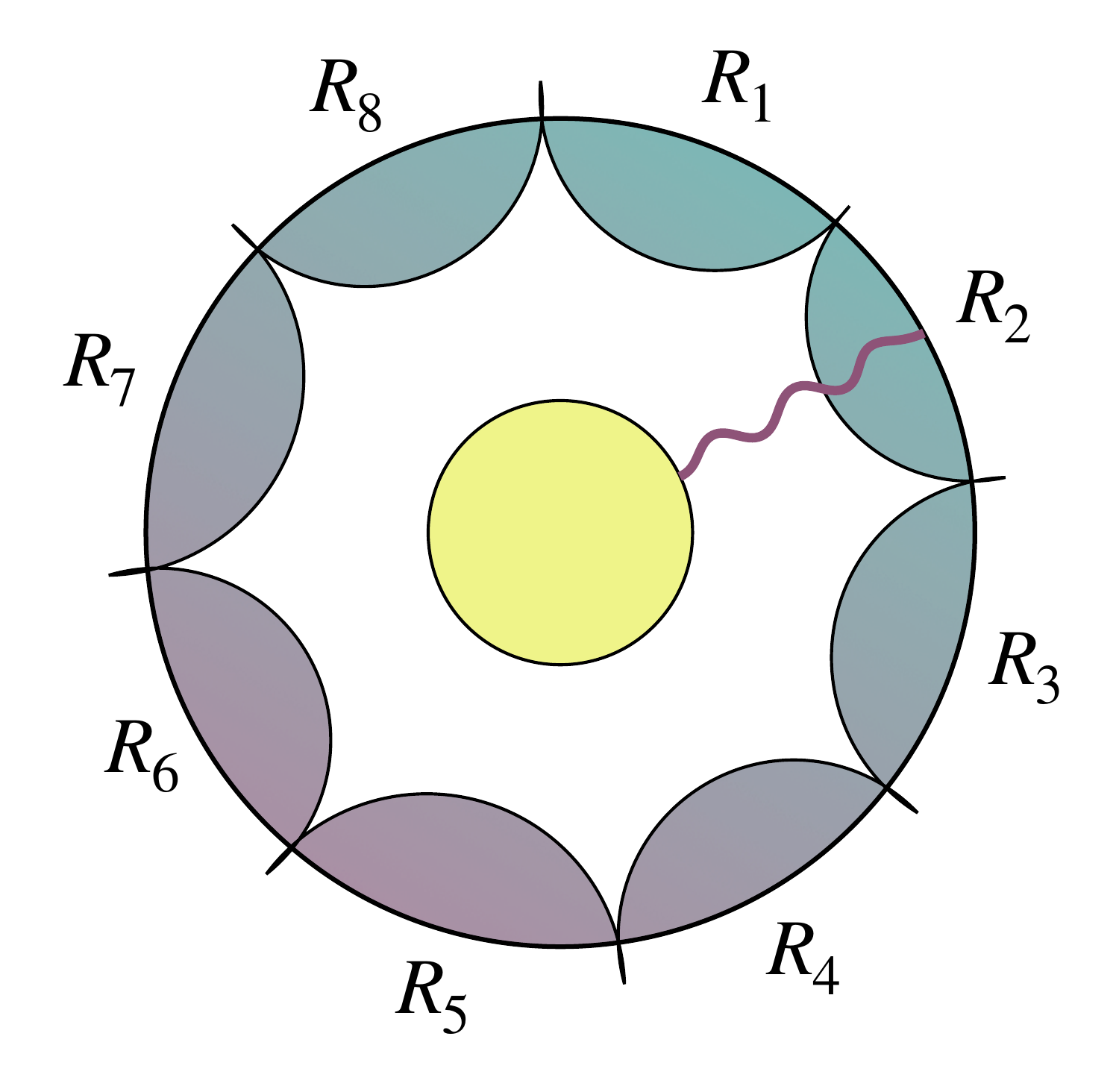}
    \caption{A time slice of the bulk. Suppose that a charged operator is localized in a certain region of the interior, highlighted in yellow here. By considering a sufficiently fine boundary cover $\{R_i\}_i$, we can make this region lie in the complement of all the entanglement wedges, shown in blue/purple. If the operator is charged under a global symmetry, this prompts a contradiction with entanglement wedge reconstruction. If however the symmetry is gauged, no contradiction arises as there will be a Wilson line ---the wavy line in the picture--- that extends from the operator to the boundary.}
    \label{fig:Harlow-Ooguri_arg}
\end{figure}

\paragraph{The argument.} Say that a global symmetry were present in the gravitational bulk theory. In AdS/CFT, every CFT operator of definite conformal dimension can be obtained as the appropriately rescaled boundary limit of a bulk operator;\footnote{In a gravitational EFT in asymptotically-AdS spacetime, physical observables must be invariant under diffeomorphisms which become trivial at the boundary (but are allowed to transform under diffeomorphisms which approach isometries of AdS, since those preserve the boundary conditions). Strictly local operators are not invariant under those diffeomorphisms and therefore do not correspond to physical observables, whose location must instead be specified in relation to the boundary through the gravitational analogue of Wilson lines. A more precise version of the Harlow-Ooguri argument than the one given here ---see \cite{Harlow:2018tng}--- involves local operators dressed with these gravitational Wilson lines, which transform in representations of the asymptotic conformal group. This is related to one of the limitations of the argument, to be discussed below, which is that the gravitational dressing is only under control semiclassically.
\label{fn:HO_dressing}} therefore, a bulk global symmetry also induces a global symmetry in the boundary.  As we have just argued, we expect any global symmetry of the boundary theory, a CFT on spatial $S^{d-1}$, to be splittable. Under that assumption, the operators which implement the global symmetry on a boundary time slice $\partial\Sigma$ may be rewritten as
\begin{equation}\label{eq:splittability_arg}
    U(g) = U(g, R_1) U(g, R_2) \cdots U(g,R_n) U_{\mathrm{edge}} \;,
\end{equation}
where $\{R_i\}_{i=1}^n$ is a disjoint cover of $\partial\Sigma$ and the $U(g, R_i)$ act trivially on operators localized on $\partial\Sigma \setminus R_i$. The last operator on the rhs has been introduced to fix the ambiguity on how the product of the $U(g, R_i)$ acts on operators with support on the boundaries of the $R_i$'s; it only acts on operators localized on those boundaries. Now, an important property of what we call a global symmetry is that at least one operator which is charged under the symmetry must exist. Suppose that the charged operator (under the boundary symmetry) is localized in $R_i$. Entanglement wedge reconstruction then tells us that its bulk image $\Phi$, which must also be charged under the symmetry, will be localized in the entanglement wedge of $R_i$. Next, replace the $\{R_i\}_{i=1}^n$ by a finer cover $\{\tilde{R}_j\}_{j=1}^m$ such that $\Phi$ lies in the complement of all the wedges, and split $U(g)$ again. The bulk image can no longer correspond to a boundary operator localized in any of the $\tilde{R}_j$ (or its boundaries), so it must commute with $U(g) = \prod_j U(g, \tilde{R}_j)\, \tilde{U}_{\text{edge}}$, yielding a contradiction. We thus conclude that global symmetries may not exist in the bulk.

Introducing the concept of splittability ---as opposed to assuming the existence of a conserved current--- allowed Harlow and Ooguri to develop an argument that not only applies to continuous global symmetries, but also to discrete ones.  Although here we have restricted to internal global symmetries, there is a slightly more general version that excludes spacetime global symmetries ---see Section 7 of \cite{Harlow:2018tng}--- and another one for (no) global higher-form symmetries ---\textit{ibid.}, Section 8.3.

\paragraph*{} A possibly disturbing feature of this argument is that it would seem to exclude \emph{boundary} global symmetries, which are a known feature of most top-down holographic pairs. How do these symmetries arise if not through a bulk global symmetry? There is actually another kind of bulk symmetry that can induce a global symmetry at the boundary: in the language of \cite{Harlow:2018tng}, \emph{long-range gauge symmetries}.

\paragraph*{Long-range gauge symmetries.} It is often said that holographic dualities map boundary global symmetries to gauge redundancies in the bulk, with the symmetry group on the field theory side matching the gauge group of the gravitational theory. Although in some sense true \cite{Witten:1998qj}, this seems to be in tension with the widespread lore that gauge redundancies are an artifact of our description without any physical meaning. A more precise statement would be that boundary global symmetries are dual to what Harlow and Ooguri have called `long-range gauge symmetries', which we define below.\footnote{The motivation for this definition came from lattice gauge theory, see \cite[Sec.~3.2-3.3]{Harlow:2018tng}.}

A quantum field theory on an infinite-volume spatial manifold $\Sigma$ with asymptotic boundary $\partial\Sigma$ and fall-off boundary conditions has a \emph{long-range gauge symmetry} with gauge group $G$ if the following conditions are satisfied: 
\begin{enumerate}[label=(\roman*)]
    \item For any (finite-dimensional, irreducible) representation $\alpha$ of $G$, there exist two sets of directed line operators:
    \begin{itemize}[]
        \item The \emph{Wilson loops} $W_\alpha(C)$, where $C$ is any closed curve in spacetime.
        \item The \emph{Wilson lines} $W_{\alpha,\, ij}(C')$, where $C'$ is any open curve in spacetime with both endpoints on $\partial\Sigma$ and $i$, $j$ are representation indices.
    \end{itemize}
    Taking the adjoint (defined here to exchange any representation indices) reverses the orientation of the loop/line, i.e., $W_\alpha^\dagger(C)  = W_\alpha(-C)$ and $W_{\alpha,\,ij}^\dagger(C)  = W_{\alpha,\,ij}(-C)$.
    In addition, Wilson lines must obey
    \begin{equation}
        \sum_k W_{\alpha,\, ik}(-C) W_{\alpha,\, kj}(C) = \delta_{ij} \,.
    \end{equation}
    Bringing the endpoints of a Wilson line together, taking the trace and deforming away from the boundary produces a Wilson loop.
    
    \item For every spatial region $R \subset \partial\Sigma$ and every $g \in G$, there exists a unitary operator $U(g,R)$ which commutes with all operators localized in the interior of $\Sigma$, and also with their boundary limits (provided that they are spacelike-separated from $\partial R$), but whose action on a Wilson line $W_\alpha$ starting at $x \in \partial\Sigma$ and ending at $y \in \partial\Sigma$ is 
    \begin{equation}
        U^\dagger(g,R) W_\alpha U(g,R) = \left\{\begin{array}{ll}
            D_\alpha(g) W_\alpha D_\alpha(g^{-1})  &\quad x,y \in R \\
            W_\alpha D_\alpha(g^{-1}) &\quad x \in R, y \notin R \\
            D_\alpha(g) W_\alpha &\quad x \notin R, y \in R \\
            W_\alpha &\quad x,y \notin R
        \end{array}\right.\;, 
    \end{equation}
    where $D_\alpha(g)$ denotes the representation matrix of $g$ in the representation $\alpha$, acting on the representation indices of the Wilson line, which we have suppressed. When $R$ is a connected component of $\partial\Sigma$, $U(g,R)$ commutes with the Hamiltonian, so we can call it an \emph{asymptotic symmetry operator}. In all other cases, we call it a \emph{localized asymptotic symmetry operator}.
    \item The ground state is invariant under $U(g,\partial\Sigma)$ and, moreover, the dynamics of the theory allow finite-energy states charged under $U(g,\partial\Sigma)$.
\end{enumerate}
Not all gauge symmetries are long-range; the above definition also prescribes the phase that the theory is in, mainly via condition (iii). The theory must be in what is sometimes called a `free charge' phase (Coulomb or topological), which resembles a weakly-coupled gauge field. It cannot be in a confining or Higgs phase because no finite-energy states with asymptotic charge exist in either case; in the first one because of the formation of flux tubes between free charges (whose energy diverges as one of their ends approaches the boundary), and in the second one because of screening of charge by the Higgs condensate (which gives mass to the gauge boson). The basic example of a theory with a long-range gauge symmetry is Maxwell electrodynamics (with or without dynamical charges), which has $G = \mathrm{U}(1)$. QCD in AdS$_4$ also has a long-range symmetry, with $G = \mathrm{SU}(3)$, but only when $\Lambda_{\text{QCD}} \ll L_{\text{AdS}}^{-1}$, not in the confining regime $\Lambda_{\text{QCD}} \gg L_{\text{AdS}}^{-1}$. 

In the case of a bulk long-range gauge symmetry, the preceding argument does not lead to a contradiction because any operator charged under this symmetry needs to be attached to the asymptotic boundary by a gauge Wilson line, which will always intersect one of the entanglement wedges independently of how small these are. This situation is depicted in Figure \ref{fig:Harlow-Ooguri_arg}. Interestingly, assuming completeness of gauge representations,\footnote{\label{fn:completeness_AdS-CFT}Completeness of gauge representations is the conjectured feature of quantum gravity that there must be dynamical objects transforming in all irreducible representations of any internal gauge symmetry. Harlow and Ooguri \cite[Sec.~5]{Harlow:2018tng} prove this conjecture for AdS/CFT assuming that the bulk gauge group is compact, which in turn relies on the assumption that the CFT dual is finitely generated, i.e., that there is a finite set of primary operators whose operator product expansion (OPE) recursively generates all other primaries.} one can show that any long-range gauge symmetry in the bulk necessarily gives rise to a splittable global symmetry in the boundary. And, perhaps less rigorously, one can also argue that the converse is true. So, as we anticipated earlier, long-range gauge symmetries in the bulk are dual to splittable global symmetries in the boundary CFT.

\paragraph{Limitations of the Harlow-Ooguri argument.} The domain of validity of the Harlow-Ooguri argument is delimited by the following observations:
\begin{itemize}
    \item It is only valid for holographic theories of quantum gravity with a conformal field theory dual. In fewer than four spacetime dimensions, there are known examples of quantum gravity theories (e.g., oriented pure 3d gravity) with global symmetries (e.g., spatial reflection and time reversal\footnote{For a discussion on C, P and T in the context of the Cobordism Conjecture, see \cite{McNamara:2022lrw}.}), but in general they are also known to be non-unitary and/or not holographic to a single CFT \cite{Maloney:2007ud, Harlow:factorization_JT}. The current interpretation is that the gravitational path integral in these theories computes an ensemble average over boundary theories \cite{Saad:2019lba, Cotler:2020ugk}, and that the global symmetry may just arise from this averaging \cite{McNamara:2020uza,Benini:2022hzx}. For example, if the ensemble contains each CFT together with its parity image, then the average is P-invariant (and T-invariant) even if no single member is. Presumably, in a single instance of the ensemble, the Harlow-Ooguri argument would still apply. Note in addition that the argument implicitly assumes spatial locality of the boundary theory, which is only true for a bulk dimension greater than or equal to three. On top of that, there are also holographic theories that are dual to non-conformal field theories, see for example \cite{Itzhaki:1998dd, Polchinski:2000uf, Klebanov:2000hb, Pilch:2000ue}. It would be thus interesting to find an argument that does not rely on the conformal invariance of the boundary theory.\footnote{The extension of the argument to the non-conformal D$p$-brane setup of \cite{Itzhaki:1998dd} could be relatively straightforward due to the \emph{generalized} conformal invariance of the boundary theory \cite{Kanitscheider:2008kd}. Note, however, that this structure is not a general feature of non-conformal holography; it is not present in any of the other examples cited above.}
    \item It does not apply to approximate global symmetries, which are commonplace among the low-energy effective actions of string theory. By approximate global symmetries we mean transformations that are only symmetries of some low-energy subspace of the theory. This is as opposed to proper global symmetries, which should act on the entire Hilbert space of quantum gravity, including for example black hole states.
    \item It relies on ingredients that are only under control semiclassically. Localized bulk operators, dressed gravitationally as in footnote \ref{fn:HO_dressing}, are well understood perturbatively in $G_N$ (i.e., in the $1/N$ expansion of the CFT), but it is not clear whether they make sense non-perturbatively. Similarly, entanglement wedge reconstruction is established for semiclassical states, perturbatively around a fixed bulk geometry \cite{Dong:2016eik, Harlow:EWR_proof}. This is a concern because the absence of global symmetries is a statement about the full, non-perturbative theory: as the axion shift symmetries discussed in the previous section illustrate, perturbative considerations alone can be misleading. 
    \item More importantly, precisely because it only uses semiclassical ingredients, the argument goes through even when gravity is switched off, $G_N \to 0$. Consider for instance a free QFT in AdS, dual to a theory of generalized free fields (GFFs) on the boundary. In this setup, local bulk operators are perfectly well defined and entanglement wedge reconstruction works (see Section 3.1 of \cite{Harlow:2018fse} and references therein)---the free bulk field can be written explicitly as a (highly non-local) smeared integral of the GFFs---, so the argument applies. Yet a free QFT in AdS is certainly allowed to have global symmetries. The resolution is that the corresponding boundary symmetry is non-splittable, which is not surprising since GFF theories are badly non-local (in particular, they have no stress tensor). What the argument actually establishes is therefore that having a global symmetry in AdS leads to a non-splittable global symmetry in the dual CFT. To conclude that quantum gravity in AdS has no global symmetries, one additionally needs that every global symmetry of a local CFT ---with a stress tensor, and possibly further conditions such as modular invariance in $d=2$--- is splittable. This is precisely the splittability assumption made above. As discussed, it is supported by results in algebraic quantum field theory on $\R^{d-1}$, but these rely on a rather restrictive set of axioms and may be weaker than what is actually needed. A proof based on the standard axioms of local CFT, in the spirit of the conformal bootstrap, would be desirable.
\end{itemize}

\subsection{Connections to other Swampland Conjectures}\label{sec:relation}

The absence of exact global symmetries is deeply intertwined with several other Swampland conjectures. For instance, the Weak Gravity Conjecture \cite{ArkaniHamed:2006dz, Harlow:2022ich} may be viewed as a refinement of this principle for gauge symmetries: it ensures that charged states exist so that extremal black holes can decay, preventing an emergent global symmetry in the limit of vanishing gauge coupling. Similarly, as we saw in section \ref{sec:heuristic}, the Completeness Hypothesis asserts that all allowed charges are realized in the spectrum, again obstructing the appearance of conserved global charges \cite{Heidenreich:2021xpr,McNamara:2021cuo}. More broadly, many Swampland criteria can be understood as mechanisms that prevent global symmetries from arising. For instance, the Swampland Distance Conjecture \cite{Ooguri:2006in} also fits naturally into this picture. In many situations, sending a gauge coupling $g \to 0$ would appear to restore a global symmetry in the infrared, since the gauge field effectively decouples and charged states become non-interacting. However, the Distance Conjecture asserts that such limits correspond to infinite-distance points in field space, where an infinite tower of states becomes exponentially light. As a result, the effective field theory description breaks down precisely before an exact global symmetry can be recovered. In this way, the Distance Conjecture provides a dynamical obstruction to taking the $g \to 0$ limit inside a consistent theory of quantum gravity. Rather than obtaining a theory with an exact global symmetry, one instead encounters a proliferation of light states that invalidates the low-energy description, preventing the symmetry from ever becoming exact.  These connections can be made more quantitative in terms of approximate symmetries \cite{Cordova:2022rer}.

A particularly sharp and modern formulation of these ideas is provided by the Cobordism Conjecture, which proposes a precise topological criterion for the absence of global symmetries in quantum gravity. Before turning to the Cobordism Conjecture, we first introduce the mathematical machinery that we will need to define it: bordism groups.

\section{Bordisms 101}\label{sec:Bordisms101}

The main point of these lectures is to get to the Cobordism Conjecture. Roughly speaking, the idea is that possible global symmetries in quantum gravity are detected by cobordism groups. As these groups are central to the discussion, we now pause to introduce bordism groups themselves. We first introduce bordism groups for the non-expert in Section~\ref{sec:bordismgeneralities}, in the hope of clarifying what kind of information these groups contain. Then, we go a bit deeper into spin bordism, which will be especially relevant later on, when discussing the Cobordism Conjecture.

We will assume a basic understanding of fibre bundles and (co)homology theories, which can be found in e.g.~\cite{nakahara2003geometry,taubes2011differential,hamilton2017mathematical}. Readers primarily interested in the physics may wish to skip ahead to the next section. 

In an effort to try to make this section more accessible to the casual reader, let us devote a couple of lines to the mathematical concept of a fibre bundle, which will play a central role in the following discussion. A fibre bundle is a geometric structure constructed by interweaving a family of copies of a model fibre $F$ over a base space $M$ such that
\begin{enumerate}
    \item each point $x$ of $M$ is assigned a copy of $F$ (a `fibre' of $x$), and
    \item the local structure of the resulting space resembles that of the trivial product $M \times F$.
\end{enumerate}
A great virtue of this concept is its flexibility: locally, a fibre bundle is a trivial product, but that does not prevent a more complex global structure from emerging from the interweaving of the fibres. Fibre bundles come in a variety of flavors that depend on which type of object the model fibre is and what kind of maps are used to glue the fibres. For example, vector fibre bundles are those bundles which have a general vector space as model fibre and in which fibres are interwoven using linear automorphisms. Meanwhile, in a principal fibre bundle, the model fibre is a Lie group $G$ and fibres are assembled using an action of $G$ itself on the fibres. The group $G$ is often referred to as the \emph{structure group} of the principal fibre bundle.\footnote{Mathematically, principal bundles can be defined for any topological group, but in physics we are interested in smooth principal bundles over smooth base manifolds. This forces the transition functions to be smooth maps, which only makes sense if $G$ is itself a smooth manifold, i.e., a Lie group.}

\subsection{Generalities on Bordism Groups}\label{sec:bordismgeneralities}
In this section we provide a brief introduction to bordism groups for readers that have little (or no) knowledge of algebraic topology. For a mathematical introduction on bordism groups see \cite{Stong+1968}. For more in-depth discussions with a special focus on physical applications, see \cite{McNamara:2019rup,Garcia-Etxebarria:2018ajm,Debray:2023yrs}.

\begin{figure}[h!]
\centering
\begin{subfigure}{.5\textwidth}
  \centering
  \includegraphics[width=0.9\linewidth]{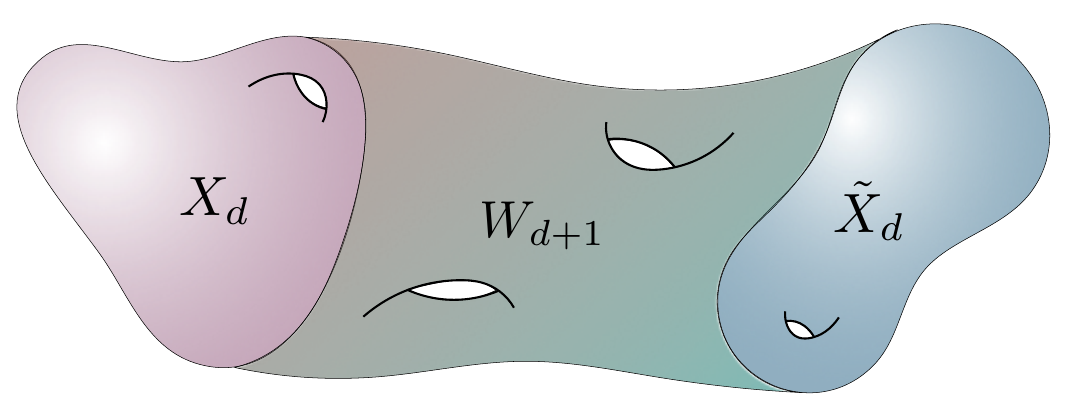}
  \caption{A bordism between two manifolds}
  \label{fig:bordism}
\end{subfigure}%
\begin{subfigure}{.5\textwidth}
  \centering
  \includegraphics[width=.9\linewidth]{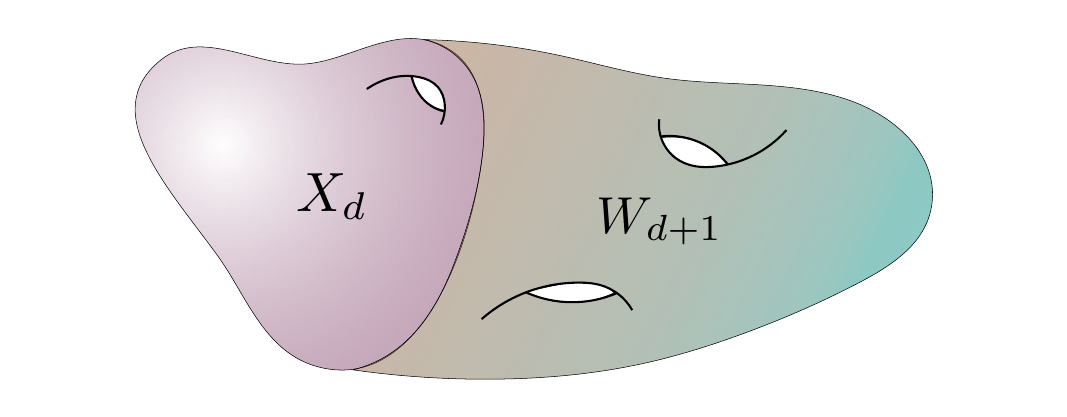}
  \caption{A bordism to nothing}
  \label{fig:bordismtonothing}
\end{subfigure}
\caption{a) $X_d$ and $\tilde X_d$ are cobordant or bordant to each other: they are in the same bordism equivalence class. b) $X_d$ is in the same equivalence class as the neutral element. $X_d$ is said to be bordant to nothing.  }
\label{fig:17}
\end{figure}

Bordism groups are elaborate mathematical objects used in algebraic topology to classify closed manifolds up to a certain \hl{equivalence relation called bordism.} Put simply, a bordism group collects equivalence classes of \hl{closed manifolds} of a given dimension, considering two manifolds $X_d$ and $\tilde{X}_d$ to be equivalent if there exists a higher-dimensional manifold $W_{d+1}$ whose boundary is the disjoint union of $X_d$ and $\tilde{X}_d$, denoted by $X_d \sqcup \tilde{X}_d$. One says that $X_d \,\sqcup \,\tilde{X}_d$ \textit{bounds} $W_{d+1}$ and that $X_d$ and $\tilde{X}_d$ are \textit{cobordant}. This is depicted schematically in Figure \ref{fig:bordism}. This higher dimensional manifold parametrizes the deformation of $X_d$ into $\tilde{X}_d$. 

\vspace{0.3cm}

Schematically, the bordism group $\Omega_d$ is the set of equivalence classes of $d$-dimensional manifolds, where two manifolds are considered equivalent if they are cobordant: 
\begin{equation}
    \Omega_d \cong \left. \{\text{$d$-dimensional manifolds }X_d\} \middle/ \text{bordism} \right..
\end{equation}

\begin{figure}[]
\centering
\begin{subfigure}{.5\textwidth}
  \centering
  \includegraphics[width=1.2\linewidth]{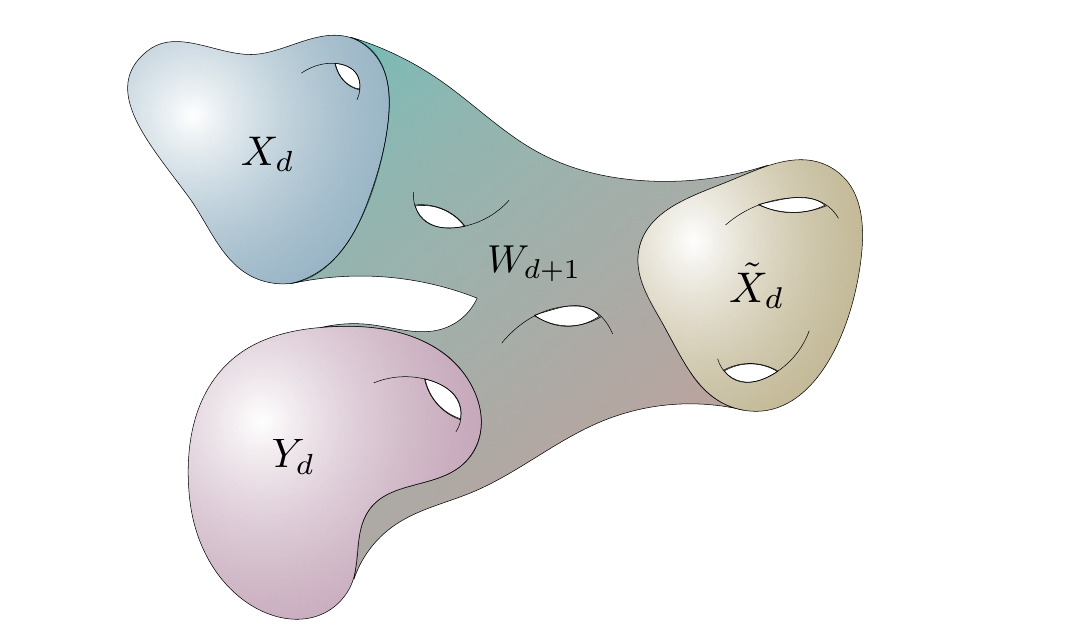}
  \caption{The bordism of a disjoint union}
  \label{fig:directsum}
\end{subfigure}%
\begin{subfigure}{.5\textwidth}
  \centering
  \includegraphics[width=1.2\linewidth]{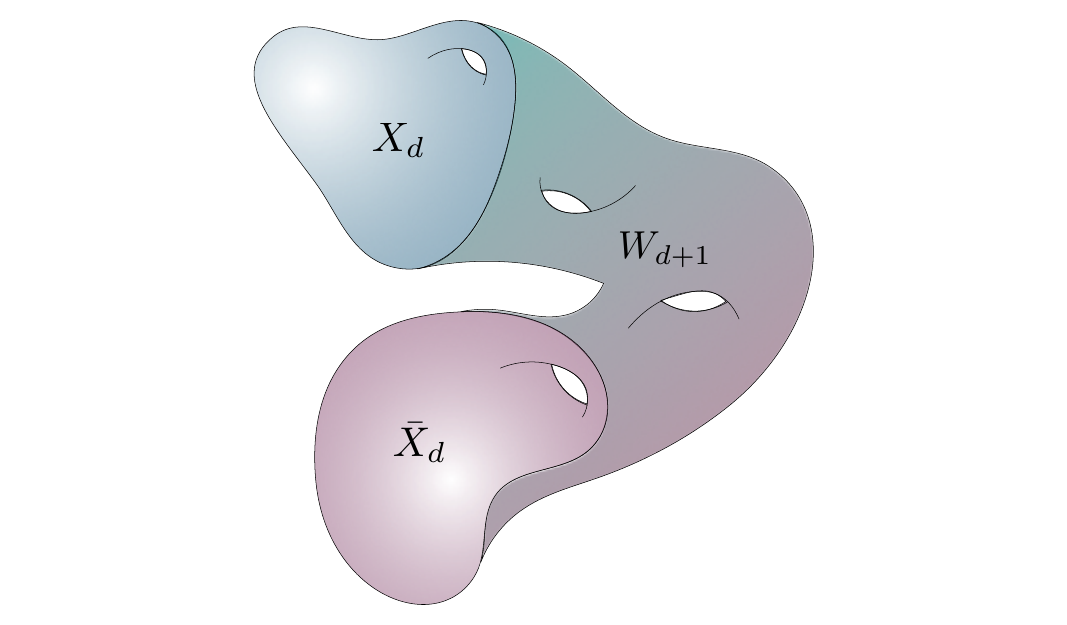}
  \caption{The inverse}
  \label{fig:inverse}
\end{subfigure}
\caption{a) The disjoint union $X_d\sqcup Y_d$ is bordant to $\tilde {X}_d$. b) The inverse of $X_d$ is defined as the manifold that is such that the disjoint union of $X_d$ and its inverse is bordant to nothing. }
\label{fig:test}
\end{figure}

\paragraph*{Group Structure.}Bordism groups $\Omega_d$ have an \hl{abelian group structure,} with addition under the disjoint union of manifolds. A bordism for a disjoint union of manifolds is depicted in Figure \ref{fig:directsum}. There is therefore a neutral element, the equivalence class of ``nothing'', the empty manifold. This means that some manifolds $X_d$ can be bordant to nothing, which is to say that they are boundaries. In that case, there is a higher dimensional manifold $W_{d+1}$ that parametrizes how $X_d$ can be shrunk into nothing, as depicted in Figure \ref{fig:bordismtonothing}. The inverse is given by an appropriate orientation-reversed version of $X_d$ such that their disjoint union is bordant to nothing, as depicted in Figure \ref{fig:inverse}.\footnote{If there is no orientation (in unoriented bordism), all manifolds are their own inverse since one can always write $M\sqcup M = \partial ( M\times [0,1])$, which implies that $M\sqcup M$ is bordant to nothing.}

\paragraph*{$\xi$-Structure.} Things get a bit more complicated when you want to consider bordisms between manifolds that satisfy extra conditions. So far, we have not specified anything about $X_d$ except for the fact that it is a manifold; this is the most basic form of bordism, known as \textit{unoriented bordism}. We can also consider the bordism of oriented manifolds in which case we call it \textit{oriented bordism}. More generally, we can consider the bordisms of manifolds with a specific \textit{structure}. When there is additional structure, the notion of cobordism must be formulated more precisely: the structure has to extend to the $(d+1)$-dimensional manifold $W_{d+1}$ that defines the bordism. 

Defining a structure on a manifold $X_d$ could be described as equipping its tangent bundle $TX_d$\footnote{More generally, $TX_d$ could be replaced by any other vector bundle $E$ over $X_d$.} with some additional geometric data. For instance, an orientable structure on a Riemannian manifold $X_d$ consists of a reduction of the frame bundle of $TX_d$ ---which is a principal $G$-bundle--- from $G = \mathrm{O}(n)$ to $G = \mathrm{SO}(n)$. This is equivalent to choosing an orientation of $X_d$. Meanwhile, a spin structure amounts to finding a lift of the same orthonormal frame bundle to a principal $\mathrm{Spin}(n)$-bundle.
Let us denote a generic structure by a $\xi$-structure. When rigorously defined, a $\xi$-structure on $W_{d+1}$ induces a structure on its boundary. In that way, we can define bordisms between manifolds $X_d$ and $\tilde{X}_d$ with $\xi$-structure through a manifold $W_{d+1}$ with a $\xi$-structure. This leads to e.g.~oriented bordism groups $\Omega_d^{\SO}$ and spin bordism groups $\Omega_d^{\Spin}$ (we will describe the latter in detail in section \ref{sec:spinbord}).

It can help to think of adding structure as trivializing a new class of topological invariants. For instance, for a $d$-manifold $X_d$ to be orientable, the first Stiefel-Whitney class of its tangent bundle $w_1(TX_d)$ \cite{milnor} must vanish, and on top of that, for it to be a spin manifold, the second Stiefel-Whitney class $w_2(TX_d)$ must vanish as well. Stiefel-Whitney classes are topological invariants that measure intrinsic obstructions to consistently choosing directions, orientations, or frames on a manifold, detecting `twists' in the tangent bundle that prevent such data from being chosen globally. Once the vanishing of these two classes is checked, the manifold under consideration can be shown to admit a spin structure. The point is that manifolds with non-trivial topology can admit various spin structures for each dimension $d$. Some of these spin structures can bound a $(d+1)$-dimensional manifold with spin structure, others cannot. The former are trivial elements and the latter are non-trivial elements in  $\Omega_d^{\Spin}$.

\paragraph*{Generators and Bordism Invariants.} If the bordism group under consideration $\Omega_d^\xi$ vanishes, then all $d$-manifolds with $\xi$-structure can be deformed into one another and into nothing. However, when it is not trivial, it generically takes the form:
\begin{equation}\label{eq:bordismex}
    \Omega^\xi _ d\cong \underbrace{\mathbb{Z}\times \cdots \times \mathbb{Z}}_{r\;\text{times}}  \times \;\underbrace{\mathbb{Z}_{p_1} \times \mathbb{Z}_{p_2} \cdots}_{s\;\text{times}}
\end{equation} 
for $p_i$ non-negative integers. This decomposition is possible whenever the bordism group is finitely generated, which is always the case in these lecture notes (but is not proven in general).  The first part corresponds to the \hl{free part} of the group, whilst the second part corresponds to \hl{torsion classes}. Manifolds that are part of a torsion class indexed by an element of $\Z_{p_i}$ are such that only the disjoint union of $p_i$ copies of them is bordant to nothing. Equation \eqref{eq:bordismex} means that any $d$-manifold with $\xi$-structure is in an equivalence class of bordism that can be indexed by $r$ integers and $s$ integers mod $p_i$.

To see where these components are coming from, we need to consider \hl{bordism invariants}. These are topological invariants that, when evaluated on a manifold, yield (one of) the aforementioned indices. With a complete set of $r + s$ bordism invariants, one can identify precisely which bordism equivalence class a manifold belongs to. In particular, these invariants help in identifying the \hl{generators of the bordism group:} the set of $d$-manifolds that are representative of the bordism class under consideration. The generators need not be unique, all that matters is that they carry the right bordism invariant. For instance, Pontryagin numbers and Stiefel-Whitney numbers \cite{milnor} are oriented cobordism invariants; they determine an oriented manifold's oriented cobordism class completely.

Determining the bordism invariant is in general easier said than done; they are not always simple integrals of characteristic classes of some tangent or gauge bundle. For instance, some of the spin bordism invariants are more complicated quantities known as $\eta$-invariants of Dirac operators \cite{Atiyah_Patodi_Singer_1975}. In such cases, finding the relevant generator may not be straightforward.

\paragraph*{Bordisms and Cohomology.} To a bordism theory $\Omega^\xi_d$, we can associate a generalized homology theory, which assigns a group $\Omega^\xi_d (Y)$ to each topological space $Y$. $\Omega^\xi_d (Y)$ describes the set of equivalence classes of $d$-manifolds $X_d$ equipped with a map into $Y$. If $Y$ is a point `$\pt$' then $\Omega^\xi_d (Y)$ reduces to the bordism group described above $\Omega_d^{\Spin} (\pt) \coloneq \Omega_d^{\Spin} $.\footnote{Note that the point is the classifying space of the trivial group: $B\mathbf{1}=\pt$, where $\mathbf{1}$ is the identity.} 

Generalized homology theories share all of the (Eilenberg--Steenrod) axioms of a homology theory \cite{1945PNAS...31..117E} except for the dimension axiom: the bordism groups of a point are not trivial for $d> 0$. In physical setups, we often have to describe the bordisms of $d$-manifolds with $\xi$-structure and with a non-trivial $G$-bundle, which one can imagine to be some sort of gauge bundle. In this case, the bordism groups one has to consider are $\Omega^\xi _ d( BG)$ where $BG$ is the \hl{classifying space of $G$} which classifies the various physically distinct $G$-bundles. Instead of just a manifold $X_d$, one is studying pairs $(X_d,f)$, where $f:X_d\rightarrow BG$ is a map into the classifying space of $G$. This map $f$ is exactly the same thing as a $G$-bundle over $X_d$. Two pairs $(X_d^{1},f_{1})$ and $(X_d^{2},f_{2})$ are bordant if there exists a $(d+1)$-dimensional manifold $W_{d+1}$ and a map $F:W_{d+1}\rightarrow BG$ such that the boundary of $W_{d+1}$ is the union of $X_d^{1}$ and $X_d^{2}$, and the map $F$ restricts to $f_{1}$ and $f_{2}$ on the ends. Importantly, even if we know all of the $\Omega_d^\xi= \Omega_d^\xi(\pt)$ and the homology of $BG$, this does not mean that we can immediately compute $\Omega^\xi_d (BG)$. The \hl{Atiyah-Hirzebruch spectral sequence}, however, gives a starting point for these calculations by reducing the computation of $\Omega^\xi_d (BG)$ to that of homology groups of $BG$ with coefficients in $\Omega^\xi_q(\pt)$ (more on this in Section~\ref{sec:spectralsequences}).

\paragraph*{Final Remarks.}
\begin{itemize}
    \item On top of the abelian group structure described above, many types of $\xi$-structure (though not all of them) allow for multiplication under the Cartesian product of two manifolds. In that case we may define the cobordism \hl{ring} $\Omega^\xi_ {*} = \bigoplus _ d \Omega^\xi _ d$, which is $\mathbb{Z}$-graded by manifold dimension. This is for example possible for oriented bordism, spin bordism and string bordism, but not for pin$^{\pm}$ bordism.
    \item Given the discussion above, one might wonder what is the difference between considering $\Omega^\xi_d(BG)$ and considering $\Omega_d ^{\xi+BG} (\pt)$ where the $G$-bundle is added to the structure. The difference is that $\Omega_d ^{\xi+BG} (\pt)$ can involve an additional relationship between the $G$-bundle and (a frame bundle of) $TX_d$, whereas in $\Omega^\xi_d(BG)$ the two are independent. Intuitively, the difference is the same as the difference between considering a manifold with spin structure and considering an arbitrary $\Spin(n)$-bundle on the same manifold. One is a property of the tangent bundle, and the other is an auxiliary bundle unrelated to it. If the action of the gauge group $G$ starts mixing with the structure, then this changes the structure. For instance, consider the bordism groups of manifolds with spin structure $\Omega_d^{\Spin}$. A quantum field theory with fermions always has a $\Z_2$ symmetry operator known as fermion parity $(-1)^F$, that assigns $-1$ to a fermion and $1$ to a boson. The holonomies of this $\Z_2$ gauge field classify spin structures. If we then consider a $\UU(1)$ gauge bundle that is completely independent of fermion parity, we would consider $\Omega_d^{\Spin}(B\UU(1))$. However, if we decide to identify the $-1$ in $\UU(1)$ with the $-1$ in fermion parity, we obtain a spin$^c$-structure,  a spin structure twisted by a line bundle: \begin{equation}
    \Spin^c(n) = \frac{\Spin(n) \times \UU(1)}{\Z_2}\,,
\end{equation}
where the $\Z_2$ acts as $(x,y) \sim (-x, -y)$.
\end{itemize}

\subsection{Spin Bordism}\label{sec:spinbord}
To put in context what we discussed in the previous section and provide specific examples, we now turn to a more detailed discussion of spin bordism groups $\Omega_d^{\Spin}(BG)$, which describe equivalence classes of $d$-dimensional spin manifolds equipped with a $G$-bundle under the bordism equivalence relation described in Figure \ref{fig:17}.

\subsubsection{Spin bordism of the Point}
Let us first consider the bordism theory of spin manifolds with trivial $G$-bundle: the spin bordism groups of the point $\Omega_d^{\Spin}(\pt)$ (recall the point is the classifying space of the trivial group). The bordism groups and generators for the first few degrees are given by:

\begin{figure}[h!]
    \centering
    \begin{tabular}{c c c c c c c}
        \toprule
        $d$ & 0 & 1 & 2 & 3 & 4 & 5 \\
        \midrule
        $\Omega_d^{\Spin}(\pt)$ 
            & $\mathbb{Z}$ 
            & $\mathbb{Z}_2$ 
            & $\mathbb{Z}_2$ 
            & $0$ 
            & $\mathbb{Z}$ 
            & $0$ \\
        gen. 
            & $\pt_+$ 
            & $S^1_{\mathrm{p}}$ 
            & $S^1_{\mathrm{p}} \times S^1_{\mathrm{p}}$ 
            & $-$ 
            & K3
            & $-$ \\
        \bottomrule
    \end{tabular}
    \caption{Spin bordism groups of a point for degrees $d = 0,\dots,5$ and representative generators.} 
    \label{fig:spin-bordism}
\end{figure}
\noindent  We see that there are various entries that are non-zero. As explained above, this is the hallmark of a \textit{generalized} homology theory: spin bordism groups detect non-trivial topology in degrees $d\geq 1$, even for a point. Let us go through them one by one and try to understand what these non-trivial classes are.

\begin{figure}
    \centering
    \includegraphics[width=0.7\linewidth]{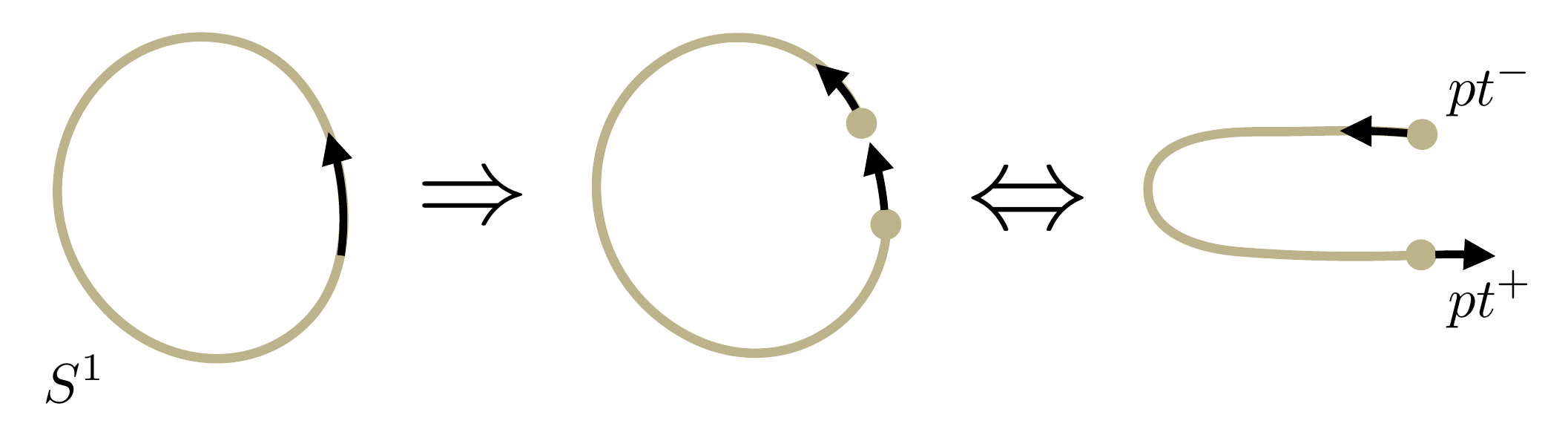}
    \caption{Cutting open a circle (with its intrinsic orientation), we obtain a line bounded by two points with opposite orientation. Deforming this, we obtain the figure on the right which shows how the disjoint union of these two points is the boundary of a 1d manifold (the line). The negative oriented point is therefore the inverse of the positively oriented one.}
    \label{fig:omega0}
\end{figure}

\begin{figure}
    \centering
    \includegraphics[width=0.7\linewidth]{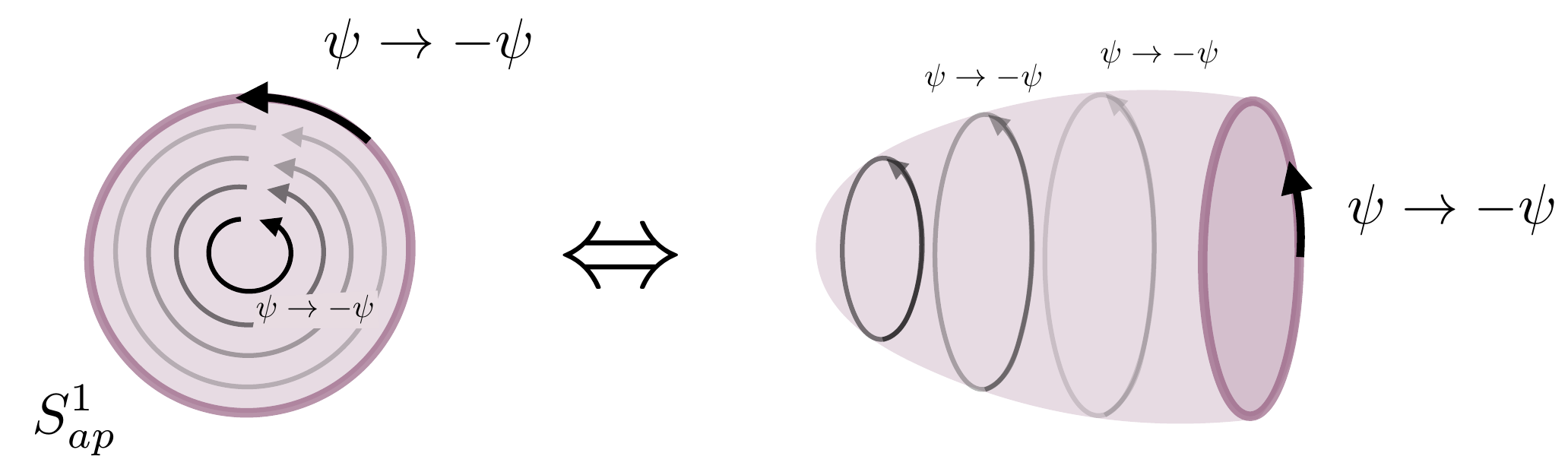}
    \caption{ The disc admits a single spin structure: the one by which fermions acquire a minus sign after going around a small loop. The circle with anti-periodic boundary conditions can therefore be the boundary of a disc. Therefore, we can deform this into the figure on the right, which shows that $S^1_{\mathrm{ap}}$ is trivial in spin bordism, since it can be bounded by a 2d spin manifold.}
    \label{fig:omega1}
\end{figure}

\begin{figure}
    \centering
    \includegraphics[width=0.7\linewidth]{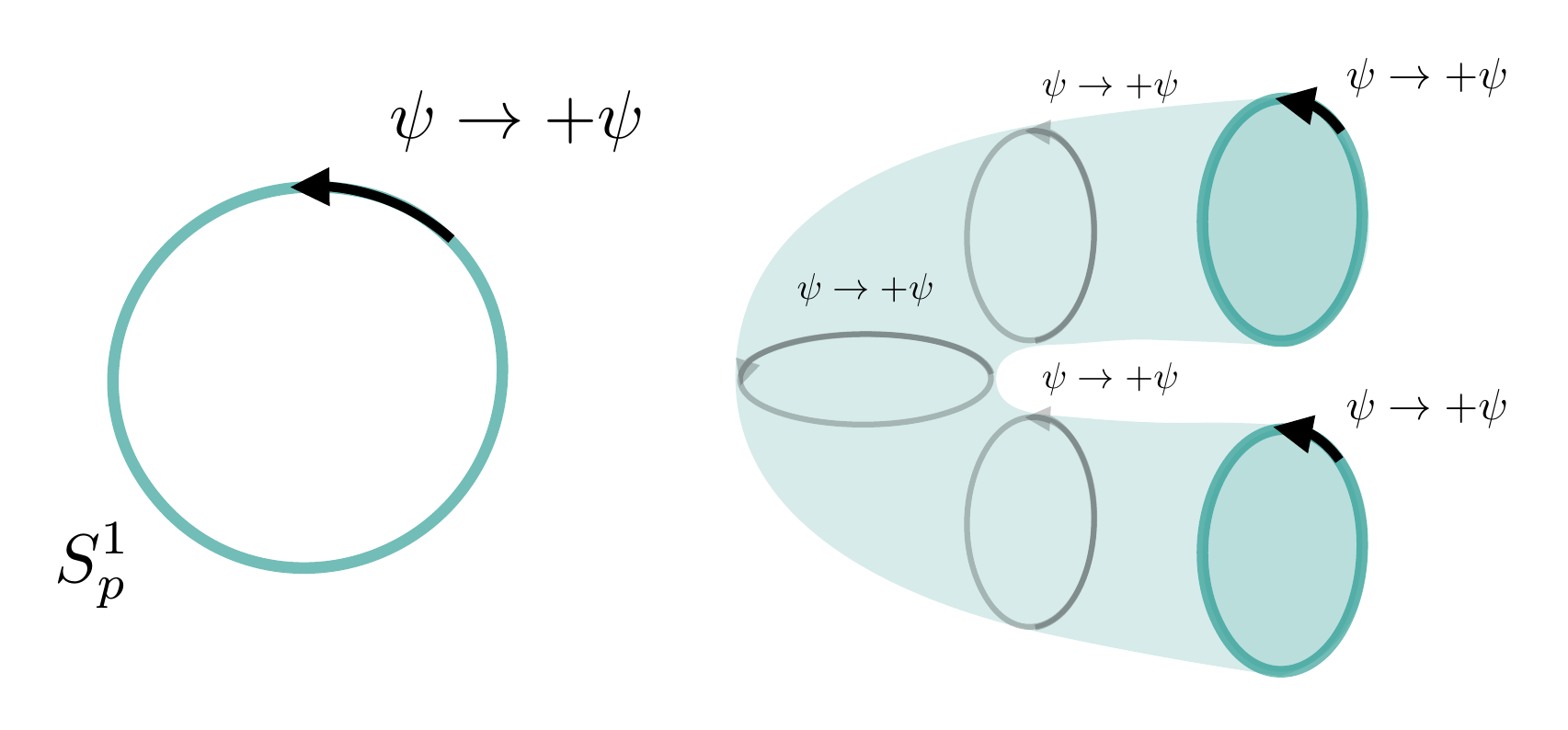}
    \caption{The circle with periodic boundary conditions cannot be the boundary of any 2d spin manifold. Nevertheless, the disjoint union of two copies of it is trivial in spin bordism as they can be connected by a cylinder with a periodic spin structure, showing that $S^1_p$ generates an order 2 element in spin bordism. }
    \label{fig:omega2}
\end{figure}

\begin{itemize}[leftmargin=*]
  \item \textbf{For degree} $d=0$: the spin bordism group is generated by the positively oriented point. It seems odd to assign an orientation to a point, since a point has no directions to rotate. To clarify, imagine we cut open an oriented circle to obtain a segment. The points at the end of this segment will have opposite orientation: one pointing outwards, and one pointing inwards. The fact that these two points bound a 1d spin manifold means that the negatively oriented point is just the inverse of the positively oriented one. The positively oriented point is therefore the generator of $\Omega_0^{\Spin}(\pt) = \Z$. This is depicted in Figure \ref{fig:omega0}.

  \item \textbf{For degree} $d=1$: the only closed manifold at our disposal is a circle. A circle can have two spin structures: one such that the fermions pick up a minus sign as one winds around the circle $S^1_{\mathrm{ap}}$ (anti-periodic (NS) boundary conditions) and one such that the fermions remain unchanged as one goes around the circle $S^1_{\mathrm{p}}$ (periodic (R) boundary conditions). To determine which of these two corresponds to a non-trivial class in bordism, let us consider the 2-dimensional spin manifolds that can be bounded by our circle. Let us try to ``fill in'' our circle: we obtain a disc. A disc admits a single spin structure, which is the usual one, where fermions pick up a minus sign as one takes them along a loop inside the disc. Imagine making this loop larger and larger such that it is almost on top of the circular boundary of the disc. As the fermions move around this large loop, they pick up a minus sign, due to the spin structure on the disc. It becomes clear that the circle with NS boundary conditions is the boundary of a disc with a spin structure. Therefore, $S^1_{\mathrm{ap}}$ is trivial in spin bordism, since it is the boundary of a 2d spin manifold. This is shown in Figure \ref{fig:omega1}. We are therefore left with the circle with R boundary conditions $S^1_{\mathrm{p}}$ for the fermions, which generates the non-trivial spin bordism class of degree 1. Indeed, there is no analog of the disc for this circle: there is no spin manifold that can fill it. The reason why the group is a $\Z_2$ is because two circles with periodic boundary conditions \textbf{can} together bound a 2d spin manifold. When you have two such circles, you aren't trying to fill them in with disks. Instead, you are connecting them with a cylinder $S^{1}\times I$. The cylinder has two boundary components, both with periodic boundary conditions. You can define a consistent spin structure on the entire cylinder where every cross-section $S^{1}\times \{t\}$ is a circle with periodic boundary conditions. The spin structure is simply constant along the cylinder: fermions are periodic around every circle $S^1\times\{t\}$, and no sign flip ever has to be introduced or removed from one end to the other. This is depicted in Figure \ref{fig:omega2}. The bordism invariant that detects this is the mod-2 index of the Dirac operator, which assigns $1$ to $S^1_{\mathrm{p}}$ and $0$ to $S^1_{\mathrm{ap}}$.
  
  \item \textbf{For degree} $d=2$: we have to consider 2d closed spin manifolds. A closed, oriented 2d surface of genus $g$ has $2^{2g}$ distinct spin structures, corresponding to choosing whether fermions transform with a $+1$ or a $-1$ around each of the $2g$ fundamental 1-cycles (two for each handle). Let's consider the torus $g=1$. The torus admits 4 different spin structures which we can denote $S^1_{\mathrm{p}}\times S^1_{\mathrm{p}}$, $S^1_{\mathrm{ap}}\times S^1_{\mathrm{p}}$, $S^1_{\mathrm{p}}\times S^1_{\mathrm{ap}}$ and $S^1_{\mathrm{ap}}\times S^1_{\mathrm{ap}}$. Following the same reasoning as above, any time we have an $S^1_{\mathrm{ap}}$, we can fill it in with a disk and the resulting ``full'' torus provides the 3d spin manifold bounded by the corresponding 2-torus. We are therefore left with $S^1_{\mathrm{p}}\times S^1_{\mathrm{p}}$, for the same reasons as the 1d case above; we cannot fill up any of the two fundamental cycles of this spin manifold with a disk, and so it constitutes a generator of $\Omega_2^{\Spin}(\pt)$. The bordism invariant in this case is called the \hl{Arf invariant}. To construct it, consider a symplectic basis of cycles $\{a_i,b_i\}^g_{i=1}$ on your genus $g$ surface. Let $q(a_i)$ be a $\mathbb{Z}_2$-valued function that assigns $1$ to cycles that have R spin structure, and $0$ to those with NS spin structure. Then the Arf invariant of a genus $g$ spin surface $\Sigma_g$ is given by:
  \begin{equation}
     \text{Arf}( \Sigma_g) = \sum_{i=1}^g q(a_i)q(b_i) \mod 2\,.
  \end{equation}
  It is clear that $\text{Arf}(S^1_{\mathrm{p}}\times S^1_{\mathrm{p}}) = 1\times 1 = 1$. This proves that $S^1_{\mathrm{p}}\times S^1_{\mathrm{p}}$ is a generator. As before, the disjoint union of this manifold with itself is a boundary, hence $\Omega_2^{\Spin}(\pt) = \Z_2$. 
  There are infinitely many equally good choices of generator with $g>1$: any genus-$g$ spin surface with Arf invariant $1$ is bordant to $S^1_{\mathrm{p}}\times S^1_{\mathrm{p}}$ \cite{KirbyTaylor:1990}.\footnote{In order to see this, note first that a genus-$g$ surface is a connected sum $\Sigma_g = T_1^2\#T_2^2\#\cdots\#T_g^2$. Since the connected sum $M\# N$ of two spin manifolds is spin bordant to $M\sqcup N$ (the bordism is obtained by attaching a 1-handle to $(M\sqcup N)\times[0,1]$), the bordism class of the surface can be expressed as $  [\Sigma_g] = \sum_{i=1}^g\, [T_i^2] = n_{\rm pp} \cdot [S_{\rm p}^1\times S_{\rm p}^1]\,,$  where $n_{\rm pp}$ is the number of RR torii ($T_i^2 = S_{\rm p}^1\times S_{\rm p}^1$) in $\Sigma_g$. Now, the Arf invariant is additive under connected sums (the union of the bases of cycles of $M$ and $N$ is a basis of cycles for $M\# N$), meaning that $  \operatorname{Arf}(\Sigma_g) = \sum_{i=1}^g \operatorname{Arf}(T_i^2) = n_{\rm pp} \,.$ It then follows that $[\Sigma_g] = [S_{\rm p}^1\times S_{\rm p}^1] \iff \operatorname{Arf}(\Sigma_g)=1$, in which case $\Sigma_g$ generates $\Omega_2^{\Spin}(\pt)$.} This is depicted in Figure \ref{fig:omega3}.
  
  \paragraph{Interlude on the Arf invariant.}We stop here to give a little bit more information on this invariant that we have just introduced. In particular, we want to better understand how one can define the map $q(a_i)$. As alluded to above, a spin structure amounts to a choice of boundary conditions around each 1-cycle. For this reason, a spin structure assigns a $0$ or a $1$ to every loop, and this assignment descends to a function on $H_{1}(\Sigma_g,\mathbb{Z}_{2})$, the mod-$2$ homology of a Riemann surface of genus $g$. On this space, we can add cycles together $x+y$ with mod $2$ coefficients (in the homological sense, i.e., as formal sums of cycles, not by concatenating paths), and we have a bilinear pairing $(x,y)$ which tells you whether the two cycles intersect. In our previous symplectic basis,  $(a_i,b_j) = \delta _{ij}$. How can we then define the desired map $q: H_1(\Sigma_g; \Z_2)\to \Z_2$? Clearly, if we want $q$ to give us $1$ for cycles that have R spin structure and $0$ to those with NS spin structure, it must be that:
  \begin{equation}\label{eq:quadrule}
      q(x + y) = q(x) + q(y) + (x,y) \;.
  \end{equation}
  Indeed, for two cycles that do not intersect, $q$ is simply additive mod 2, so combining two cycles with R spin structure gives one with NS spin structure; when the cycles intersect, resolving each intersection point to obtain an embedded curve contributes an additional unit. For this reason, $q$ is often called a quadratic form (more precisely, a quadratic refinement of the intersection form), since it follows the quadratic rule \eqref{eq:quadrule}. Note that $q$ is not linear, so spin structures are not canonically elements of $H^1(\Sigma_g;\Z_2)$; rather, they form an affine space over it. For a closed oriented surface, spin structures are in one-to-one correspondence with such quadratic forms \cite{Atiyah:1971,Johnson:1980}.

  \item \textbf{For degree} $d=3$: the bordism group is trivial. In other words, every 3d spin manifold bounds a 4d spin manifold. This may seem surprising: extrapolating from the last two examples, can't we argue that $S^1_{\mathrm{p}}\times S^1_{\mathrm{p}} \times S^1_{\mathrm{p}} $ cannot be the boundary of a spin manifold? Indeed, the 4-manifold $\mathrm{Disc}\times S^1_{\mathrm{p}} \times S^1_{\mathrm{p}}$ can only fill in 3-tori that have at least one cycle with NS spin structure, but as it turns out there is another 4d spin manifold whose boundary \textbf{can} be $S^1_{\mathrm{p}}\times S^1_{\mathrm{p}} \times S^1_{\mathrm{p}} $. It is obtained by filling in one of the circles and letting the remaining torus fiber degenerate at 12 isolated points in a way that the whole manifold remains smooth; see e.g.~\cite{GarciaEtxebarria:2020xsr,Scorpan2005TheWW} for this construction (the complement of a neighbourhood of a regular fibre in the rational elliptic surface). This manifold is sometimes called half of a K3 surface, because gluing two of them along their $S^1_{\mathrm{p}}\times S^1_{\mathrm{p}} \times S^1_{\mathrm{p}} $ boundary gives exactly the K3 surface. It follows that all subsequent tori $\mathbb{T}^n$ with R boundary conditions on all cycles will be trivial in bordism, since we can always write them as $S^1_{\mathrm{p}}\times S^1_{\mathrm{p}} \times S^1_{\mathrm{p}} \times \mathbb{T}^{n-3}$ and ``fill in'' $S^1_{\mathrm{p}}\times S^1_{\mathrm{p}} \times S^1_{\mathrm{p}}$ with half a K3.

    \item \textbf{For degree} $d=4$: the spin bordism group is $\Z$, and is generated by the $\mathrm{K3}$ surface. There is no easy way to see this. The simple fact is that the bordism invariant is in $d=4$ given by the signature: 
    \begin{equation}
       \sigma (M_4) =  \frac{1}{3}\int_{M_4} p_1 = -\frac{1}{24 \pi ^2}\int_{M_4} \mathrm{tr} (R\wedge R) \;,
    \end{equation}
    where $p_1$ is the first Pontryagin number. The index of the Dirac operator is in turn given by $-\sigma /8$. Rokhlin's theorem \cite{Rokhlin_52} says that all 4d spin manifolds have a signature that is a multiple of 16.  It so happens that $\sigma(\mathrm{K3}) = -16$ and so in that sense K3 is the minimal choice and \textbf{the} generator of $\Omega_4^{\Spin}(\pt)$. In other words, $\sigma/16$ defines an isomorphism $\Omega_4^{\Spin}(\pt) \cong \Z$, so $\Omega_4^{\Spin}(\pt)$ is generated by a manifold with signature $\pm16$, such as K3. 
\end{itemize}

\begin{figure}
    \centering
    \includegraphics[width=0.5\linewidth]{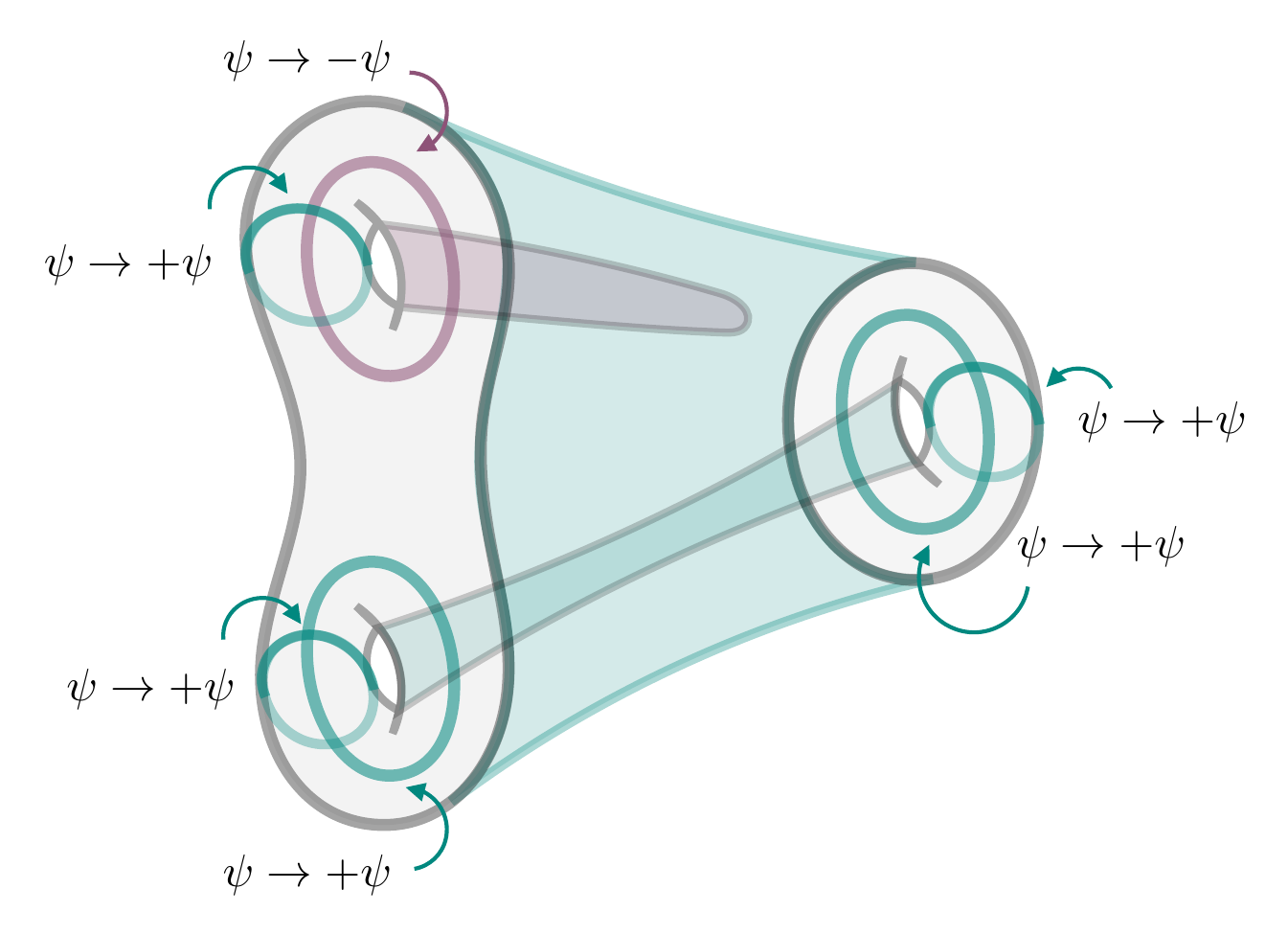}
    \caption{A schematic depiction of the fact that any genus-$g$ spin surface with Arf invariant $1$ is bordant to $S^1_{\mathrm{p}} \times S^1_{\mathrm{p}}$.}
    \label{fig:omega3}
\end{figure}

With these examples in hand, we hope to have built some intuition surrounding bordism groups and their invariants. Even if somewhat abstract, the above discussion has important physical implications when applied to spacetime backgrounds in the presence of fermions: topologies that correspond to nontrivial spin bordism classes are protected under smooth spacetime deformations. This establishes a selection rule for dynamical processes in a candidate theory of quantum gravity. In contrast with the local formulation of conventional symmetries, in the case of bordism invariants the obstruction is coming from global topological features of the spacetime manifold. As we will see in Section~\ref{sec:coco}, this ``conserved'', ``irremovable'' topology in the background gives rise to a (higher-form) global symmetry. Before that, now that we know $\Omega^{\Spin}_d(\pt)$, let us briefly sketch the tools involved in computing $\Omega^{\Spin}_d(BG)$.

\subsubsection{Computing Spin Bordism Groups $\Omega_d^{\Spin}(BG)$}\label{sec:spectralsequences} Bordism groups can be computed from tools in algebraic topology such as the \hl{Atiyah-Hirzebruch spectral sequence} (AHSS) or the Adams spectral sequence, which apply to generalized homology theories such as bordism (see Section~\ref{sec:bordismgeneralities}). A proper description of these tools is well beyond the scope of these lectures;\footnote{For an introduction to spectral sequences see for instance \cite{McCleary:2001}, and for physics-oriented treatments of the AHSS for spin bordism see \cite{Garcia-Etxebarria:2018ajm,Debray:2023yrs}.} thus here we will limit ourselves to discussing a few basic examples that illustrate the general philosophy behind these methods, focusing on AHSS. 

Previously, we discussed the spin bordism groups of the point. Those correspond to bordism equivalence classes of manifolds on which fermions can propagate (spin manifolds). They are in some sense purely gravitational, as they only involve the topology (and spin structure) of spacetime itself, and no gauge fields. We now want to consider spacetime backgrounds with a non-trivial gauge background for some gauge symmetry group $G$. For now there is nothing dynamical, it is pure topology.
As sketched in section \ref{sec:bordismgeneralities}, this comes down to considering pairs $(X_d,f)$, where $X_d$ is a spin manifold and $f:X_d\rightarrow BG$ is a map into the classifying space of $G$. This map $f$ is exactly the same thing as a $G$-bundle over $X_d$. Two pairs $(X_d^{1},f_{1})$ and $(X_d^{2},f_{2})$ are bordant if there exists a $(d+1)$-dimensional spin manifold $W_{d+1}$ and a map $F:W_{d+1}\rightarrow BG$ such that the boundary of $W_{d+1}$ is the union of $X_d^{1}$ and $X_d^{2}$, and the map $F$ restricts to $f_{1}$ and $f_{2}$ on the ends. As we will see, under this new bordism equivalence relation, there will be new non-trivial classes on top of the gravitational ones we discussed in the previous section. 

The point is that there are two sources of data: the topology of the spin manifold itself (the bordism invariants of which are encoded in $\Omega^{\Spin}_d(\pt)$) and the topology coming from the gauge bundle $f: X_d\to BG$ (dictated by the homology of $BG$). The question is to determine how these two kinds of data combine, and which combinations actually survive the aforementioned bordism relation.

At the most naive level, a $d$-manifold mapping into $BG$ determines a homology class in $H_d(BG;\Z)$. If one ignored the presence of spin structures, it would be tempting to identify bordism classes with ordinary homology classes. This viewpoint captures only the ``shape'' of how manifolds sit inside $BG$, but completely misses the additional constraints and invariants imposed by fermions. A better approximation is to think of a spin bordism class as a homology cycle in $BG$ that is \emph{decorated} by intrinsic spin bordism data living on the manifold itself. At this stage, many combinations of cycles and decorations appear to be allowed. The AHSS can be understood as a systematic refinement of this naive homological picture. Indeed, not every such decorated cycle can arise from a genuine spin manifold mapping into $BG$. When attempting to glue local data into a global spin manifold, the spin structure may fail to extend, even though the underlying homology class is perfectly well defined. The role of the AHSS is to identify and remove precisely those decorated cycles that are obstructed in this way. What remains after all such consistency conditions are imposed is exactly the true spin bordism group $\Omega^{\Spin}_d(BG)$. {Bordism therefore captures the ordinary homology of $BG$, and more}: it also records the intrinsic spin bordism data carried by the manifolds and how the two are glued together. 

The AHSS for spin bordism starts by writing down the \hl{``second page''}, which contains all of the data of the homology of $BG$ and the spin bordism groups of the point:\footnote{There is also a first page, $E^1_{p,q} = C_p(BG;\Omega_q^{\Spin}(\pt))$, given by cellular chains of $BG$ with coefficients in $\Omega_q^{\Spin}(\pt)$, whose differential $d^1$ is the cellular boundary map. Taking its homology gives the $E^2$ page. In practice, one always starts from $E^2$.}
\begin{equation}
E^2_{p,q}
\;=\;
H_p\!\left(BG;\Omega_q^{\Spin}(\mathrm{pt})\right).
\end{equation}
One may visualize the $E^2$ page as a grid indexed by $(p,q)$, where the horizontal direction $p$ records the homological degree in $BG$ (roughly, the dimension of the cycle that a manifold is mapped onto), and the vertical direction $q$ records intrinsic spin bordism data. Schematically, the $E^2$ page has the form

\begin{equation}
\renewcommand{\arraystretch}{1.25}
\begin{array}{@{}c cccc@{}}
\toprule
\addlinespace[-0.1ex]
q \backslash p
  & 0
  & 1
  & 2
  & \cdots
\\
\midrule
\vdots
  & \vdots
  & \vdots
  & \vdots
  & \ddots
\\
2
  & \Omega^{\mathrm{Spin}}_2(\mathrm{pt})
  & H_1\bigl(BG;\Omega^{\mathrm{Spin}}_2(\mathrm{pt})\bigr)
  & H_2\bigl(BG;\Omega^{\mathrm{Spin}}_2(\mathrm{pt})\bigr)
  & \cdots
\\
1
  & \Omega^{\mathrm{Spin}}_1(\mathrm{pt})
  & H_1\bigl(BG;\Omega^{\mathrm{Spin}}_1(\mathrm{pt})\bigr)
  & H_2\bigl(BG;\Omega^{\mathrm{Spin}}_1(\mathrm{pt})\bigr)
  & \cdots
\\
0
  & \Omega^{\mathrm{Spin}}_0(\mathrm{pt})
  & H_1\bigl(BG;\Omega^{\mathrm{Spin}}_0(\mathrm{pt})\bigr)
  & H_2\bigl(BG;\Omega^{\mathrm{Spin}}_0(\mathrm{pt})\bigr)
  & \cdots
\\
\bottomrule
\end{array}
\end{equation}
The $E^2$ page makes all potential contributions to the spin bordism group
$\Omega^{\Spin}_{d}(BG)$ visible at once. 
Indeed, for fixed total degree $d=p+q$, the groups $E^2_{p,q} $ lie along a diagonal in the $(p,q)$-plane, and collectively encode all ways in which a $d$-dimensional spin manifold may contribute to bordism data. Reading along such a diagonal therefore lists all \emph{a priori} possible contributions to an element of $\Omega^{\Spin}_{d}(BG)$.

However, this diagonal data is only provisional. 
The entries of the $E^2$ page represent candidate bordism classes, not guaranteed ones.
As explained above, the purpose of the Atiyah--Hirzebruch spectral sequence is precisely to \hl{eliminate those combinations of data that cannot arise from a genuine spin manifold}.
This elimination is implemented through a sequence of differentials.

Starting on the second page, there are differentials
\begin{equation}
d_2 \colon E^2_{p,q} \longrightarrow E^2_{p-2,\,q+1} \;.    
\end{equation}
A class that lies in the image of $d_2$ disappears on the third page, reflecting the fact that it is obstructed and does not define a true bordism invariant.
The resulting $E^3$ page therefore contains fewer elements than the $E^2$ page, and provides a better approximation to the spin bordism group.
However, the process does not stop there.
The $E^3$ page itself carries further differentials, and iterating this procedure produces a sequence of increasingly refined approximations.
At each stage, additional incompatible data is removed until the spectral sequence stabilizes and the remaining classes assemble into the true spin bordism group. The differentials on the $r$-th page of the Atiyah--Hirzebruch spectral sequence act on 
\begin{equation}
d_r \colon E^r_{p,q} \longrightarrow E^r_{p-r,\,q+r-1} \;.    
\end{equation}
Some of the differentials in the Atiyah--Hirzebruch spectral sequence are known explicitly.
For example, the first potentially nontrivial differential $d_2$ can often be described in cohomological terms as a reduction modulo $2$ combined with the dual of the second Steenrod square. Explicitly, $d_2\colon E^2_{p,1}=H_p(BG;\Z_2)\to E^2_{p-2,2}=H_{p-2}(BG;\Z_2)$ is the dual of $\mathrm{Sq}^2$, while $d_2\colon E^2_{p,0}=H_p(BG;\Z)\to E^2_{p-2,1}=H_{p-2}(BG;\Z_2)$ is reduction mod 2 composed with the dual of $\mathrm{Sq}^2$ \cite{Garcia-Etxebarria:2018ajm}. This allows one to determine $d_2$ in many concrete cases. Beyond the second page, however, the situation becomes significantly more complicated. The higher differentials correspond to higher-order cohomological operations, whose explicit form is generally difficult to identify and depends on subtle global information.
As a result, these \hl{differentials are rarely known in full generality}.
In practice, one often restricts attention to low-dimensional bordism groups, where degree considerations force the higher differentials to vanish automatically, or to situations in which symmetry or simplicity of the space $BG$ severely constrains the possible differentials.

When all differentials have been taken into account, the spectral sequence stabilizes at a \hl{final page}, denoted $E^\infty$.
At this stage, no further classes are removed, and the surviving entries represent all bordism data that is free of obstructions.
Nevertheless, the $E^\infty$ page does not yet give the spin bordism group itself in a completely explicit form.
Rather, it determines $\Omega^{\Spin}_d(BG)$ only up to extension.

Indeed, the AHSS does not directly produce the spin bordism group $\Omega^{\Spin}_d(BG)$ as a single object.
Instead, it computes a \hl{filtration} of this group. By a filtration of an abelian group $A$ one means a nested sequence of subgroups
\begin{equation}
0 = F_{-1}A \subset F_0A \subset F_1A \subset \cdots \subset F_dA = A,
\end{equation}
which organizes the elements of $A$ according to increasing levels of complexity.
The associated graded pieces
\begin{equation}
\operatorname{gr}_p A := F_pA / F_{p-1}A
\end{equation}
record the genuinely new information that appears at each stage of the filtration.
Knowing all graded pieces determines the group $A$ only up to extension, since different groups may share the same associated graded structure. A simple example illustrates why knowledge of a filtration does not uniquely determine a group.
Consider the abelian groups $\mathbb{Z}_4$ and $\mathbb{Z}_2 \oplus \mathbb{Z}_2$.
Both admit a filtration
\begin{equation}
0 \subset F_0 A\subset F_1A = A,
\qquad
F_0 A\cong \mathbb{Z}_2,  
\end{equation}
such that the associated graded pieces are
\begin{equation}
\operatorname{gr}_0 A \cong \mathbb{Z}_2,
\qquad
\operatorname{gr}_1 A \cong \mathbb{Z}_2.    
\end{equation}
From the perspective of the associated graded group, these two possibilities are indistinguishable.
Nevertheless, the resulting groups are not isomorphic: in $\mathbb{Z}_4$ the two $\mathbb{Z}_2$ layers are nontrivially glued together, while in $\mathbb{Z}_2 \oplus \mathbb{Z}_2$ they remain independent. More specifically, $\Z_4$ has a single subgroup of order 2, whereas $\Z_2 \oplus \Z_2$ has three. This distinction is precisely what is meant by an \hl{extension problem}.

The $E^\infty$ page describes the graded pieces associated to a natural filtration on $\Omega^{\Spin}_d(BG)$.
The groups appearing along the diagonal $p+q=d$ record how much new bordism data enters at each stage of the filtration, but they do not specify how these pieces fit together into a single abelian group.
Different ways of assembling the same graded pieces can lead to an extension problem. Concretely,
\begin{equation}\label{eq:E^infty-page}
E^\infty_{p,q}
\;\cong\;
\frac{F_p \Omega^{\Spin}_{p+q}(BG)}
     {F_{p-1} \Omega^{\Spin}_{p+q}(BG)}.
\end{equation}
For fixed total degree $d=p+q$, the groups $E^\infty_{p,q}$ therefore describe how $\Omega^{\Spin}_d(BG)$ is built up from successive layers of bordism data.
Reconstructing the full group from these layers may involve extension ambiguities, which are not resolved by the spectral sequence itself. In favourable cases, such as when all groups involved are free abelian or when degree considerations force a unique outcome, no nontrivial extensions can occur, and the $E^\infty$ page determines the bordism group uniquely:
\begin{equation}
\Omega^{\Spin}_{d}(BG)
\;\cong\;
\bigoplus_{p+q=d} E^\infty_{p,\,q}.
\end{equation}
In general, however, there will be extension problems. Resolving these ambiguities typically requires identifying \hl{explicit generators} of $\Omega^{\Spin}_d(BG)$ and determining their orders.

\paragraph{Example: $G=\mathbb{Z}$.}
As a simple illustrative example, consider the case $BG = B\mathbb{Z}=  S^1$. The relevant homology groups of the circle are given by
\begin{alignat}{2}
H_0(S^1;\mathbb{Z})   &\cong \mathbb{Z}\,,   \qquad & H_1(S^1;\mathbb{Z})   &\cong \mathbb{Z}\,,\\
H_0(S^1;\mathbb{Z}_2) &\cong \mathbb{Z}_2\,, \qquad & H_1(S^1;\mathbb{Z}_2) &\cong \mathbb{Z}_2\,,
\end{alignat}
with all higher homology groups vanishing.
The second page of the Atiyah--Hirzebruch spectral sequence for spin bordism is therefore
\begin{equation}
E^2_{p,q} = H_p(S^1;\Omega_q^{\Spin}(\mathrm{pt}))\,.
\end{equation}
Since only $p=0,1$ contribute, the $E^2$ page consists of two nonzero columns:
\begin{equation}
E^2_{0,q} \cong \Omega_q^{\Spin}(\mathrm{pt})\,,
\qquad
E^2_{1,q} \cong \Omega_q^{\Spin}(\mathrm{pt})\,,
\end{equation}
and $E^2_{p,q}=0$ for $p\ge 2$. The $E^2$ page has the form
\begin{equation}
\begin{array}{c@{\qquad}cccc}
\toprule
\addlinespace[-0.1ex]
q \backslash p
    & 0 & 1 & 2 & \cdots \\
\midrule
\vdots
    & \vdots & \vdots & \vdots & \\
4   & \mathbb{Z}   & \mathbb{Z}   & 0 & \cdots \\
3   & 0            & 0            & 0 & \cdots \\
2   & \mathbb{Z}_2 & \mathbb{Z}_2 & 0 & \cdots \\
1   & \mathbb{Z}_2 & \mathbb{Z}_2 & 0 & \cdots \\
0   & \mathbb{Z}   & \mathbb{Z}   & 0 & \cdots \\
\bottomrule
\end{array}
\end{equation}
with all entries outside these two columns equal to zero. For degree reasons, all differentials vanish identically ---notice that there are no possible targets for any differential $d_2$ starting from the nonzero entries of the $E^2$ page.
As a result, the spectral sequence collapses at the second page, so that $E^\infty = E^2$.
Reading along the diagonal $p+q=d$, one finds
\begin{equation}\label{eq:spin-bordism-BZ-explicit}
\Omega_d^{\Spin}(S^1)
\cong
\Omega_d^{\Spin}(\mathrm{pt})
\oplus
\Omega_{d-1}^{\Spin}(\mathrm{pt}) \,.
\end{equation}
The possible extension ambiguity is resolved because the short exact sequence
\begin{equation}
  0 \,\rightarrow\, E^\infty_{0,d} = \Omega^{\mathrm{Spin}}_d(\pt) \,\xrightarrow{\iota}\, \Omega^{\mathrm{Spin}}_d(S^1) \,\xrightarrow{\pi}\, E^\infty_{1,d-1} = \Omega^{\mathrm{Spin}}_{d-1}(\pt) \,\rightarrow\, 0
\end{equation}
with
\begin{equation}
\begin{aligned}
    \iota : \Omega^{\mathrm{Spin}}_d(\pt) \;&\longrightarrow\; \Omega^{\mathrm{Spin}}_d(S^1)\\
    [M] \;&\longmapsto\; [M,\, \text{constant map to a fixed point } * \in S^1]
\end{aligned}
\end{equation}
is split. Indeed, the map
\begin{equation}
\begin{aligned}
    r : \Omega^{\mathrm{Spin}}_d(S^1) \;&\longrightarrow\; \Omega^{\mathrm{Spin}}_d(\pt)\\
    [M,f] \;&\longmapsto\; [M]
\end{aligned}
\end{equation}
is such that $r\circ\iota = \operatorname{id}$. The splitting lemma then gives \eqref{eq:spin-bordism-BZ-explicit}. The same argument works for any space $X$ with a basepoint. The bordism of a point always splits off as a direct summand, and the reduced bordism
$\widetilde{\Omega}^{\mathrm{Spin}}_d(X) = \Omega^{\mathrm{Spin}}_d(X)/\iota\big(\Omega^{\mathrm{Spin}}_d(\pt)\big)$ is built from the columns $p \geq 1$. For $S^1$ there is only one such column, so $\widetilde{\Omega}^{\mathrm{Spin}}_d(S^1)$ is a single graded piece. For spaces with more columns, extension problems can remain among the $p \geq 1$ pieces.

This example is particularly simple: there are no nontrivial differentials and no extension ambiguities, so the Atiyah--Hirzebruch spectral sequence collapses immediately. In general, however, the situation can become much more complicated. For instance, replacing $\mathbb{Z}$ by a finite group such as $\mathbb Z_2$ already leads to a substantially richer structure.

The underlying reason is that
\begin{equation}
B\mathbb Z_2 = \mathbb{RP}^{\infty}
\end{equation}
has a very large mod-$2$ cohomology ring. In fact,
\begin{equation}
H^*(\mathbb{RP}^{\infty};\mathbb Z_2) \cong \mathbb Z_2[x] \,, \qquad |x|=1\;,
\end{equation}
so there are nontrivial cohomology classes in every degree. Consequently, the $E^2$ page of the AHSS for $\Omega_*^{\Spin}(B\mathbb Z_2)$ contains many $\mathbb Z_2$ summands, and several nontrivial differentials occur. These differentials are closely related to Steenrod operations, which act on mod-$2$ cohomology and relate classes in different degrees. Computations of the groups $\Omega_d^{\Spin}(B\mathbb Z_2)$ using the AHSS appear, for example, in \cite{Garcia-Etxebarria:2018ajm}. 

In contrast, for cyclic groups $\mathbb Z_k$ with $k$ odd, the spectral sequence becomes considerably simpler. In this case the mod-$2$ cohomology of $B\mathbb Z_k$ is trivial in positive degrees, since
\begin{equation}
\widetilde{H}^*(B\mathbb Z_k;\mathbb Z_2)=0 \qquad (k \text{ odd}) \,,
\end{equation}
reflecting the fact that $\mathbb Z_2 \otimes \mathbb Z_k =0$. As a result, the $E^2$ page contains far fewer torsion classes and the AHSS is much easier to analyze. This case was also studied in \cite{Garcia-Etxebarria:2018ajm}.

More generally, the complexity of these computations is largely governed by the structure of the mod-$2$ cohomology of the space $BG$, since spin structures are closely tied to mod-$2$ phenomena.

\paragraph{Summary.}Having reviewed the basic mathematical structure of bordism groups and the type of topological information that they encode, we now return to physics. In particular, we will see that these seemingly abstract invariants play the role of detecting global symmetries in the background of quantum gravity theories. This powerful connection is made precise by the \emph{Cobordism Conjecture}, the central topic of these lectures, to which we finally turn. As we will see, the conjecture turns the abstract classification of manifolds developed in this section into sharp physical predictions: every non-trivial bordism class signals a would-be global symmetry, which quantum gravity must remove by gauging or breaking it, often by predicting new, sometimes non-supersymmetric, objects in string theory.

\section{The Cobordism Conjecture}\label{sec:coco}
In the previous section, we defined (spin) bordism groups abstractly, as equivalence classes of manifolds related by bordisms. We now turn to physics. We interpret these manifolds as compact backgrounds $X_d$ on which a quantum gravity theory is compactified, and a bordism between two of them as a physical, topology-changing process that deforms one background into the other.

\begin{figure}[h!]
\centering
\begin{subfigure}{0.5\textwidth}
  \centering
  \includegraphics[width=0.85\linewidth]{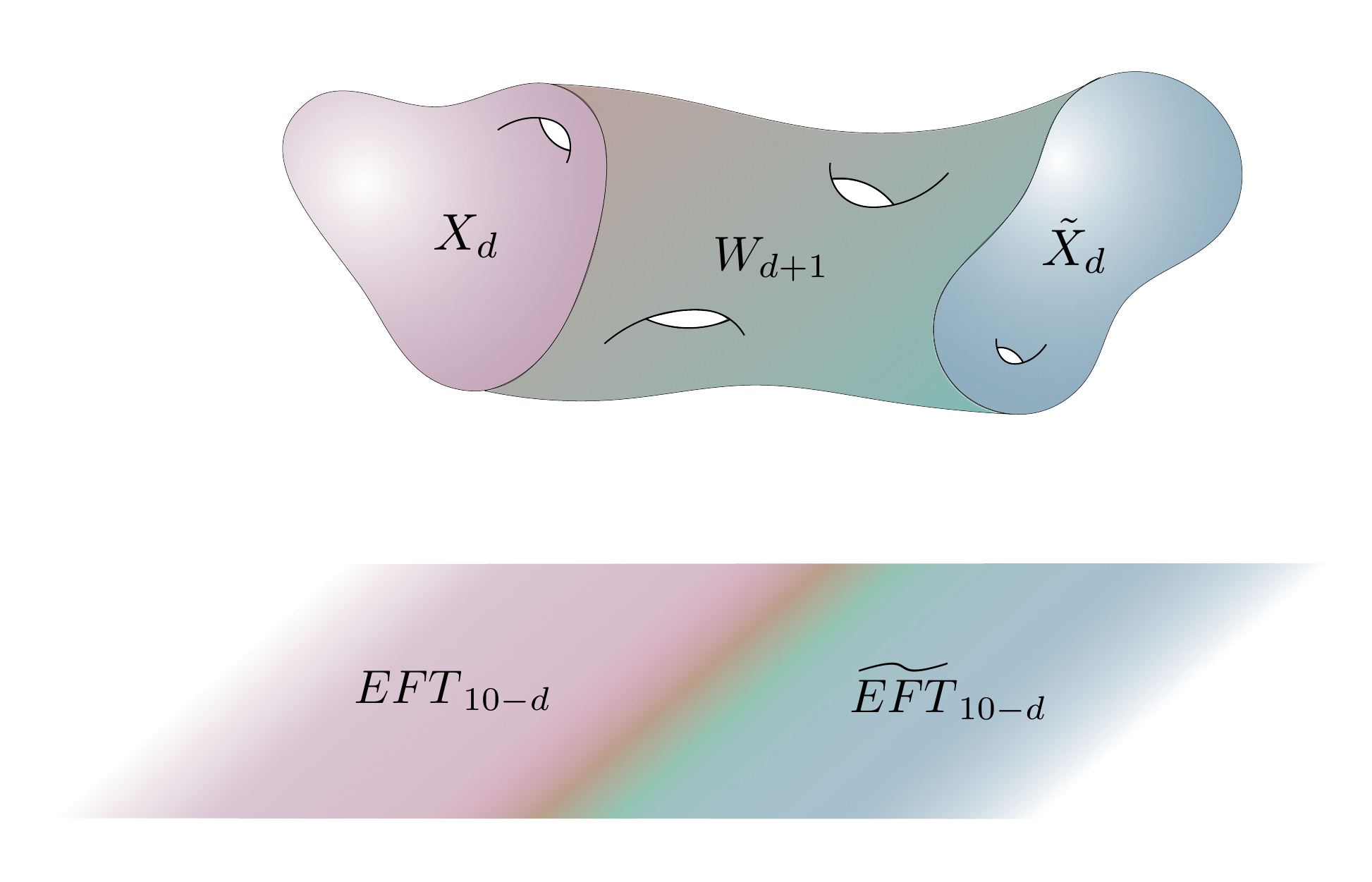}
  \caption{A bordism between two manifolds}
  \label{fig:sub1}
\end{subfigure}%
\begin{subfigure}{0.5\textwidth}
  \centering
\includegraphics[width=0.65 \linewidth]{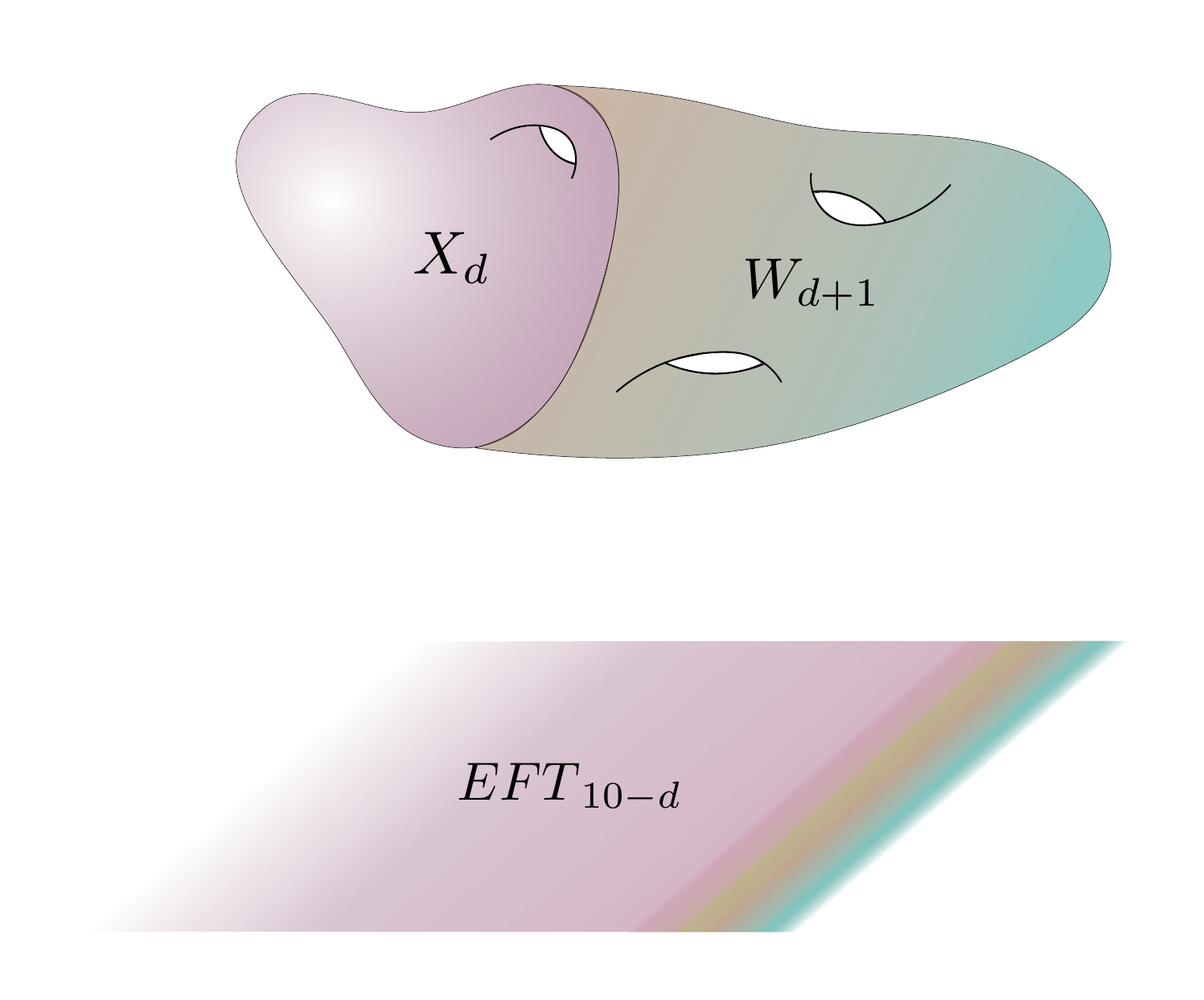}
  \caption{A bordism to nothing}
  \label{fig:sub2}
\end{subfigure}
\caption{Bordisms from the perspective of the lower dimensional EFTs one gets after compactifying on $X_d$ and $\tilde{X}_d$. (a) A bordism between $X_d$ and $\tilde{X}_d$ looks like a domain wall; (b) a bordism to nothing looks like a boundary.}
\label{fig:eftbordism}
\end{figure}

The Cobordism Conjecture \cite{McNamara:2019rup} states that such a process always exists: in quantum gravity, any background can be deformed into any other. This is a sharp formulation of \hl{background independence}, since the theory should not split into disconnected sectors labelled by the topology of the background. In the language of bordism, all $d$-dimensional backgrounds must lie in a single class, and since the empty manifold is one of them, this must be the trivial class, the class of ``nothing''. From the perspective of the lower-dimensional EFTs obtained by compactifying on $X_d$, any two such theories can then be connected by a \hl{domain wall}, as in Figure~\ref{fig:sub1}, and any such theory admits a \hl{boundary} or End-of-the-World wall, as in Figure~\ref{fig:sub2}. The conjecture can therefore be stated as:
\begin{symmetrybox}{dustyPlum}
    \textbf{The Swampland Cobordism Conjecture:}
Given a $D$-dimensional QG theory compactified on a $d$-dimensional compact manifold, all bordism classes must be trivial:
\begin{equation}\label{eq:COCO}
    \forall\, d \;\; \text{such that}\; d \leq D,\;\; \Omega _ d ^{\rm QG} = 0\,.
\end{equation}
\end{symmetrybox}

\noindent Here, $\Omega^{\rm QG}_d$ is the bordism group of quantum gravity backgrounds with every ingredient the theory contains: all the structure its backgrounds carry, all the defects it admits, and all the topology-changing processes it allows. This group is not known a priori, and determining it is precisely the aim of the program. In practice, the conjecture is applied through successive approximations to $\Omega^{\rm QG}_d$. A natural first one comes from the fact that the low-energy EFTs obtained from string theory generically contain fermions, arising as KK modes of higher-dimensional fermionic fields. Compactification manifolds must then admit a spin structure, which suggests spin bordism $\Omega_d^{\Spin}$ as a starting point. In any given approximation, a non-trivial class $[X_d]$ is a background that cannot be deformed to nothing. It carries a global topological charge, measured by the corresponding bordism invariant, which labels the background in exactly the way background independence forbids. Such a charge signals a global symmetry, which must be either gauged, meaning the backgrounds carry additional structure, or broken by a defect, a new object that the theory must contain and whose inclusion trivializes the class. Either way, a non-trivial class tells us that the approximation is too coarse, and the conjecture predicts the new ingredients needed to trivialize it:
\begin{equation}
    \Omega_d^{\rm approx.} \neq 0 \quad\Longrightarrow\quad \Omega_d^{\rm approx.\,+\,new\ ingredients} = 0\,.
\end{equation}
Step by step, each new ingredient brings us closer to the full set of structures, defects and processes of quantum gravity, and the sequence of approximations is meant to asymptote to $\Omega^{\rm QG}_d$.\footnote{For a possible organizing principle of this sequence of refinements see \cite{Blumenhagen:2022bvh}.} Before discussing the consequences of the conjecture, let us make precise why a non-trivial bordism class signals a global symmetry in the language of Section~\ref{sec:GlobalSYms101}.

\paragraph*{From bordism classes to global symmetries.} Consider a $D$-dimensional theory whose backgrounds carry a $\xi$-structure, and suppose that $\Omega^\xi_d\neq0$, detected by a bordism invariant $I$ (an integer, an integer mod $n$, or more generally a phase such as an $\eta$-invariant, see Section~\ref{sec:bordismgeneralities}).
\begin{itemize}
    \item \textbf{Symmetry operators.} For any closed $d$-dimensional submanifold $M_d$ of spacetime, equipped with the structure induced from spacetime, we can evaluate the invariant on the background restricted to $M_d$ and define
    \begin{equation}
        U_\alpha(M_d) = \exp\left(i\alpha\, I[M_d]\right)\,.
    \end{equation}
    If $M_d$ is deformed into $M'_d$ without crossing any defect, the region swept out is precisely a bordism between them, so $I[M_d]=I[M'_d]$: $U_\alpha$ is \emph{topological}. It is also invertible, and it composes according to the group law of the Pontryagin dual of $\Omega^\xi_d$ ($\mathrm{U}(1)$ for a $\Z$ summand, $\Z_n$ for a $\Z_n$ summand). By construction, $I[M_d]$ is non-trivial if $M_d$ is bordant to the generator $X_d$ of $\Omega^\xi_d$.
    \item \textbf{Form degree.} $U_\alpha$ is supported on a submanifold of dimension $d$. By the definitions of Section~\ref{ssec:higher_form}, a codimension-$(p+1)$ topological symmmetry operator generates a $p$-form symmetry, so a non-trivial $\Omega^\xi_d$ gives a \hl{$(D-d-1)$-form global symmetry}. Note that if the corresponding bordism element is $\mathbb{Z}$, then the Pontryagin dual is U(1) and $I[M_d]$ can be thought of as the integral of  $\star \mathbf{j}_{D-d}$ over $M_d$.
    \item \textbf{Charged objects.} The objects charged under this symmetry are $(D-d-1)$-dimensional defects, around which the linking $d$-dimensional manifold carries non-trivial invariant $I$. Equivalently, if the generator of $\Omega^\xi_d$, $X_d$ cannot be filled in smoothly, one can still fill it in \emph{singularly}: the transverse $(d+1)$-dimensional space of the defect is a ``cone'' over $X_d$, and the singular locus is the defect.
    \item \textbf{Breaking and gauging.} Following Section~\ref{sec:gaugebreaking}, the symmetry is \emph{broken} if these defects exist as dynamical objects, since then $U_\alpha$ is no longer topological. It is \emph{gauged} if $I$ can be trivialized by introducing a dynamical gauge field. In both cases, some new ingredient beyond the naive $\xi$-structure must be present: the defect, or the gauge field and its couplings.
\end{itemize}
As a concrete example, take $\Omega_1^{\Spin}=\Z_2$, generated by $S^1_{\rm p}$, with invariant the mod-2 index. The operator $U(M_1)=(-1)^{I[M_1]}$ is supported on closed curves and measures whether fermions are periodic around them. It is topological, of codimension $D-1$, and hence generates a $\Z_2$ $(D-2)$-form symmetry. Its charged objects are codimension-two defects called ``spin defects'' around which fermions pick up an extra sign, i.e., defects with a $(-1)^F$ monodromy; in $D=10$ they have eight-dimensional worldvolumes. From the viewpoint of the $(D-1)$-dimensional theory obtained by compactifying on a circle, the symmetry separates the sector with periodic fermions from the one with antiperiodic fermions, and its charged objects are domain walls between them. We return to this example below; see also \cite{McNamara:2019rup}.

\vspace{0.6em}

\noindent We conclude this introduction to Section \ref{sec:coco} with a few words on the case $d=D$. A closed $D$-dimensional manifold is an entire closed universe, and its bordism class is measured by a spacetime-filling topological operator \cite{Montero:2022vva}. It corresponds to a \hl{$(-1)$-form symmetry} (see Example~4 in Section~\ref{ssec:globsym_examples}). The triviality of this bordism class is closely tied to the Baby Universe Hypothesis, discussed in Section~\ref{sssec:BabyUniv}. 

\subsection{Gauging \& Breaking Symmetries}
Once a global symmetry is identified by a non-trivial bordism class, one has to determine whether the symmetry is gauged or broken in the quantum gravity theory under consideration. This is determined in a case-by-case manner, and a global symmetry could be gauged in one theory and broken in another. Following the mechanisms for gauging and breaking global symmetries introduced in Section \ref{sec:gaugebreaking}, we now describe how global symmetries are gauged or broken, with an explicit example in each case.  

\paragraph{Breaking the symmetry}
amounts to introducing a defect that is charged under this symmetry and can thus absorb the uncancelled topological charge in the compactification. The objects that are charged under the $(D-d-1)$-form global symmetry generated by this charge are $(D-d-1)$-dimensional defects.

A simple example of this was discussed in \cite{McNamara:2019rup}. Consider the one-dimensional spin bordism group
\begin{equation}
\Omega^{\Spin}_1=\mathbb{Z}_2 \,,
\end{equation}
which we just discussed. A fermion with periodic boundary conditions on $S^1$ creates a configuration that you cannot deform to a point. In the case of the ten-dimensional heterotic string theories, their compactification on the circle with periodic spin structure should be perfectly consistent, so this symmetry has to be broken by an eight-dimensional spin defect, which has yet to be identified.

\paragraph{Gauging the symmetry} amounts to introducing a dynamical $(D-d)$-form gauge field $B_{D-d}$ that couples to the conserved $(D-d)$-form current $\mathbf{j}$ of the symmetry through $\int B_{D-d}\wedge\star\mathbf{j}$, where $\star\mathbf{j}$ is the closed $d$-form whose integral defines the bordism invariant. The equation of motion of $B_{D-d}$ then makes $\star\mathbf{j}$ exact, so the bordism invariant must vanish on every compact background unless sources are present. Gauging the symmetry consistently therefore imposes new constraints on the theory.

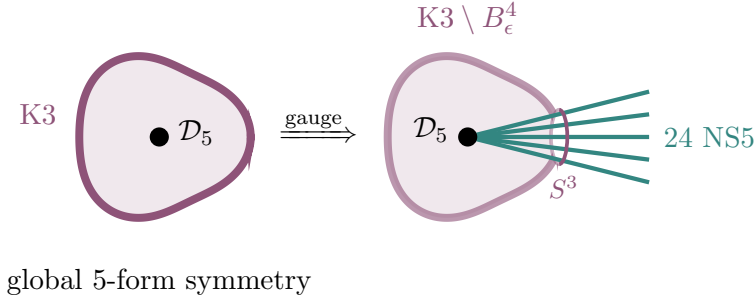
\begin{figure}[htbp]
\vspace{0.5cm}
\centering
\begin{tikzpicture}[scale=1.2]

  \def\rK{0.90}     
  \def\yB{-0.60}    
  \pgfmathsetmacro\rE{1.12*\rK}  

  \def\xL{-1.4}
  \def\xM{2.0}

  \pgfmathsetmacro\xGauge{0.5*((\xL+\rE)+(\xM-\rK))}


  \def\yL{1.0}

  \fill[dustyPlum, opacity=0.13]
    plot[domain=0:360, samples=120, smooth cycle]
    ({\xL+\rK*(1+0.07*cos(3*\x)+0.05*cos(2*\x)-0.04*sin(2*\x)*sin(\x))*cos(\x)},
     {\yL+\rK*(1+0.07*cos(3*\x)+0.05*cos(2*\x)-0.04*sin(2*\x)*sin(\x))*sin(\x)});

  \draw[dustyPlum, line width=3]
    plot[domain=0:360, samples=120, smooth cycle]
    ({\xL+\rK*(1+0.07*cos(3*\x)+0.05*cos(2*\x)-0.04*sin(2*\x)*sin(\x))*cos(\x)},
     {\yL+\rK*(1+0.07*cos(3*\x)+0.05*cos(2*\x)-0.04*sin(2*\x)*sin(\x))*sin(\x)});

  \fill[black]
    (\xL,\yL) circle (3pt);

  \node[black,anchor=west]
    at ({\xL+0.10},{\yL+0.05})
    {$\mathcal{D}_5$};

  \node[dustyPlum,anchor=east]
    at ({\xL-\rK-0.12},{\yL+0.25})
    {$\mathrm{K3}$};

  \node at (\xL,\yB)
    {$\text{global }5\text{-form symmetry}$};


  \node at (\xGauge,1)
    {$\xRightarrow{\text{gauge}}$};


  \def\yM{1.0}
  \def\hy{0.30}     
  \def\hx{0.09}     

  \fill[dustyPlum, opacity=0.13]
    plot[domain=0:360, samples=120, smooth cycle]
    ({\xM+\rK*(1+0.07*cos(3*\x)+0.05*cos(2*\x)-0.04*sin(2*\x)*sin(\x))*cos(\x)},
     {\yM+\rK*(1+0.07*cos(3*\x)+0.05*cos(2*\x)-0.04*sin(2*\x)*sin(\x))*sin(\x)});

  \draw[dustyPlum, line width=3, opacity=0.6]
    plot[domain=0:360, samples=120, smooth cycle]
    ({\xM+\rK*(1+0.07*cos(3*\x)+0.05*cos(2*\x)-0.04*sin(2*\x)*sin(\x))*cos(\x)},
     {\yM+\rK*(1+0.07*cos(3*\x)+0.05*cos(2*\x)-0.04*sin(2*\x)*sin(\x))*sin(\x)});

  \fill[white]
    ({\xM+\rE},\yM) ellipse [x radius=\hx, y radius=\hy];

  \draw[dustyPlum, line width=1.4pt, opacity=0.45]
    ({\xM+\rE},{\yM+\hy})
    arc[start angle=90, end angle=270, x radius=\hx, y radius=\hy];

  \foreach \d in {-0.50,-0.25,0,0.25,0.50}{
    \draw[deepAqua, line width=1.6]
      (\xM,\yM) -- ({\xM+2.0},{\yM+\d});
  }

  \draw[dustyPlum, line width=1.4pt]
    ({\xM+\rE},{\yM-\hy})
    arc[start angle=-90, end angle=90, x radius=\hx, y radius=\hy];

  \fill[black]
    (\xM,\yM) circle (3pt);

  \node[black,anchor=east]
    at ({\xM-0.10},{\yM+0.08})
    {$\mathcal{D}_5$};

  \node[deepAqua,anchor=west]
    at ({\xM+2.05},\yM)
    {$24\ \text{NS5}$};

  \node[dustyPlum,anchor=north]
    at ({\xM+\rE+0.05},{\yM-\hy-0.02})
    {\small $S^3$};

  \node[dustyPlum,anchor=south]
    at (\xM,{\yM+\rK+0.12})
    {$\mathrm{K3}\setminus B^4_\epsilon$};

\end{tikzpicture}

\caption{\small
Left: a 5-dimensional defect $\mathcal{D}_5$ charged under the would-be $5$-form symmetry associated with $[\mathrm{K3}]\in\Omega_4^{\Spin}$ is detected by the topological operator $U_\alpha(\mathrm{K3})$ on a K3 surrounding it.
Right: making the symmetry consistent with heterotic string theory requires $\mathcal{D}_5$ to be the boundary of 24 NS5-branes, which pierce the K3. The configuration is only consistent once a small ball $B^4_\epsilon$ is excised around the piercing, leaving $K3\setminus B^4_\epsilon$, whose boundary is an $S^3$ linking the 24 NS5-branes. This is to be compared with Figure \ref{fig:gauged-zero-form}.}
\label{fig:K3-NS5}
\end{figure}

A simple example of this was also discussed in \cite{McNamara:2019rup}. As reviewed in Section~\ref{sec:spinbord}, the four-dimensional spin bordism group
\begin{equation}
\Omega^{\Spin}_4=\mathbb{Z}
\end{equation}
is generated by the K3 surface, with bordism invariant $I(X_4)=\tfrac{1}{48}\int_{X_4} p_1$ and $I(\mathrm{K3})=-1$. The associated would-be 5-form symmetry has a 6-form current with $\star\mathbf{j}_6 = \tfrac{1}{48}p_1$, so gauging it requires a dynamical 6-form gauge field coupled to $p_1$. Remarkably, the low-energy effective action of heterotic string theory contains exactly such a coupling:
\begin{equation}\label{eq:hetactioncoup}
 S_{\rm het} \supset 24 \int B_6 \wedge \star\mathbf{j}_6 = \frac12 \int B_6 \wedge p_1 \,,
\end{equation}
where $B_6$ is the NSNS 6-form, the magnetic dual of the B-field $B_2$. Writing $H_7 = dB_6$ and $H_3 = \star H_7$, the equation of motion of $B_6$ becomes the Bianchi identity
\begin{equation}\label{eq:H3bianchi}
    d H_3 = \tfrac{1}{2} p_1\,.
\end{equation}
Hence $\tfrac12 p_1$ is not only closed but exact, and its integral over any compact $X_4$ must vanish. For K3, $\int_{\mathrm{K3}}\tfrac12 p_1 = -24$: the background carries $-24$ units of NS5-brane charge, which is the well-known statement that heterotic string theory on K3 is inconsistent on its own. The charge must be cancelled by including 24 NS5-branes, which act as magnetic sources for $H_3$ on the right-hand side of \eqref{eq:H3bianchi}. Therefore the 5-dimensional defect that is charged under the would-be 5-form global symmetry detected by $I(\mathrm{K3})$ is actually the end-point of 24 six-dimensional NS5 branes. This is depicted in Figure~\ref{fig:K3-NS5}.

This is an instance of the refinement procedure described above. Once $B_6$ is included, consistent backgrounds without sources satisfy $\tfrac12 p_1 = dH_3$, i.e., they carry a spin structure together with a trivialization of $\tfrac12 p_1$. This is a \emph{string structure}, and string bordism $\Omega_d^{\rm String}$ is the next approximation to $\Omega_d^{\rm QG}$. In four dimensions the K3 class is gone, $\Omega_4^{\rm String}=0$, but a new class appears in three dimensions,
\begin{equation}
\Omega_3^{\rm String}=\mathbb{Z}_{24}\,, \qquad I(S^3) = \int_{S^3} H_3 \mod 24\,,
\end{equation}
generated by an $S^3$ carrying one unit of $H_3$-flux. The corresponding would-be global symmetry is broken precisely by the NS5-branes, which are magnetic sources for $H_3$: an $S^3$ with flux can end on them.

So far we have ignored one ingredient of heterotic string theory that also enters the Bianchi identity: the gauge field. The full Bianchi identity reads
\begin{equation}\label{eq:fullbianchi}
    d H_3 = \tfrac12 p_1 + c_2(F)\,,
\end{equation}
where $c_2(F)$ is the second Chern class of the heterotic gauge bundle. The $p_1$ charge of K3 can therefore also be cancelled by gauge instantons instead of NS5-branes, e.g.\ by embedding 24 units of instanton number in the gauge bundle. Consistent heterotic backgrounds are then those with $\tfrac12 p_1 + c_2(F)$ trivialized, which defines a \emph{twisted string structure}: a spin structure, a gauge bundle, and a trivialization of $\tfrac12 p_1 + c_2(F)$. The associated twisted string bordism groups provide a further refinement towards $\Omega^{\rm QG}_d$ for heterotic string theory. Note that, as one can read off from \eqref{eq:fullbianchi}, the heterotic action also contains a coupling of $c_2(F)$ to $B_6$, analogous to \eqref{eq:hetactioncoup}. Here $c_2(F)$ is the current of a Chern--Weil global symmetry, and this Green--Schwarz-type coupling is one of the ways in which quantum gravity removes such symmetries, as discussed in detail in \cite{Heidenreich:2020pkc}.

\subsection{Applying the Conjecture}\label{sec:applying}

Having discussed what the Cobordism Conjecture is, we now turn to discussing how one applies it. In particular, what kinds of bordism groups are appropriate for each of the different string theories?

In applying the Cobordism Conjecture, the choice of bordism group is not universal but depends on the theory under consideration.  An EFT of quantum gravity does not sum over arbitrary smooth manifolds: it sums over manifolds equipped with whatever additional structures are required for the fields in the theory to be consistently defined.  For example, fermions require a spin (or spin$^c$) structure, orientifolds may require non-orientable manifolds with a specified Pin$^\pm$ structure, heterotic strings require a principal $G$-bundle satisfying anomaly cancellation conditions, and background $B$-fields introduce twists.  Accordingly, the relevant classification of backgrounds is given not by a single bordism group, but by the bordism theory appropriate to the tangential and auxiliary data required by the model:
\begin{equation}
\Omega_*^{\mathrm{SO}}, \quad
\Omega_*^{\mathrm{Spin}}, \quad
\Omega_*^{\mathrm{Pin}^\pm}, \quad
\Omega_*^{\mathrm{Spin}}(BG), \quad
\Omega_*^{\mathrm{String}}, \quad
\Omega_*^{\mathrm{twisted-String}}, \quad \text{etc.},
\end{equation}
where the $*$ is a placeholder for the degree of the corresponding bordism group. Each of these bordism groups is sensitive to different kinds of geometric and topological data. Refining the tangential structure on manifolds accordingly changes the associated bordism theory. Passing, for example, from oriented to spin bordism excludes manifolds with $w_2 \neq 0$; refining further to String bordism excludes spin manifolds with nontrivial $\tfrac12 p_1$. Conversely, imposing additional structure can split previously equivalent bordism classes into distinct ones, since the extra data may obstruct bordisms that were allowed in the coarser theory. Thus refinement can cause some classes to disappear while others newly emerge.

From the physical point of view underlying the Cobordism Conjecture, this behaviour has a natural interpretation. Classes that disappear under refinement correspond to would-be topological charges that no longer generate a global symmetry once the additional structure of the improved theory is taken into account: the refined structure either forces the vanishing of the charge with a Gauss-law constraint, outright forbidding classes with nonzero charge (gauging), or permits dynamical processes that connect the bordism generator to nothing (breaking). On the other hand, classes that appear only after refinement detect more delicate topological obstructions that are invisible in the coarser description; these may signal potential global symmetries that become meaningful only in a more microscopic formulation of the theory. Coarser bordism groups do not become obsolete once a refinement is available, however. The sequence of refinements is rarely canonical, and every non-trivial class, at any level, is a would-be global symmetry that the full theory must gauge or break. Identifying how it does so is precisely what tells us which structures and defects the theory must contain, and therefore which refinements of the description are appropriate.

For example, consider the effective field theory of type IIB string theory. We can start by noting that there are fermions in this theory, which means that our spacetime background should admit a spin structure. One could then notice that there is an SL$(2,\Z)$ duality group that acts on the bosonic degrees of freedom. Therefore, one can consider spin bordism groups with a duality group bundle $\Omega_*^{\mathrm{Spin}}\!\left(B\mathrm{SL}(2,\mathbb{Z})\right)$. But this is not yet the full story: the fermions also transform under the duality and so we have to consider the fermionic lift  Mp$(2,\Z)$. In this case, the duality group mixes with the spin structure, so we have to consider  $\Omega_*^{\mathrm{Spin}-\mathrm{Mp}(2,\mathbb{Z})}$. We can go one step further by taking into account the worldsheet symmetries of Type IIB string theory: worldsheet fermion parity $(-1)^{F_L}$ and worldsheet orientation reversal $\Omega$. These symmetries act on spacetime fields and lead one to consider the bordism groups $\Omega_*^{\mathrm{Pin}^{+}(\mathrm{GL}(2,\mathbb{Z}))}$. One can in principle keep identifying more structure in type IIB string theory and adding it into the bordism group computation \cite{Debray:2023yrs}.

In Table \ref{tab:bordismtable}, we list a few ten-dimensional string theories and the bordism groups that are appropriate for them, with varying amounts of structure. The ellipses indicate that each of these theories carries additional geometric or symmetry data that could, in principle, be captured by further refinements of the relevant bordism groups, many of which have not yet been computed. 

We now turn to the objects that are needed to break/gauge the symmetries detected by bordism groups. We will not aim to be exhaustive, and will only give a few examples. For further examples of direct applications of the Cobordism Conjecture, the interested reader can turn to \cite{Dierigl:2022reg,Kaidi:2023tqo,Andriot:2022mri,GarciaEtxebarria:2020xsr,Ooguri:2020sua,Montero:2020icj,Hamada:2021bbz,Blumenhagen:2021nmi,Dierigl:2020lai,Dierigl:2023jdp,Debray:2023yrs}.

\begin{table}
    \centering
    \renewcommand{\arraystretch}{1.3}
    \begin{tabular}{
        @{}
        >{\raggedright\arraybackslash}p{1.8cm}
        >{\raggedright\arraybackslash}p{12.8cm}
        @{}
    }
        \toprule
        \textbf{Theory / context}
        & \textbf{Required structure and corresponding bordism group} \\
        \midrule

        M-theory\newline
        Type IIA
        & Pin structure:
          $\Omega_*^{\mathrm{Pin}^{+}}$
          \cite{McNamara:2019rup}
         \par\smallskip
          Including trivialization of $W_7(TM)$ \cite{Diaconescu:2000wy}: $\Omega_{*}^{\mathrm{String}^h} $\cite{Debray:2024czl}
          \par\smallskip
          $\cdots$ \\

        \addlinespace
        Type IIB
        & Spin structure:
          $\Omega_*^{\mathrm{Spin}}$
          \cite{McNamara:2019rup}
          \par\smallskip
          F-theory compactifications:
          $\Omega_*^{\mathrm{Spin}^c}$
          \cite{McNamara:2019rup}
          \par\smallskip
          Duality group bundle:
          $\Omega_*^{\mathrm{Spin}}
            \bigl(B\mathrm{SL}(2,\mathbb{Z})\bigr)$
          \cite{Debray:2023yrs}
          \par\smallskip
          Including fermion transformation under
          $\mathrm{SL}(2,\mathbb{Z})$:
          $\Omega_*^{\mathrm{Spin}\text{-}\mathrm{Mp}(2,\mathbb{Z})}$
          \cite{Debray:2023yrs}
          \par\smallskip
          Accounting for worldsheet symmetries
          $(-1)^{F_L}$ and $\Omega$:
          $\Omega_*^{\mathrm{Pin}^{+}
            (\mathrm{GL}(2,\mathbb{Z}))}$
          \cite{Debray:2023yrs}
          \par\smallskip
          $\cdots$ \\
          
       \addlinespace
IIA/IIB \newline on $\mathbb{T}^n$ & Spin structure $+$ U-duality: $\Omega_*^{\mathrm{Spin}}(BG_U)$
\cite{Braeger:2025kra} \par\smallskip Including fermion transformation under U-duality: $\Omega_*^{\mathrm{Spin}-\widetilde{G_U}^{(+)}}$ \cite{Chakrabhavi:2025bfi} 
          \par\smallskip
                    $\cdots$ \\

        \addlinespace
        Type I/\newline
        Het O
        & Spin structure $+$ $SO(32)$ bundle:
          $\Omega_*^{\mathrm{Spin}}(B\mathrm{Ss}(32))$
          \cite{Kneissl:2024zox}
          \par\smallskip
          $\cdots$ \\

        \addlinespace
        Het E
        & Spin structure $+$ Bianchi identity $+$ $E_8$ swap:
          $\Omega_*^{\xi_{E_8\times E_8}}$
          \cite{Debray:2023rlx}
          \par\smallskip
          CHL string:
          $\Omega_*^{\xi_{\mathrm{CHL}}}$
          \cite{Debray:2023rlx}
          \par\smallskip
          $\cdots$ \\

        \addlinespace
        Non-SUSY\newline
        Hets
        & Spin structure $+$ $G$-bundle $+$ Bianchi identities:
          \par\smallskip
          $\Omega_*^{\mathrm{String}\text{-}\mathrm{Spin}(16)^2}$,
          $\Omega_*^{\mathrm{String}\text{-}\mathrm{Sp}(32)}$,
          $\Omega_*^{\mathrm{String}\text{-}
            \mathrm{U}(32)\langle c_3\rangle}$
          \cite{Basile:2023knk}
          \par\smallskip
          $\cdots$ \\

        \bottomrule
    \end{tabular}
    \caption{Some 10d string theories and the bordism groups
        detecting their relevant structures at varying levels of refinement.} 
    \label{tab:bordismtable}
\end{table}

\paragraph{D-branes as Cobordism Defects.}One of the first examples of extended objects predicted by the Cobordism Conjecture are defects that carry magnetic charge under the various $p$-form gauge symmetries of string theory, namely D-branes. In Type II string theory, a D$p$-brane is linked by a sphere $S^{8-p}$ on which one measures an integer-quantized Ramond--Ramond (RR) flux $F_{8-p}$. The associated bordism charge is given by
\begin{equation}\label{eq:rrcharge}
\left[ \int_{S^{8-p}} F_{8-p} \right]\!
\;\in\; \Omega^{\mathrm{Spin}}_{8-p}\!\left(K(\mathbb{Z},\,8-p)\right),
\end{equation}
which generates an integer-valued bordism group \cite{McNamara:2019rup,Nevoa:2025xiq}.\footnote{Here $K(\mathbb{Z},8-p)$ is the Eilenberg--MacLane space that classifies degree $(8-p)$ cohomology classes. The flux $\int_{S^{8-p}}F_{8-p}$ determines a map from the spacetime (or linking sphere) to $K(\mathbb{Z},8-p)$, making the appropriate bordism group $\Omega^{\mathrm{Spin}}_{8-p}(K(\mathbb{Z},8-p))$. } Physically, this charge detects the leading top-form Chern--Simons coupling on the D-brane worldvolume. 

This description, however, does not capture the full structure of D-brane charge. In Type II string theory the stable charge of D-branes is more precisely classified by $K$-theory (or twisted $K$-theory), reflecting the fact that configurations differing by the addition of brane--antibrane pairs represent the same conserved charge \cite{Witten:1998cd,Moore:1999gb}. The cohomological charge \eqref{eq:rrcharge} should therefore be viewed as the leading approximation to the full $K$-theoretic classification. This refinement can also be formulated in the language of bordism groups; see e.g.\ \cite{Nevoa:2025xiq}.

The fact that the global symmetry in \eqref{eq:rrcharge} is broken is due to the fact that \hl{D-branes are \textit{dynamical} objects} in string theory. In other words, they can \emph{end} on their creation operator. \begin{figure}
    \centering
    \includegraphics[width=0.5\linewidth]{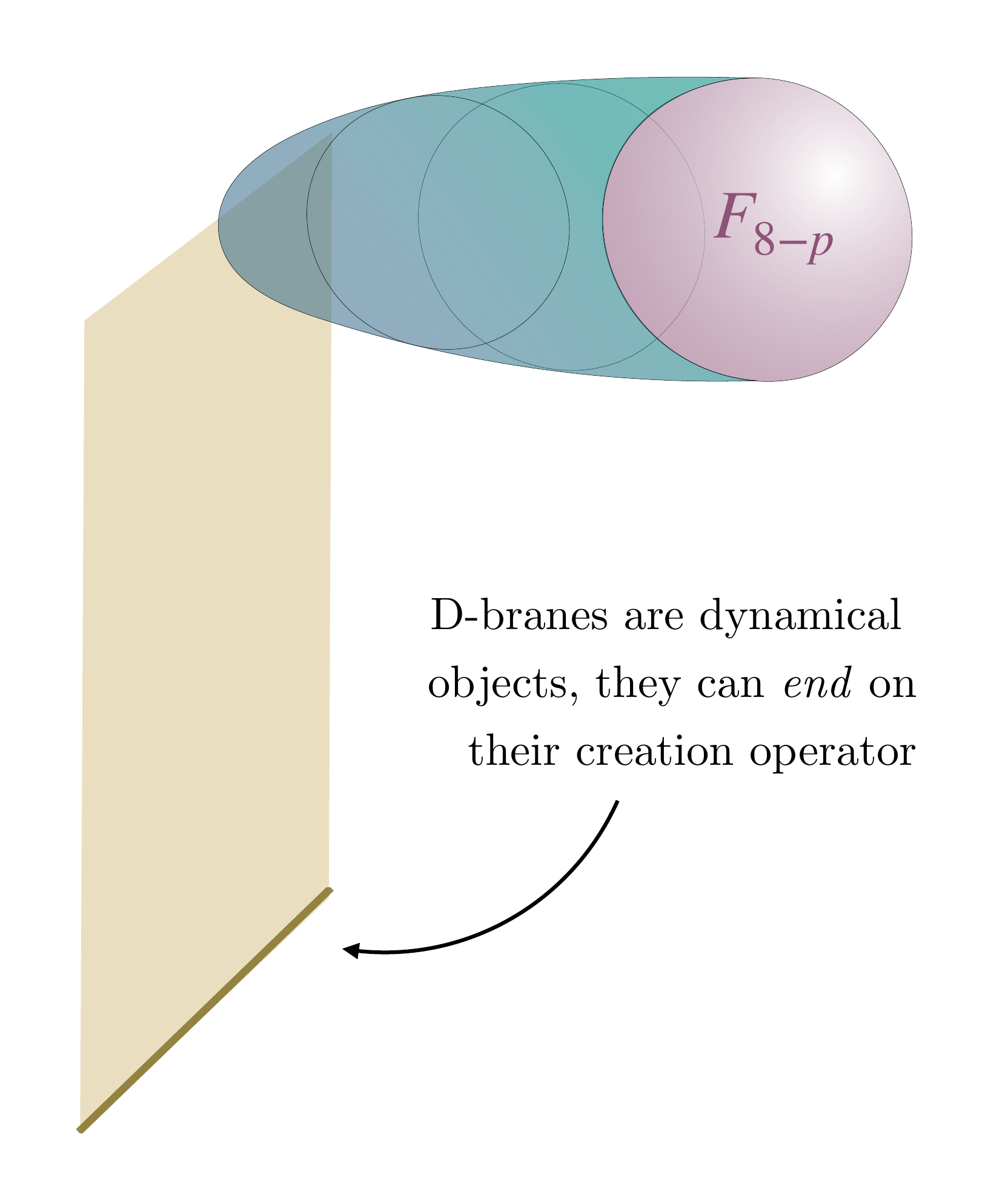}
    \caption{The global symmetry in \eqref{eq:rrcharge} is broken in string theory because D-branes are dynamical: the D-brane can absorb the charge and then disappear into the vacuum. In other words, there are creation operators in string theory that can end a D-brane.}
    \label{fig:dbranedynamical}
\end{figure}
\paragraph{Boundaries.}The Cobordism Conjecture implies that all theories compatible with quantum gravity admit boundaries. Indeed, $\Omega^{\rm QG}_d = 0$ implies in particular that every compactification manifold $X_d$ is bordant to nothing, i.e., there is a configuration in which the internal space smoothly shrinks away. From the point of view of the lower-dimensional EFT, this is an end-of-the-world boundary, see Figure~\ref{fig:sub2}. This may seem surprising at first, but they already occur in very concrete corners of string/M-theory.  The classic example is the Hořava--Witten description of strongly coupled $E_8\times E_8$ heterotic theory as M-theory on an orbifold $\mathbb{R}^{10}\times S^1/\mathbb{Z}_2$, whose fixed loci are genuine ten-dimensional end-of-the-world walls supporting $E_8$ gauge fields; from the eleven-dimensional viewpoint the spacetime is effectively a manifold with boundary, and consistency (notably anomaly cancellation) forces nontrivial boundary degrees of freedom \cite{Horava:1995qa,Horava:1996ma}. A closely related example appears in type IIA/type I$'$ constructions with O8-planes: the orientifold 8-plane is a codimension-one object, and in the familiar type I$'$  \cite{Polchinski:1995df} interval picture one has O8-planes at the ends together with stacks of D8-branes whose distribution cancels tadpoles; in the low-energy description this is literally a ``wall''/boundary-like configuration (a domain wall of massive IIA supergravity) rather than some wildly non-geometric endpoint.

\paragraph{New objects.} The Cobordism Conjecture does not only post-dict the existence of known extended objects in string theory; it also predicts new ones. One example is the R7-brane (``reflection 7-brane'') of type IIB, which trivializes a $\Z_2$ class in $\Omega_1^{\Spin\text{-GL}^+(2,\Z)}(\pt)$ \cite{Dierigl:2022reg, Dierigl:2023jdp}. It is a \emph{non-supersymmetric} codimension-two defect around which one picks up a $(-1)^{F_L}$ or $\Omega$ monodromy rather than an ordinary $\SL(2,\mathbb{Z})$ monodromy. In the type IIB effective action, this is a monodromy of the axio-dilaton that reflects the F-theory torus, $\tau \to -\bar{\tau}$. Another striking example is \cite{Montero:2020icj}: trivializing $\Omega_2^{\mathrm{Pin}^-}=\mathbb{Z}_8$ in 9d and 8d theories with 16 supercharges predicts \emph{I-folds}, a generalization of orientifolds. Anomaly cancellation in compactifications on $T^3/\mathbb{Z}_2$ with I-folds at the fixed points then restricts the gauge group rank to $r\equiv 1 \bmod 8$ in 9d and $r\equiv 2 \bmod 8$ in 8d. These are exactly the ranks realized in string theory, which is strong evidence for the String Lamppost Principle: the conjecture that all consistent quantum gravity EFTs can arise from string theory. Many other cobordism defects remain less well understood. Examples include non-supersymmetric NS5-branes in non-supersymmetric heterotic string theories \cite{Basile:2023knk} and other non-supersymmetric heterotic branes \cite{Kaidi:2023tqo,Kaidi:2024cbx,Hamada:2024cdd,Hamada:2025cpe}. Perhaps the simplest, and most mysterious, is the spin defect that should trivialize the class of the periodic circle $S^1_p$ in spin bordism in heterotic string theory.

\subsection{Dynamical Aspects}\label{sec:dynamicalaspects}
In the previous section we saw that the Cobordism Conjecture predicts the existence of dynamical objects carrying specific types of topological charges. One of the simplest and most familiar examples is provided by D-branes. The conjecture predicts that there should exist objects charged under Ramond--Ramond (RR) fields. However, a D-brane is much more than an object characterized by its topological couplings to RR gauge potentials. Dynamically, a D-brane possesses a rich structure: its low-energy dynamics are governed by a Dirac--Born--Infeld (DBI) action, while in the ultraviolet it supports open string degrees of freedom. Thus the topological charge is only one aspect of a much more elaborate physical object.

This naturally raises the question of whether similar dynamical structures exist for the other objects predicted by the Cobordism Conjecture. More generally, we would like to develop tools that allow us to study the dynamical aspects of these defects, beyond their purely topological characterization. Several related questions arise. What degrees of freedom live on their worldvolume? How do they interact? How do they couple to the bulk fields? Can these objects be realized in supergravity as solitonic solutions? If so, how do they backreact on the spacetime? As we will see, these questions are difficult in that the object is in general \hl{non-supersymmetric} and can involve strong string coupling effects.

The Cobordism Conjecture itself is \hl{entirely silent about these dynamical aspects.} In particular, it may happen that a single physical object trivializes several distinct bordism classes, indicating that the relationship between topological charge and physical realization need not be one-to-one. Nonetheless, understanding the dynamics of cobordism defects is essential if one hopes to extract phenomenological consequences from the conjecture. One of the crucial questions is to identify the tension or mass of these objects, which dictates the scale at which the global symmetry is broken. At present, it is entirely possible that these cobordism defects have tensions of order the string or quantum gravity scale. If this were the case, they would be effectively decoupled from low-energy physics and would leave little observable imprint in the infrared. On the other hand, if the tensions of some of these objects were parametrically lower than the string scale, they could play an important role at low energies for model building. In other words, the Cobordism Conjecture only states that the symmetry must be broken (or gauged) \emph{at some scale}; it does not forbid it from being an approximate symmetry in the IR. If the relevant cobordism defects are heavy, processes that change the would-be topological charge, such as nucleating a defect, are suppressed, and the symmetry is effectively exact at low energies. Determining the breaking scale therefore requires dynamical information (the tensions, couplings and backreaction of the defects) which the topological analysis alone cannot provide. This motivates the rest of this section.
 
More broadly, the Cobordism Conjecture asserts that it should always be possible either to end spacetime or to connect two quantum gravity EFTs by means of a domain wall. Even when the relevant bordism classes are trivial, so that no additional topological defect is required, providing a dynamical description of such a transition within the regime of validity of the EFT appears difficult. In Lorentzian spacetimes, topology-changing processes are typically associated with \hl{spacetime singularities} (see e.g.~\cite{Gibbons:2011dh}), indicating that intrinsically stringy effects must become important. Similarly, when two EFTs are connected by a domain wall, one generically expects the wall itself to lie outside the regime of validity of either EFT. In particular, the two sides of the wall may correspond to completely different string constructions with different spectra and degrees of freedom, and the domain wall must interpolate between them, effectively converting the degrees of freedom of one theory into those of the other. The difficulty in doing so is particularly acute when interpolating between a chiral theory and a non-chiral one, since the chiral degrees of freedom must become gapped along the interpolation \cite{Angius:2024pqk}. From the perspective of the low-energy EFT, such a process is therefore expected to be singular or involve \hl{strong string coupling effects}. 

In the remainder of this section we make these statements more precise and illustrate them with concrete examples. In Section~\ref{sec:charact}, we study brane-like defects in EFTs of quantum gravity and attempt to extract general lessons about how bordism defects backreact on spacetime. In Section~\ref{sec:etw}, we turn to the more general question of how bordism transitions (such as ending spacetime or connecting two theories via a domain wall) can be described on-shell both from the perspective of the low-energy EFT and from that of the full string theory. Finally, in Section \ref{sec:unknownBdef}, we outline recent progress in understanding the dynamical properties of new, non-supersymmetric cobordism defects.

\subsubsection{Dynamics of Cobordism Defects: a Case Study}\label{sec:charact}
We now discuss what is known about the dynamical aspects of cobordism defects. We will start with D-branes as an example, which will serve as a template for discussing what can be understood about new non-supersymmetric defects predicted by the Cobordism Conjecture in the next section. 

\subsubsubsection{Couplings and Worldvolume theory }At most, the Cobordism Conjecture determines which topological couplings a defect must possess in order to carry the appropriate bordism charge. This, however, is only part of the story: the worldvolume theory of a given brane may contain many additional couplings. To begin with, a single defect may trivialize several bordism classes, and in that case it can carry more than one topological coupling. Beyond these required topological terms, much less is known in general about the full set of worldvolume interactions. Since the brane is dynamical, it necessarily has finite tension and couples to the spacetime metric. But it may also couple to other bulk fields, and, moreover, it can support its own intrinsic worldvolume degrees of freedom. To illustrate these points, we will use D-branes as examples.

\paragraph{D-brane DBI action.}In the UV, a D-brane is defined by the boundary conditions imposed on open strings. The endpoints of an open string are constrained to lie on the D-brane, satisfying Neumann boundary conditions along directions tangent to the brane and Dirichlet boundary conditions along directions transverse to it. One may quantize these open strings and compute their scattering amplitudes directly. At energies much smaller than the string scale, the excitations of these open strings give rise to fields localized on the brane worldvolume. For example, consider a single D$p$-brane in type II string theory. The massless spectrum of open strings ending on the brane contains
\begin{itemize}
    \item a U$(1)$ gauge field $A_a$ living on the $(p+1)$-dimensional worldvolume,
    \item scalar fields $\Phi^i$ describing fluctuations of the brane in the transverse directions,
    \item and their supersymmetric fermionic partners.
\end{itemize}
The leading-order effective action governing these worldvolume fields (at leading order in $g_s$, and in $\alpha' $) is the Dirac--Born--Infeld (DBI) action,
\begin{equation}
S_{\mathrm{DBI}}
=
-T_p
\int_{\mathcal W_{p+1}} d^{p+1}\xi \,
e^{-\phi}
\sqrt{
-\det\!\left(
P[G+B]_{ab}
+
2\pi\alpha' F_{ab}
\right)
},
\end{equation}
where $T_p$ is the D$p$-brane tension, $G_{\mu\nu}$ is the spacetime metric, $B_{\mu\nu}$ is the NS--NS two-form, $F_{ab}$ is the worldvolume field strength, and $P[\cdots]$ denotes the pullback of bulk fields onto the brane worldvolume. Expanding the DBI action for small field strengths reproduces ordinary Maxwell theory together with kinetic terms for the transverse scalar fields,
\begin{equation}
S_{\mathrm{DBI}}
=
-T_p
\int_{\mathcal W_{p+1}} d^{p+1}\xi
\left[
1
+
\frac{(2\pi\alpha')^2}{4}F_{ab}F^{ab}
+
\frac{1}{2}\partial_a\Phi^i\,\partial^a\Phi^i
+\cdots
\right].
\end{equation}
The DBI action therefore provides the infrared (IR) effective description of the dynamics of open-string endpoints. More generally, a stack of $N$ coincident D-branes supports a non-Abelian $U(N)$ gauge theory, since open strings can begin and end on different branes. In this way, gauge fields and matter living on D-branes emerge directly from the dynamics of open strings, while the DBI action captures their low-energy interactions.

\paragraph{Topological couplings on D-branes.}In addition to the DBI action, D-branes possess topological couplings to the Ramond--Ramond (RR) gauge potentials of the bulk theory. These couplings are described by a Chern--Simons (CS), or Wess--Zumino, action of the form 
\begin{equation}
S_{\mathrm{CS}}
=
\mu_p
\int_{\mathcal W_{p+1}}
P\!\left[
\sum_q C_q
\right]
\wedge
e^{B+2\pi\alpha' F},
\end{equation}
where $\mu_p$ is the RR charge of the D$p$-brane, $C_q$ denotes the collection of RR gauge potentials, $B$ is the NS--NS two-form, and $F$ is the worldvolume gauge field strength. Expanding the exponential generates a tower of topological couplings,
\begin{equation}\label{eq:CSaction}
S_{\mathrm{CS}}
=
\mu_p
\int_{\mathcal W_{p+1}}
\left(
C_{p+1}
+
(2\pi\alpha')\, C_{p-1}\wedge F
+
\frac{(2\pi\alpha')^2}{2}
C_{p-3}\wedge F\wedge F
+\cdots
\right).
\end{equation}
The first term expresses the fact that a D$p$-brane is electrically charged under the RR $(p+1)$-form potential $C_{p+1}$. The higher-order terms reveal a remarkable feature: gauge-field configurations on the brane can themselves carry lower-dimensional D-brane charges. For example, an instanton configuration with non-vanishing
\begin{equation}
\int F\wedge F
\end{equation}
induces D$(p-4)$-brane charge through the coupling to $C_{p-3}$. All of these charges are encoded in the K-theoretic classification of D-brane charges \cite{Witten:1998cd}.\footnote{For an open-closed duality interpretation of the relation between Cobordisms and K-theory, see \cite{Blumenhagen:2021nmi,Blumenhagen:2022bvh}.}

\paragraph{Summary.}Together, the DBI and CS terms furnish the leading-order low-energy effective action for a D-brane,
\begin{equation}
S_{\mathrm{D}p}
=
S_{\mathrm{DBI}}
+
S_{\mathrm{CS}}\,,
\end{equation}
combining the dynamical worldvolume degrees of freedom with their couplings to the bulk supergravity fields. 
A natural goal is to develop an analogous understanding for more general extended objects appearing in quantum gravity, including the exotic defects predicted by the Cobordism Conjecture. For D-branes, the existence of a microscopic open-string description provides a powerful handle on both the propagating worldvolume degrees of freedom and their topological couplings. In contrast, for a generic cobordism defect, one may not have access to such a UV description, even if it exists. The Cobordism Conjecture may constrain topological couplings of these objects, but it does not by itself determine the full analogue of the DBI$+$CS action.

\subsubsubsection{Backreaction on supergravity}
With the full worldvolume action in hand, we know all the couplings of the defect to the bulk fields. We can then determine the defect's spacetime profile and its backreaction on the metric and other bulk fields simply by solving the coupled equations of motion. However, even without a full microscopic description of cobordism defects and their worldvolume action and couplings, one can still attempt to characterize these objects from the perspective of the low-energy gravitational theory. After all, many of the basic properties of D-branes can be inferred without detailed knowledge of open-string dynamics. In particular, D-branes admit classical supergravity descriptions as BPS black brane solutions carrying Ramond--Ramond charges. The existence of such solutions follows from the bulk equations of motion and does not require explicit access to the microscopic open-string sector. We turn to this now, taking once more D-branes as a working example. 

\paragraph{D-branes as solitons in supergravity.}In the low-energy limit, D$p$-branes can be described as solitonic objects in supergravity, magnetically charged under RR ${F}_{8-p}$ (and electrically under $\tilde{F}_{p+2}$) given by the following Einstein-frame solution (\hl{for $p \leq 6$}):
\begin{align}
    ds_{\rm E}^2 &=H(r)^{\frac{p-7}{8}}\,dx_{\parallel}^2+H(r)^{\frac{p+1}{8}}\,dx_{\perp}^2\\
    \tilde{F}_{m0\cdots p} &=\partial_m H(r)^{-1} \label{eq:flux}\\
    e^{\phi(r)}&=H(r)^{\frac{3-p}{4}}\,,\label{eq:dil}
\end{align}
where the harmonic function is given by 
\begin{equation}
    H(r)=1+\frac{\alpha}{r^{7-p}}\,.
\end{equation}
Here $r$ is the radial distance in the space transverse to the brane and $\alpha$ is a dimensionful constant,
\begin{align}
    \alpha
=
(4\pi)^{\frac{5-p}{2}}
\Gamma\!\left(\frac{7-p}{2}\right)
(\alpha')^{\frac{7-p}{2}}
g_s N \,.
\end{align}
\begin{figure}
    \centering
    \includegraphics[width=0.5\linewidth]{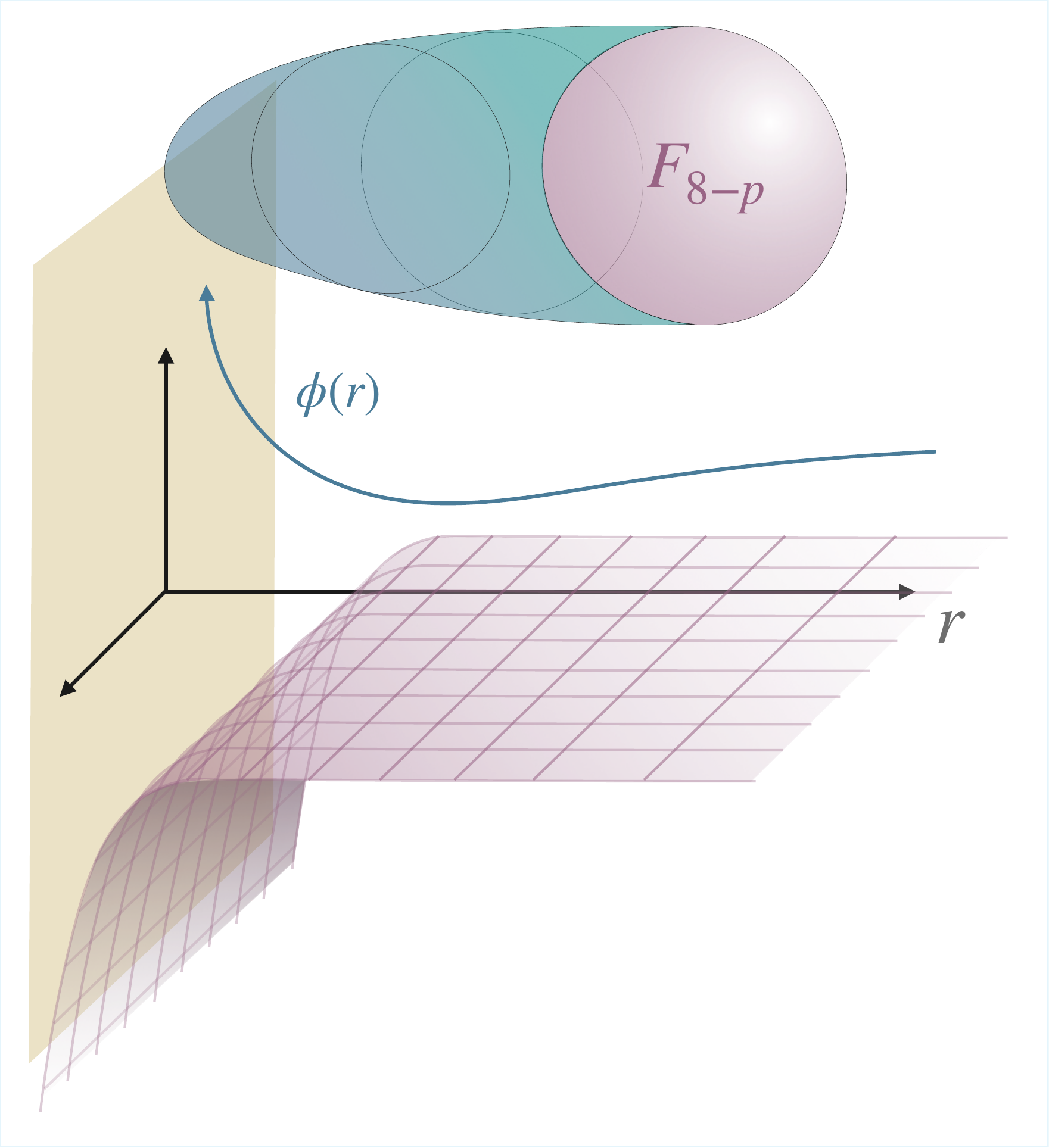}
    \caption{An illustration of the backreaction of a D$p$-brane. The brane is a magnetic source for $F_{8-p}$-flux, and is the locus of a spacetime singularity where the dilaton blows up (except for $p=3$).}
    \label{fig:dynamicalcobordism}
\end{figure}
From this solution, one can infer that there exists a transverse sphere surrounding the D$p$-brane which carries $N$ units of $(8-p)$-form flux. This demonstrates that D-branes act as magnetic sources for the corresponding Ramond--Ramond flux.

\paragraph{Singularities.}At the origin of the transverse space, $r = 0$, the D$p$-brane (for $p \neq 3$) sits at a curvature singularity where the dilaton diverges ($\phi \to \pm\infty$), signaling a breakdown of the supergravity description. This is depicted schematically in Figure \ref{fig:dynamicalcobordism}. The D3-brane case is exceptional: it is the unique D-brane solution for which the locus $r = 0$ corresponds instead to a regular geometric horizon, and the dilaton remains constant. This reflects the well-known \cite{Itzhaki:1998dd} fact that the D3-brane is the only solution in the family whose near-horizon limit ---$\rm{AdS}_5 \times \rm{S}^5$--- is holographically dual to a conformally invariant quantum field theory. In all other cases, the dilaton runs with the transverse radial coordinate, breaking the dilatation isometry of AdS space, which implies that the boundary dual is not conformal.\footnote{In holography, radial position in the bulk is roughly identified with boundary energy scale; see e.g.~\cite{Peet:1998wn,Polchinski:2010hw}.} For $p\neq3$, the type II near-horizon solution is only to be trusted in a specific range of the radial coordinate, which corresponds to an intermediate region of the scaling RG flow followed by the QFT dual.

The singular nature of D-brane solutions within the low-energy effective field theory is not unexpected. D-branes are intrinsically stringy objects, defined as loci where open strings can end. As such, their full description necessarily involves degrees of freedom that lie beyond the supergravity approximation, which captures only the massless modes of the string.

It is also instructive to examine the behavior of the dilaton near the core of the D$p$-brane solutions. For $p \geq 4$, the dilaton approaches $\phi \to -\infty$, corresponding to a weakly coupled regime. In contrast, for $p \leq 2$, one finds $\phi \to +\infty$, indicating strong coupling at the core. This behavior suggests a possible connection with the Swampland Distance Conjecture \cite{Buratti:2021fiv}, which asserts that effective field theories coupled to gravity break down when scalar fields traverse infinite distances in field space, typically accompanied by the emergence of an infinite tower of light states. In the present case, the breakdown of the effective field theory is indeed evident: the dilaton runs off to the boundary of its moduli space, and the Einstein-frame  geometry develops a curvature singularity.\footnote{One should be cautious in directly applying the Swampland Distance Conjecture in this context. The conjecture is usually formulated for vacuum expectation values of scalar fields in backgrounds without significant spacetime warping \cite{Raucci:2026fzp}. Here, the dilaton profile is position-dependent, and the geometry is highly curved. It is therefore not immediately clear that an infinite tower of states becomes parametrically light as one approaches the singularity. }

There are several additional examples in string theory in which bordism defects can be explicitly identified and characterized within supergravity. In the cases studied in \cite{Buratti:2021yia,Buratti:2021fiv,Angius:2022aeq,Mourad:2024dur,Angius:2024zjv,Angius:2023uqk,Raucci:2022jgw}, one finds that the bordism defect is typically localized at a spacetime singularity where scalar fields traverse an infinite distance in moduli space. More concretely, many of the known solutions take a form closely analogous to extremal black brane geometries, in which scalar fields exhibit logarithmic running near the defect \cite{Angius:2022aeq}. This behavior can be understood from the fact that the presence of a non-trivial bordism charge induces an effective potential for the scalars—a so-called dynamical tadpole \cite{Buratti:2021yia}. When expressed in terms of canonically normalized fields, this potential is exponential, leading generically to logarithmic profiles for the scalars. This feature is consistent with broader expectations from the Swampland Program, where scalar potentials in quantum gravity are often exponential in the canonically normalized moduli. We will discuss this in more detail in section \ref{sec:unknownBdef}. 

\paragraph{Low codimension defects.}It is important to emphasize, however, that the picture just described primarily applies to defects of sufficiently high codimension. Lower-codimension branes exhibit qualitatively different behavior. A prototypical example is the D8-brane, whose solution in Einstein frame takes the form \cite{Polchinski:1995df,Bergshoeff:1995vh}:
\begin{equation}
ds_{\rm E}^2 = H(x)^{\frac{1}{8}} \, dx_{||}^2 + H(x)^{\frac{9}{8}} \, dx^2, 
\qquad
e^{\phi} = H(x)^{-5/4},
\qquad
H(x) = c + m x \, .
\end{equation}
Here $x$ denotes the single transverse direction. Unlike higher-codimension branes, \hl{the dilaton remains finite} at the location of the D8-brane, and the geometry does not develop a curvature singularity at its core. More generally, as observed in \cite{Buratti:2021fiv}, domain wall defects can produce only finite discontinuities in physical quantities. However, their distinctive feature is that they drastically modify the global structure of spacetime: one cannot move arbitrarily far from a D8-brane without encountering a singularity at finite proper distance, where the dilaton diverges and the effective description breaks down.

A related phenomenon occurs for codimension-two branes, such as the D7-brane. In Einstein frame, its solution reads \cite{Greene:1989ya,Gibbons:1995vg}
\begin{equation}
ds_{\rm E}^2 = \eta_{\mu\nu} dx^\mu dx^\nu + \tau_2\, |\eta(\tau)|^4\, |z|^{-1/6}\, |dz|^2, 
\qquad 
\tau(z) = C_0 + i e^{-\phi} = \frac{1}{2\pi i}\log \frac{z}{z_0} \, ,
\end{equation}
where $z$ is the complex coordinate on the transverse plane, $\tau$ is the axio-dilaton with $\tau_2 = \mathrm{Im}\,\tau = e^{-\phi}$, $\eta(\tau)$ is the Dedekind eta function, and the expression for $\tau$ holds near the brane. Going around the brane, $z\to e^{2\pi i}z$, the axio-dilaton undergoes the monodromy $\tau\to\tau+1$, reflecting the unit of D7-brane charge. Near the core, $|\eta(\tau)|^4 \simeq |z/z_0|^{1/6}$, so the metric reduces to $ds_{\rm E}^2 \simeq \eta_{\mu\nu}dx^\mu dx^\nu + e^{-\phi}|dz|^2$ up to a constant rescaling. At the core, $z\to0$, one has $e^{-\phi} = \frac{1}{2\pi}\log|z_0/z|\to\infty$, i.e., weak coupling, but the curvature diverges. Since the core lies at finite proper distance, this is a genuine singularity of the solution rather than an asymptotic region. Moving away from the brane, $e^{-\phi}$ decreases and vanishes at $|z|=|z_0|$, where the solution breaks down \hl{at finite spacetime distance}, much as for the D8-brane. This is characteristic of codimension-two defects: in two transverse dimensions, localized sources produce \hl{logarithmic profiles and a deficit angle}, which prevents the recovery of asymptotically flat space. Indeed, at large $z= r e^{i \varphi}$, the metric in the transverse space becomes: 
\begin{equation}
    ds^2=  \, r^{-1/12}\, (dr ^2 + r^2 d\varphi^2), 
\end{equation}
After a change of variables, this becomes: 
\begin{equation}
    ds^2 = d\rho^2 + \rho^2 d\theta^2 \text{  with } \theta \in [0,\pi/6]\;.
\end{equation}
This means that every D7-brane produces a deficit angle of $\pi/6$. Globally consistent configurations require combining several 7-branes, e.g., 24 of them for a compact transverse space, as in F-theory on K3.

\paragraph{Summary.}Even without a microscopic description, the low-energy gravitational theory captures many properties of a defect. For D-branes, these are BPS black brane solutions of supergravity that magnetically source RR flux through a surrounding transverse sphere. For codimension three or higher, the core is a curvature singularity where the dilaton runs to infinite distance in moduli space (except for the D3-brane, which has a regular horizon). This running is consistent with the logarithmic scalar profiles that dynamical tadpoles induce for other bordism defects \cite{Angius:2022aeq}. Low-codimension branes behave differently. The D8-brane has a regular core but hits a singularity at finite distance. The D7-brane produces a deficit angle, so the EFT breaks down at finite distance from the core and a globally consistent configuration needs several 7-branes. All of this is the case of D-branes, which are supersymmetric and well understood. We will see what we can say about other, non-supersymmetric defects in Section~\ref{sec:unknownBdef}.

\subsubsection{Ending Spacetime in the EFT and in String Theory}\label{sec:etw}
The Cobordism Conjecture predicts that there should always be a (finite-energy) quantum gravity process that can end spacetime, regardless of the need for introducing a cobordism defect to absorb some global charge. In other words, even if all bordism classes are trivial, there should exist finite energy processes that allow one to end spacetime. It is therefore of interest to understand, more generally, how spacetimes can end, before even including singular defects.

Importantly, the Cobordism Conjecture predicts that this should be a \emph{quantum gravity} process. The existence of such an on-shell process need not be visible within the low-energy EFT: it is a statement about the full string theory, and the transition may well involve regions of strong coupling or high curvature where the EFT breaks down. 

We already saw in Section \ref{sec:applying} that in highly supersymmetric cases, some string theories (and M-theory) are known to admit boundaries. For instance, Type IIA can end on a stack of D8 branes and O8 planes. Without supersymmetry, one expects boundaries to correspond to more drastic ``end of the world'' transition, where the EFT completely breaks down. This raises the question: how can spacetime end in string theory? There are two complementary ways to approach this issue. From the low-energy effective field theory (EFT) perspective, one may ask what it means for spacetime to terminate as a solution to the equations of motion, and how such an endpoint can arise dynamically, for instance through the nucleation of a bubble of nothing \cite{Witten:1981gj}. From the ultraviolet (UV) perspective of string theory, where the number of extended spacetime dimensions can change, one may instead ask what it means for spacetime to disappear from the low-energy description altogether. In this sense, the endpoint of spacetime may correspond not merely to a boundary in a fixed-dimensional geometry, but to a transition in which certain dimensions cease to admit a geometric interpretation.

We will discuss these two viewpoints separately: first analyzing the EFT description of spacetime-ending configurations, and then turning to the UV interpretation in string theory, where the notion of spacetime itself becomes dynamical. 

\subsubsubsection{Bordism transitions in the EFT.}Solutions of the Einstein--matter equations in which spacetime literally ``ends'' have a long history, and the canonical example is Witten's bubble of nothing \cite{Witten:1981gj}.  The setup is five-dimensional Einstein gravity with one spatial direction compactified on a circle, so that asymptotically the geometry approaches the Kaluza--Klein vacuum $\mathbb{R}^{3,1}\times S^1$:
\begin{equation}
ds^2 = -dt^2 + dr^2 + r^2 d\Omega_2^2 + d\chi^2,
\qquad \chi \sim \chi + 2\pi R .
\end{equation} Witten exhibited a semiclassical decay channel of this vacuum mediated by a Euclidean gravitational instanton: one starts from the five-dimensional Euclidean Schwarzschild solution and interprets the Euclidean time circle as the Kaluza--Klein circle \cite{Witten:1981gj}:\begin{equation}
ds_{\rm Eucl}^2
= f(r)\, d\chi^2 + f(r)^{-1} dr^2 + r^2 d\Omega_3^2,
\qquad
f(r)=1-\frac{r_0^2}{r^2},
\qquad
r\ge r_0 .
\end{equation}  Smoothness at the ``bolt'' (the minimal-radius locus) forces the would-be circle to shrink to zero size there, fixing its periodicity so that no conical singularity appears:
\begin{equation}
\Delta\chi
= \frac{4\pi}{f'(r_0)}
= 2\pi r_0,
\qquad\Rightarrow\qquad
r_0 = R .
\end{equation}In Lorentzian signature, the analytic continuation of the instanton using $d\Omega_3^2 = d\psi^2 + \sin^2\psi\, d\Omega_2^2\,$
with
$\psi = \frac{\pi}{2} + i\tau $ gives the Lorentzian bubble-of-nothing metric
\begin{equation}
ds^2
= f(r)\, d\chi^2 + f(r)^{-1} dr^2
+ r^2\left(-d\tau^2 + \cosh^2\tau\, d\Omega_2^2\right),
\qquad
f(r)=1-\frac{R^2}{r^2}.
\end{equation}
This metric describes a spherical bubble wall (a timelike hypersurface) on which the $S^1$ pinches off. Beyond this wall there is no interior region at all.  In this precise sense, the end of spacetime occurs because the metric is only defined for radii outside the bubble: the circle degenerates smoothly at the bubble, and the manifold terminates.

\begin{figure}
    \centering
    \includegraphics[width=0.4\linewidth]{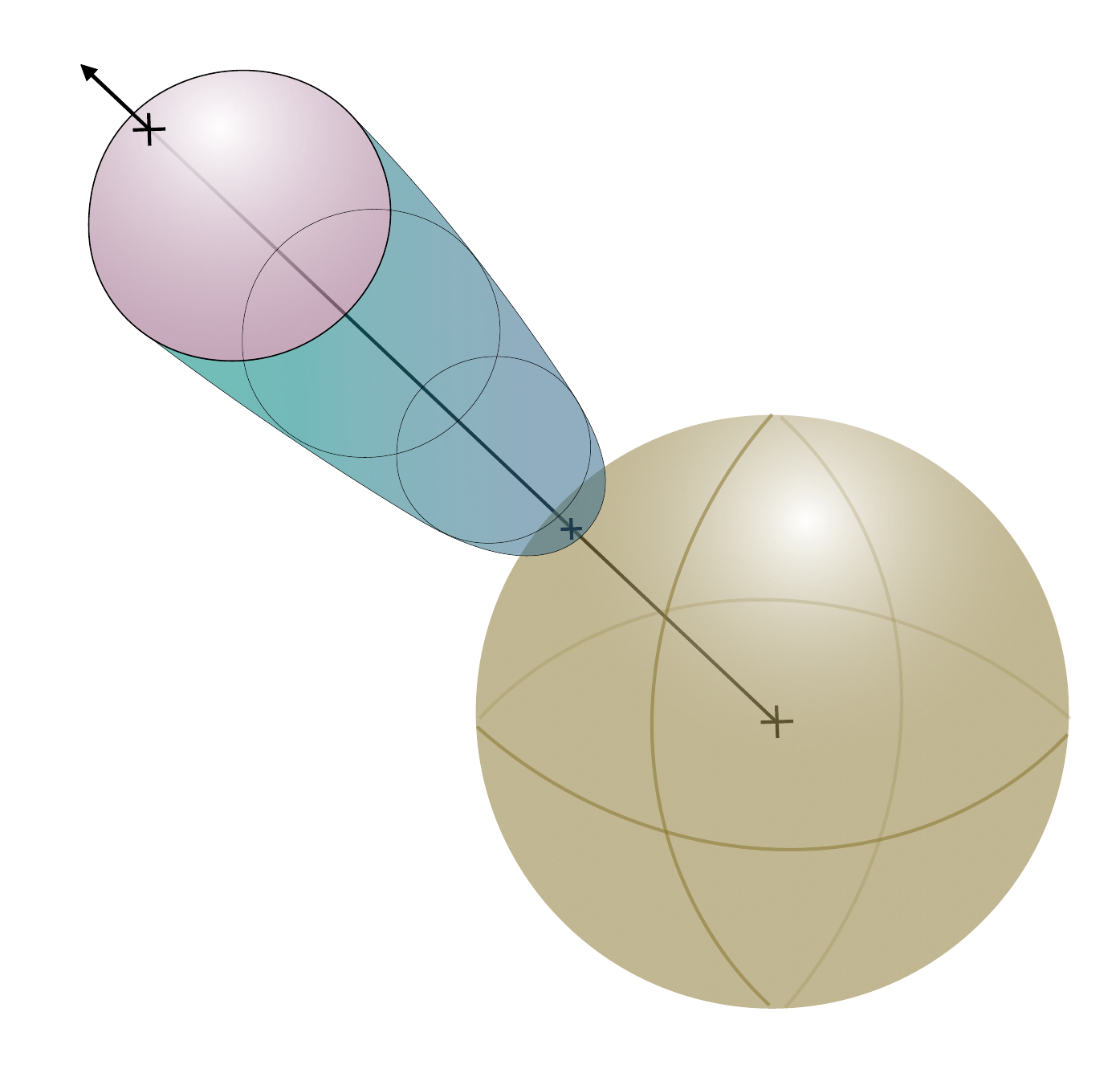}
    \caption{Depiction of the Bubble of Nothing geometry. The $\chi$-circle shrinks and eventually pinches off on the surface of the bubble, in a smooth way.}
    \label{fig:bubbleofnothing}
\end{figure}

The subsequent evolution is dramatic.  After nucleation, the bubble expands outward; in the standard slicing one finds that the bubble wall accelerates and asymptotically approaches the speed of light, so that it ultimately consumes the entire asymptotic spacetime \cite{Witten:1981gj}.  From the four-dimensional effective-field-theory viewpoint, the bubble wall is a finite-action end-of-the-world (ETW) surface where the extra dimension caps off.  This provides a concrete EFT realization of ``spacetime ending'' without introducing curvature singularities at the endpoint.

However, the bubble of nothing is not universally allowed.  A key point is that the Kaluza--Klein circle becomes contractible in the instanton geometry.  For spinors, this places a strong constraint on the spin structure: a contractible circle cannot support periodic fermions without encountering a global inconsistency, and Witten argued that the presence of elementary fermions (equivalently, a spin structure compatible with supersymmetry) obstructs the decay \cite{Witten:1981gj}.  This is the standard sense in which supersymmetric Kaluza--Klein vacua are protected against the original bubble-of-nothing instability: the topology required for the instanton is incompatible with the required fermion boundary conditions. In other words, the decay to nothing is topologically obstructed. 

In EFTs compatible with quantum gravity, the Cobordism Conjecture implies that no such topological obstructions exist, so decay to nothing should always be allowed. This makes it worthwhile to study whether bubbles of nothing are viable in theories that come from string compactifications. These theories contain many more ingredients than pure gravity on $\mathbb{R}^{3,1}\times S^1$: fluxes, branes, higher-derivative corrections, moduli, and so on. This question was studied in \cite{GarciaEtxebarria:2020xsr}, which highlighted two different kinds of obstruction. The first kind is \emph{topological}: these obstructions are encoded in bordism groups, and the Cobordism Conjecture requires all of them to be trivialized. The second kind is \emph{dynamical}, of Coleman--De~Luccia type: gravitational effects can suppress a decay even when topology allows it \cite{Coleman:1980aw}. So even though the Cobordism Conjecture guarantees that decay to nothing is always topologically allowed, the decay can still be a high-energy process that is heavily suppressed at low energies. The dynamical aspects of generalizations of the bubble of nothing as a cobordism to nothing have been further discussed under dualities in \cite{Delgado:2023uqk} and with non-trivial gauge bundles in the extra dimension in \cite{lasHeras:2026xdl}.

A different, but related, notion of ``nothing'' comes from viewing the endpoint as an infinitely deep anti--de Sitter region.  It was proposed in \cite{Brown:2011gt} that the absence of spacetime can be understood as a limit of AdS in which the curvature radius $\ell$ is taken to zero (equivalently $\Lambda\to -\infty$), so that the interior becomes more and more negatively curved until no low-energy degrees of freedom remain.  This proposal has been used as a conceptual bridge between bubbles of nothing (as endpoints of tunneling) and creation-from-nothing processes \cite{Cespedes:2020xpn}. This AdS$_{\ell\to 0}$ interpretation raises an apparent tension with the bubble-of-nothing intuition: naively, Israel junction conditions suggest that taking $\Lambda\to -\infty$ on one side of a domain wall forces the wall tension to become infinite (and even negative) to support the gluing.  A recent analysis in \cite{MuntzPadillaSaffin:2025} resolves this puzzle by emphasizing that the divergence is an artifact of using an \emph{unrenormalized} gravitational description in the $\ell\to 0$ limit. Implementing holographic renormalization, they show that the would-be infinite contribution can be isolated and cancelled by the appropriate counterterms.

\subsubsubsection{Bordism transitions in String Theory.}
When discussing the Cobordism Conjecture, one often hears statements that ``dimensions disappear'' or that compact spaces ``cap off.'' Strictly speaking, this should be interpreted as a statement about the \emph{low-energy effective description}.  In a UV complete framework such as string theory, what we call a ``dimension'' is not absolute: it is encoded in the spectrum of light worldsheet modes.

In perturbative string theory, a spacetime dimension corresponds to a free
worldsheet scalar field $X^\mu(\sigma,\tau)$ with action
\begin{equation}
S = \frac{1}{4\pi \alpha'} \int d^2\sigma \, \partial_a X^\mu \partial^a X_\mu .
\end{equation}
If the corresponding worldsheet degrees of freedom become infinitely massive,
they decouple from the low-energy spectrum, and the associated spacetime
dimension disappears from the effective description.

This phenomenon is familiar in compactification.  For example, compactifying the bosonic string on a circle of radius $R$ yields momentum and winding excitations with masses:
\begin{equation}
M^2 = \frac{n^2}{R^2} + \frac{w^2 R^2}{\alpha'^2} + \frac{2}{\alpha'}(N+\tilde N-2).
\end{equation}
When $R\to 0$, momentum modes become heavy, and when $R\to\infty$, winding modes
become heavy.  In both limits, the effective spacetime description changes,
demonstrating that the notion of spacetime dimension is dynamical in string
theory.

More dramatic examples arise in tachyon condensation, where entire spacetime directions can disappear from the low-energy theory.

\paragraph*{Winding tachyons.}
In non-supersymmetric backgrounds with a shrinking circle along which fermions are antiperiodic (Scherk--Schwarz boundary conditions), the lightest winding modes ($w=1$, $n=0$) of the type II superstring can become tachyonic,
\begin{equation}\label{eq:winding_tachyons}
M^2_{\text{winding}} = \frac{R^2}{\alpha'^2} - \frac{2}{\alpha'} < 0
\quad \text{for } R<\sqrt{2\alpha'} .
\end{equation}
Condensation of this winding tachyon caps off the circle, removing the
associated dimension from the infrared geometry.  The endpoint of the
condensation is a smooth geometry in which the circle shrinks to zero size,
analogous to the ``bubble of nothing'' picture, but realized within a fully
UV-complete worldsheet description \cite{Adams:2001sv,Adams:2005rb}.

\paragraph*{Hellerman--Swanson Bulk Tachyons.}
Hellerman and Swanson \cite{Hellerman:2006nx, Hellerman:2006hf,Hellerman:2007fc,Hellerman:2006ff} (see also \cite{Kaidi:2020jla}) considered backgrounds with a \emph{bulk} tachyon profile that depends on a lightlike spacetime coordinate,
\begin{equation}
T(X) \sim \mu \, e^{\beta X^+},
\end{equation}
on top of a lightlike linear dilaton. The dilaton slope fixes $\beta$, since the tachyon vertex operator must be marginal. This profile induces a worldsheet potential that grows exponentially along a lightlike direction. Tachyon condensation then dynamically removes the region of spacetime where the tachyon diverges, so that region drops out of the late-time description. This is a realization of a dynamical cobordism to nothing \cite{Angius:2022mgh}. The mechanism is easiest to see in the bosonic string, where the tachyon enters the worldsheet action as a potential,
\begin{equation}
S \supset \int d^2\sigma \, \sqrt{-h}\, T(X) ,
\end{equation}
with $h_{ab}$ the worldsheet metric. As $X^+ \to \infty$ the string becomes infinitely massive and all of its massless modes disappear. The resulting configuration is a bubble that interpolates between a smooth spacetime with a lightlike dilaton at $X^+ \to -\infty$ and, at $X^+ \to \infty$, a region with no notion of a low-energy spacetime \cite{Hellerman:2006nx}. The construction is exact in $\alpha'$, so it can be trusted wherever the string coupling is weak. The setup has been generalized to superstrings and to dimension-changing bubbles, in which only \emph{some} of the spacetime dimensions disappear. In that case the tachyon also depends on additional fields $X^i$, which become massive as $X^+ \to \infty$. The central charge lost when these fields decouple is compensated by a change in the slope of the linear dilaton, so the total worldsheet central charge is preserved.

These examples illustrate that ``dimension disappearance'' is not mysterious in a UV complete theory: it corresponds to worldsheet degrees of freedom becoming massive and decoupling.  Tachyon condensation provides explicit mechanisms by which compact directions cap off or spacetime regions are removed.  In this sense, such processes can be viewed as concrete realizations of cobordism transitions within string theory, where the full worldsheet description tracks how spacetime dimensions dynamically change.

\paragraph{Summary.}
From the low-energy perspective, spacetime can end smoothly, as in the bubble of nothing, where a compact direction caps off on an ``end of the world'' wall. The Cobordism Conjecture guarantees that such decays are topologically allowed, although dynamical obstructions may still suppress them. In string theory, a dimension disappears when the worldsheet degrees of freedom that describe it become massive and decouple. Tachyon condensation provides concrete realizations of this, in which compact directions cap off or entire spacetime regions are removed. Admittedly, this discussion reads as a list of examples, since there is, as of now, no universal understanding of how bordism transitions can be realized dynamically. Nevertheless, these examples already show that such transitions are not exotic, but a natural feature of string theory, and understanding them more systematically is the subject of ongoing research.

\subsubsection{Towards Understanding Unknown Cobordism Defects} \label{sec:unknownBdef}
What can be said about the dynamical aspects of the new, non-supersymmetric branes predicted by the Cobordism Conjecture? In this final part, we follow the template of Section~\ref{sec:charact}. We first discuss what can be said about the worldvolume theory of these defects, and then their backreaction on spacetime. We then review other approaches that have been used to study them.

\paragraph{Worldvolume theory.} We start with the worldvolume theory, since knowing it would determine everything about the brane, including how it couples to the bulk. In general, the only thing one can say about the worldvolume theory of a generic bordism defect is that it \hl{contains a topological term identifying the bordism class it trivializes}. Indeed, consider a $D$-dimensional theory and a class $[M_d]\in\Omega_d^{\xi}$ trivialized by a defect $\mathcal{D}$ with $(D-d-1)$-dimensional worldvolume $\mathcal{W}_{D-d-1}$, so that a representative $M_d$ arises as the boundary of a small neighbourhood of $\mathcal{W}_{D-d-1}$. Suppose that $[M_d]$ generates a free summand $\mathbb{Z}\subset\Omega_d^{\xi}$,\footnote{The torsion case is more subtle, since no closed differential form detects it. The corresponding invariant is valued in $\mathbb{Z}_n$ or $\mathbb{R}/\mathbb{Z}$ (it can be, e.g., an $\eta$-invariant), and the broken symmetry is a discrete $\mathbb{Z}_n$ $(D-d-1)$-form symmetry. The defect can then at most couple to a discrete gauge field $b_{D-d-1}$.} and that the corresponding bordism invariant is the integral of a closed $d$-form $\omega_d$ built out of bulk fields,
\begin{equation}
  I(M_d)=\frac{1}{2\pi}\int_{M_d}\omega_d \in \mathbb{Z}\,.
\end{equation}
The operators $U_\alpha(\Sigma_d)=\exp\big(\tfrac{i\alpha}{2\pi}\int_{\Sigma_d}\omega_d\big)$ generate a $(D-d-1)$-form U$(1)$ global symmetry, and the defect breaks it by violating the Bianchi identity,
\begin{equation}
  d\omega_d = 2\pi q\,\delta_{d+1}(\mathcal{W}_{D-d-1})\,.
\end{equation}
Writing $\omega_d=F_d=dA_{d-1}$ away from $\mathcal{W}_{D-d-1}$, we see that the defect is a magnetic source for $A_{d-1}$. Equivalently, introducing the dual potential $\tilde{A}_{D-d-1}$ with $d\tilde{A}_{D-d-1}\propto\star F_d$, the topological term on the worldvolume is simply the electric coupling
\begin{equation}\label{eq:generalizeCS}
  S_{\mathcal{W}} \supset q\int_{\mathcal{W}_{D-d-1}} \tilde{A}_{D-d-1}\,.
\end{equation}
This is the analogue of the first term in the CS action for D-branes \eqref{eq:CSaction}. In general, nothing more is known: the analogues of the DBI action and of the higher CS couplings, as well as the intrinsic worldvolume degrees of freedom, are not fixed by the topological analysis. In some cases, anomaly inflow can shed further light on the worldvolume theory, since the degrees of freedom living on the brane must cancel the anomalies induced by the bulk couplings (see e.g.~\cite{Basile:2023knk,Basile:2025mnj}).

\paragraph{Backreaction on spacetime.}It is tempting to describe these unknown objects in the same way as the brane solutions of Section~\ref{sec:charact}. Indeed, many other cobordism defects have been found to share features with D-branes, a pattern explored under the name of Dynamical Cobordism~\cite{Buratti:2021fiv}. In string models with tadpoles or scalar potentials, one always finds spacetime-dependent solutions that evolve along one direction and terminate at a finite spacetime distance, at a ``wall of nothing'' that realizes the cobordism defect dynamically \cite{Angius:2022aeq}. As in the D-brane case, the scalars run off to infinite distance $D$ in field space as one approaches the defect. Across all examples studied, the spacetime distance $\Delta$ to the defect and the scalar curvature $R$ obey universal scalings near the location of the defect:
\begin{equation}
\Delta \sim e^{-\frac{\delta}{2} D}\,, \qquad |R| \sim e^{\delta D}\,,
\end{equation}
where $\delta$ is a model-dependent $\mathcal{O}(1)$ coefficient. In fact, it was argued in \cite{Angius:2022aeq} that this behaviour is generic in string EFTs, close enough to a finite distance singularity. These relations tie the end of spacetime directly to the infinite-distance limits of the Swampland Distance Conjecture. Applying these ideas to unknown defects, one can for instance try to resolve the singularity of a backreacted solution with a localized, delta-function source carrying the appropriate tension and dilaton coupling~\cite{Blumenhagen:2022mqw,Blumenhagen:2023abk,Delgado:2023uqk}. This approach starts from an ansatz based on the expected form and symmetries of the solution, and then solves the resulting equations of motion. In this way, one can obtain generalizations of the brane-type solutions described above \cite{Damian:2026jsz,Basile:2022ypo}. In the absence of supersymmetry, however, the system is often underconstrained. Several families of solutions may satisfy the same basic requirements, with no clear criterion for identifying the physically relevant one. In some cases, enough prior knowledge of the defect allows one to come close to a complete solution, as for the R7-brane studied in \cite{Heckman:2025wqd,Cavusoglu:2026xiv}, or for the non-supersymmetric heterotic branes of \cite{Kaidi:2023tqo,Kaidi:2024cbx,Fukuda:2024pvu}. Beyond case-by-case analyses, general principles of quantum gravity can be used to argue for nontrivial constraints on the tensions of cobordism defects, even without supersymmetry \cite{Nevoa:2025xiq}. Moreover, from the bottom-up, it may be possible to constrain the space of \emph{allowed} singularities \cite{Calderon-Infante:2026ymy,Makridou:2026jzy}.

More generally, there is no general argument that a cobordism defect should be a localized brane at all: all the Cobordism Conjecture predicts is something that carries the right topological coupling \eqref{eq:generalizeCS}. One can sometimes argue that the topological magnetic charge it carries forces the configuration to shrink, so that the defect becomes localized in spacetime \cite{Kaidi:2023tqo,Kaidi:2024cbx}, but this need not always be the case. In some cases, the required cobordism defect can be identified as the intersection of higher-dimensional objects rather than as a standalone object \cite{Dierigl:2026psj}. One can also imagine that a cobordism to nothing proceeds step by step, removing part of the topology at each stage \cite{Ruiz:2024jiz}.

\paragraph{Holography.}A complementary handle on end-of-the-world (ETW) branes is provided by holography. When the ETW brane is embedded in an asymptotically AdS spacetime, its properties can be studied from the perspective of the dual field theory, where the brane corresponds to a boundary condition or defect. This has been used to study the dynamics of ETW branes in \cite{Huertas:2023syg,Huertas:2024mvy,Anastasi:2025puv,Hassfeld:2025hjx}.

\paragraph{Domain walls between different theories.}One of the most striking predictions of the Cobordism Conjecture is the existence of a domain wall interpolating between type IIA and type IIB string theory. While type IIA degrees of freedom can terminate on a Ho\v{r}ava--Witten wall, as discussed in Section~\ref{sec:applying}, the situation is more subtle for type IIB. Its chiral degrees of freedom cannot simply be gapped. Understanding how chiral theories can end is a general challenge for the Cobordism Conjecture \cite{Angius:2024pqk}, and presumably requires intrinsically strongly coupled stringy effects. Several complementary approaches have been used to investigate the properties of such a domain wall, including worldsheet methods \cite{Heckman:2025wqd}, non-compact CFTs \cite{Anastasi:2026cus}, and matrix theory \cite{Torres:2026vxx}. Junctions between more than two theories have also been discussed \cite{Hellerman:2010dv,Altavista:2026edv,Tachikawa:2026jaj,Altavista:2026muc}. A simpler realization arises for domain walls between theories that differ only by some amount of flux. These can be realized using D-branes, which change the flux as one crosses them \cite{Buratti:2021fiv,Shiu:2023bay}. 

\paragraph{Generalized Bubbles of Nothing.}Finally, generalizations of the bubble of nothing (see Section~\ref{sec:etw}) have been used to describe dynamical cobordisms \cite{Blanco-Pillado:2023aom,Blanco-Pillado:2023hog,Ookouchi:2024tfz}, and to argue that gravitational instantons of high genus carry a large tension \cite{lasHeras:2026xdl}.

\paragraph{Summary.}Unlike for D-branes, for generic non-supersymmetric cobordism defects one typically knows only the topological coupling that identifies the bordism class they trivialize. Their worldvolume theory, tension and backreaction must be inferred by other means: anomaly inflow, ans\"atze for supergravity solutions, quantum gravity constraints on tensions and singularities, holography, and worldsheet or matrix theory techniques. Beyond their intrinsic interest, cobordism defects and dynamical cobordism solutions can be used to connect to other Swampland conjectures (e.g.~\cite{Aparici:2026svo,Bedroya:2025ltj,Calderon-Infante:2023ler,Velazquez:2022eco}). For instance, domain walls can be used to define notions of distance between vacua in the quantum gravity landscape \cite{Basile:2023rvm,Cribiori:2026btb}. Understanding the dynamics of these objects is the subject of ongoing research. Several steps remain before one can connect directly to realistic universes,\footnote{For instance, dynamical cobordism solutions should be described in the presence of the stabilizing scalar potentials of string phenomenology models \cite{Angius:2023xtu,Delgado:2025crl}.} but such an understanding could have implications for global symmetries in our own universe \cite{Wang:2021ayd,Hamada:2025duq}.

\subsection{Further Topics}

In this final part, we briefly discuss several further topics that are closely related to the Cobordism Conjecture and to the role of bordism groups in the Swampland Program, but that we did not have the chance to elaborate on in the main text. These include the Baby Universe Hypothesis, the nucleation of universes with boundaries in quantum cosmology, and the use of bordism groups in classifying global anomalies. Our aim is not to give a complete treatment, but rather to highlight how the ideas developed so far connect to these topics, and to point the interested reader to the relevant literature.

\subsubsection{The Baby Universe Hypothesis}\label{sssec:BabyUniv}

\begin{figure}[htpb!]
    \centering
    \includegraphics[width=0.6\linewidth]{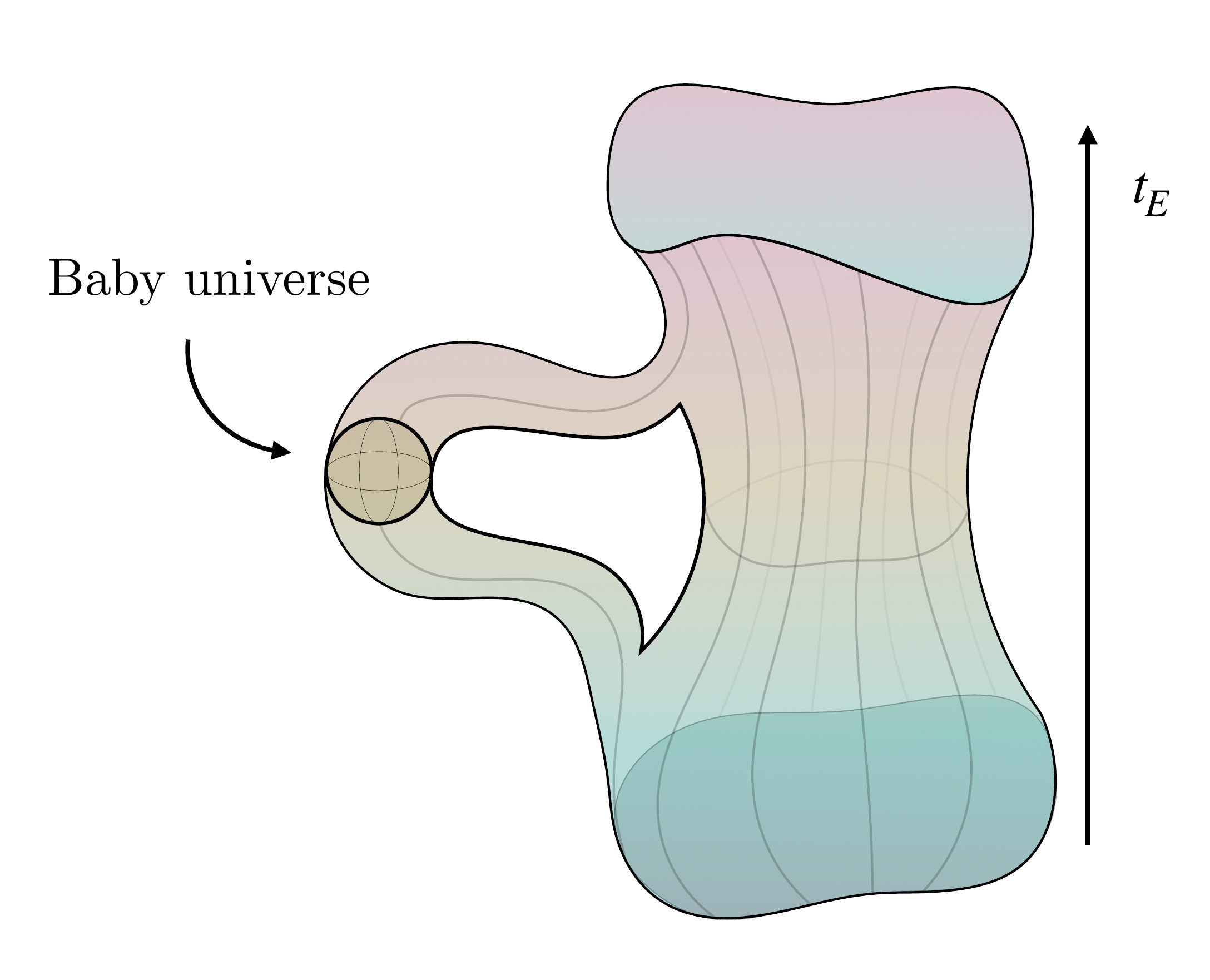}
    \caption{Formation and absorption of a baby universe along the Euclidean time direction.}
    \label{fig}
\end{figure}

A \emph{baby universe} is a closed component of spacetime which branches off from a larger universe and may later be reabsorbed, as illustrated in Figure~\ref{fig}. Since it has no asymptotic boundary, it carries no conserved energy or momentum measured at infinity. We define the \hl{baby-universe Hilbert space $\mathcal{H}_{\rm BU}$} as the space of quantum-gravity states whose spatial boundary is empty.

At first sight, $\mathcal{H}_{\rm BU}$ could contain many states, corresponding for instance to different internal geometries or topologies. This would have observable consequences for the parent universe. In the Euclidean path integral, the emission and absorption of baby universes produces effective couplings of the form \cite{COLEMAN1988867,Hebecker:2018ofv,Marolf:2020xie}
\begin{equation}
    S_{\rm eff}
    =
    S_0+\sum_i \alpha_i\int \mathcal{L}_i .
    \label{eq:alpha-parameters}
\end{equation}
The parameters $\alpha_i$ are eigenvalues of baby-universe creation and annihilation operators, which commute with all observables in the parent universe. Different values of $\alpha_i$ therefore label different superselection sectors,\footnote{Let $\hat A_i$ be the (Hermitian) baby-universe operator with eigenvalue $\alpha_i$, and suppose $[\hat A_i,\mathcal{O}]=0$ for every observable $\mathcal{O}$ of the parent universe. For eigenstates $\hat A_i|\alpha\rangle=\alpha_i|\alpha\rangle$ and $\hat A_i|\alpha'\rangle=\alpha_i'|\alpha'\rangle$, one has $(\alpha_i-\alpha_i')\langle\alpha'|\mathcal{O}|\alpha\rangle=0$. Hence $\langle\alpha'|\mathcal{O}|\alpha\rangle=0$ whenever $\alpha_i\neq\alpha_i'$, meaning that no observable can connect different values of $\alpha_i$. So $\alpha_i$ is a conserved, superselected label.} or equivalently different effective theories seen by observers in the parent universe.

This is in tension with the absence of global symmetries. A fixed coupling constant can be viewed as a background for a $(-1)$-form symmetry (see Example~4 in Section~\ref{ssec:globsym_examples}). In quantum gravity such a symmetry cannot remain global: the coupling must arise as the vacuum expectation value of a dynamical field, so that finite-energy processes can change its value locally. The $\alpha_i$, however, are independent of position and cannot be changed by any local process. Non-trivial $\alpha$-parameters would therefore amount to $(-1)$-form global symmetries.

The Cobordism Conjecture gives a complementary perspective on the same obstruction. It implies that any two closed spatial manifolds can be connected by a finite-energy process or domain wall. Consequently, closed universes distinguished only by their topology cannot define independent superselection sectors: they are connected by dynamical cobordisms and must be regarded as gauge equivalent in the full theory. After quotienting by these redundancies, \hl{only one physical baby-universe state remains}.

The absence of global symmetries and the Cobordism Conjecture therefore lead to

\begin{symmetrybox}{dustyPlum}
    \textbf{The Baby Universe Hypothesis \cite{McNamara:2020uza}}\;
In a unitary theory of quantum gravity in $d>3$ spacetime dimensions, $
    \dim \mathcal{H}_{\rm BU}=1 
$.
\end{symmetrybox}

\noindent Thus, closed universes do not carry independent quantum information and do not introduce free $\alpha$-parameters. Equivalently, the full theory admits no superselection sectors associated with finite, closed components of spacetime. The restriction to $d>3$ is important \cite{McNamara:2020uza}: in lower-dimensional theories, such as JT gravity, ensemble interpretations and non-trivial baby-universe sectors may arise \cite{Saad:2019lba,Marolf:2020xie}. In $d=2$, massless scalars cannot acquire vacuum expectation values, so couplings need not be set by dynamical fields, while in $d=3$ gravity has no local propagating degrees of freedom. Consequently, the higher-dimensional Swampland arguments for the absence of free parameters, global symmetries, and non-trivial cobordism need not apply in the same form.


\subsubsection{On Nucleating Universes with Boundaries}

The Cobordism Conjecture implies the availability of end-of-the-world (ETW) branes (or analogous objects) that can serve as genuine, dynamical spacetime boundaries. In other words, any quantum gravity vacuum admits a boundary. In \cite{Hassfeld:2023kpu,Hassfeld:2024hnx}, the authors take this slogan seriously and ask what it could mean in \emph{quantum cosmology}: if dynamical boundaries are allowed/required by cobordism, then Euclidean saddles with genuine boundary components should be placed on essentially the same footing as the usual compact ``no-boundary'' saddles.

They propose a concrete alternative to the Hartle--Hawking (HH) no-boundary saddle for de Sitter creation: instead of the round $S^4$ instanton, take a Euclidean four-sphere with two small four-balls excised around the north and south poles, producing an instanton with \emph{two} $S^3$ boundary components, interpreted as ETW branes (the ``Boundary Proposal''). One then cuts at the equator and glues the remaining half to an expanding Lorentzian de Sitter region. Crucially, this boundary saddle is a legitimate competitor in the semiclassical sum, and the authors argue that the boundary creation process can even \emph{dominate} over the traditional HH saddle. 

More generally, the existence of boundaries and ETW branes plays an important role in quantum cosmology, as they participate in the rates of creating universes from nothing \cite{Hassfeld:2024ake}. 

\subsubsection{Bordism Groups and Anomalies}
Bordism groups also play an important role in identifying global gauge and gravitational anomalies in EFTs of quantum gravity. Recall that an anomaly is a gauge transformation or diffeomorphism of the spacetime $X_d$ under which the path integral is not invariant. If the anomalous transformation can be continuously deformed to the identity, the anomaly is called \emph{local}. Local anomalies are the most severe, since they arise for infinitesimal transformations. In four dimensions they arise from triangle Feynman diagrams, and they are captured by the anomaly polynomial in $d+2$ dimensions \cite{Alvarez-Gaume:1984zlq}.

Once local anomalies cancel, there can still be \emph{global} anomalies, associated with transformations that cannot be deformed to the identity. A famous example is Witten's $\SU(2)$ anomaly \cite{Witten:1982fp}. In the modern approach, anomalies are encoded in an \emph{anomaly theory}, an invertible field theory in $d+1$ dimensions whose boundary is $X_d$ (see \cite{Dai:1994kq,Witten:2015aba,Yonekura:2016wuc} and \cite{Garcia-Etxebarria:2018ajm} for a review). The anomaly theory is not a physical theory in one dimension more. Rather, it is constructed to have exactly the opposite anomaly to the $d$-dimensional theory, so that the coupled system is anomaly-free.

The advantage of this approach is that, once local anomalies cancel, the partition function of the anomaly theory is a bordism invariant of the $(d+1)$-dimensional manifold $Y_{d+1}$ \cite{Atiyah_Patodi_Singer_1975}. Detecting global anomalies therefore reduces to computing the relevant bordism groups $\Omega_{d+1}$. This is not always easy, depending on the structure (spin, gauge bundle, etc.) one has to impose. For instance, Witten's $\SU(2)$ anomaly is detected by $\Omega_5^{\rm Spin}(B\SU(2)) = \mathbb{Z}_2$. In a theory with dynamical gravity, anomaly cancellation must hold on \emph{all} closed manifolds $Y_{d+1}$, including those that describe topology-changing transitions of $X_d$. Evaluating the path integral directly across such transitions would require a non-perturbative description of quantum gravity, since they are typically singular in the EFT. Bordism groups nevertheless characterize them topologically. This makes global anomaly cancellation a sharp consistency condition on any EFT coupled to gravity and, in this sense, a bottom-up tool for the Swampland Program.

\section*{Acknowledgements}
MD thanks the participants of the 21st Modave Summer School in Mathematical Physics and the KMPB School on D-branes in Physics and Mathematics, whose questions and discussions greatly improved these notes. MD is also grateful to the organizers of both schools for the invitation to lecture and for creating such a welcoming and stimulating environment. We thank Jos\'e Calder\'on-Infante, Jacob McNamara, Miguel Montero, Angel M.~Uranga, and Cumrun Vafa for helpful discussions and valuable comments on the manuscript.
EVA thanks Pablo Saura-Bastida and Michelangelo Tartaglia for helpful discussions on higher-form symmetries and topology and Fangyi Guo for pointing out relevant references, as well as Jos\'e Barb\'on for his guidance and support.
GMA would also like to thank Nikolay Bobev for encouraging him to participate in the writing of these notes, and Joel Karlsson for helpful discussions.
GMA is partially supported by the FWO projects G003523N, G094523N, and G0E2723N, and the KU Leuven C1 project C16/25/01. GMA is also a PhD Fellow of the Research Foundation -- Flanders (FWO) with project number 1141426N.
CE acknowledges travel support from the Chair of Theoretical Physics 3 at JMU Würzburg through the German Research Foundation (DFG) through the Collaborative Research Center ToCoTronics, Project ID 258499086 - SFB 1170, as well as Germany's Excellence Strategy through the Würzburg-Dresden Cluster of Excellence on Complexity and Topology in Quantum Matter - ctd.qmat (EXC 2147, Project-ID 390858490).
This publication has been funded within the framework of the R\&D\&I Projects PID2024-156043NB-I00 and CEX2025-001574-S, funded by MICIU/AEI/10.13039/501100011033. The research presented in this publication falls within the research line Strings and Quantum Gravity. The work was also partially supported by the grants PID2022-137127NB-I00, funded by MCIN/AEI/10.13039/501100011033/ FEDER, UE, and `ERDF A way of making Europe'. EVA also acknowledges support from the European Union’s Horizon 2020 research and innovation programme under the Marie Sklodowska-Curie grant agreement No 860881-HIDDeN. The work of MD is supported in part by a grant from
the Simons Foundation (602883,CV) and the DellaPietra Foundation.

\newpage

\bibliographystyle{JHEP}
\bibliography{references_v2}

@article{Benini:2022hzx,
    author = "Benini, Francesco and Copetti, Christian and Di Pietro, Lorenzo",
    title = "{Factorization and global symmetries in holography}",
    eprint = "2203.09537",
    archivePrefix = "arXiv",
    primaryClass = "hep-th",
    reportNumber = "SISSA 05/2022/FISI, SISSA 05/2022/FISI",
    doi = "10.21468/SciPostPhys.14.2.019",
    journal = "SciPost Phys.",
    volume = "14",
    number = "2",
    pages = "019",
    year = "2023"
}

@article{Maloney:2007ud,
    author = "Maloney, Alexander and Witten, Edward",
    title = "{Quantum Gravity Partition Functions in Three Dimensions}",
    eprint = "0712.0155",
    archivePrefix = "arXiv",
    primaryClass = "hep-th",
    doi = "10.1007/JHEP02(2010)029",
    journal = "JHEP",
    volume = "02",
    pages = "029",
    year = "2010"
}

@article{Cotler:2020ugk,
    author = "Cotler, Jordan and Jensen, Kristan",
    title = "{AdS$_{3}$ gravity and random CFT}",
    eprint = "2006.08648",
    archivePrefix = "arXiv",
    primaryClass = "hep-th",
    doi = "10.1007/JHEP04(2021)033",
    journal = "JHEP",
    volume = "04",
    pages = "033",
    year = "2021"
}

@article{Klebanov:2000hb,
    author = "Klebanov, Igor R. and Strassler, Matthew J.",
    title = "{Supergravity and a confining gauge theory: Duality cascades and chi SB resolution of naked singularities}",
    eprint = "hep-th/0007191",
    archivePrefix = "arXiv",
    reportNumber = "IASSNS-HEP-00-56, PUPT-1944",
    doi = "10.1088/1126-6708/2000/08/052",
    journal = "JHEP",
    volume = "08",
    pages = "052",
    year = "2000"
}

@article{Polchinski:2000uf,
    author = "Polchinski, Joseph and Strassler, Matthew J.",
    title = "{The String dual of a confining four-dimensional gauge theory}",
    eprint = "hep-th/0003136",
    archivePrefix = "arXiv",
    reportNumber = "IAS-TH-00-18, NSF-ITP-00-16",
    month = "3",
    year = "2000"
}

@article{Pilch:2000ue,
    author = "Pilch, Krzysztof and Warner, Nicholas P.",
    title = "{N=2 supersymmetric RG flows and the IIB dilaton}",
    eprint = "hep-th/0004063",
    archivePrefix = "arXiv",
    reportNumber = "CITUSC-00-018, USC-00-02",
    doi = "10.1016/S0550-3213(00)00656-8",
    journal = "Nucl. Phys. B",
    volume = "594",
    pages = "209--228",
    year = "2001"
}

@article{Kanitscheider:2008kd,
    author = "Kanitscheider, Ingmar and Skenderis, Kostas and Taylor, Marika",
    title = "{Precision holography for non-conformal branes}",
    eprint = "0807.3324",
    archivePrefix = "arXiv",
    primaryClass = "hep-th",
    reportNumber = "ITFA-2008-25",
    doi = "10.1088/1126-6708/2008/09/094",
    journal = "JHEP",
    volume = "09",
    pages = "094",
    year = "2008"
}

@article{MuntzPadillaSaffin:2025,
    author = "Muntz, Benjamin and Padilla, Antonio and Saffin, Paul M.",
    title = "{Do we live on the End of the World?}",
    eprint = "2411.05912",
    archivePrefix = "arXiv",
    primaryClass = "hep-th",
    doi = "10.1007/JHEP05(2025)006",
    journal = "JHEP",
    volume = "05",
    pages = "006",
    year = "2025"
}

@article{Hellerman:2006nx,
    author = "Hellerman, Simeon and Swanson, Ian",
    title = "{Cosmological solutions of supercritical string theory}",
    eprint = "hep-th/0611317",
    archivePrefix = "arXiv",
    doi = "10.1103/PhysRevD.77.126011",
    journal = "Phys. Rev. D",
    volume = "77",
    pages = "126011",
    year = "2008"
}

@ARTICLE{1945PNAS...31..117E,
       author = {{Eilenberg}, Samuel and {Steenrod}, Norman E.},
        title = "{Axiomatic Approach to Homology Theory}",
      journal = {Proceedings of the National Academy of Science},
         year = 1945,
        month = apr,
       volume = {31},
       number = {4},
        pages = {117-120},
          doi = {10.1073/pnas.31.4.117},
       adsurl = {https://ui.adsabs.harvard.edu/abs/1945PNAS...31..117E}
}

@inproceedings{Scorpan2005TheWW,
  title={The wild world of 4-manifolds},
  author={Alexandru Scorpan},
  year={2005},
  url={https://api.semanticscholar.org/CorpusID:117823549}
}

@article{Becker:2024ayh,
    author = "Becker, Katrin and Brady, Nathan and Gra{\~n}a, Mariana and Morros, Miguel and Sengupta, Anindya and You, Qi",
    title = "{Tadpole conjecture in non-geometric backgrounds}",
    eprint = "2407.16758",
    archivePrefix = "arXiv",
    primaryClass = "hep-th",
    doi = "10.1007/JHEP10(2024)021",
    journal = "JHEP",
    volume = "10",
    pages = "021",
    year = "2024"
}

@article{Hellerman:2006tx,
    author = "Hellerman, Simeon and Walcher, Johannes",
    title = "{Worldsheet CFTs for Flat Monodrofolds}",
    eprint = "hep-th/0604191",
    archivePrefix = "arXiv",
    month = "4",
    year = "2006"
}

@article{Chen:2025rkb,
    author = "Chen, Shi and van de Heisteeg, Damian and Vafa, Cumrun",
    title = "{Symmetries and M-theory-like vacua in four dimensions}",
    eprint = "2503.16599",
    archivePrefix = "arXiv",
    primaryClass = "hep-th",
    doi = "10.1007/JHEP07(2025)258",
    journal = "JHEP",
    volume = "07",
    pages = "258",
    year = "2025"
}

@article{Rajaguru:2024emw,
    author = "Rajaguru, Muthusamy and Sengupta, Anindya and Wrase, Timm",
    title = "{Fully stabilized Minkowski vacua in the 2$^{6}$ Landau-Ginzburg model}",
    eprint = "2407.16756",
    archivePrefix = "arXiv",
    primaryClass = "hep-th",
    doi = "10.1007/JHEP10(2024)095",
    journal = "JHEP",
    volume = "10",
    pages = "095",
    year = "2024"
}

@article{Narain:1986qm,
    author = "Narain, K. S. and Sarmadi, M. H. and Vafa, C.",
    title = "{Asymmetric Orbifolds}",
    reportNumber = "HUTP-86-A089",
    doi = "10.1016/0550-3213(87)90228-8",
    journal = "Nucl. Phys. B",
    volume = "288",
    pages = "551",
    year = "1987"
}

@article{Baykara:2024tjr,
    author = "Baykara, Zihni Kaan and Tarazi, Houri-Christina and Vafa, Cumrun",
    title = "{New Nonsupersymmetric Tachyon-Free Strings}",
    eprint = "2406.00185",
    archivePrefix = "arXiv",
    primaryClass = "hep-th",
    doi = "10.1103/3c5v-yv16",
    journal = "Phys. Rev. Lett.",
    volume = "136",
    number = "6",
    pages = "061602",
    year = "2026"
}

@article{Perez:2012wv,
    author = "Perez, Alejandro",
    title = "{The Spin Foam Approach to Quantum Gravity}",
    eprint = "1205.2019",
    archivePrefix = "arXiv",
    primaryClass = "gr-qc",
    doi = "10.12942/lrr-2013-3",
    journal = "Living Rev. Rel.",
    volume = "16",
    pages = "3",
    year = "2013"
}

@article{Basile:2025zjc,
    author = "Basile, Ivano and Knorr, Benjamin and Platania, Alessia and Schiffer, Marc",
    title = "{Asymptotic safety, quantum gravity, and the swampland: A conceptual assessment}",
    eprint = "2502.12290",
    archivePrefix = "arXiv",
    primaryClass = "hep-th",
    reportNumber = "MPP-2025-37",
    doi = "10.21468/SciPostPhys.20.2.027",
    journal = "SciPost Phys.",
    volume = "20",
    number = "2",
    pages = "027",
    year = "2026"
}

@article{Emparan:2001wn,
    author = "Emparan, Roberto and Reall, Harvey S.",
    title = "{A Rotating black ring solution in five-dimensions}",
    eprint = "hep-th/0110260",
    archivePrefix = "arXiv",
    reportNumber = "CERN-TH-2001-294",
    doi = "10.1103/PhysRevLett.88.101101",
    journal = "Phys. Rev. Lett.",
    volume = "88",
    pages = "101101",
    year = "2002"
}

@article{Hellerman:2010dv,
    author = "Hellerman, Simeon and Kleban, Matthew",
    title = "{Dynamical Cobordisms in General Relativity and String Theory}",
    eprint = "1009.3277",
    archivePrefix = "arXiv",
    primaryClass = "hep-th",
    reportNumber = "IPMU-10-0156",
    doi = "10.1007/JHEP02(2011)022",
    journal = "JHEP",
    volume = "02",
    pages = "022",
    year = "2011"
}

@article{Tachikawa:2026jaj,
    author = "Tachikawa, Yuji",
    title = "{On the trivalent junction of three non-tachyonic heterotic string theories}",
    eprint = "2603.28133",
    archivePrefix = "arXiv",
    primaryClass = "hep-th",
    doi = "10.1007/JHEP06(2026)194",
    journal = "JHEP",
    volume = "06",
    pages = "194",
    year = "2026"
}

@article{Dai:1994kq,
    author = "Dai, Xian-zhe and Freed, Daniel S.",
    title = "{eta invariants and determinant lines}",
    eprint = "hep-th/9405012",
    archivePrefix = "arXiv",
    doi = "10.1063/1.530747",
    journal = "J. Math. Phys.",
    volume = "35",
    pages = "5155--5194",
    year = "1994",
    note = "[Erratum: J.Math.Phys. 42, 2343--2344 (2001)]"
}

@article{Witten:2015aba,
    author = "Witten, Edward",
    title = "{Fermion Path Integrals And Topological Phases}",
    eprint = "1508.04715",
    archivePrefix = "arXiv",
    primaryClass = "cond-mat.mes-hall",
    doi = "10.1103/RevModPhys.88.035001",
    journal = "Rev. Mod. Phys.",
    volume = "88",
    number = "3",
    pages = "035001",
    year = "2016"
}

@article{Yonekura:2016wuc,
    author = "Yonekura, Kazuya",
    title = "{Dai-Freed theorem and topological phases of matter}",
    eprint = "1607.01873",
    archivePrefix = "arXiv",
    primaryClass = "hep-th",
    reportNumber = "IPMU-16-0094",
    doi = "10.1007/JHEP09(2016)022",
    journal = "JHEP",
    volume = "09",
    pages = "022",
    year = "2016"
}

@article{Atiyah_Patodi_Singer_1975, title={Spectral asymmetry and Riemannian Geometry. I}, volume={77}, DOI={10.1017/S0305004100049410}, number={1}, journal={Mathematical Proceedings of the Cambridge Philosophical Society}, author={Atiyah, M. F. and Patodi, V. K. and Singer, I. M.}, year={1975}, pages={43–69}}

@article{Harlow:2018fse,
    author = "Harlow, Daniel",
    title = "{TASI Lectures on the Emergence of Bulk Physics in AdS/CFT}",
    eprint = "1802.01040",
    archivePrefix = "arXiv",
    primaryClass = "hep-th",
    doi = "10.22323/1.305.0002",
    journal = "PoS",
    volume = "TASI2017",
    pages = "002",
    year = "2018"
}

@article{Grana:2021zvf,
    author = "Gra{\~n}a, Mariana and Herr{\'a}ez, Alvaro",
    title = "{The Swampland Conjectures: A Bridge from Quantum Gravity to Particle Physics}",
    eprint = "2107.00087",
    archivePrefix = "arXiv",
    primaryClass = "hep-th",
    doi = "10.3390/universe7080273",
    journal = "Universe",
    volume = "7",
    number = "8",
    pages = "273",
    year = "2021"
}

@article{Cheng:2026abk,
    author = "Cheng, Peng and De Freitas, Hector Parra",
    title = "{Dai-Freed anomalies and level matching in heterotic asymmetric orbifolds}",
    eprint = "2604.19634",
    archivePrefix = "arXiv",
    primaryClass = "hep-th",
    month = "4",
    year = "2026"
}

@article{Freed:1987qk,
    author = "Freed, Daniel S. and Vafa, Cumrun",
    title = "{GLOBAL ANOMALIES ON ORBIFOLDS}",
    reportNumber = "HUTP-86/A090",
    doi = "10.1007/BF01212418",
    journal = "Commun. Math. Phys.",
    volume = "110",
    pages = "349",
    year = "1987",
    note = "[Addendum: Commun.Math.Phys. 117, 349 (1988)]"
}

@article{Delgado:2026qvy,
    author = "Delgado, Matilda and Eberhardt, Lorenz and Toma{\v{s}}evi{\'c}, Marija",
    title = "{What makes spacetime spin in string theory?}",
    eprint = "2606.18380",
    archivePrefix = "arXiv",
    primaryClass = "hep-th",
    reportNumber = "CERN-TH-2026-134, MPP-2026-109",
    month = "6",
    year = "2026"
}

@article{PhysRevLett.37.8,
  title = {Symmetry Breaking through Bell-Jackiw Anomalies},
  author = {'t Hooft, G.},
  journal = {Phys. Rev. Lett.},
  volume = {37},
  issue = {1},
  pages = {8--11},
  numpages = {0},
  year = {1976},
  month = {Jul},
  publisher = {American Physical Society},
  doi = {10.1103/PhysRevLett.37.8},
  url = {https://link.aps.org/doi/10.1103/PhysRevLett.37.8}
}

@article{PhysRevLett.43.1566,
  title = {Baryon- and Lepton-Nonconserving Processes},
  author = {Weinberg, Steven},
  journal = {Phys. Rev. Lett.},
  volume = {43},
  issue = {21},
  pages = {1566--1570},
  numpages = {0},
  year = {1979},
  month = {Nov},
  publisher = {American Physical Society},
  doi = {10.1103/PhysRevLett.43.1566},
  url = {https://link.aps.org/doi/10.1103/PhysRevLett.43.1566}
}

@article{McNamara:2022lrw,
    author = "McNamara, Jacob and Reece, Matthew",
    title = "{Reflections on Parity Breaking}",
    eprint = "2212.00039",
    archivePrefix = "arXiv",
    primaryClass = "hep-th",
    month = "11",
    year = "2022"
}

@article{Debray:2023rlx,
    author = "Debray, Arun",
    title = "{Bordism for the 2-group symmetries of the heterotic and CHL strings}",
    eprint = "2304.14764",
    archivePrefix = "arXiv",
    primaryClass = "math.AT",
    doi = "10.1090/conm/802/16079",
    journal = "Contemp. Math.",
    volume = "802",
    pages = "227--98",
    year = "2024"
}

@article{Lake:2018dqm,
    author = "Lake, Ethan",
    title = "{Higher-form symmetries and spontaneous symmetry breaking}",
    eprint = "1802.07747",
    archivePrefix = "arXiv",
    primaryClass = "hep-th",
    month = "2",
    year = "2018"
}

@article{Angius:2023uqk,
    author = "Angius, Roberta and Makridou, Andriana and Uranga, Angel M.",
    title = "{Intersecting end of the world branes}",
    eprint = "2312.16286",
    archivePrefix = "arXiv",
    primaryClass = "hep-th",
    doi = "10.1007/JHEP03(2024)110",
    journal = "JHEP",
    volume = "03",
    pages = "110",
    year = "2024"
}

@article{Kaidi:2020jla,
    author = "Kaidi, Justin",
    title = "{Stable Vacua for Tachyonic Strings}",
    eprint = "2010.10521",
    archivePrefix = "arXiv",
    primaryClass = "hep-th",
    doi = "10.1103/PhysRevD.103.106026",
    journal = "Phys. Rev. D",
    volume = "103",
    number = "10",
    pages = "106026",
    year = "2021"
}

@article{Wang:2021ayd,
    author = "Wang, Juven and Wan, Zheyan and You, Yi-Zhuang",
    title = "{Cobordism and deformation class of the standard model}",
    eprint = "2112.14765",
    archivePrefix = "arXiv",
    primaryClass = "hep-th",
    doi = "10.1103/PhysRevD.106.L041701",
    journal = "Phys. Rev. D",
    volume = "106",
    number = "4",
    pages = "L041701",
    year = "2022"
}

@article{Raucci:2022jgw,
    author = "Raucci, Salvatore",
    title = "{On codimension-one vacua and string theory}",
    eprint = "2206.06399",
    archivePrefix = "arXiv",
    primaryClass = "hep-th",
    doi = "10.1016/j.nuclphysb.2022.116002",
    journal = "Nucl. Phys. B",
    volume = "985",
    pages = "116002",
    year = "2022"
}

@article{Blumenhagen:2021nmi,
    author = "Blumenhagen, Ralph and Cribiori, Niccol{\`o}",
    title = "{Open-closed correspondence of K-theory and cobordism}",
    eprint = "2112.07678",
    archivePrefix = "arXiv",
    primaryClass = "hep-th",
    reportNumber = "MPP-2021-201",
    doi = "10.1007/JHEP08(2022)037",
    journal = "JHEP",
    volume = "08",
    pages = "037",
    year = "2022"
}

@article{Blumenhagen:2022bvh,
    author = "Blumenhagen, Ralph and Cribiori, Niccol{\`o} and Kneissl, Christian and Makridou, Andriana",
    title = "{Dimensional Reduction of Cobordism and K-theory}",
    eprint = "2208.01656",
    archivePrefix = "arXiv",
    primaryClass = "hep-th",
    reportNumber = "MPP-2022-95",
    doi = "10.1007/JHEP03(2023)181",
    journal = "JHEP",
    volume = "03",
    pages = "181",
    year = "2023"
}

@article{Montero:2022vva,
    author = "Montero, Miguel and Parra de Freitas, Hector",
    title = "{New supersymmetric string theories from discrete theta angles}",
    eprint = "2209.03361",
    archivePrefix = "arXiv",
    primaryClass = "hep-th",
    doi = "10.1007/JHEP01(2023)091",
    journal = "JHEP",
    volume = "01",
    pages = "091",
    year = "2023"
}

@article{Basile:2022ypo,
    author = "Basile, Ivano and Raucci, Salvatore and Thom{\'e}e, Sylvain",
    title = "{Revisiting Dudas-Mourad Compactifications}",
    eprint = "2209.10553",
    archivePrefix = "arXiv",
    primaryClass = "hep-th",
    doi = "10.3390/universe8100544",
    journal = "Universe",
    volume = "8",
    number = "10",
    pages = "544",
    year = "2022"
}

@article{Velazquez:2022eco,
    author = "Vel{\'a}zquez, David Mart{\'\i}n and De Biasio, Davide and Lust, Dieter",
    title = "{Cobordism, singularities and the Ricci flow conjecture}",
    eprint = "2209.10297",
    archivePrefix = "arXiv",
    primaryClass = "hep-th",
    doi = "10.1007/JHEP01(2023)126",
    journal = "JHEP",
    volume = "01",
    pages = "126",
    year = "2023"
}

@article{Blumenhagen:2023abk,
    author = "Blumenhagen, Ralph and Kneissl, Christian and Wang, Chuying",
    title = "{Dynamical Cobordism Conjecture: solutions for end-of-the-world branes}",
    eprint = "2303.03423",
    archivePrefix = "arXiv",
    primaryClass = "hep-th",
    reportNumber = "MPP-2023-33",
    doi = "10.1007/JHEP05(2023)123",
    journal = "JHEP",
    volume = "05",
    pages = "123",
    year = "2023"
}

@article{Delgado:2025crl,
    author = "Delgado, Matilda and Reymond, S{\'e}bastien and Van Riet, Thomas",
    title = "{Black Holes, Moduli Stabilisation and the Swampland}",
    eprint = "2504.02645",
    archivePrefix = "arXiv",
    primaryClass = "hep-th",
    reportNumber = "MPP-2025-62",
    month = "4",
    year = "2025"
}

@article{Angius:2023xtu,
    author = "Angius, Roberta and Huertas, Jesus and Uranga, Angel M.",
    title = "{Small black hole explosions}",
    eprint = "2303.15903",
    archivePrefix = "arXiv",
    primaryClass = "hep-th",
    reportNumber = "IFT-UAM/CSIC-23-31",
    doi = "10.1007/JHEP06(2023)070",
    journal = "JHEP",
    volume = "06",
    pages = "070",
    year = "2023"
}

@article{Cribiori:2026btb,
    author = "Cribiori, Niccol{\`o} and Paraskevopoulou, Antonia and Van Riet, Thomas",
    title = "{A domain wall bound on anti-de Sitter vacua}",
    eprint = "2603.08779",
    archivePrefix = "arXiv",
    primaryClass = "hep-th",
    reportNumber = "MPP-2026-25",
    doi = "10.21468/SciPostPhys.20.6.157",
    journal = "SciPost Phys.",
    volume = "20",
    pages = "157",
    year = "2026"
}

@article{Basile:2023rvm,
    author = "Basile, Ivano and Montella, Carmine",
    title = "{Domain walls and distances in discrete landscapes}",
    eprint = "2309.04519",
    archivePrefix = "arXiv",
    primaryClass = "hep-th",
    doi = "10.1007/JHEP02(2024)227",
    journal = "JHEP",
    volume = "02",
    pages = "227",
    year = "2024"
}

@article{Shiu:2023bay,
    author = "Shiu, Gary and Tonioni, Flavio and Van Hemelryck, Vincent and Van Riet, Thomas",
    title = "{Connecting flux vacua through scalar field excursions}",
    eprint = "2311.10828",
    archivePrefix = "arXiv",
    primaryClass = "hep-th",
    reportNumber = "UUITP-32/23",
    doi = "10.1103/PhysRevD.109.066017",
    journal = "Phys. Rev. D",
    volume = "109",
    number = "6",
    pages = "066017",
    year = "2024"
}

@article{Angius:2024zjv,
    author = "Angius, Roberta",
    title = "{End of the world brane networks for infinite distance limits in CY moduli space}",
    eprint = "2404.14486",
    archivePrefix = "arXiv",
    primaryClass = "hep-th",
    reportNumber = "IFT-UAM/CSIC-24-62",
    doi = "10.1007/JHEP09(2024)178",
    journal = "JHEP",
    volume = "09",
    pages = "178",
    year = "2024"
}

@article{Ookouchi:2024tfz,
    author = "Ookouchi, Yutaka and Sato, Ryota and Tsukahara, Sohei",
    title = "{Decay of Kaluza-Klein vacuum via singular instanton}",
    eprint = "2404.13917",
    archivePrefix = "arXiv",
    primaryClass = "hep-th",
    doi = "10.1007/JHEP10(2025)040",
    journal = "JHEP",
    volume = "10",
    pages = "040",
    year = "2025"
}

@article{Mourad:2024dur,
    author = "Mourad, J. and Raucci, S. and Sagnotti, A.",
    title = "{Brane-like solutions and other non-supersymmetric vacua}",
    eprint = "2406.14926",
    archivePrefix = "arXiv",
    primaryClass = "hep-th",
    doi = "10.1007/JHEP10(2024)054",
    journal = "JHEP",
    volume = "10",
    pages = "054",
    year = "2024"
}

@article{Fukuda:2024pvu,
    author = "Fukuda, Masaki and Kobayashi, Shun K. and Watanabe, Kento and Yonekura, Kazuya",
    title = "{Black p-branes in heterotic string theory}",
    eprint = "2412.02277",
    archivePrefix = "arXiv",
    primaryClass = "hep-th",
    reportNumber = "TU-1247",
    doi = "10.1007/JHEP05(2025)043",
    journal = "JHEP",
    volume = "05",
    pages = "043",
    year = "2025"
}

@article{Hamada:2025duq,
    author = "Hamada, Yu and Hamada, Yuta and Kimura, Hayate",
    title = "{Black string in the standard model}",
    eprint = "2501.05678",
    archivePrefix = "arXiv",
    primaryClass = "hep-th",
    reportNumber = "KEK-TH-2679, DESY-24-222",
    doi = "10.1103/x3sh-45j7",
    journal = "Phys. Rev. D",
    volume = "111",
    number = "12",
    pages = "125009",
    year = "2025"
}

@article{Hassfeld:2025hjx,
    author = "Hassfeld, Bjoern and Hebecker, Arthur and Schiller, Daniel",
    title = "{Localized gravity, de Sitter, and the Horizon Criterion}",
    eprint = "2505.07934",
    archivePrefix = "arXiv",
    primaryClass = "hep-th",
    doi = "10.1007/JHEP10(2025)151",
    journal = "JHEP",
    volume = "10",
    pages = "151",
    year = "2025"
}

@article{Basile:2025mnj,
    author = "Basile, Ivano and Larotonda, Vittorio",
    title = "{Non-supersymmetric branes and discrete topological terms}",
    eprint = "2507.11610",
    archivePrefix = "arXiv",
    primaryClass = "hep-th",
    doi = "10.1007/JHEP10(2025)030",
    journal = "JHEP",
    volume = "10",
    pages = "030",
    year = "2025"
}

@article{Aparici:2026svo,
    author = "Aparici, Miquel and Herr{\'a}ez, Alvaro and Masias, Joaquin",
    title = "{Black Hole-Tower Correspondence: Species backreaction in minimal black hole limits}",
    eprint = "2609.00119",
    archivePrefix = "arXiv",
    primaryClass = "hep-th",
    month = "8",
    year = "2026"
}

@article{Calderon-Infante:2023ler,
    author = "Calder{\'o}n-Infante, Jos{\'e} and Castellano, Alberto and Herr{\'a}ez, Alvaro and Ib{\'a}{\~n}ez, Luis E.",
    title = "{Entropy bounds and the species scale distance conjecture}",
    eprint = "2306.16450",
    archivePrefix = "arXiv",
    primaryClass = "hep-th",
    reportNumber = "CERN-TH-2023-113",
    doi = "10.1007/JHEP01(2024)039",
    journal = "JHEP",
    volume = "01",
    pages = "039",
    year = "2024"
}

@article{Bedroya:2025ltj,
    author = "Bedroya, Alek and Steinhardt, Paul J.",
    title = "{Holography vs. Scale Separation}",
    eprint = "2509.25313",
    archivePrefix = "arXiv",
    primaryClass = "hep-th",
    month = "9",
    year = "2025"
}

@article{Casini:2008cr,
    author = "Casini, H.",
    title = "{Relative entropy and the Bekenstein bound}",
    eprint = "0804.2182",
    archivePrefix = "arXiv",
    primaryClass = "hep-th",
    doi = "10.1088/0264-9381/25/20/205021",
    journal = "Class. Quant. Grav.",
    volume = "25",
    pages = "205021",
    year = "2008"
}

@article{Blanco-Pillado:2023aom,
    author = "Blanco-Pillado, J. J. and Espinosa, J. R. and Huertas, J. and Sousa, K.",
    title = "{Bubbles of nothing: the tunneling potential approach}",
    eprint = "2312.00133",
    archivePrefix = "arXiv",
    primaryClass = "hep-th",
    doi = "10.1088/1475-7516/2024/03/029",
    journal = "JCAP",
    volume = "03",
    pages = "029",
    year = "2024"
}

@article{Blanco-Pillado:2023hog,
    author = "Blanco-Pillado, J. J. and Espinosa, J. R. and Huertas, J. and Sousa, K.",
    title = "{Tunneling potentials to nothing}",
    eprint = "2311.18821",
    archivePrefix = "arXiv",
    primaryClass = "hep-th",
    doi = "10.1103/PhysRevD.109.084057",
    journal = "Phys. Rev. D",
    volume = "109",
    number = "8",
    pages = "084057",
    year = "2024"
}

@article{Huertas:2024mvy,
    author = "Huertas, Jes{\'u}s and Uranga, Angel M.",
    title = "{End of the world brane dynamics in holographic 4d $ \mathcal{N} $ = 4 SU(N) with 3d $ \mathcal{N} $ = 2 boundary conditions}",
    eprint = "2410.05368",
    archivePrefix = "arXiv",
    primaryClass = "hep-th",
    doi = "10.1007/JHEP01(2025)002",
    journal = "JHEP",
    volume = "01",
    pages = "002",
    year = "2025"
}

@article{Huertas:2023syg,
    author = "Huertas, Jes{\'u}s and Uranga, Angel M.",
    title = "{Aspects of dynamical cobordism in AdS/CFT}",
    eprint = "2306.07335",
    archivePrefix = "arXiv",
    primaryClass = "hep-th",
    doi = "10.1007/JHEP08(2023)140",
    journal = "JHEP",
    volume = "08",
    pages = "140",
    year = "2023"
}

@article{Anastasi:2025puv,
    author = "Anastasi, Edoardo and Angius, Roberta and Huertas, Jes{\'u}s and Uranga, Angel M. and Wang, Chuying",
    title = "{Relative quantum gravity: localized gravity and the swampland}",
    eprint = "2501.03310",
    archivePrefix = "arXiv",
    primaryClass = "hep-th",
    doi = "10.1007/JHEP08(2025)107",
    journal = "JHEP",
    volume = "08",
    pages = "107",
    year = "2025"
}

@article{Chakrabhavi:2025bfi,
    author = "Chakrabhavi, Vivek and Debray, Arun and Dierigl, Markus and Heckman, Jonathan J.",
    title = "{Reflection branes, bordisms, and U dualities}",
    eprint = "2509.03573",
    archivePrefix = "arXiv",
    primaryClass = "hep-th",
    reportNumber = "CERN-TH-2025-180",
    doi = "10.1103/99d8-hyhs",
    journal = "Phys. Rev. D",
    volume = "113",
    number = "6",
    pages = "066015",
    year = "2026"
}

@article{Hamada:2025cpe,
    author = "Hamada, Yuta and Ishige, Arata and Koga, Yuichi",
    title = "{More on 8D nonsupersymmetric branes and heterotic strings}",
    eprint = "2505.15144",
    archivePrefix = "arXiv",
    primaryClass = "hep-th",
    reportNumber = "KEK-TH-2723,KYUSHU-HET-323",
    doi = "10.1103/yb7p-425g",
    journal = "Phys. Rev. D",
    volume = "113",
    number = "4",
    pages = "046012",
    year = "2026"
}

@article{Hassfeld:2024ake,
    author = "Hassfeld, Bjoern and Hebecker, Arthur and Westphal, Alexander",
    title = "{End-of-the-world branes and inflationary predictions for rocky and swampy landscapes}",
    eprint = "2411.11944",
    archivePrefix = "arXiv",
    primaryClass = "hep-th",
    doi = "10.1007/JHEP03(2025)196",
    journal = "JHEP",
    volume = "03",
    pages = "196",
    year = "2025"
}

@article{Ruiz:2024jiz,
    author = "Ruiz, Ignacio",
    title = "{Morse-Bott inequalities, topology change and cobordisms to nothing}",
    eprint = "2410.21372",
    archivePrefix = "arXiv",
    primaryClass = "hep-th",
    reportNumber = "IFT-UAM/CSIC-24-154",
    doi = "10.1007/JHEP06(2025)030",
    journal = "JHEP",
    volume = "06",
    pages = "030",
    year = "2025"
}

@article{Hamada:2024cdd,
    author = "Hamada, Yuta and Ishige, Arata",
    title = "{Investigating 9d/8d non-supersymmetric branes and theories from supersymmetric heterotic strings}",
    eprint = "2409.04770",
    archivePrefix = "arXiv",
    primaryClass = "hep-th",
    reportNumber = "KEK-TH-2652",
    doi = "10.1007/JHEP01(2025)141",
    journal = "JHEP",
    volume = "01",
    pages = "141",
    year = "2025"
}

@article{Altavista:2026muc,
    author = "Altavista, Chiara and Anastasi, Edoardo and Angius, Roberta and Uranga, Angel M. and Wang, Chuying",
    title = "{The Art of Networking: Networks of Trivalent 10d Heterotic Junctions}",
    eprint = "2607.01321",
    archivePrefix = "arXiv",
    primaryClass = "hep-th",
    month = "7",
    year = "2026"
}

@article{Dierigl:2026psj,
    author = "Dierigl, Markus and Ruiz, Ignacio",
    title = "{Brane networks and the cobordism conjecture}",
    eprint = "2605.18952",
    archivePrefix = "arXiv",
    primaryClass = "hep-th",
    reportNumber = "CERN-TH-2026-108",
    doi = "10.1103/dk2b-rgkf",
    journal = "Phys. Rev. D",
    volume = "114",
    number = "6",
    pages = "066012",
    year = "2026"
}

@article{lasHeras:2026xdl,
    author = "las Heras, Camilo and Ruiz, Ignacio",
    title = "{Bordisms between 9d type IIB supergravities and commutator widths of duality groups}",
    eprint = "2605.15276",
    archivePrefix = "arXiv",
    primaryClass = "hep-th",
    reportNumber = "CERN-TH-2026-107",
    month = "5",
    year = "2026"
}

@article{Damian:2026jsz,
    author = "Damian, Cesar and Loaiza-Brito, Oscar and L{\'o}pez-Ramos, V{\'\i}ctor M.",
    title = "{Double fibration in G-theory and the cobordism conjecture}",
    eprint = "2605.09868",
    archivePrefix = "arXiv",
    primaryClass = "hep-th",
    month = "5",
    year = "2026"
}

@article{Altavista:2026edv,
    author = "Altavista, Chiara and Anastasi, Edoardo and Angius, Roberta and Uranga, Angel M.",
    title = "{The art of branching: Cobordism junctions of 10d string theories}",
    eprint = "2603.24667",
    archivePrefix = "arXiv",
    primaryClass = "hep-th",
    doi = "10.1007/JHEP07(2026)211",
    journal = "JHEP",
    volume = "07",
    pages = "211",
    year = "2026"
}

@article{Braeger:2025kra,
    author = "Braeger, Noah and Debray, Arun and Dierigl, Markus and Heckman, Jonathan J. and Montero, Miguel",
    title = "{Cobordism Utopia: U-Dualities, Bordisms, and the Swampland}",
    eprint = "2505.15885",
    archivePrefix = "arXiv",
    primaryClass = "hep-th",
    reportNumber = "CERN-TH-2025-103, IFT-025-51",
    month = "5",
    year = "2025"
}

@article{Angius:2022mgh,
    author = "Angius, Roberta and Delgado, Matilda and Uranga, Angel M.",
    title = "{Dynamical Cobordism and the beginning of time: supercritical strings and tachyon condensation}",
    eprint = "2207.13108",
    archivePrefix = "arXiv",
    primaryClass = "hep-th",
    doi = "10.1007/JHEP08(2022)285",
    journal = "JHEP",
    volume = "08",
    pages = "285",
    year = "2022"
}

@article{Hellerman:2006ff,
    author = "Hellerman, Simeon and Swanson, Ian",
    title = "{Dimension-changing exact solutions of string theory}",
    eprint = "hep-th/0612051",
    archivePrefix = "arXiv",
    doi = "10.1088/1126-6708/2007/09/096",
    journal = "JHEP",
    volume = "09",
    pages = "096",
    year = "2007"
}

@article{Hellerman:2006hf,
    author = "Hellerman, Simeon and Swanson, Ian",
    title = "{Cosmological unification of string theories}",
    eprint = "hep-th/0612116",
    archivePrefix = "arXiv",
    doi = "10.1088/1126-6708/2008/07/022",
    journal = "JHEP",
    volume = "07",
    pages = "022",
    year = "2008"
}

@article{Hellerman:2007fc,
    author = "Hellerman, Simeon and Swanson, Ian",
    title = "{Charting the landscape of supercritical string theory}",
    eprint = "0705.0980",
    archivePrefix = "arXiv",
    primaryClass = "hep-th",
    doi = "10.1103/PhysRevLett.99.171601",
    journal = "Phys. Rev. Lett.",
    volume = "99",
    pages = "171601",
    year = "2007"
}

@article{Makridou:2026jzy,
    author = "Makridou, Andriana and G{\'o}mez, Alejandro Javier Puga",
    title = "{Sharpened Dynamical Cobordism}",
    eprint = "2605.06793",
    archivePrefix = "arXiv",
    primaryClass = "hep-th",
    reportNumber = "IFT-UAM/CSIC-26-45",
    month = "5",
    year = "2026"
}

@article{Calderon-Infante:2026ymy,
    author = "Calder{\'o}n-Infante, Jos{\'e} and Cheng, Gongrui and Herr{\'a}ez, Alvaro and Van Riet, Thomas",
    title = "{End-of-the-World Singularities: The Good, the Bad, and the Heated-up}",
    eprint = "2603.18133",
    archivePrefix = "arXiv",
    primaryClass = "hep-th",
    month = "3",
    year = "2026"
}

@article{Eardley:2002re,
    author = "Eardley, Douglas M. and Giddings, Steven B.",
    title = "{Classical black hole production in high-energy collisions}",
    eprint = "gr-qc/0201034",
    archivePrefix = "arXiv",
    reportNumber = "NSF-ITP-02-01",
    doi = "10.1103/PhysRevD.66.044011",
    journal = "Phys. Rev. D",
    volume = "66",
    pages = "044011",
    year = "2002"
}

@article{Gubser:1998bc,
    author = "Gubser, S. S. and Klebanov, Igor R. and Polyakov, Alexander M.",
    title = "{Gauge theory correlators from noncritical string theory}",
    eprint = "hep-th/9802109",
    archivePrefix = "arXiv",
    reportNumber = "PUPT-1767",
    doi = "10.1016/S0370-2693(98)00377-3",
    journal = "Phys. Lett. B",
    volume = "428",
    pages = "105--114",
    year = "1998"
}

@article{Kaidi:2024cbx,
    author = "Kaidi, Justin and Tachikawa, Yuji and Yonekura, Kazuya",
    title = "{On non-supersymmetric heterotic branes}",
    eprint = "2411.04344",
    archivePrefix = "arXiv",
    primaryClass = "hep-th",
    reportNumber = "TU-1248, KYUSHU-HET-294",
    doi = "10.1007/JHEP03(2025)211",
    journal = "JHEP",
    volume = "03",
    pages = "211",
    year = "2025"
}

@article{Harlow:2022ich,
    author = "Harlow, Daniel and Heidenreich, Ben and Reece, Matthew and Rudelius, Tom",
    title = "{Weak gravity conjecture}",
    eprint = "2201.08380",
    archivePrefix = "arXiv",
    primaryClass = "hep-th",
    reportNumber = "ACFI-T22-01",
    doi = "10.1103/RevModPhys.95.035003",
    journal = "Rev. Mod. Phys.",
    volume = "95",
    number = "3",
    pages = "035003",
    year = "2023"
}

@article{Agmon:2022thq,
    author = "Agmon, Nathan Benjamin and Bedroya, Alek and Kang, Monica Jinwoo and Vafa, Cumrun",
    title = "{Lectures on the string landscape and the Swampland}",
    eprint = "2212.06187",
    archivePrefix = "arXiv",
    primaryClass = "hep-th",
    month = "12",
    year = "2022"
}

@book{Ammon:2015wua,
    author = "Ammon, Martin and Erdmenger, Johanna",
    title = "{Gauge/gravity duality}: {Foundations and applications}",
    doi = "10.1017/CBO9780511846373",
    isbn = "978-1-107-01034-5, 978-1-316-23594-2",
    publisher = "Cambridge University Press",
    address = "Cambridge",
    month = "4",
    year = "2015"
}

@article{Heidenreich:2021xpr,
    author = "Heidenreich, Ben and McNamara, Jacob and Montero, Miguel and Reece, Matthew and Rudelius, Tom and Valenzuela, Irene",
    title = "{Non-invertible global symmetries and completeness of the spectrum}",
    eprint = "2104.07036",
    archivePrefix = "arXiv",
    primaryClass = "hep-th",
    reportNumber = "ACFI-T21-03",
    doi = "10.1007/JHEP09(2021)203",
    journal = "JHEP",
    volume = "09",
    pages = "203",
    year = "2021"
}

@ARTICLE{bekensetinbound,
       author = {{Bekenstein}, Jacob D.},
        title = "{Universal upper bound on the entropy-to-energy ratio for bounded systems}",
      journal = "Phys. Rev. D",
         year = 1981,
        month = jan,
       volume = {23},
       number = {2},
        pages = {287-298},
          doi = {10.1103/PhysRevD.23.287},
       adsurl = {https://ui.adsabs.harvard.edu/abs/1981PhRvD..23..287B}
}

@article{bekensetin2,
    author = "Bekenstein, Jacob D.",
    title = "{Black holes and the second law}",
    journal = "Lett. Nuovo Cim.",
    volume = "4",
    pages = "737--740",
    doi = "10.1007/BF02757029",
    year = "1972"
}

@article{MISNER1957525,
title = {Classical physics as geometry},
journal = {Annals of Physics},
volume = {2},
number = {6},
pages = {525-603},
year = {1957},
issn = {0003-4916},
doi = {https://doi.org/10.1016/0003-4916(57)90049-0},
url = {https://www.sciencedirect.com/science/article/pii/0003491657900490},
author = {Charles W Misner and John A Wheeler}
}

@article{Anastasi:2026cus,
    author = "Anastasi, Edoardo and Montero, Miguel and Uranga, Angel M. and Wang, Chuying",
    title = "{What IIB looks IIA string: string cobordisms via non-compact CFTs}",
    eprint = "2603.00225",
    archivePrefix = "arXiv",
    primaryClass = "hep-th",
    reportNumber = "IFT-26-016",
    doi = "10.1007/JHEP07(2026)182",
    journal = "JHEP",
    volume = "07",
    pages = "182",
    year = "2026"
}

@article{Torres:2026vxx,
    author = "Torres, Ethan",
    title = "{A Matrix Theory Construction of the IIA/IIB Wall}",
    eprint = "2603.02199",
    archivePrefix = "arXiv",
    primaryClass = "hep-th",
    reportNumber = "CERN-TH-2026-030",
    month = "3",
    year = "2026"
}

@article{Heckman:2025wqd,
    author = "Heckman, Jonathan J. and McNamara, Jacob and Parra-Martinez, Julio and Torres, Ethan",
    title = "{Gliozzi-Scherk-Olive defects: IIA/IIB walls and the surprisingly stable R7-brane}",
    eprint = "2507.21210",
    archivePrefix = "arXiv",
    primaryClass = "hep-th",
    reportNumber = "CERN-TH-2025-136",
    doi = "10.1103/wqys-zsyf",
    journal = "Phys. Rev. D",
    volume = "113",
    number = "6",
    pages = "066021",
    year = "2026"
}

@article{Angius:2024pqk,
    author = "Angius, Roberta and Uranga, Angel Maria and Wang, Chuying",
    title = "{End of the world boundaries for chiral quantum gravity theories}",
    eprint = "2410.07322",
    archivePrefix = "arXiv",
    primaryClass = "hep-th",
    doi = "10.1007/JHEP03(2025)064",
    journal = "JHEP",
    volume = "03",
    pages = "064",
    year = "2025"
}

@article{Cavusoglu:2026xiv,
    author = "{\c{C}}avu{\c{s}}o{\u{g}}lu, Atakan and Cveti{\v{c}}, Mirjam and Heckman, Jonathan J. and Kuntz, Jeffrey and Murdia, Chitraang",
    title = "{Gravitational background of Alice vortices and R7-branes}",
    eprint = "2602.13196",
    archivePrefix = "arXiv",
    primaryClass = "hep-th",
    doi = "10.1103/scv8-1gj8",
    journal = "Phys. Rev. D",
    volume = "114",
    number = "2",
    pages = "026031",
    year = "2026"
}

@article{Peet:1998wn,
    author = "Peet, Amanda W. and Polchinski, Joseph",
    title = "{UV / IR relations in AdS dynamics}",
    eprint = "hep-th/9809022",
    archivePrefix = "arXiv",
    reportNumber = "NSF-ITP-98-086",
    doi = "10.1103/PhysRevD.59.065011",
    journal = "Phys. Rev. D",
    volume = "59",
    pages = "065011",
    year = "1999"
}

@book{Schwartz:2014sze,
    author = "Schwartz, Matthew D.",
    title = "{Quantum Field Theory and the Standard Model}",
    doi = "10.1017/9781139540940",
    isbn = "978-1-107-03473-0, 978-1-107-03473-0",
    publisher = "Cambridge University Press",
    month = "3",
    year = "2014"
}

@book{Weinberg:QM, place={Cambridge}, edition={2}, title={Lectures on Quantum Mechanics}, publisher={Cambridge University Press}, author={Weinberg, Steven}, year={2015}}

@article{Delgado:2023uqk,
    author = "Delgado, Matilda",
    title = "{The bubble of nothing under T-duality}",
    eprint = "2312.09291",
    archivePrefix = "arXiv",
    primaryClass = "hep-th",
    reportNumber = "IFT-UAM/CSIC-23-146",
    doi = "10.1007/JHEP05(2024)333",
    journal = "JHEP",
    volume = "05",
    pages = "333",
    year = "2024"
}

@article{Blumenhagen:2022mqw,
    author = "Blumenhagen, Ralph and Cribiori, Niccol{\`o} and Kneissl, Christian and Makridou, Andriana",
    title = "{Dynamical cobordism of a domain wall and its companion defect 7-brane}",
    eprint = "2205.09782",
    archivePrefix = "arXiv",
    primaryClass = "hep-th",
    reportNumber = "MPP-2022-57",
    doi = "10.1007/JHEP08(2022)204",
    journal = "JHEP",
    volume = "08",
    pages = "204",
    year = "2022"
}

@article{Raucci:2026fzp,
    author = "Raucci, Salvatore and Ruiz, Ignacio and Valenzuela, Irene",
    title = "{Alice in Warpland: KK modes, Warped Compactifications and the Swampland}",
    eprint = "2603.11163",
    archivePrefix = "arXiv",
    primaryClass = "hep-th",
    reportNumber = "IFT-UAM/CSIC-26-22, CERN-TH-2026-040",
    month = "3",
    year = "2026"
}

@article{Adams:2001sv,
    author = "Adams, A. and Polchinski, J. and Silverstein, Eva",
    title = "{Don't panic! Closed string tachyons in ALE space-times}",
    eprint = "hep-th/0108075",
    archivePrefix = "arXiv",
    reportNumber = "SLAC-PUB-8955, NSF-ITP-01-75",
    doi = "10.1088/1126-6708/2001/10/029",
    journal = "JHEP",
    volume = "10",
    pages = "029",
    year = "2001"
}

@book{hamilton2017mathematical,
  title={Mathematical Gauge Theory: With Applications to the Standard Model of Particle Physics},
  author={Hamilton, M.J.D.},
  isbn={9783319684390},
  series={Universitext},
  url={https://books.google.be/books?id=8zVCDwAAQBAJ},
  year={2017},
  publisher={Springer International Publishing}
}

@article{Heckman:2024obe,
    author = "Heckman, Jonathan J. and McNamara, Jacob and Montero, Miguel and Sharon, Adar and Vafa, Cumrun and Valenzuela, Irene",
    title = "{Fate of stringy noninvertible symmetries}",
    eprint = "2402.00118",
    archivePrefix = "arXiv",
    primaryClass = "hep-th",
    reportNumber = "CERN-TH-2024-019",
    doi = "10.1103/PhysRevD.110.106001",
    journal = "Phys. Rev. D",
    volume = "110",
    number = "10",
    pages = "106001",
    year = "2024"
}

@book{taubes2011differential,
  title={Differential Geometry: Bundles, Connections, Metrics and Curvature},
  author={Taubes, C.},
  isbn={9780199605880},
  lccn={2011499898},
  series={Differential Geometry: Bundles, Connections, Metrics and Curvature},
  url={https://books.google.be/books?id=hinInCKHwwMC},
  year={2011},
  publisher={OUP Oxford}
}

@article{Dixon:1986jc,
    author = "Dixon, Lance J. and Harvey, Jeffrey A. and Vafa, C. and Witten, Edward",
    title = "{Strings on Orbifolds. 2.}",
    reportNumber = "PRINT-86-0246 (PRINCETON)",
    doi = "10.1016/0550-3213(86)90287-7",
    journal = "Nucl. Phys. B",
    volume = "274",
    pages = "285--314",
    year = "1986"
}

@book{nakahara2003geometry,
  title={Geometry, Topology and Physics, Second Edition},
  author={Nakahara, M.},
  isbn={9780750306065},
  lccn={2003282202},
  series={Graduate student series in physics},
  url={https://books.google.be/books?id=cH-XQB0Ex5wC},
  year={2003},
  publisher={Taylor \& Francis}
}

@article{Debray:2024czl,
    author = "Debray, Arun and Yu, Matthew",
    title = "{Type IIA String Theory and tmf with Level Structure}",
    eprint = "2411.07299",
    archivePrefix = "arXiv",
    primaryClass = "math.AT",
    month = "11",
    year = "2024"
}

@article{Diaconescu:2000wy,
    author = "Diaconescu, Duiliu-Emanuel and Moore, Gregory W. and Witten, Edward",
    title = "{E(8) gauge theory, and a derivation of K theory from M theory}",
    eprint = "hep-th/0005090",
    archivePrefix = "arXiv",
    reportNumber = "IASSNS-HEP-00-39",
    doi = "10.4310/ATMP.2002.v6.n6.a2",
    journal = "Adv. Theor. Math. Phys.",
    volume = "6",
    pages = "1031--1134",
    year = "2003"
}

@article{Burgess:2008ri,
    author = "Burgess, C. P. and Conlon, J. P. and Hung, L-Y. and Kom, C. H. and Maharana, Anshuman and Quevedo, F.",
    title = "{Continuous Global Symmetries and Hyperweak Interactions in String Compactifications}",
    eprint = "0805.4037",
    archivePrefix = "arXiv",
    primaryClass = "hep-th",
    reportNumber = "DAMTP-2008-45, CAV-HEP-08-08",
    doi = "10.1088/1126-6708/2008/07/073",
    journal = "JHEP",
    volume = "07",
    pages = "073",
    year = "2008"
}

@article{Witten:1998cd,
    author = "Witten, Edward",
    title = "{D-branes and K-theory}",
    eprint = "hep-th/9810188",
    archivePrefix = "arXiv",
    reportNumber = "IASSNS-HEP-98-82",
    doi = "10.1088/1126-6708/1998/12/019",
    journal = "JHEP",
    volume = "12",
    pages = "019",
    year = "1998"
}

@article{Gibbons:2011dh,
    author = "Gibbons, G. W.",
    title = "{Topology change in classical and quantum gravity}",
    eprint = "1110.0611",
    archivePrefix = "arXiv",
    primaryClass = "gr-qc",
    month = "10",
    year = "2011"
}

@article{Moore:1999gb,
    author = "Moore, Gregory W. and Witten, Edward",
    title = "{Selfduality, Ramond-Ramond fields, and K theory}",
    eprint = "hep-th/9912279",
    archivePrefix = "arXiv",
    doi = "10.1088/1126-6708/2000/05/032",
    journal = "JHEP",
    volume = "05",
    pages = "032",
    year = "2000"
}

@article{Harlow:TASI,
    author = "Harlow, Daniel",
    title = "{TASI Lectures on the Emergence of Bulk Physics in AdS/CFT}",
    eprint = "1802.01040",
    archivePrefix = "arXiv",
    primaryClass = "hep-th",
    doi = "10.22323/1.305.0002",
    journal = "PoS",
    volume = "TASI2017",
    pages = "002",
    year = "2018"
}

@article{Nevoa:2025xiq,
    author = "Nevoa, Vinicius and Raman, Sanjay and Vafa, Cumrun",
    title = "{Elementary Constituents Conjecture}",
    eprint = "2511.13813",
    archivePrefix = "arXiv",
    primaryClass = "hep-th",
    month = "11",
    year = "2025"
}

@article{McNamara:2021cuo,
    author = "McNamara, Jacob",
    title = "{Gravitational Solitons and Completeness}",
    eprint = "2108.02228",
    archivePrefix = "arXiv",
    primaryClass = "hep-th",
    month = "8",
    year = "2021"
}

@article{Horava:1995qa,
    author = "Horava, Petr and Witten, Edward",
    title = "{Heterotic and Type I string dynamics from eleven dimensions}",
    eprint = "hep-th/9510209",
    archivePrefix = "arXiv",
    reportNumber = "IASSNS-HEP-95-86, PUPT-1571A",
    doi = "",
    journal = "Nucl. Phys. B",
    volume = "460",
    pages = "506--524",
    year = "1996"
}

@article{Dierigl:2022reg,
    author = "Dierigl, Markus and Heckman, Jonathan J. and Montero, Miguel and Torres, Ethan",
    title = "{IIB string theory explored: Reflection 7-branes}",
    eprint = "2212.05077",
    archivePrefix = "arXiv",
    primaryClass = "hep-th",
    reportNumber = "LMU-ASC 31/22",
    doi = "10.1103/PhysRevD.107.086015",
    journal = "Phys. Rev. D",
    volume = "107",
    number = "8",
    pages = "086015",
    year = "2023"
}

@article{Basile:2023knk,
    author = "Basile, Ivano and Debray, Arun and Delgado, Matilda and Montero, Miguel",
    title = "{Global anomalies {\&} bordism of non-supersymmetric strings}",
    eprint = "2310.06895",
    archivePrefix = "arXiv",
    primaryClass = "hep-th",
    reportNumber = "IFT-23-129",
    doi = "10.1007/JHEP02(2024)092",
    journal = "JHEP",
    volume = "02",
    pages = "092",
    year = "2024"
}

@article{Kneissl:2024zox,
    author = "Kneissl, Christian",
    title = "{Spin cobordism and the gauge group of type I/heterotic string theory}",
    eprint = "2407.20333",
    archivePrefix = "arXiv",
    primaryClass = "hep-th",
    reportNumber = "MPP-2024-159",
    doi = "10.1007/JHEP01(2025)181",
    journal = "JHEP",
    volume = "01",
    pages = "181",
    year = "2025"
}

@article{Hassfeld:2023kpu,
    author = "Hassfeld, Bjoern and Hebecker, Arthur and Walcher, Johannes",
    title = "{Cobordism and bubbles of anything in the string landscape}",
    eprint = "2310.06021",
    archivePrefix = "arXiv",
    primaryClass = "hep-th",
    doi = "10.1007/JHEP02(2024)127",
    journal = "JHEP",
    volume = "02",
    pages = "127",
    year = "2024"
}

@article{Belenchia:2018szb,
    author = "Belenchia, Alessio and Wald, Robert M. and Giacomini, Flaminia and Castro-Ruiz, Esteban and Brukner, {\v{C}}aslav and Aspelmeyer, Markus",
    title = "{Quantum Superposition of Massive Objects and the Quantization of Gravity}",
    eprint = "1807.07015",
    archivePrefix = "arXiv",
    primaryClass = "quant-ph",
    doi = "10.1103/PhysRevD.98.126009",
    journal = "Phys. Rev. D",
    volume = "98",
    number = "12",
    pages = "126009",
    year = "2018"
}

@article{Hassfeld:2024hnx,
    author = "Hassfeld, Bjoern and Hebecker, Arthur",
    title = "{The boundary proposal}",
    eprint = "2403.18892",
    archivePrefix = "arXiv",
    primaryClass = "hep-th",
    doi = "10.1016/j.physletb.2024.138946",
    journal = "Phys. Lett. B",
    volume = "856",
    pages = "138946",
    year = "2024"
}

@article{Dong:2016eik,
    author = "Dong, Xi and Harlow, Daniel and Wall, Aron C.",
    title = "{Reconstruction of Bulk Operators within the Entanglement Wedge in Gauge-Gravity Duality}",
    eprint = "1601.05416",
    archivePrefix = "arXiv",
    primaryClass = "hep-th",
    reportNumber = "NSF-KITP-16-005, NSF-KITP-16-005",
    doi = "10.1103/PhysRevLett.117.021601",
    journal = "Phys. Rev. Lett.",
    volume = "117",
    number = "2",
    pages = "021601",
    year = "2016"
}

@article{Harlow:EWR_proof,
    author = "Harlow, Daniel",
    title = "{The Ryu{\textendash}Takayanagi Formula from Quantum Error Correction}",
    eprint = "1607.03901",
    archivePrefix = "arXiv",
    primaryClass = "hep-th",
    doi = "10.1007/s00220-017-2904-z",
    journal = "Commun. Math. Phys.",
    volume = "354",
    number = "3",
    pages = "865--912",
    year = "2017"
}

@article{Witten:BH_lectures,
    author = "Witten, Edward",
    title = "{Introduction to black hole thermodynamics}",
    eprint = "2412.16795",
    archivePrefix = "arXiv",
    primaryClass = "hep-th",
    doi = "10.1140/epjp/s13360-025-06288-y",
    journal = "Eur. Phys. J. Plus",
    volume = "140",
    number = "5",
    pages = "430",
    year = "2025"
}

@article{Harlow:factorization_JT,
    author = "Harlow, Daniel and Jafferis, Daniel",
    title = "{The Factorization Problem in Jackiw-Teitelboim Gravity}",
    eprint = "1804.01081",
    archivePrefix = "arXiv",
    primaryClass = "hep-th",
    doi = "10.1007/JHEP02(2020)177",
    journal = "JHEP",
    volume = "02",
    pages = "177",
    year = "2020"
}

@article{Itzhaki:1998dd,
    author = "Itzhaki, Nissan and Maldacena, Juan Martin and Sonnenschein, Jacob and Yankielowicz, Shimon",
    title = "{Supergravity and the large N limit of theories with sixteen supercharges}",
    eprint = "hep-th/9802042",
    archivePrefix = "arXiv",
    reportNumber = "TAUP-2474-98, HUTP-98-A003",
    doi = "10.1103/PhysRevD.58.046004",
    journal = "Phys. Rev. D",
    volume = "58",
    pages = "046004",
    year = "1998"
}

@article{Bousso1999,
  author  = {Bousso, Raphael},
  title   = {A Covariant Entropy Conjecture},
  journal = {Journal of High Energy Physics},
  year    = {1999},
  volume  = {1999},
  number  = {07},
  pages   = {004},
  doi     = {10.1088/1126-6708/1999/07/004},
  eprint  = {hep-th/9905177},
  archivePrefix = {arXiv}
}

@article{Bousso2002,
  author  = {Bousso, Raphael},
  title   = {The Holographic Principle},
  journal = {Reviews of Modern Physics},
  year    = {2002},
  volume  = {74},
  number  = {3},
  pages   = {825--874},
  doi     = {10.1103/RevModPhys.74.825},
  eprint  = {hep-th/0203101},
  archivePrefix = {arXiv}
}

@book{Heusler1996,
  author    = {Heusler, Markus},
  title     = {Black Hole Uniqueness Theorems},
  publisher = {Cambridge University Press},
  year      = {1996},
  series    = {Cambridge Lecture Notes in Physics},
  volume    = {6},
  address   = {Cambridge},
  isbn      = {978-0521567355}
}

@article{Reece:2023TASI,
  author        = {Reece, Matthew},
  title         = {TASI Lectures: (No) Global Symmetries to Axion Physics},
  journal       = {Proceedings of Science},
  year          = {2024},
  volume        = {TASI2022},
  pages         = {008},
  eprint        = {2304.08512},
  archivePrefix = {arXiv},
  primaryClass  = {hep-ph},
  doi           = {10.22323/1.439.0008}
}

@article{Hofman:2018lfz,
    author = "Hofman, Diego M. and Iqbal, Nabil",
    title = "{Goldstone modes and photonization for higher form symmetries}",
    eprint = "1802.09512",
    archivePrefix = "arXiv",
    primaryClass = "hep-th",
    doi = "10.21468/SciPostPhys.6.1.006",
    journal = "SciPost Phys.",
    volume = "6",
    number = "1",
    pages = "006",
    year = "2019"
}

@article{Polchinski2003Monopoles,
  author        = {Polchinski, Joseph},
  title         = {Monopoles, Duality, and String Theory},
  journal       = {Nuclear Physics B - Proceedings Supplements},
  year          = {2003},
  volume        = {117},
  pages         = {255--269},
  doi           = {},
  eprint        = {hep-th/0304042},
  archivePrefix = {arXiv},
  primaryClass  = {hep-th}
}

@article{McNamara:2020uza,
    author = "McNamara, Jacob and Vafa, Cumrun",
    title = "{Baby Universes, Holography, and the Swampland}",
    eprint = "2004.06738",
    archivePrefix = "arXiv",
    primaryClass = "hep-th",
    month = "4",
    year = "2020"
}

@article{Saad:2019lba,
    author = "Saad, Phil and Shenker, Stephen H. and Stanford, Douglas",
    title = "{JT gravity as a matrix integral}",
    eprint = "1903.11115",
    archivePrefix = "arXiv",
    primaryClass = "hep-th",
    month = "3",
    year = "2019"
}

@article{COLEMAN1988867,
title = {Black holes as red herrings: Topological fluctuations and the loss of quantum coherence},
journal = {Nuclear Physics B},
volume = {307},
number = {4},
pages = {867-882},
year = {1988},
issn = {0550-3213},
doi = {https://doi.org/10.1016/0550-3213(88)90110-1},
url = {https://www.sciencedirect.com/science/article/pii/0550321388901101},
author = {Sidney Coleman}
}

@article{Marolf:2020xie,
    author = "Marolf, Donald and Maxfield, Henry",
    title = "{Transcending the ensemble: baby universes, spacetime wormholes, and the order and disorder of black hole information}",
    eprint = "2002.08950",
    archivePrefix = "arXiv",
    primaryClass = "hep-th",
    doi = "10.1007/JHEP08(2020)044",
    journal = "JHEP",
    volume = "08",
    pages = "044",
    year = "2020"
}

@article{Rokhlin_52,
    author = {Vladimir Abramovich Rokhlin},
    title ={New results in the theory of four-dimensional manifolds},
    journal = {Doklady Akademii Nauk SSSR},
    volume = {84},
    pages = {221–224},
    year = {1952}
}

@article{tHooft:1993dmi,
    author = "'t Hooft, Gerard",
    title = "{Dimensional reduction in quantum gravity}",
    eprint = "gr-qc/9310026",
    archivePrefix = "arXiv",
    reportNumber = "THU-93-26",
    journal = "Conf. Proc. C",
    volume = "930308",
    pages = "284--296",
    year = "1993"
}

@book{milnor,
    author = "Milnor, John and Stasheff, James D",
    title = "{Characteristic Classes (AM-76), Volume 76}", 
    publisher =  "Princeton: Princeton University Press", 
    year =" 1974"
}

@article{Sen:1994kx,
    author = "Sen, Ashoke and Zwiebach, Barton",
    title = "{Background independent algebraic structures in closed string field theory}",
    eprint = "hep-th/9408053",
    archivePrefix = "arXiv",
    reportNumber = "MIT-CTP-2346",
    doi = "10.1007/BF02101895",
    journal = "Commun. Math. Phys.",
    volume = "177",
    pages = "305--326",
    year = "1996"
}

@article{Brandenberger:1988aj,
    author = "Brandenberger, Robert H. and Vafa, C.",
    title = "{Superstrings in the Early Universe}",
    reportNumber = "HUTP-88-A035, BROWN-HET-673",
    doi = "10.1016/0550-3213(89)90037-0",
    journal = "Nucl. Phys. B",
    volume = "316",
    pages = "391--410",
    year = "1989"
}

@article{Witten:1982fp,
    author = "Witten, Edward",
    editor = "Shifman, Mikhail A.",
    title = "{An SU(2) Anomaly}",
    doi = "10.1016/0370-2693(82)90728-6",
    journal = "Phys. Lett. B",
    volume = "117",
    pages = "324--328",
    year = "1982"
}

@article{Witten:1992qy,
    author = "Witten, Edward",
    title = "{On background independent open string field theory}",
    eprint = "hep-th/9208027",
    archivePrefix = "arXiv",
    reportNumber = "IASSNS-HEP-92-53",
    doi = "10.1103/PhysRevD.46.5467",
    journal = "Phys. Rev. D",
    volume = "46",
    pages = "5467--5473",
    year = "1992"
}

@article{Alvarez-Gaume:1984zlq,
    author = "Alvarez-Gaume, Luis and Ginsparg, Paul H.",
    editor = "Salam, A. and Sezgin, E.",
    title = "{The Structure of Gauge and Gravitational Anomalies}",
    reportNumber = "HUTP-84/A016",
    doi = "10.1016/0003-4916(85)90087-9",
    journal = "Annals Phys.",
    volume = "161",
    pages = "423",
    year = "1985",
    note = "[Erratum: Annals Phys. 171, 233 (1986)]"
}

@article{Montero:2020icj,
    author = "Montero, Miguel and Vafa, Cumrun",
    title = "{Cobordism Conjecture, Anomalies, and the String Lamppost Principle}",
    eprint = "2008.11729",
    archivePrefix = "arXiv",
    primaryClass = "hep-th",
    doi = "10.1007/JHEP01(2021)063",
    journal = "JHEP",
    volume = "01",
    pages = "063",
    year = "2021"
}

@article{Dierigl:2023jdp,
    author = "Dierigl, Markus and Heckman, Jonathan J. and Montero, Miguel and Torres, Ethan",
    title = "{R7-branes as charge conjugation operators}",
    eprint = "2305.05689",
    archivePrefix = "arXiv",
    primaryClass = "hep-th",
    reportNumber = "LMU-ASC 16/23, IFT-UAM/CSIC-23-50",
    doi = "10.1103/PhysRevD.109.046004",
    journal = "Phys. Rev. D",
    volume = "109",
    number = "4",
    pages = "046004",
    year = "2024"
}

@book{Stong+1968,
url = {https://doi.org/10.1515/9781400879977},
title = {Notes on Cobordism Theory},
author = {Robert E. Stong},
publisher = {Princeton University Press},
address = {Princeton},
doi = {doi:10.1515/9781400879977},
isbn = {9781400879977},
year = {1968},
lastchecked = {2024-03-12}
}

@article{Garcia-Etxebarria:2018ajm,
    author = "Garc\'\i{}a-Etxebarria, I\~naki and Montero, Miguel",
    title = "{Dai-Freed anomalies in particle physics}",
    eprint = "1808.00009",
    archivePrefix = "arXiv",
    primaryClass = "hep-th",
    reportNumber = "MPP-2018-188",
    doi = "10.1007/JHEP08(2019)003",
    journal = "JHEP",
    volume = "08",
    pages = "003",
    year = "2019"
}

@article{Kaidi:2023tqo,
    author = "Kaidi, Justin and Ohmori, Kantaro and Tachikawa, Yuji and Yonekura, Kazuya",
    title = "{Nonsupersymmetric Heterotic Branes}",
    eprint = "2303.17623",
    archivePrefix = "arXiv",
    primaryClass = "hep-th",
    doi = "10.1103/PhysRevLett.131.121601",
    journal = "Phys. Rev. Lett.",
    volume = "131",
    number = "12",
    pages = "121601",
    year = "2023"
}

@article{Andriot:2022mri,
    author = "Andriot, David and Carqueville, Nils and Cribiori, Niccol\`o",
    title = "{Looking for structure in the cobordism conjecture}",
    eprint = "2204.00021",
    archivePrefix = "arXiv",
    primaryClass = "hep-th",
    reportNumber = "MPP-2022-11",
    doi = "10.21468/SciPostPhys.13.3.071",
    journal = "SciPost Phys.",
    volume = "13",
    number = "3",
    pages = "071",
    year = "2022"
}

@inproceedings{Sagnotti:1995ga,
    author = "Sagnotti, Augusto",
    title = "{Some properties of open string theories}",
    booktitle = "{International Workshop on Supersymmetry and Unification of Fundamental Interactions (SUSY 95)}",
    eprint = "hep-th/9509080",
    archivePrefix = "arXiv",
    reportNumber = "ROM2F-95-18",
    pages = "473--484",
    month = "9",
    year = "1995"
}

@article{Dixon:1986iz,
    author = "Dixon, Lance J. and Harvey, Jeffrey A.",
    editor = "Schellekens, B.",
    title = "{String Theories in Ten-Dimensions Without Space-Time Supersymmetry}",
    reportNumber = "PRINT-86-0244 (PRINCETON)",
    doi = "10.1016/0550-3213(86)90619-X",
    journal = "Nucl. Phys. B",
    volume = "274",
    pages = "93--105",
    year = "1986"
}

@article{Alvarez-Gaume:1986ghj,
    author = "Alvarez-Gaume, Luis and Ginsparg, Paul H. and Moore, Gregory W. and Vafa, C.",
    title = "{An O(16) x O(16) Heterotic String}",
    reportNumber = "HUTP-86/A013",
    doi = "10.1016/0370-2693(86)91524-8",
    journal = "Phys. Lett. B",
    volume = "171",
    pages = "155--162",
    year = "1986"
}

@article{Sagnotti:1996qj,
    author = "Sagnotti, Augusto",
    editor = "Lust, D. and Otto, H. J. and Weigt, G.",
    title = "{Surprises in open string perturbation theory}",
    eprint = "hep-th/9702093",
    archivePrefix = "arXiv",
    reportNumber = "ROM2F-97-4",
    doi = "10.1016/S0920-5632(97)00344-7",
    journal = "Nucl. Phys. B Proc. Suppl.",
    volume = "56",
    pages = "332--343",
    year = "1997"
}

@article{Sugimoto:1999tx,
    author = "Sugimoto, Shigeki",
    title = "{Anomaly cancellations in type I D-9 - anti-D-9 system and the USp(32) string theory}",
    eprint = "hep-th/9905159",
    archivePrefix = "arXiv",
    reportNumber = "YITP-99-25",
    doi = "10.1143/PTP.102.685",
    journal = "Prog. Theor. Phys.",
    volume = "102",
    pages = "685--699",
    year = "1999"
}

@book{Polchinski:1998rr,
    author = "Polchinski, J.",
    title = "{String theory. Vol. 2: Superstring theory and beyond}",
    doi = "10.1017/CBO9780511618123",
    isbn = "978-0-511-25228-0, 978-0-521-63304-8, 978-0-521-67228-3",
    publisher = "Cambridge University Press",
    series = "Cambridge Monographs on Mathematical Physics",
    month = "12",
    year = "2007"
}

@article{Susskind:1994vu,
    author = "Susskind, Leonard",
    title = "{The World as a hologram}",
    eprint = "hep-th/9409089",
    archivePrefix = "arXiv",
    reportNumber = "SU-ITP-94-33",
    doi = "10.1063/1.531249",
    journal = "J. Math. Phys.",
    volume = "36",
    pages = "6377--6396",
    year = "1995"
}

@article{Debray:2023yrs,
    author = "Debray, Arun and Dierigl, Markus and Heckman, Jonathan J. and Montero, Miguel",
    title = "{The Chronicles of IIBordia: Dualities, Bordisms, and the Swampland}",
    eprint = "2302.00007",
    archivePrefix = "arXiv",
    primaryClass = "hep-th",
    reportNumber = "LMU-ASC 06/23, IFT-UAM/CSIC-23-7",
    month = "1",
    year = "2023"
}

@article{Buratti:2021fiv,
    author = "Buratti, Ginevra and Calder\'on-Infante, Jos\'e and Delgado, Matilda and Uranga, Angel M.",
    title = "{Dynamical Cobordism and Swampland Distance Conjectures}",
    eprint = "2107.09098",
    archivePrefix = "arXiv",
    primaryClass = "hep-th",
    doi = "10.1007/JHEP10(2021)037",
    journal = "JHEP",
    volume = "10",
    pages = "037",
    year = "2021"
}

@article{Angius:2022aeq,
    author = "Angius, Roberta and Calder\'on-Infante, Jos\'e and Delgado, Matilda and Huertas, Jes\'us and Uranga, Angel M.",
    title = "{At the End of the World: Local Dynamical Cobordism}",
    eprint = "2203.11240",
    archivePrefix = "arXiv",
    primaryClass = "hep-th",
    reportNumber = "IFT-UAM/CSIC-22-31",
    month = "3",
    year = "2022"
}

@article{LIGOScientific:2016aoc,
    author = "Abbott, B. P. and others",
    collaboration = "LIGO Scientific, Virgo",
    title = "{Observation of Gravitational Waves from a Binary Black Hole Merger}",
    eprint = "1602.03837",
    archivePrefix = "arXiv",
    primaryClass = "gr-qc",
    reportNumber = "LIGO-P150914",
    doi = "10.1103/PhysRevLett.116.061102",
    journal = "Phys. Rev. Lett.",
    volume = "116",
    number = "6",
    pages = "061102",
    year = "2016"
}

@article{Bekenstein:1973ur,
    author = "Bekenstein, Jacob D.",
    title = "{Black holes and entropy}",
    doi = "10.1103/PhysRevD.7.2333",
    journal = "Phys. Rev. D",
    volume = "7",
    pages = "2333--2346",
    year = "1973"
}

@article{Hawking:1974rv,
    author = "Hawking, S. W.",
    title = "{Black hole explosions}",
    doi = "10.1038/248030a0",
    journal = "Nature",
    volume = "248",
    pages = "30--31",
    year = "1974"
}

@article{Hawking:1975vcx,
    author = "Hawking, S. W.",
    editor = "Gibbons, G. W. and Hawking, S. W.",
    title = "{Particle Creation by Black Holes}",
    doi = "10.1007/BF02345020",
    journal = "Commun. Math. Phys.",
    volume = "43",
    pages = "199--220",
    year = "1975",
    note = "[Erratum: Commun.Math.Phys. 46, 206 (1976)]"
}

@article{Maldacena:1997re,
    author = "Maldacena, Juan Martin",
    title = "{The Large N limit of superconformal field theories and supergravity}",
    eprint = "hep-th/9711200",
    archivePrefix = "arXiv",
    reportNumber = "HUTP-97-A097, HUTP-98-A097",
    doi = "10.1023/A:1026654312961",
    journal = "Adv. Theor. Math. Phys.",
    volume = "2",
    pages = "231--252",
    year = "1998"
}

@article{Susskind:1998dq,
    author = "Susskind, Leonard and Witten, Edward",
    title = "{The Holographic bound in anti-de Sitter space}",
    eprint = "hep-th/9805114",
    archivePrefix = "arXiv",
    reportNumber = "SU-ITP-98-39, IASSNS-HEP-98-44",
    month = "5",
    year = "1998"
}

@article{Witten:1998qj,
    author = "Witten, Edward",
    title = "{Anti-de Sitter space and holography}",
    eprint = "hep-th/9802150",
    archivePrefix = "arXiv",
    reportNumber = "IASSNS-HEP-98-15",
    doi = "10.4310/ATMP.1998.v2.n2.a2",
    journal = "Adv. Theor. Math. Phys.",
    volume = "2",
    pages = "253--291",
    year = "1998"
}

@article{Vafa:2005ui,
    author = "Vafa, Cumrun",
    title = "{The String landscape and the swampland}",
    eprint = "hep-th/0509212",
    archivePrefix = "arXiv",
    reportNumber = "HUTP-05-A043",
    month = "9",
    year = "2005"
}

@article{Gaiotto:2014kfa,
    author = "Gaiotto, Davide and Kapustin, Anton and Seiberg, Nathan and Willett, Brian",
    title = "{Generalized Global Symmetries}",
    eprint = "1412.5148",
    archivePrefix = "arXiv",
    primaryClass = "hep-th",
    doi = "10.1007/JHEP02(2015)172",
    journal = "JHEP",
    volume = "02",
    pages = "172",
    year = "2015"
}

@article{Hamada:2021bbz,
    author = "Hamada, Yuta and Vafa, Cumrun",
    title = "{8d supergravity, reconstruction of internal geometry and the Swampland}",
    eprint = "2104.05724",
    archivePrefix = "arXiv",
    primaryClass = "hep-th",
    doi = "10.1007/JHEP06(2021)178",
    journal = "JHEP",
    volume = "06",
    pages = "178",
    year = "2021"
}

@article{Hebecker:2018ofv,
    author = "Hebecker, Arthur and Mikhail, Thomas and Soler, Pablo",
    title = "{Euclidean wormholes, baby universes, and their impact on particle physics and cosmology}",
    eprint = "1807.00824",
    archivePrefix = "arXiv",
    primaryClass = "hep-th",
    doi = "10.3389/fspas.2018.00035",
    journal = "Front. Astron. Space Sci.",
    volume = "5",
    pages = "35",
    year = "2018"
}

@article{Harlow:2018tng,
      author         = "Harlow, Daniel and Ooguri, Hirosi",
      title          = "{Symmetries in quantum field theory and quantum gravity}",
      year           = "2018",
      eprint         = "1810.05338",
      archivePrefix  = "arXiv",
      primaryClass   = "hep-th",
      SLACcitation   = "%%CITATION = ARXIV:1810.05338;%%"
}

@article{Harlow:2018jwu,
      author         = "Harlow, Daniel and Ooguri, Hirosi",
      title          = "{Constraints on symmetry from holography}",
      year           = "2018",
      eprint         = "1810.05337",
      archivePrefix  = "arXiv",
      primaryClass   = "hep-th",
      SLACcitation   = "%%CITATION = ARXIV:1810.05337;%%"
}

@article{Ooguri:2006in,
      author         = "Ooguri, Hirosi and Vafa, Cumrun",
      title          = "{On the Geometry of the String Landscape and the
                        Swampland}",
      journal        = "Nucl. Phys.",
      volume         = "B766",
      year           = "2007",
      pages          = "21-33",
      doi            = "10.1016/j.nuclphysb.2006.10.033",
      eprint         = "hep-th/0605264",
      archivePrefix  = "arXiv",
      primaryClass   = "hep-th",
      reportNumber   = "CALT-68-2600, HUTP-06-A017",
      SLACcitation   = "%%CITATION = HEP-TH/0605264;%%"
}

@inproceedings{Polchinski:2010hw,
    author = "Polchinski, Joseph",
    title = "{Introduction to Gauge/Gravity Duality}",
    booktitle = "{Theoretical Advanced Study Institute in Elementary Particle Physics}: {String theory and its Applications: From meV to the Planck Scale}",
    eprint = "1010.6134",
    archivePrefix = "arXiv",
    primaryClass = "hep-th",
    doi = "10.1142/9789814350525_0001",
    pages = "3--46",
    month = "10",
    year = "2010"
}

@article{Witten:1998zw,
    author = "Witten, Edward",
    editor = "Bergstrom, L. and Lindstrom, U.",
    title = "{Anti-de Sitter space, thermal phase transition, and confinement in gauge theories}",
    eprint = "hep-th/9803131",
    archivePrefix = "arXiv",
    reportNumber = "IASSNS-HEP-98-21",
    doi = "10.4310/ATMP.1998.v2.n3.a3",
    journal = "Adv. Theor. Math. Phys.",
    volume = "2",
    pages = "505--532",
    year = "1998"
}

@article{tHooft:1978pzo,
    author = "'t Hooft, Gerard",
    editor = "Lannutti, J. E. and Williams, P. K.",
    title = "{Non-Abelian gauge theories and quark confinement}",
    doi = "10.1063/1.31600",
    journal = "AIP Conf. Proc.",
    volume = "48",
    pages = "1--8",
    year = "1978"
}

@article{tHooft:1977nqb,
    author = "'t Hooft, Gerard",
    title = "{On the Phase Transition Towards Permanent Quark Confinement}",
    reportNumber = "Print-78-0099 (UTRECHT)",
    doi = "10.1016/0550-3213(78)90153-0",
    journal = "Nucl. Phys. B",
    volume = "138",
    pages = "1--25",
    year = "1978"
}

@article{Wilson:1974sk,
    author = "Wilson, Kenneth G.",
    editor = "Taylor, J. C.",
    title = "{Confinement of Quarks}",
    reportNumber = "CLNS-262",
    doi = "10.1103/PhysRevD.10.2445",
    journal = "Phys. Rev. D",
    volume = "10",
    pages = "2445--2459",
    year = "1974"
}

@article{ArkaniHamed:2006dz,
      author         = "Arkani-Hamed, Nima and Motl, Lubos and Nicolis, Alberto
                        and Vafa, Cumrun",
      title          = "{The String landscape, black holes and gravity as the
                        weakest force}",
      journal        = "JHEP",
      volume         = "06",
      year           = "2007",
      pages          = "060",
      doi            = "10.1088/1126-6708/2007/06/060",
      eprint         = "hep-th/0601001",
      archivePrefix  = "arXiv",
      primaryClass   = "hep-th",
      reportNumber   = "HUTP-05-A0057",
      SLACcitation   = "%%CITATION = HEP-TH/0601001;%%"
}

@article{Rakic:2023vhv,
    author = "Rakic, Ilija and Rangamani, Mukund and Turiaci, Gustavo J.",
    title = "{Thermodynamics of the near-extremal Kerr spacetime}",
    eprint = "2310.04532",
    archivePrefix = "arXiv",
    primaryClass = "hep-th",
    doi = "10.1007/JHEP06(2024)011",
    journal = "JHEP",
    volume = "06",
    pages = "011",
    year = "2024"
}

@article{Iliesiu:2022onk,
    author = "Iliesiu, Luca V. and Murthy, Sameer and Turiaci, Gustavo J.",
    title = "{Revisiting the logarithmic corrections to the black hole entropy}",
    eprint = "2209.13608",
    archivePrefix = "arXiv",
    primaryClass = "hep-th",
    doi = "10.1007/JHEP07(2025)058",
    journal = "JHEP",
    volume = "07",
    pages = "058",
    year = "2025"
}

@article{Turiaci:2023wrh,
    author = "Turiaci, Gustavo J.",
    title = "{New insights on near-extremal black holes}",
    eprint = "2307.10423",
    archivePrefix = "arXiv",
    primaryClass = "hep-th",
    month = "7",
    year = "2023"
}

@article{Coleman:1980aw,
      author         = "Coleman, Sidney R. and De Luccia, Frank",
      title          = "{Gravitational Effects on and of Vacuum Decay}",
      journal        = "Phys. Rev.",
      volume         = "D21",
      year           = "1980",
      pages          = "3305",
      doi            = "10.1103/PhysRevD.21.3305",
      reportNumber   = "SLAC-PUB-2463",
      SLACcitation   = "%%CITATION = PHRVA,D21,3305;%%"
}

@article{Banks:2010zn,
      author         = "Banks, Tom and Seiberg, Nathan",
      title          = "{Symmetries and Strings in Field Theory and Gravity}",
      journal        = "Phys. Rev.",
      volume         = "D83",
      year           = "2011",
      pages          = "084019",
      doi            = "10.1103/PhysRevD.83.084019",
      eprint         = "1011.5120",
      archivePrefix  = "arXiv",
      primaryClass   = "hep-th",
      SLACcitation   = "%%CITATION = ARXIV:1011.5120;%%"
}

@article{Polchinski:1995df,
	Archiveprefix = {arXiv},
	Author = {Polchinski, Joseph and Witten, Edward},
	Doi = {10.1016/0550-3213(95)00614-1},
	Eprint = {hep-th/9510169},
	Journal = {Nucl. Phys.},
	Pages = {525-540},
	Primaryclass = {hep-th},
	Reportnumber = {IASSNS-HEP-95-81, NSF-ITP-95-135},
	Slaccitation = {%%CITATION = HEP-TH/9510169;%%},
	Title = {{Evidence for heterotic - type I string duality}},
	Volume = {B460},
	Year = {1996}}

@article{Aharony:1999ti,
	Archiveprefix = {arXiv},
	Author = {Aharony, Ofer and Gubser, Steven S. and Maldacena, Juan Martin and Ooguri, Hirosi and Oz, Yaron},
	Doi = {10.1016/S0370-1573(99)00083-6},
	Eprint = {hep-th/9905111},
	Journal = {Phys. Rept.},
	Pages = {183-386},
	Primaryclass = {hep-th},
	Reportnumber = {CERN-TH-99-122, HUTP-99-A027, LBNL-43113, RU-99-18, UCB-PTH-99-16, LBL-43113},
	Slaccitation = {%%CITATION = HEP-TH/9905111;%%},
	Title = {{Large N field theories, string theory and gravity}},
	Volume = {323},
	Year = {2000}}

@article{Banks:1988yz,
	Author = {Banks, Tom and Dixon, Lance J.},
	Doi = {10.1016/0550-3213(88)90523-8},
	Journal = {Nucl. Phys.},
	Pages = {93-108},
	Reportnumber = {PUPT-1086, SCIPP-8805},
	Slaccitation = {%%CITATION = NUPHA,B307,93;%%},
	Title = {{Constraints on String Vacua with Space-Time Supersymmetry}},
	Volume = {B307},
	Year = {1988}}

@article{Krauss:1988zc,
	Author = {Krauss, Lawrence M. and Wilczek, Frank},
	Doi = {10.1103/PhysRevLett.62.1221},
	Journal = {Phys. Rev. Lett.},
	Pages = {1221},
	Reportnumber = {YCTP-P26-88, NSF-ITP-88-187},
	Slaccitation = {%%CITATION = PRLTA,62,1221;%%},
	Title = {{Discrete Gauge Symmetry in Continuum Theories}},
	Volume = {62},
	Year = {1989}}

@article{Coleman:1991ku,
	Archiveprefix = {arXiv},
	Author = {Coleman, Sidney R. and Preskill, John and Wilczek, Frank},
	Doi = {10.1016/0550-3213(92)90008-Y},
	Eprint = {hep-th/9201059},
	Journal = {Nucl. Phys.},
	Pages = {175-246},
	Primaryclass = {hep-th},
	Reportnumber = {IASSNS-HEP-91-64, CALT-68-1764, HUTP-92-A003},
	Slaccitation = {%%CITATION = HEP-TH/9201059;%%},
	Title = {{Quantum hair on black holes}},
	Volume = {B378},
	Year = {1992}}

@article{Palti:2019pca,
	Archiveprefix = {arXiv},
	Author = {Palti, Eran},
	Doi = {10.1002/prop.201900037},
	Eprint = {1903.06239},
	Journal = {Fortsch. Phys.},
	Number = {6},
	Pages = {1900037},
	Primaryclass = {hep-th},
	Reportnumber = {MPP-2019-53},
	Slaccitation = {%%CITATION = ARXIV:1903.06239;%%},
	Title = {{The Swampland: Introduction and Review}},
	Volume = {67},
	Year = {2019}}

@article{Buratti:2021yia,
    author = "Buratti, Ginevra and Delgado, Matilda and Uranga, Angel M.",
    title = "{Dynamical tadpoles, stringy cobordism, and the SM from spontaneous compactification}",
    eprint = "2104.02091",
    archivePrefix = "arXiv",
    primaryClass = "hep-th",
    doi = "10.1007/JHEP06(2021)170",
    journal = "JHEP",
    volume = "06",
    pages = "170",
    year = "2021"
}

@article{Witten:1981gj,
    author = "Witten, Edward",
    title = "{Instability of the Kaluza-Klein Vacuum}",
    reportNumber = "PRINT-81-0441 (PRINCETON)",
    doi = "10.1016/0550-3213(82)90007-4",
    journal = "Nucl. Phys. B",
    volume = "195",
    pages = "481--492",
    year = "1982"
}

@article{McNamara:2019rup,
    author = "McNamara, Jacob and Vafa, Cumrun",
    title = "{Cobordism Classes and the Swampland}",
    eprint = "1909.10355",
    archivePrefix = "arXiv",
    primaryClass = "hep-th",
    month = "9",
    year = "2019"
}

@article{Dierigl:2020lai,
    author = "Dierigl, Markus and Heckman, Jonathan J.",
    title = "{Swampland cobordism conjecture and non-Abelian duality groups}",
    eprint = "2012.00013",
    archivePrefix = "arXiv",
    primaryClass = "hep-th",
    doi = "10.1103/PhysRevD.103.066006",
    journal = "Phys. Rev. D",
    volume = "103",
    number = "6",
    pages = "066006",
    year = "2021"
}

@article{Bergshoeff:1995vh,
    author = "Bergshoeff, E. and Green, Michael B. and Papadopoulos, G. and Townsend, P. K.",
    title = "{The IIA supereight-brane}",
    eprint = "hep-th/9511079",
    archivePrefix = "arXiv",
    reportNumber = "DAMTP-R-95-55, UG-12-95",
    month = "11",
    year = "1995"
}

@article{GarciaEtxebarria:2020xsr,
    author = "Garc\'\i{}a Etxebarria, I\~naki and Montero, Miguel and Sousa, Kepa and Valenzuela, Irene",
    title = "{Nothing is certain in string compactifications}",
    eprint = "2005.06494",
    archivePrefix = "arXiv",
    primaryClass = "hep-th",
    doi = "10.1007/JHEP12(2020)032",
    journal = "JHEP",
    volume = "12",
    pages = "032",
    year = "2020"
}

@article{Horava:1996ma,
    author = "Horava, Petr and Witten, Edward",
    title = "{Eleven-dimensional supergravity on a manifold with boundary}",
    eprint = "hep-th/9603142",
    archivePrefix = "arXiv",
    reportNumber = "IASSNS-HEP-96-17, PUPT-1597",
    doi = "10.1016/0550-3213(96)00308-2",
    journal = "Nucl. Phys. B",
    volume = "475",
    pages = "94--114",
    year = "1996"
}

@article{vanBeest:2021lhn,
    author = "van Beest, Marieke and Calder\'on-Infante, Jos\'e and Mirfendereski, Delaram and Valenzuela, Irene",
    title = "{Lectures on the Swampland Program in String Compactifications}",
    eprint = "2102.01111",
    archivePrefix = "arXiv",
    primaryClass = "hep-th",
    month = "2",
    year = "2021"
}

@article{Ooguri:2020sua,
    author = "Ooguri, Hirosi and Takayanagi, Tadashi",
    title = "{Cobordism Conjecture in AdS}",
    eprint = "2006.13953",
    archivePrefix = "arXiv",
    primaryClass = "hep-th",
    month = "6",
    year = "2020"
}

@inproceedings{NabilR,
    author = "Iqbal, Nabil",
    title = "{Jena lectures on generalized global symmetries: principles and applications}",
    eprint = "2407.20815",
    archivePrefix = "arXiv",
    primaryClass = "hep-th",
    month = "7",
    year = "2024"
}

@article{NamekiR,
    author = "Schafer-Nameki, Sakura",
    title = "{ICTP lectures on (non-)invertible generalized symmetries}",
    eprint = "2305.18296",
    archivePrefix = "arXiv",
    primaryClass = "hep-th",
    doi = "10.1016/j.physrep.2024.01.007",
    journal = "Phys. Rept.",
    volume = "1063",
    pages = "1--55",
    year = "2024"
}

@article{BottiniR,
    author = "Bhardwaj, Lakshya and Bottini, Lea E. and Fraser-Taliente, Ludovic and Gladden, Liam and Gould, Dewi S. W. and Platschorre, Arthur and Tillim, Hannah",
    title = "{Lectures on generalized symmetries}",
    eprint = "2307.07547",
    archivePrefix = "arXiv",
    primaryClass = "hep-th",
    doi = "10.1016/j.physrep.2023.11.002",
    journal = "Phys. Rept.",
    volume = "1051",
    pages = "1--87",
    year = "2024"
}

@inproceedings{Cordova:2022ruw,
    author = "Cordova, Clay and Dumitrescu, Thomas T. and Intriligator, Kenneth and Shao, Shu-Heng",
    title = "{Snowmass White Paper: Generalized Symmetries in Quantum Field Theory and Beyond}",
    booktitle = "{Snowmass 2021}",
    eprint = "2205.09545",
    archivePrefix = "arXiv",
    primaryClass = "hep-th",
    month = "5",
    year = "2022"
}

@article{McGreevy:2022oyu,
    author = "McGreevy, John",
    title = "{Generalized Symmetries in Condensed Matter}",
    eprint = "2204.03045",
    archivePrefix = "arXiv",
    primaryClass = "cond-mat.str-el",
    doi = "10.1146/annurev-conmatphys-040721-021029",
    journal = "Ann. Rev. Condensed Matter Phys.",
    volume = "14",
    pages = "57--82",
    year = "2023"
}

@article{Brennan:2023mmt,
    author = "Brennan, T. Daniel and Hong, Sungwoo",
    title = "{Introduction to Generalized Global Symmetries in QFT and Particle Physics}",
    eprint = "2306.00912",
    archivePrefix = "arXiv",
    primaryClass = "hep-ph",
    month = "6",
    year = "2023"
}

@article{Gomes:2023ahz,
    author = "Gomes, Pedro R. S.",
    title = "{An introduction to higher-form symmetries}",
    eprint = "2303.01817",
    archivePrefix = "arXiv",
    primaryClass = "hep-th",
    doi = "10.21468/SciPostPhysLectNotes.74",
    journal = "SciPost Phys. Lect. Notes",
    volume = "74",
    pages = "1",
    year = "2023"
}

@article{Bhardwaj:2023wzd,
    author = "Bhardwaj, Lakshya and Schafer-Nameki, Sakura",
    title = "{Generalized charges, part I: Invertible symmetries and higher representations}",
    eprint = "2304.02660",
    archivePrefix = "arXiv",
    primaryClass = "hep-th",
    doi = "10.21468/SciPostPhys.16.4.093",
    journal = "SciPost Phys.",
    volume = "16",
    number = "4",
    pages = "093",
    year = "2024"
}

@article{Bartsch:2023pzl,
    author = "Bartsch, Thomas and Bullimore, Mathew and Grigoletto, Andrea",
    title = "{Higher representations for extended operators}",
    eprint = "2304.03789",
    archivePrefix = "arXiv",
    primaryClass = "hep-th",
    month = "4",
    year = "2023"
}

@article{Brown:2011gt,
    author = "Brown, Adam R. and Dahlen, Alex",
    title = "{On `nothing' as an infinitely negatively curved spacetime}",
    eprint = "1111.0301",
    archivePrefix = "arXiv",
    primaryClass = "hep-th",
    doi = "10.1103/PhysRevD.85.104026",
    journal = "Phys. Rev. D",
    volume = "85",
    pages = "104026",
    year = "2012"
}

@article{Cespedes:2020xpn,
    author = "Cespedes, Sebastian and de Alwis, Senarath P. and Muia, Francesco and Quevedo, Fernando",
    title = "{Lorentzian vacuum transitions: Open or closed universes?}",
    eprint = "2011.13936",
    archivePrefix = "arXiv",
    primaryClass = "hep-th",
    journal = "Phys. Rev. D",
    volume = "104",
    pages = "026013",
    year = "2021"
}

@article{Adams:2005rb,
    author = "Adams, Allan and Liu, Xiao and McGreevy, John and Saltman, Alex and Silverstein, Eva",
    title = "{Things fall apart: Topology change from winding tachyons}",
    eprint = "hep-th/0502021",
    archivePrefix = "arXiv",
    journal = "JHEP",
    volume = "10",
    pages = "033",
    year = "2005"
}

@article{Heidenreich:2020pkc,
    author = "Heidenreich, Ben and McNamara, Jacob and Montero, Miguel and Reece, Matthew and Rudelius, Tom and Valenzuela, Irene",
    title = "{Chern-Weil global symmetries and how quantum gravity avoids them}",
    eprint = "2012.00009",
    archivePrefix = "arXiv",
    primaryClass = "hep-th",
    year = "2020"
}

@article{Greene:1989ya,
    author = "Greene, Brian R. and Shapere, Alfred D. and Vafa, Cumrun and Yau, Shing-Tung",
    title = "{Stringy Cosmic Strings and Noncompact Calabi-Yau Manifolds}",
    journal = "Nucl. Phys. B",
    volume = "337",
    pages = "1",
    year = "1990"
}

@article{Gibbons:1995vg,
    author = "Gibbons, G. W. and Green, Michael B. and Perry, Malcolm J.",
    title = "{Instantons and seven-branes in type IIB superstring theory}",
    eprint = "hep-th/9511080",
    archivePrefix = "arXiv",
    journal = "Phys. Lett. B",
    volume = "370",
    pages = "37",
    year = "1996"
}

@article{Atiyah:1971,
    author = "Atiyah, Michael F.",
    title = "{Riemann surfaces and spin structures}",
    journal = "Ann. Sci. \'Ecole Norm. Sup. (4)",
    volume = "4",
    pages = "47--62",
    year = "1971"
}

@article{Johnson:1980,
    author = "Johnson, Dennis",
    title = "{Spin structures and quadratic forms on surfaces}",
    journal = "J. London Math. Soc. (2)",
    volume = "22",
    pages = "365--373",
    year = "1980"
}

@incollection{KirbyTaylor:1990,
    author = "Kirby, Robion C. and Taylor, Laurence R.",
    title = "{Pin structures on low-dimensional manifolds}",
    booktitle = "{Geometry of Low-Dimensional Manifolds 2}",
    series = "London Math. Soc. Lecture Note Ser.",
    volume = "151",
    pages = "177--242",
    publisher = "Cambridge University Press",
    year = "1990"
}

@book{McCleary:2001,
    author = "McCleary, John",
    title = "{A User's Guide to Spectral Sequences}",
    edition = "2nd",
    series = "Cambridge Studies in Advanced Mathematics",
    volume = "58",
    publisher = "Cambridge University Press",
    year = "2001"
}

@article{Cordova:2022rer,
    author = "Cordova, Clay and Ohmori, Kantaro and Rudelius, Tom",
    title = "{Generalized symmetry breaking scales and weak gravity conjectures}",
    eprint = "2202.05866",
    archivePrefix = "arXiv",
    primaryClass = "hep-th",
    journal = "JHEP",
    volume = "11",
    pages = "154",
    year = "2022"
}

@article{Kapustin:2014gua,
    author = "Kapustin, Anton and Seiberg, Nathan",
    title = "{Coupling a QFT to a TQFT and Duality}",
    eprint = "1401.0740",
    archivePrefix = "arXiv",
    primaryClass = "hep-th",
    journal = "JHEP",
    volume = "04",
    pages = "001",
    year = "2014"
}

@article{AharonovBohm,
  title = {Significance of Electromagnetic Potentials in the Quantum Theory},
  author = {Aharonov, Y. and Bohm, D.},
  journal = {Phys. Rev.},
  volume = {115},
  issue = {3},
  pages = {485--491},
  numpages = {0},
  year = {1959},
  month = {Aug},
  publisher = {American Physical Society},
  doi = {10.1103/PhysRev.115.485},
  url = {https://link.aps.org/doi/10.1103/PhysRev.115.485}
}

\newpage

\end{document}